\documentclass[14pt,twoside]{article}
\usepackage[cp866]{inputenc}
\usepackage[russian]{babel}
\usepackage{euscript,graphicx}
\usepackage{amsmath,amsfonts,amssymb}
\usepackage{euscript}

\usepackage{fancyhdr} 
\date{}
\title{}
\begin{document}
\thispagestyle{empty} \fontsize{14}{21}\selectfont $$$$
\thispagestyle{empty}

  \begin{center}
  \Large {\bf Г.Г. Михайличенко }
  \end{center}

\vspace{1mm}

  \begin{center}
  \Large {\bf Математические Основы \\ и Результаты \\ Теории
  Физических Структур }
  \end{center}

\vspace{7mm}

\begin{center}
\Large   {\it Посвящается моим сыновьям \\
Илье и Виктору, \\ а также моим Коллегам \\ по Горно-Алтайской школе \\
теории физических структур}
\end{center}

\vspace{7mm}

\begin{center}
\Large { \bf $ \infty \ \bullet \ \infty \ \bullet \ \infty $}
\end{center}

\vspace{7mm}

  \begin{center}
  \Large {\bf G.G. Mikhailichenko }
  \end{center}

\vspace{1mm}

  \begin{center}
  \Large {\bf The Mathematical Basics \\and Results of the \\
  Theory of Physical Structures  }
  \end{center}

\vspace{7mm}

  \begin{center}
   \Large {\it  Dedicated to my sons \\
Iliya and Victor, \\
and to my Colleagues \\ at Gorno-Altaisk school of \\
theory of physical structures }
\end{center}

\newpage
\thispagestyle{empty}
\begin{center}
{\bf Минобрнауки России \\
Федеральное государственное бюджетное образовательное \\
учреждение высшего профессионального образования} \\
{\bf Горно-Алтайский государственный университет}
\end{center}

\vspace{35mm}

\begin{center}
\Large{\bf Г.Г. МИХАЙЛИЧЕНКО}
\end{center}

\vspace{10mm}

\begin{center}
\Huge{ \bf

МАТЕМАТИЧЕСКИЕ \ ОСНОВЫ \\
\vspace {2mm}
 И \ РЕЗУЛЬТАТЫ  \ ТЕОРИИ \\
 \vspace{2mm}
   ФИЗИЧЕСКИХ \ СТРУКТУР}
\end{center}

\vspace{20mm}

\begin{flushright}
{\bf Приложение \hspace{25mm} \\
 А.Н. Бородина}
\end{flushright}

\vspace{40mm}

\begin{center}
{\bf Горно-Алтайский госуниверситет \\
 Горно-Алтайск,  Россия \\
2016}
\end{center}

\newpage
\thispagestyle{empty}
     \begin{center}
   {\bf  Ministry of Education and Science of Russia \\
     Federal state-financed educational \\
     institution of higher professional education\\
     Gorno-Altaisk State University}
     \end{center}

     \vspace{35mm}

     \begin{center}
     \Large{\bf G.G. MIKHAILICHENKO}
     \end{center}

     \vspace{10mm}

     \begin{center}
     \Huge{ \bf

     THE \ MATHEMATICAL \ BASICS  \\
     \vspace{2mm}
     AND \ RESULTS \ OF \ THE \ THEORY \\
     \vspace{2mm}
     \ OF \ PHYSICAL \ STRUCTURES}
     \end{center}

     \vspace{20mm}

     \begin{flushright}
     {\bf An appendix by \hspace{25mm} \\
      A.N. Borodin} \phantom{a}
     \end{flushright}

     \vspace{5mm}

     \begin{flushright}
    {\bf  Translation by \hspace{25mm} \\
      N.V. Faranosov}
     \end{flushright}

     \vspace{15mm}

     \begin{center}
      {\bf Gorno-Altaisk State University \\
       Gorno-Altaisk,   Russia \\
    2016}
     \end{center}

\newpage

  \begin{center}
  \Large{\bf В \ Е \ Х \ И}
  \end{center}

\vspace{5mm}

{\bf 1965 --} формулировка принципа феноменологической симметрии \\

{\bf 1970 --} исследование физических структур ранга (2,2) и (3,2) \\

{\bf 1975 --} классификация физических структур произвольного ранга  \\

 {\bf 1980 --} классификация феноменологически симметричных
 двумер- \\
 \phantom{aaaaaaaaa} ных геометрий ранга 4 на одном множестве\\

{\bf 1985 --} исследование групповых свойств физических структур \\

{\bf 1990 --} классификация феноменологически симметричных
трехмер-
\\
\phantom{aaaaaaaaa} ных геометрий ранга 5 на одном множестве \\

{\bf 1995 --} определение полиметрических физических структур и их  \\
\phantom {aaaaaaaaa} классификация для некоторых рангов \\

{\bf 2000 --} исследование алгебраических свойств абстрактных физиче- \\
\phantom{aaaaaaaaa} ских структур на множествах произвольной природы \\

{\bf 2005 --} выход в свет монографии \ "Теория физических структур"  \\
\phantom{aaaaaaaaa} Кулакова Ю.И.\
 (847 стр. с ил.) \\

{\bf 2010 --} исследование гипотезы вложения физических структур \\
\phantom{aaaaaaaaa} в структуры более высокого ранга \\

{\bf 2015 --} вывод уравнения феноменологической симметрии для неко- \\
\phantom{aaaaaaaaa} торых трехмерных геометрий \\

{\bf P.S. --} более полная информация о теории физических структур \\
\phantom{aaaaaaaa} размещена  на сайте  ТФС  \ ( {\bf
http://www.tphs.info} )

\newpage

\begin{center}
\Large{\bf  L \  A \ N \ D \ M \ A \ R \ K \ S}
\end{center}

\vspace{5mm}

{\bf 1965 --} the principle of phenomenological symmetry formulated   \\

{\bf 1970 --} research of the physical structures of rank (2,2) and (3,2)    \\

{\bf 1975 --} classification of physical structures of arbitrary rank     \\

{\bf 1980 --} classification of phenomenologically symmetric two-
\\  \phantom{aaaaaaaaa}dimensional  geometries of rank 4 on one set    \\

{\bf 1985 --} research of the group properties of physical structures      \\

{\bf 1990 --} classification of phenomenologically symmetric three- \\
 \phantom{aaaaaaaaa} dimensional geometries of rank 5 on one set    \\

{\bf 1995 --} definition of polymetric physical structures and their \\
\phantom{aaaaaaaaa} classification for some ranks     \\

{\bf 2000 --} research of some algebraic properties of abstract physical \\
\phantom{aaaaaaaaa} structures on sets of arbitrary nature      \\

{\bf 2005 --} publication of the monograph \ "Theory of Physical Structures"  \\
\phantom{aaaaaaaaa} by Yu. I. Kulakov (847 pp. with il., Russian) \\

{\bf 2010 --} research of the hypothesis of enclosure of physical structures \\
\phantom{aaaaaaaaa} in structures of higher ranks    \\

{\bf 2015 --} equation of phenomenological symmetry for some \\
\phantom{aaaaaaaaa} three-dimensional geometries derived    \\

{\bf P.S. --} more information about the theory of physical structures \\
\phantom{aaaaaaaa} is to be found on the site TPS \ ( {\bf
http://www.tphs.info} )

    \newpage

\begin{center}
Печатается по решению редакционно-издательского совета \\
Горно-Алтайского госуниверситета
\end{center}

УДК 514.1 \hspace{5mm} ББК 22.3 \hspace{5mm} М 69

\vspace{1mm}

{\bf Г.Г. Михайличенко

Математические Основы и Результаты Теории Физических Структур:}
Монография / Г.Г. Михайличенко -- Горно-Алтайск: \ РИО \ ГАГУ,
2016, - 297 с.

\vspace{1mm}

{\bf ISBN 978-5-91425-079-6}

\vspace{1mm}

Р е ц е н з е н т ы: доктор физ.-мат. наук, профессор Ю.С.
Владимиров (МГУ); доктор физ.-мат. наук, в.н.с. Е.Е. Витяев (ИМ СО
РАН); кандидат физ.-мат. наук, доцент В.А. Кыров (ГАГУ)

\vspace{5mm}

Теория физических структур (ТФС) была предложена профессором Ю.И.
Кулаковым для классификации законов физики. История возникновения
и развития этой теории достаточно подробно изложена в его
монографии [1]. Физическая структура представляет собой геометрию
одного или двух множеств, метрическая функция которой паре точек
сопоставляет число. Ее феноменологическая симметрия по Кулакову
означает, что для любой совокупности некоторого конечного числа
точек все их взаимные расстояния функционально связаны. Такие
геометрии наделены групповой симметрией по Клейну, которая
эквивалентна феноменологической симметрии, и многие из них имеют
содержательную физическую интерпретацию. Поэтому, прежде всего,
они должны быть точно определены и подробно изучены как чисто
математические объекты. В данной книге представлены математические
основы и полученные к настоящему времени классификационные
результаты ТФС. Книга адресована научным сотрудникам и
преподавателям, аспирантам и студентам старших курсов, всем тем,
чьи интересы лежат в области алгебры, геометрии и теоретической
физики, которые хотели бы использовать ТФС в своих исследованиях
или внести свой вклад в развитие ее математического аппарата ( см. на сайте
arXiv по адресу \ {\bf https://arxiv.org/pdf/1602.02795} ).

\vspace{2mm}

\begin{flushright}
\copyright \ Г.Г. Михайличенко, 2016
\end{flushright}

\newpage

     \begin{center}
     Printed by the Editorial and Printing Council of Gorno-Altaisk \\
     state university order
     \end{center}

     UDK 514.1 \hspace{5mm} BBL 22.3 \hspace{5mm} М 69

     \vspace{1mm}

     {\bf G.G. Mikhailichenko

      The Mathematical Basics and Results of the Theory of \ Physical Structures:}
     Monograph / G.G. Mikhailichenko -- Gorno-Altaisk: \ EPD \ GASU,
     2016, - 297 p.

     \vspace{1mm}

     {\bf ISBN 978-5-91425-079-6}

     \vspace{1mm}

     R e v i e w e d \ by: Doctor of Physical and Mathematical Sciences Professor Yu. S.
     Vladimirov (MSU); Doctor of Physical and Mathematical Sciences E.E. Vitiaev (out-of-staff
     member IM SD RAS); Candidate of Physical and Mathematical Sciences associate
     professor V.A. Kyrov (GASU).

     \vspace{5mm}

     The theory of Physical Structures (TPS) was put forward by Professor Yu. I. Kulakov
     for the sake of classifying the laws of Physics. The history of the development of
     that theory is given in his monograph [1]. A physical structure is a geometry of
     one or two sets whose metric function assigns a number to every pair
     of points. Its phenomenological symmetry, under Kulakov,
     means that for every collection of some finite number of
     points all of their reciprocal distances are functionally related. Such
     geometries are endowed with a group symmetry under Klein, which is
     equivalent to the phenomenological symmetry, and many of them have
     an essential physical interpretation. That is why they are to be defined
     precisely and explored as purely mathematical objects. In this monograph
     we treat the mathematical basics of the TPS and present the results of
     attempts at classification that have been obtained by now. We hope
     that the monograph will be of interest for research workers and
     teachers, senior and post graduate students, as well as to all
     those interested in algebra, geometry and theoretical
     physics who would like to use the TPS in their research projects
     or could want to contribute to the development of its mathematical apparatus.
    ( see on the site arXiv on the address \ {\bf https://arxiv.org/pdf/1602.02795} ).

     \vspace{5mm}

     \begin{flushright}
     \copyright \ G.G. Mikhailichenko, 2016
     \end{flushright}

\newpage

\begin{center}
{\bf  \Large СОДЕРЖАНИЕ}
\end{center}

\vspace{4mm}

ВВЕДЕНИЕ \dotfill\ 10

\vspace{4mm}

ГЛАВА I. {\bf Геометрия как физическая структура на одном \phantom{a} \linebreak
\phantom{aaaaaaaaaa} множестве} \dotfill\ 32

\vspace{4mm}

\S1. Феноменологическая и групповая симметрии в геометрии  \dotfill\ 32

\S2. Классификация одномерных, двумерных и трехмерных \phantom{aaaaaaa}
\linebreak \phantom{aaaaa} геометрий  \dotfill\  48

\S3. Двуметрические геометрии на плоскости и триметрические
\phantom{aaa} \linebreak \phantom{aaaaa} геометрии в пространстве \dotfill\ 76

\S4. К вопросу о симметрии расстояния в геометрии  \dotfill\ 102

\S5. Бинарные и тернарные геометрии \dotfill\  108

\S6. Функциональные уравнения в геометрии  \dotfill\ 124

\S7. Вопросы классификации феноменологически симметричных
\phantom{aaa} \linebreak \phantom{aaaaa} геометрий \dotfill\ 134

\vspace{4mm}

ГЛАВА II.  {\bf Физическая структура как геометрия двух \phantom{aaa}
\linebreak \phantom{aaaaaaaaaaa} множеств} \dotfill\ 144

\vspace{4mm}

\S8. Феноменологическая и групповая симметрии физических \phantom{aaaaaa}
\linebreak \phantom{aaaaa} структур \dotfill\ 144

\S9. Классификация однометрических физических структур
\dotfill\ 158

\S10. Двуметрические и триметрические физические структуры \dotfill\ 188

\S11. Групповая симметрия произвольных физических структур \dotfill\ 216

\S12. Функциональные уравнения в теории физических структур \dotfill\ 232

\S13. Интерпретации физических структур \dotfill\ 252

\S14. Нерешенные задачи в теории физических структур \dotfill\ 266

\vspace{4mm}

ЗАКЛЮЧЕНИЕ  \dotfill\ 270

\vspace{4mm}

Л и т е р а т у р а \dotfill\ 272

\vspace{4mm}

{\it Приложение.} \ А.Н. Бородин. Груда и физическая структура \phantom{aaaaaaa}
 \linebreak \phantom{aaaaaaaaaaaaaaa} ранга (2,2)\dotfill\  280

\vspace{4mm}

{\it Сведения об авторах} \dotfill\ 294

 \newpage

     \begin{center}

     {\bf  \Large CONTENTS}
     \end{center}

     \vspace{4mm}

     INTRODUCTION \dotfill\ 11

     \vspace{4mm}

    CHAPTER I. {\bf Geometry as a physical structure on one \phantom{a} \linebreak
    \phantom{aaaaaaaaaa} set} \dotfill\ 33

     \vspace{4mm}

     \S1. Phenomenological and group symmetries in geometry  \dotfill\ 33

     \S2. Classification of one-, two- and three-dimensional
      \phantom{aaaaaaa} \linebreak \phantom{aaaaa} geometries  \dotfill\  49

     \S3. Dimetric geometries on a plane and trimetric
     \phantom{aaa} \linebreak \phantom{aaaaa} geometries
      in space \dotfill\ 77

     \S4. The symmetry of distance in geometry \dotfill\ 103

     \S5. Binary and ternary geometries\dotfill\ 109

     \S6. Functional equations in geometry  \dotfill\ 125

     \S7. Problems of classification of phenomenologically symmetric\phantom{aaa}
      \linebreak \phantom{aaaaa} geometries  \dotfill\ 135

     \vspace{4mm}

     CHAPTER II.  {\bf A physical structure as a geometry of two \phantom{aaa}
     \linebreak \phantom{aaaaaaaaaaa} sets} \dotfill\ 145

     \vspace{4mm}

    \S8. The phenomenological and group symmetry of physical \phantom{aaaaaa}
    \linebreak \phantom{aaaaa} structures \dotfill\ 145

     \S9. A classification of unimetric physical structures
     \dotfill\ 159

     \S10. Dimetric and trimetric physical structures\dotfill\ 189

     \S11. The group symmetry of arbitrary physical structures \dotfill\  217

     \S12. Functional equations in the theory of physical structures \dotfill\   233

     \S13. Interpretations of physical structures \dotfill\ 253

     \S14. The unresolved problems in the theory of physical structures  \dotfill\  267

     \vspace{4mm}

     CONCLUSION  \dotfill\ 271
     \vspace{4mm}

     B i b l i o g r a p h y  \dotfill\ 273

     \vspace{4mm}

     {\it Appendix.} \ A.N. Borodin. A heap and physical structures
     \phantom{aaaaaaa} \linebreak \phantom{aaaaaaaaaaaaaaa} of rank (2,2)\dotfill\  281

     \vspace{4mm}

     {\it About the authors} \dotfill\ 295

\newpage

\begin{center}
{\bf \Large ВВЕДЕНИЕ}
\end{center}

\vspace{5mm}

Для иллюстрации феноменологической и групповой симметрий в \linebreak обычной
геометрии, а также связи между ними, рассмотрим сначала
плоскость Евклида. В  декартовой  прямоугольной  системе
координат $(x,y)$  квадрат  расстояния  $\rho(ij)$  между любыми
двумя ее точками $i = (x_i, y_i)$ и $j = (x_j,y_j)$ задается функцией
\begin{equation}
\label{a1}
f(ij) = \rho ^2(ij) = (x_i - x_j) ^2 + (y_i - y_j) ^2.
\tag{В.1}
\end{equation}

Возьмем четыре произвольные точки $i$, $j$, $k$, $l$ и запишем для них шесть
значений метрической функции (В.1): $f(ij)$, $f(ik)$,
$f(il)$, $f(jk)$, $f(jl)$, $f(kl)$. Хорошо известно, что шесть
взаимных расстояний между любыми четырьмя точками евклидовой
плоскости функционально связаны, обращая в нуль определитель
Кэли-Менгера пятого порядка:
\begin{equation}
\label{a2}
\left|
\begin{array}[tcccb]{ccccc}
0 & 1 & 1 & 1 & 1 \\
1 & 0 & f(ij) & f(ik) & f(il) \\
1 & f(ij) & 0 & f(jk) & f(jl) \\
1 & f(ik) & f(jk) & 0 & f(kl) \\
1 & f(il) & f(jl) & f(kl) & 0 \\
\end{array}
\right| = 0.
\tag{В.2}
\end{equation}

Геометрический смысл соотношения (В.2) состоит в том,
что объем тетраэдра с вершинами, лежащими на плоскости, равен нулю.

В соответствии с терминологией Ю.И. Кулакова [2] соотношение (В.2),
справедливое для любой четверки $\langle ijkl \rangle$, выражает
феноменологическую симметрию евклидовой плоскости.

По метрической функции (В.1) можно найти множество движений плоскости
Евклида, то есть таких гладких и обратимых ее преобразований
\begin{equation}
\label{a3}
x' = \lambda(x,y), \quad   y' = \sigma(x,y),
\tag{В.3}
\end{equation}
относительно которых эта функция сохраняется: $f(i'j')=f(ij)$.

Действительно, если преобразование (В.3) является
движением, то для его функций $\lambda$ и $\sigma$ получаем
функциональное уравнение
\begin{displaymath}
(\lambda(i) - \lambda(j))^2 + (\sigma(i) - \sigma(j))^2
= (x_i - x_j)^2 + (y_i - y_j)^2,
\end{displaymath}

 \newpage

     \begin{center}
     {\bf \Large INTRODUCTION}
     \end{center}

     \vspace{5mm}

     To illustrate phenomenological and group symmetries in ordinary geometry as well
     as their relation, let us first take Euclidean plane. In the Cartesian rectangular coordinate
     system $(x,y)$  the squared distance $\rho(ij)$  between any
     two points $i = (x_i, y_i)$ and $j = (x_j,y_j)$ of it is determined by the function
     \begin{equation}
     \label{a1}
     f(ij) = \rho ^2(ij) = (x_i - x_j) ^2 + (y_i - y_j) ^2.
     \tag{В.1}
     \end{equation}

     We shall take four arbitrary points $i$, $j$, $k$, $l$ and write six values
     of the metric function  (В.1): $f(ij)$, $f(ik)$,
     $f(il)$, $f(jk)$, $f(jl)$, $f(kl)$ for them. It is well known that the six
     reciprocal distances among any four points of an Euclidean plane
     are functionally related, turning into zero the Cayly-Menger determinant
     of the fifth order:
     \phantom{aaaaa} \\
     \begin{equation}
     \label{a2}
     \left|
     \begin{array}[tcccb]{ccccc}
     0 & 1 & 1 & 1 & 1 \\
     1 & 0 & f(ij) & f(ik) & f(il) \\
     1 & f(ij) & 0 & f(jk) & f(jl) \\
     1 & f(ik) & f(jk) & 0 & f(kl) \\
     1 & f(il) & f(jl) & f(kl) & 0 \\
     \end{array}
     \right| = 0.
     \tag{В.2}
     \end{equation}
     \phantom{aaaaa}

     The geometrical meaning of the relation (В.2) is in that the volume of
     a tetrahedron with all the apices in one plane is equal to zero.

     Under  Yu.I. Kulakov's terms [2], the relation (В.2),
     that is true for any quadruple of points $\langle ijkl \rangle$, expresses
     the phenomenological symmetry of Euclidean plane.

     By the metric function (В.1), it is possible to find the set of motions of
     Euclidean plane, i.e. the set of such smooth and invertible transformations
     \begin{equation}
     \label{a3}
     x' = \lambda(x,y), \quad   y' = \sigma(x,y),
     \tag{В.3}
     \end{equation}
     of it with respect to which the function is preserved: $f(i'j')=f(ij)$.

     Indeed, if the transformation (В.3) is a
     motion, then for the functions $\lambda$ and $\sigma$ of it we have
     the functional relation
     \begin{displaymath}
     (\lambda(i) - \lambda(j))^2 + (\sigma(i) - \sigma(j))^2
     = (x_i - x_j)^2 + (y_i - y_j)^2,
     \end{displaymath}

\newpage

\noindent
где, например, $\lambda(i)=\lambda(x_i,y_i)$. Сводя это уравнение
к системе функцио- нально-дифференциальных соотношений, можно
найти все его решения:
\begin{equation}
\label{a4}
\left.\begin{array}{rcl}
\lambda(x,y) = ax - \varepsilon by + c,\\
\sigma(x,y) = bx + \varepsilon ay + d\;,
\end{array}\right\}
\tag{B.4}
\end{equation}
где $\varepsilon = \pm 1$; $a^2 + b^2 = 1$, $c$, $d$ -- произвольные постоянные.

Множество всех движений (В.3) с функциями (В.4)
является группой, определяющей групповую симметрию
плоскости Евклида. С другой стороны, эта трехпараметрическая группа
преобразований координатной плоскости $(x,y)$
задает на ней по Ф.Клейну [3] евклидову геометрию. В частности, метрическая
функция $f(ij)$ может быть найдена решением функционального уравнения
$$
f(x'_i,y'_i,x'_j,y'_j)=f(x_i,y_i,x_j,y_j)
$$
как ее двухточечный инвариант. Общее решение этого уравнения
совпадает с метрической функцией (В.1) с точностью до масштабного
преобразования:
$$
f(ij)=\psi((x_i-x_j)^2+(y_i-y_j)^2),
$$
где $\psi$ -- произвольная функция одной переменной.

Выясним, имеется ли связь феноменологической и групповой
симметрий для произвольной плоской геометрии, задаваемой метрической функцией
\begin{equation}
\label{a5}
f(ij) = f(x_i,y_i,x_j,y_j),
\tag{В.5}
\end{equation}
обобщающей выражение (В.1).

Жесткая фигура на плоскости при любом разумном определении понятия
движения имеет три степени свободы. Возьмем четырехточечную фигуру
$\langle  ijkl  \rangle$. Каждая ее точка задается двумя
координатами, а вся фигура -- восемью. Шесть значений функции
(В.5) для этой фигуры должны быть зависимы, так как, иначе, число
ее степеней свободы будет равно только двум: $8 - 6 = 2$. Таким
образом, для любой четверки $\langle  ijkl  \rangle$ должна
существовать функциональная связь
\begin{equation}
\label{a6}
\Phi(f(ij),f(ik),f(il),f(jk),f(jl),f(kl)) = 0,
\tag{В.6}
\end{equation}

\newpage

\noindent
     where, for example, $\lambda(i)=\lambda(x_i,y_i)$.
     By way of reducing that equation to a system of functional differential relations,
     it is possible to find all its solutions:
     \begin{equation}
     \label{a4}
     \left.\begin{array}{rcl}
     \lambda(x,y) = ax - \varepsilon by + c,\\
     \sigma(x,y) = bx + \varepsilon ay + d\;,
     \end{array}\right\}
     \tag{B.4}
    \end{equation}
     where $\varepsilon = \pm 1$; $a^2 + b^2 = 1$, $c$, $d$ are arbitrary constants.

     The set of all the motions (В.3) with the functions (В.4)
     is a group that determines the group symmetry
     of Euclidean plane. On the other hand, that three-parameter group
     of transformations of the coordinate plane $(x,y)$
     defines on it, under F. Klein [3], an Euclidean geometry. In particular, the metric function
     $f(ij)$ may be found by way of solving the functional equation
     $$
     f(x'_i,y'_i,x'_j,y'_j)=f(x_i,y_i,x_j,y_j)
     $$ \\
     as its two-point invariant. The general solution of that equation
     coincides with the metric function (В.1) with an accuracy up to a scaling
     transformation:
     $$
     f(ij)=\psi((x_i-x_j)^2+(y_i-y_j)^2),
     $$
     where $\psi$ is an arbitrary function of one variable.

     Let us check whether there exists any relation of the phenomenological and group
     symmetries of an arbitrary planar geometry defined by the metric function
     \begin{equation}
     \label{a5}
     f(ij) = f(x_i,y_i,x_j,y_j),
     \tag{В.5}
     \end{equation}
     that generalizes the expression (В.1).

     A rigid figure on a plane, under any reasonable definition of the notion of
     motion has three degrees of freedom. Let us consider a four-point figure
     $\langle  ijkl  \rangle$. Every point of it is defined by two coordinates, and the figure on the whole
     -- by eight. The six values of the function (В.5) for that figure
     must be dependent, because otherwise the number of its degrees of freedom
     will only be as few as two: $8 - 6 = 2$. Thus, for any
     quadruple $\langle  ijkl  \rangle$ there must exist a functional relation
     \begin{equation}
     \label{a6}
     \Phi(f(ij),f(ik),f(il),f(jk),f(jl),f(kl)) = 0,
     \tag{В.6}
     \end{equation}

\newpage

 \noindent
 выражающая {\it феноменологическую симметрию} плоской
геометрии с метрической функцией (В.5).

Простые соображения приводят также к выводу о том, что
если имеет место связь (В.6), то существует трехпараметрическая
группа движений:
\begin{equation}
\label{a7}
\left.\begin{array}{rcl}
x' = \lambda(x,y;a^1,a^2,a^3),\\
y' = \sigma(x,y;a^1,a^2,a^3)\;,
\tag{В.7}
\end{array}\right\}
\end{equation}
относительно которой метрическая функция (В.5) является двухточечным
инвариантом: $f(i'j')=f(ij)$ или
\begin{equation}
\label{a8}
f(\lambda(i),\sigma(i),\lambda(j),\sigma(j)) = f(x_i,y_i,x_j,y_j),
\tag{B.8}
\end{equation}
где, например, $\lambda(i)=\lambda(x_i,y_i;a^1,a^2,a^3)$.

Множество всех движений (В.7) определяет {\it групповую симметрию}
плоской геометрии с метрической функцией (В.5).

Заметим, что приведенные выше соображения о связи феноменологической
и групповой симметрий применимы не только в отношении плоскости
Евклида, но и в отношении других двумерных геометрий (плоскости
Лобачевского, плоскости Минковского, симплектической плоскости,
обычной двумерной сферы и т.д.).

Г.Гельмгольц в своей работе \ "О фактах, лежащих в основании
геометрии" \ [4] высказал предположение, что метрическая функция
$n$-мер- ного пространства не может быть произвольной если в нем
твердое тело имеет $n(n + 1)/2$ степеней свободы. Но в таком
случае между всеми взаимными расстояниями для $n + 2$ точек
твердого тела должна существовать функциональная связь, так как
при ее отсутствии число степеней свободы $(n + 2)$-точечной
жесткой фигуры с общим расположением точек, движение которой
однозначно определяет движение всего твердого тела, уменьшится
ровно на единицу. Поэтому естественно было предположить, что и
феноменологическая симметрия $n$-мерного пространства невозможна
при произвольной метрической функции. Для $n = 1$ и $n = 2$ это
было установлено в работах автора [5] и [6], а для $n = 3$ -- в
работе В.Х.Лева [7].

Заметим еще, что задачу классификации всех плоских (двумерных)
геометрий, в которых \ "положение фигуры задается тремя
условиями"$,$

\newpage

     \noindent
      that expresses the {\it phenomenological symmetry} of the planar geometry with
     the metric function (В.5).

     By virtue of simple considerations the
     existence of a relation (В.6) implies the existence of a three-parameter
     group of motions:

     \begin{equation}
     \label{a7}
     \left.\begin{array}{rcl}
     x' = \lambda(x,y;a^1,a^2,a^3),\\

     y' = \sigma(x,y;a^1,a^2,a^3)\;,
     \tag{В.7}
     \end{array}\right\}
     \end{equation}
     \phantom{aaaaa} \\
     with respect to which the metric function  (В.5) is a two-point invariant:
     $f(i'j')=f(ij)$ or
     \begin{equation}
     \label{a8}
     f(\lambda(i),\sigma(i),\lambda(j),\sigma(j)) = f(x_i,y_i,x_j,y_j),
     \tag{B.8}
     \end{equation}
     \phantom{aaaaa} \\
     where, for example, $\lambda(i)=\lambda(x_i,y_i;a^1,a^2,a^3)$.

     The set of all the motions (В.7) defines the {\it group symmetry} of
     the planar geometry with the metric function (В.5).

     We shall note that the above considerations concerning the relation bet- \ ween the
     phenomenological and group symmetries are not only applicable to
     Euclidean plane, but to other two-dimensional geometries (the Lobachevski
     plane, the Minkowski plane, the simplectic plane, the ordinary two-dimensio- \ nal
     sphere and so on).

     H. Helmholtz in his work "On the facts underlying
     geometry" \ [4] suggested that the metric function  of an $n$-dimensional
     space cannot be an arbitrary one if in that space a rigid body
     has $n(n + 1)/2$ degrees of freedom. But then there must exist a functional
     relation among all the reciprocal distances for $n + 2$ points of the
     rigid body, because absence of such a relation would reduce by one the number of
     the degrees of freedom of a rigid body of $(n + 2)$-points with the common
     point spacing whose motion defines uniquely the motion of the whole
     solid body. So, it has been natural to suppose that the
     phenomenological symmetry of an $n$-dimensional space is impossible with
     an arbitrary function. For $n = 1$ and $n = 2$, it was established in the notes [5]
     and [6] by the author, and for $n = 3$ -- in V.H. Lev's note [7].

     We shall also note that the problem of classification of all planar (two-dimensional)
     \ geometries in which the position of a figure \ "is determined

\newpage

\noindent
 впервые сформулировал А.Пуанкаре в своей работе \
"Об основных гипотезах геометрии" \ [8].

Метрическая функция $f(ij)$ задает геометрию пространства.
Действительно, по этой функции можно найти группу движений,
относительно которой она является двухточечным инвариантом.
Групповая же симметрия лежит в основе "Эрлангенской программы" \
Ф.Клейна 1872 года [3], согласно которой геометрия пространства
есть теория инвариантов некоторой группы его преобразований.
С другой стороны, в геометрии обнаруживает себя
феноменологическая симметрия, выражаемая некоторой
функциональной связью между всеми взаимными расстояниями для определенного
числа точек пространства. На эту симметрию  впервые
особое внимание обратил Ю.И.Кулаков [2], сделав ее основным принципом
своей теории физических структур [1].

Рассмотрим множество состояний некоторой термодинамической
системы. Каждой паре состояний $\langle  ij  \rangle$ сопоставим два числа,
равные двум количествам тепла $Q^{TS}(ij)$ и
$Q^{ST}(ij)$, которые система отдает внешним телам при ее переходе
из состояния $i$ в состояние $j$ по двум различным путям $TS$ и $ST$,
состоящим из равновесных изотермического ($T$ = const) и адиабатического
($S$ = const) процессов:
\begin{equation}
\label{a9}
\left.\begin{array}{rcl}
Q^{TS}(ij) = (S_i - S_j)T_i,\\
Q^{ST}(ij) = (S_i - S_j)T_j\;,
\end{array}\right\}
\tag{B.9}
\end{equation}
где $S$ -- энтропия и $T$ -- температура системы.

Двухкомпонентная тепловая функция $Q = (Q^{TS},Q^{ST})$ с выражениями
(B.9) для ее компонент задает на плоскости $(S,T)$
геометрию, которая, подобно евклидовой геометрии на плоскости,
феноменологически симметрична, с одной стороны, и наделена
групповой симметрией -- с другой.

Возьмем произвольные три состояния $i$, $j$, $k$. Тогда
дополнительно к двум количествам тепла, задаваемым выражениями
(B.9), можно выписать еще четыре: $Q^{TS}(ik), Q^{ST}(ik)$ и
$Q^{TS}(jk)$, $Q^{ST}(jk)$ для пар состояний $\langle  ik  \rangle$ и
$\langle  jk  \rangle$. Из этих шести выражений можно исключить три энтропии
$S_i$, $S_j$, $S_k$ и три температуры $T_i$, $T_j$, $T_k$
состояний $i$, $j$, $k$, в результате чего получаются две
независимые функциональные связи

\newpage

     \noindent
     by three conditions" \     was
     formulated for the first time by J.H. Poincare
     in his work "Sur les hypotheses fondamentales de la geometrie"  \ [8].

     The metric function $f(ij)$ gives a geometry of space.
     Indeed, through that function it is possible to find the group of motions
     with respect to which it is a two-point invariant.
     Group symmetry also lies in the basis of the F. Klein's "Erlangen programme" \
     of 1872 [3], under which geometry of space
     is a theory of invariants of some group of its transformations.
     On the other hand, there appears in geometry some
     phenomenological symmetry expressed by some
     functional relation among all the reciprocal distances for a certain number
     of points of the space. For the first time that sort of symmetry  became
     an object of special attention in the works by Yu.I. Kulakov [2], who made
     it the basic principle of his theory of physical structures [1].

     Let us consider the set of states of some thermodynamic
     system. We shall assign to each pair of states $\langle  ij  \rangle$ two numbers
     equal to two quantities of heat $Q^{TS}(ij)$ and
     $Q^{ST}(ij)$ which the system gives away to other bodies in the course of the transition
     from the state $i$ to the state $j$ along two different ways, $TS$ and $ST$, that
     consist of equilibrium processes, an isothermic one ($T$ = const) and a adiabatic
     ($S$ = const) one:
     \phantom{aaaaa} \\
     \phantom{aaaaa} \\
     \begin{equation}
     \label{a9}
     \left.\begin{array}{rcl}
     Q^{TS}(ij) = (S_i - S_j)T_i,\\
     Q^{ST}(ij) = (S_i - S_j)T_j\;,
     \end{array}\right\}
     \tag{B.9}
     \end{equation}
     \phantom{aaaaa} \\
     where $S$ is the entropy and $T$ is the temperature of the system.

     A two-component thermal function $Q = (Q^{TS},Q^{ST})$ with the expressions
     (B.9) for its components gives on the plane $(S,T)$
     a geometry that, like the Euclidean geometry on a plane, is
     phenomenologically symmetric, on the one hand, and is endowed
     with a group symmetry, on the other.

     Let us take three arbitrary states, $i$, $j$, $k$. Then, in addition to
     two quantities of heat, determined by the expressions (B.9), we can write
     four more: $Q^{TS}(ik), Q^{ST}(ik)$ and $Q^{TS}(jk)$, $Q^{ST}(jk)$
     for the pairs of states $\langle  ik  \rangle$ and $\langle  jk  \rangle$.
     We may exclude of these six the three entropies, $S_i$, $S_j$, $S_k$, and
     the three temperatures, $T_i$, $T_j$, $T_k$, of the states $i$, $j$, $k$,
     which will yield as result two independent functional relations \ among \ all \ the
     \ quantities \ of

\newpage

\noindent
между всеми количествами тепла, задаваемые следующими уравнениями:
$$
\left.\begin{array}{c}
\left|
\begin{array}[tcb]{ccc}
0 & -Q^{ST}(ij) & -Q^{ST}(ik) \\
Q^{TS}(ij) & 0 & -Q^{ST}(jk) \\
Q^{TS}(ik) & Q^{TS}(jk) & 0 \\
\end{array}
\right| = 0,\\
\\
\left|
\begin{array}[tcb]{ccc}
Q^{TS}(ij) & Q^{TS}(jk) & -Q^{ST}(ik) \\
Q^{TS}(ik) & 0 & -Q^{ST}(ik) \\
Q^{TS}(ik) & -Q^{ST}(ij) & -Q^{ST}(jk) \\
\end{array}
\right| = 0.
\end{array}\right\}
\eqno(\text{B}.10)
$$

Соотношения (B.10), справедливые для любой тройки состояний \linebreak $\langle  ijk  \rangle$,
выражают феноменологическую симметрию двуметрической геометрии,
задаваемой на плоскости $(S,T)$ двухкомпонентной тепловой
функцией (B.9).
Группа движений в этой геометрии состоит из всех
тех гладких и обратимых преобразований
$$
S'=\lambda(S,T), \ \ T'=\sigma(S,T)
\eqno(\text{B}.11)
$$
плоскости $(S,T)$, которые сохраняют обе компоненты функции (В.9):

\begin{equation}
\label{a12}
\left.\begin{array}{rcl}
(\lambda(i) - \lambda(j))\sigma(i) = (S_i - S_j)T_i,\\
(\lambda(i) - \lambda(j))\sigma(j) = (S_i - S_j)T_j\;.
\end{array}\right\}
\tag{B.12}
\end{equation}

Решения этой системы функциональных
уравнений легко находятся методом разделения переменных -- координат
состояний $i$ и $j$:
\begin{equation}
\label{a13}
\lambda(S,T) = aS + b,\quad
\sigma(S,T) =T/a,
\tag{B.13}
\end{equation}
где $a\neq0$, $b$ -- произвольные постоянные.

Множество преобразований (В.11) с функциями (В.13)
является группой всех движений, которая определяет
групповую симметрию двумерной двуметрической геометрии, задаваемой на
плоскости $(S,T)$ функцией (В.9). Таким образом, к плоскости термодинамических
состояний тоже применима Эрлангенская программа Ф.Клейна [3]. В частности,
тепловая функция $Q(ij)$ может быть найдена решением функционального уравнения
$$
Q(S'_i,T'_i,S'_j,T'_j)=Q(S_i,T_i,S_j,T_j)
$$

\newpage

    \noindent
     heat, determined by the following equations:
    \phantom{aaaaa} \\
     $$
     \left.\begin{array}{c}

     \left|

     \begin{array}[tcb]{ccc}
     0 & -Q^{ST}(ij) & -Q^{ST}(ik) \\
     Q^{TS}(ij) & 0 & -Q^{ST}(jk) \\
     Q^{TS}(ik) & Q^{TS}(jk) & 0 \\
     \end{array}
     \right| = 0,\\
     \\
     \left|
     \begin{array}[tcb]{ccc}
     Q^{TS}(ij) & Q^{TS}(jk) & -Q^{ST}(ik) \\
     Q^{TS}(ik) & 0 & -Q^{ST}(ik) \\
     Q^{TS}(ik) & -Q^{ST}(ij) & -Q^{ST}(jk) \\
     \end{array}
     \right| = 0.
     \end{array}\right\}
     \eqno(\text{B}.10)
     $$
    \phantom{aaaaa} \\

     The relations (B.10), true to any triple of states \linebreak $\langle  ijk  \rangle$,
     express the phenomenological symmetry of the dimetric geometry
     defined on the plane $(S,T)$ by the two-component thermal function (B.9).
     The group of motions in that geometry consists of all
     those smooth and invertible transformations
     $$
     S'=\lambda(S,T), \ \ T'=\sigma(S,T)
     \eqno(\text{B}.11)
     $$
     of the plane $(S,T)$ that preserve the both components of the function (В.9):

     \begin{equation}
     \label{a12}
     \left.\begin{array}{rcl}
     (\lambda(i) - \lambda(j))\sigma(i) = (S_i - S_j)T_i,\\
     (\lambda(i) - \lambda(j))\sigma(j) = (S_i - S_j)T_j\;.

     \end{array}\right\}
     \tag{B.12}
     \end{equation}

     The solutions of that system of functional equations
     are readily found by the method of separating of variables, the coordinates
     of the states $i$ and $j$:
     \begin{equation}
     \label{a13}
     \lambda(S,T) = aS + b,\quad
     \sigma(S,T) =T/a,
     \tag{B.13}
     \end{equation}
     where $a\neq0$, $b$ are arbitrary  constants.

     The set of transformations (В.11) with the functions (В.13)
     is the group of all the motions that determine the group symmetry of
     the two-dimensional dimetric geometry defined on the plane $(S,T)$
     by the function (В.9). Thus, the "Erlangen programme" of F. Klein is
     applicable to the plane of \\ thermodynamic states [3]. In particular,
     the thermal function $Q(ij)$ may be found by way of solving the functional equation
     $$
     Q(S'_i,T'_i,S'_j,T'_j)=Q(S_i,T_i,S_j,T_j)
     $$

     \newpage

\noindent
как двухточечный инвариант группы преобразований (В.11) с
функциями (В.13), который совпадает с ней с точностью до
масштабного преобразования:
$$
Q(ij)=\psi((S_i-S_j)T_i,(S_i-S_j)T_j),
$$
где $\psi=(\psi^1,\psi^2)$ -- двухкомпонентная функция двух
переменных.

Легко установить, что феноменологическая симметрия геометрии,
задаваемой на плоскости $(x,y)$ некоторой двухкомпонентной
функцией
\begin{equation}
\label{a14}
f(ij) = f(x_i,y_i,x_j,y_j),
\tag{B.14}
\end{equation}
где $f=(f^1,f^2)$, выражается соотношением
\begin{equation}
\label{a15}
\Phi(f(ij),f(ik),f(jk)) = 0,
\tag{B.15}
\end{equation}
где $\Phi = (\Phi_{1},\Phi_{2})$. Групповая же симметрия этой
геометрии определяется группой всех движений:
$$
x' = \lambda(x,y;a^1,a^2),\quad
y' = \sigma(x,y;a^1,a^2),
\eqno(\text{B.16})
$$
зависящей  от двух непрерывных параметров
$a^{1},a^{2}$, относительно которой обе
компоненты метрической функции (В.14) сохраняются: $f(i'j')=f(ij)$,
являясь ее двухточечными инвариантами.
Заметим, что, как и в случае плоскости Евклида, здесь групповая
симметрия также эквивалентна феноменологической симметрии.

Ю.И.Кулаков в своих исследованиях по основаниям физики [9] предложил
математическую модель строения физического закона, рассматриваемого как
феноменологически симметричная связь между измеряемыми в опыте величинами.
Эта модель, названная им {\it физической структурой},
приложима к обычной геометрии и
представляет собой своеобразную геометрию двух множеств с метрической
функцией, сопоставляющей число паре точек, но не из одного множества, а из
двух разных.
В новой геометрии естественно вводится движение как пара таких
преобразований исходных множеств, которые сохраняют метрическую функцию.
Совокупность всех движений является группой и определяет групповую симметрию
этой геометрии.

Следуя работе [9], рассмотрим второй закон Ньютона в механике и
закон Ома в электродинамике, записав их в такой форме, которая
позволит выявить их феноменологическую симметрию.

\newpage

     \noindent
     as a two-point invariant of the set of transformations (В.11), that coincides with it with
     an accuracy up to the scaling transformation
     $$
     Q(ij)=\psi((S_i-S_j)T_i,(S_i-S_j)T_j),
     $$
     where $\psi=(\psi^1,\psi^2)$ is a two-component function of two variables.

     It is easy to establish that the phenomenological symmetry of the geometry
     defined on the plane $(x,y)$ by some two-component
     function
     \begin{equation}
     \label{a14}
     f(ij) = f(x_i,y_i,x_j,y_j),
     \tag{B.14}
     \end{equation}
     where $f=(f^1,f^2)$, is expressed by the relation
     \begin{equation}
     \label{a15}
     \Phi(f(ij),f(ik),f(jk)) = 0,
     \tag{B.15}
     \end{equation}
     where $\Phi = (\Phi_{1},\Phi_{2})$. As to the group symmetry of that
     geometry, it is defined by the group of all the motions:
     $$
     x' = \lambda(x,y;a^1,a^2),\quad
     y' = \sigma(x,y;a^1,a^2),
     \eqno(\text{B.16})
     $$
     that depends on the two continuous parameters
     $a^{1},a^{2}$, with respect to which group both components
     of the metric function (В.14) are preserved: $f(i'j')=f(ij)$,
     being its two-point invariants.
     We shall note that here, just as in case of the Euclidean plane, the group
     symmetry is also equivalent to the phenomenological one.

     Yu.I. Kulakov, in his research of physics basics [9] suggested
     a mathemati- \ cal model of the structure of a physical law considered as a
     phenomenological- \ ly symmetric relation among the quantities measured
     in the experiment. The model, he called a {\it physical structure},
     can be applied to ordinary geometry too, and is a peculiar geometry of
     two sets with a metric function assigning a number to a pair of points,
     belonging however not to one and the same but to two different sets.
     In the new geometry, naturally, a motion is introduced, as a pair of transformations
     of the original sets such that preserve the metric function.
     The totality of all the motions is a group and it determines the group symmetry
     of the geometry in question.

     Let us consider, according the principles expounded in the note [9], Newton's 2nd law
     of mechanics and Ohm's law of electrodynamics, writing them in such a form
     that would enable us to reveal their phenomenological symmetry.

\newpage

Пусть $i$ -- материальное тел о, масса которого равна $m_i$, и
$\alpha$ -- ускоритель, характеризуемый силой $F_\alpha$. Под
ускорителем подразумевается какое-то другое тело, которое при
взаимодействии с данным телом изменяет его скорость. Измеряемой в
опыте величиной является ускорение $a_{i\alpha}$, которое телу $i$
сообщает ускоритель $\alpha$. Второй закон Ньютона в его
традиционной форме утверждает, что произведение массы тела на
сообщаемое ему ускорение равно действующей силе:
$$
m_ia_{i\alpha} = F_\alpha.
\eqno(\text{B.17})
$$

Возьмем произвольные два тела $i$, $j$ и произвольные два ускорителя $\alpha$,
$\beta$. Дополнительно к соотношению (B.17) запишем еще три:
$$
m_ia_{i\beta} = F_\beta,\ \ m_ja_{j\alpha} = F_\alpha,\ \
m_ja_{j\beta} = F_\beta.
$$
Из этих четырех соотношений можно исключить массы $m_i$,$m_j$ тел
$i,j$, силы $F_\alpha$,$F_\beta$ ускорителей $\alpha,\beta$ и
получить функциональную связь только между ускорениями, задаваемую
следующим уравнением:
$$
a_{i\alpha}a_{j\beta}-a_{i\beta}a_{j\alpha}=0.
\eqno(\text{B}.18)
$$

По терминологии Ю.И.Кулакова [9] функциональная связь (B.18),
имеющая место для
любых двух тел $i,j$ и любых двух ускорителей $\alpha,\beta$,
представляет уравнение второго закона Ньютона в феноменологически
симметричной форме.

Перейдем к электродинамике. Проводнику $i$ с
сопротивлением $R_i$ и источнику тока
$\alpha$ с электродвижущей силой $\mathcal{E}_\alpha$ и внутренним
сопротивлением $r_\alpha$ сопоставим ток $I_{i\alpha}$, измеряемый
амперметром в замкнутой цепи:
$$
I_{i\alpha}=\frac{\mathcal{E}_\alpha}{R_i+r_\alpha}.
\eqno(\text{B.19})
$$

Возьмем произвольные три проводника $i,j,k$ и произвольные два источника тока
$\alpha,\beta$. Тогда дополнительно к току $I_{i\alpha}$ по выражению (B.19)
можно выписать еще пять его значений:
$$
I_{i\beta},I_{j\alpha},I_{j\beta},I_{k\alpha},I_{k\beta}.
$$
Из шести выражений для тока могут быть исключены сопротивления
$R_i,R_j,R_k$ проводников $i,j,k$, электродвижущие силы
$\mathcal{E}_\alpha,\mathcal{E}_\beta$ и внутрен-

\newpage

     Let $i$ be a body the mass of which is equal to $m_i$, and
     $\alpha$ an accelerator characterized by force $F_\alpha$. An accelerator
     means some other body that, by interacting with the given one, changes
     its speed. The quantity measured by experiment is the acceleration
     $a_{i\alpha}$, that the accelerator $\alpha$ imparts to the body
     $i$. In its traditional form, Newton's well-renowned second law
     reads that the product of the weight of the body and
     the acceleration it is imparted is equal to the force applied:
     $$
     m_ia_{i\alpha} = F_\alpha.
     \eqno(\text{B.17})
     $$

     We shall take two arbitrary bodies $i$, and $j$ and two arbitrary accelerators
     $\alpha$, and $\beta$. In addition to the relation (B.17), we shall write three more:
     $$
     m_ia_{i\beta} = F_\beta,\ \ m_ja_{j\alpha} = F_\alpha,\ \
     m_ja_{j\beta} = F_\beta.
     $$
     From the four relations, it is possible to eliminate the masses $m_i$ and $m_j$ of the bodies
     $i$ and $j$, and the forces $F_\alpha$ and $F_\beta$ of the accelerators $\alpha$ and $\beta$,
     which yields a functional relation among the accelerations only, that is defined by
     the equations as follows:
     $$
     a_{i\alpha}a_{j\beta}-a_{i\beta}a_{j\alpha}=0.
     \eqno(\text{B}.18)
     $$

     Under Yu.I. Kulakov's terms [9] the functional relation (B.18), existing for
     any two bodies $i$ and $j$ and any two accelerators  $\alpha$ and $\beta$,
     is the phenomenologically symmetric form of Newton's second law.

     Let us now look at electrodynamics. To a conductor $i$ with
     resistance $R_i$ and a source of current
     $\alpha$ with an electromotive force $\mathcal{E}_\alpha$ and the internal
     resistance $r_\alpha$ we shall assign a current $I_{i\alpha}$, measured
     by an ammeter in a closed circuit:
     $$
     I_{i\alpha}=\frac{\mathcal{E}_\alpha}{R_i+r_\alpha}.
     \eqno(\text{B.19})
     $$
     We shall take three arbitrary conductors $i,j$, and $k$ and two arbitrary current sources
     $\alpha$ and $\beta$. Then, in addition to the current $I_{i\alpha}$, under the expression (B.19)
     it is possible to write five more values of it:
     $$
     I_{i\beta},I_{j\alpha},I_{j\beta},I_{k\alpha},I_{k\beta}.
     $$
     Now, from the six expressions for the current, we can eliminate
     the resistances $R_i, \ R_j, \ R_k$ of the conductors $i, j, \ k$, the electromotive
     forces \ $\mathcal{E}_\alpha, \ \mathcal{E}_\beta$ \ and

\newpage

\noindent
ние сопротивления $r_\alpha,r_\beta$ источников тока $\alpha,
\beta$, в результате чего получается функциональная связь только
между токами, задаваемая следующим уравнением:
$$
\left|
\begin{array}{ccc}
I_{i\alpha} & I_{i\beta} & I_{i\alpha}I_{i\beta} \\
I_{j\alpha} & I_{j\beta} & I_{j\alpha}I_{j\beta} \\
I_{k\alpha} & I_{k\beta} & I_{k\alpha}I_{k\beta}
\end{array}
\right| =0.
\eqno(\text{B.20})
$$

Функциональная связь (B.20), справедливая для любых трех
проводников $i,j,k$ и любых двух источников тока $\alpha,\beta$,
представляет по терминологии Кулакова [9] закон Ома в феноменологически
симметричной форме.

Обращает на себя внимание принципиальная общность уравнений
(B.18), (B.20) и (B.2), (В.10), задающих феноменологически
симметричные функциональные связи между измеряемыми в опыте
величинами: для любых двух материальных тел $i,j$ и любых двух
ускорителей $\alpha,\beta$ четыре значения ускорения $a$ связаны
уравнением (B.18); для любых трех проводников $i,j,k$ и любых двух
источников тока $\alpha,\beta$ шесть значений тока $I$ связаны
уравнением (В.20); для любых четырех точек $i,j,k,l$ евклидовой
плоскости шесть значений квадрата расстояния $f=\rho^2$ между ними
связаны уравнением (B.2); для любых трех состояний $i,j,k$
термодинамической системы шесть количеств тепла
$Q=(Q^{TS},Q^{ST})$ связаны уравнением (В.10).

В каждом из четырех рассмотренных примеров мы имеем дело с {\it
функцией пары точек}, определяющей в некотором обобщенном смысле
расстояние между ними, то есть с {\it метрической функцией}: в
уравнении (B.18) ускорение $a_{i\alpha}$ из второго закона Ньютона
(B.17) есть такое расстояние между телом $i$ и ускорителем
$\alpha$, которые являются точками (элементами) физически
различных множеств -- множества материальных тел и множества
ускорителей; аналогично, в уравнении (В.20) ток $I_{i\alpha}$ из
закона Ома (В.19) есть расстояние между проводником $i$ и
источником тока $\alpha$, которые являются точками (элементами)
физически различных множеств -- множества проводников и множества
источников тока; в уравнении (В.2) метрическая функция $f$
сопоставляет, согласно формуле (В.1), паре точек \ $i$ \ и \ $j$ \
евклидовой плоскости число

\newpage

      \noindent
     internal resistances
     $r_\alpha$ and $r_\beta$ of the current sources $\alpha$ and $\beta$,
     which yields the functional relation among the currents
     defined by the following equation:
     \phantom{aaaaa} \\\
     \phantom{aaaaa}
     $$
     \left|
     \begin{array}{ccc}
     I_{i\alpha} & I_{i\beta} & I_{i\alpha}I_{i\beta} \\
     I_{j\alpha} & I_{j\beta} & I_{j\alpha}I_{j\beta} \\
     I_{k\alpha} & I_{k\beta} & I_{k\alpha}I_{k\beta}
     \end{array}
     \right| =0.
     \eqno(\text{B.20})
     $$
     \phantom{aaaaa} \\
     \phantom{aaaaa} \\

     The functional relation (B.20), true to any three
     conductors $i,j,k$ and any two sources of current $\alpha,\beta$,
     is, under Kulakov terminology [9], Ohm's law in the phenomenologically
     symmetric form.

      Concerning the equations (B.18), (B.20) and (B.2), (В.10) that define the
     phenomenologically symmetric  functional relations among the magnitudes
     measured by experiment, what attracts attention is their principal
     generality: for any two material bodies $i,j$ and any two
     accelerators $\alpha,\beta$ the four values of acceleration $a$ are tied by
     the equation (B.18); for any three conductors $i,j,k$ and any two
     current sources $\alpha,\beta$ the six values of current $I$ are connected
     by the equation (В.20); for any four points  $i,j,k,l$ of Euclidean plane
     the six values of the squared distances $f=\rho^2$ among them
     are tied by the equation (B.2); for any three states $i,j,k$
     of a thermodynamic system the six quantities of heat
     $Q=(Q^{TS},Q^{ST})$ are tied by the equation (В.10).

     In each of the four examples that we have considered we deal with
     a {\it function of a pair of points} that defines, in some general sense, the
     distance between them, i.e. we deal with a {\it metric function}: in the
     equation (B.18) the acceleration $a_{i\alpha}$ from Newton's 2nd law (B.17)
     is such a distance between the body $i$ and the accelerator $\alpha$ that are
     points (elements) of physically different sets - a set of bodies and a set
     of accelerators; in the equation (В.20), similarly, the current $I_{i\alpha}$
     from Ohm's law (В.19) is the distance between the conductor $i$ and
     the current source $\alpha$ that are, in their turn, points (elements) of
     physically different sets - a set of conductors and a set of sources of current;
     in the equation (В.2) the metric function $f$  assigns, according to the formula
     (В.1), to a pair of points $i$ and $j$ of an Euclidean

\newpage

\noindent
$f(ij)$, \ равное квадрату обычного расстояния $\rho(ij)$ между
ними, а в уравнениях (В.10) тепловая функция $Q$ сопоставляет паре
состояний $i$ и $j$, являющихся точками соответствующей плоскости
термодинамических состояний, два количества тепла $Q^{TS}(ij)$,
$Q^{ST}(ij)$, определяемые выражениями (В.9), которые естественно
рассматривать как два расстояния между ними.

Согласно определению Ю.И.Кулакова [9] функция ускорения (В.17)
на множестве материальных
тел и множестве ускорителей задает {\it физическую структуру ранга} (2,2),
а функция тока (В.19)
на множестве проводников и множестве источников тока задает {\it физическую
структуру ранга} (3,2). Эти физические структуры представляют собой
своеобразные геометрии двух множеств, феноменологическая симметрия которых
выражается уравнениями (B.18) и (B.20) соответственно.
Аналогично, метрическая функция (B.1)
задает на плоскости {\it физическую структуру ранга} 4, то есть
геометрию обычной евклидовой плоскости, феноменологическая симметрия которой
выражается уравнением (B.2). И наконец,
тепловая функция (В.9) задает на плоскости
термодинамических состояний {\it физическую структуру ранга} 3
как двумерную двуметрическую геометрию, феноменологическая симметрия которой
выражается уравнениями (В.10).

На примерах евклидовой плоскости и плоскости термодинамических состояний,
задаваемых метрическими функциями (B.1) и (В.9), мы
убедились в том, что их групповая симметрия, определяемая множеством всех
движений, и их феноменологическая симметрия эквивалентны друг другу.
Естественно предположить, что аналогичная ситуация имеет место
и в геометрии двух множеств -- физической структуре. Под движением в
этой геометрии будем понимать совокупность двух одновременных
преобразований каждого из множеств, сохраняющих обобщенное расстояние между
точками любой пары, для которой оно определено.

Преобразования
$$
m'=\lambda(m),\ \ F'=\sigma(F)
\eqno(B.21)
$$
множества материальных тел и множества ускорителей составляют
движение, если они сохраняют функцию ускорения $a=F/m$, определяе-

\newpage

     \noindent
     plane a number $f(ij)$
     equal to the squared distance of the ordinary distance $\rho(ij)$
     between
     them, and in the equations (В.10), the thermal function $Q$ assigns to a pair
     of states $i$ and $j$, which are points of the corresponding plane
     of thermodynamic states, two quantities of heat, $Q^{TS}(ij)$ and $Q^{ST}(ij)$,
     defined by the expressions (В.9), and it is natural to consider those expressi- \ ons
     as two distances among them.

     According to Yu.I. Kulakov's definition in [9] the function of acceleration (В.17)
     on a set of bodies and a set of accelerators gives a {\it physical structure
     of rank} (2,2), and the function of current (В.19) on a
     set of conductors and a set of current sources gives a {\it physical
     structure of rank} (3,2). These physical structures are essentially some
     peculiar geometries of two sets whose phenomenological symmetry is
     expressed by the equations (B.18) and (B.20) respectively. Similarly, the
     metric function (B.1) gives on a plane a {\it physical structure of rank} 4,
     i.e. a geometry of an ordinary Euclidean plane whose phenomenological
     symmetry is expressed by the equation (B.2). And at last, the thermal
     function (В.9) gives on a plane of thermodynamical states a {\it physical
     structure of rank} 3 as a  two-dimensional dimetric geometry whose
     phenomenological symmetry is expressed by the equations (В.10).

     The examples of an Euclidean plane and of a plane of thermodynamic
     states, defined by the metric functions (B.1) and (В.9), demonstrate
     that their group symmetry, defined by the set of all the motions,
     and their phenomenological symmetry are equivalent. It
     is natural to suppose that a similar situation takes place
     in a geometry of two sets - physical structure. Under the term 'motion'
     in that geometry we shall understand a unity of two simultaneous
     transformations of each set preserving the generalized distance between
     the points of any pair for which it has been determined.

     The transformations \\
     $$
    m'=\lambda(m),\ \ F'=\sigma(F)
     \eqno(B.21)
     $$ \\
     of the set of bodies and the set of accelerators comprise a motion if
     they preserve the function of acceleration \ $a=F/m$, \ determined by
     Newton's

\newpage

\noindent
мую вторым законом Ньютона (В.17):
$$
\frac{\sigma(F)}{\lambda(m)}=\frac{F}{m}.
$$

Относительно функций $\lambda$ и $\sigma$
получено простое функциональное уравнение, решение которого
находится методом разделения переменных:
$$
\lambda(m)=cm,\ \ \sigma(F)=cF,
\eqno(B.22)
$$
где $c\neq0$ -- произвольная постоянная. Множество всех преобразований (В.21)
с функциями (В.22) является однопараметрической группой движений,
определяющей групповую симметрию физической структуры ранга (2,2)
как феноменологически симметричной геометрии, задаваемой
на множестве материальных тел и множестве ускорителей функцией ускорения.

Если группа преобразований (В.21) известна, то функция ускорения
$a=a(m,F)$ может быть найдена решением другого функционального уравнения
$$
a(m',F') = a(m,F),
$$
как ее двухточечный инвариант, который определяет закон Ньютона (B.17)
с точностью до масштабного преобразования:
$$
a = \chi(F/m),
$$
где $\chi$ -- функция одной переменной. Ее содержательный смысл
состоит в возможности выбора шкалы
акселерометра -- прибора, измеряющего ускорение. Ясно, что физический смысл
второго закона Ньютона, его феноменологическая и групповая симметрии не
зависят от такого выбора.

Найдем преобразования множества проводников и множества источников
тока
$$
R' = \lambda(R),\ \
\mathcal{E}' = \sigma(\mathcal{E},r),\ \ r' = \rho(\mathcal{E},r),
\eqno(\text{B}.23)
$$
сохраняющих функцию тока $I=\mathcal{E}/(R+r)$, определяемую
законом Ома (В.19):
$$
\frac{\sigma(\mathcal{E},r)}{\lambda(R)+
\rho(\mathcal{E},r)} =
\frac{\mathcal{E}}{R+r}.
$$

\newpage

     \noindent
     2nd law (В.17):
     $$
     \frac{\sigma(F)}{\lambda(m)}=\frac{F}{m}.
     $$

     Concerning the functions $\lambda$ and $\sigma$, a simple functional
     equation has been obtained that is solved by the method of separating
     of variables:
     $$
     \lambda(m)=cm,\ \ \sigma(F)=cF,
     \eqno(B.22)
     $$
     where $c\neq0$ is an arbitrary constant. The set of all the transformations
      (В.21) with the functions (В.22) is a one-parameter group of motions
     that defines the group symmetry of the physical structure of rank (2,2)
     as of a phenomenologically symmetric geometry given on a set of bodies
     and a set of accelerators by a function of acceleration.

     If the group of transformations (В.21) is known the function of acceleration
     $a=a(m,F)$ may be found by way of solving another functional equation
     $$
     a(m',F') = a(m,F),
     $$
     as its two-point invariant, that determines Newton's 2nd law (B.17)
     with an accuracy up to a scaling transformation:
     $$
     a = \chi(F/m),
     $$
     where $\chi$ is a function of one variable. Its essential meaning is in
     the possibility of choosing the scale of the accelerometer - the device that
     measures \\ acceleration. It is clear that the physical meaning
     of Newton's 2nd law and its phenomenological and group symmetries do not
     depend on any such choice.

     Let us find the transformations of the set of conductors and the set of
     current sources:
     $$
     R' = \lambda(R),\ \
     \mathcal{E}' = \sigma(\mathcal{E},r),\ \ r' = \rho(\mathcal{E},r),
     \eqno(\text{B}.23)
     $$
     that preserve the function of current $I=\mathcal{E}/(R+r)$ determined by
     Ohm's law (В.19): \\
     $$
     \frac{\sigma(\mathcal{E},r)}{\lambda(R)+
     \rho(\mathcal{E},r)} =
     \frac{\mathcal{E}}{R+r}.
     $$

\newpage

Относительно функций $\lambda$, $\sigma$, $\rho$ получено
функциональное уравнение, решение которого находится методом
дифференцирования по независимым переменным $R, \mathcal{E}, r$ и
последующего их разделения:
$$
\lambda(R) = aR + b,\
\sigma(\mathcal{E},r) = a\mathcal{E},\
\rho(\mathcal{E},r) = ar - b,
\eqno(\text{B}.24)
$$
где $a\neq0$ и $b$ -- произвольные постоянные.

Преобразования (В.23) с функциями (В.24) составляют
двухпараметрическую группу движений, которая определяет групповую
симметрию физической структуры ранга (3,2) как феноменологически
симметричной геометрии, задаваемой на множестве проводников и
множестве источников тока функцией тока. С другой стороны, по
известной группе преобразований (B.23) можно, решая функциональное
уравнение
$$
I(R',\mathcal{E}',r') = I(R,\mathcal{E},r),
$$
найти функцию тока $I=I(R,\mathcal{E},r)$ как ее двухточечный
инвариант:
$$
I = \chi(\frac{\mathcal{E}}{R+r}),
$$
где $\chi$ -- функция одной переменной. Таким образом, закон Ома
(В.19) восстанавливается с точностью до масштабного преобразования,
которое не меняет физический смысл закона, его феноменологическую и
групповую симметрии, отражая возможность выбора шкалы
амперметра -- прибора для измерения тока.

Подводя итог вышеизложенному, приходим к выводу, что
для введения геометрии двух множеств
имеются не только физические, но и математические предпосылки.
Дело в том, что метрическая функция такой геометрии допускает группу движений,
которая ее однозначно определяет. Таким образом, "Эрлангенская программа" \ Ф.
Клейна (1872) с обычной геометрии на одном множестве естественно переносится
на геометрию двух множеств, причем групповая и феноменологическая симметрии
оказываются эквивалентными в каждой из них.

\newpage

    With respect to the functions $\lambda$, $\sigma$, $\rho$ \
     a functional equation has been obtained whose solution is
     found by the method of differentiating in indepen- \ dent variables
     $R, \mathcal{E}, r$ and separating them:
     $$
     \lambda(R) = aR + b,\
     \sigma(\mathcal{E},r) = a\mathcal{E},\
     \rho(\mathcal{E},r) = ar - b,
     \eqno(\text{B}.24)
     $$
     where $a\neq0$ and $b$ are arbitrary constants.

     The transformations (В.23) with the functions (В.24) comprise a
     two-parameter group of motions that defines the group
     symmetry of the physical structure of rank (3,2) as a phenomenologically
     symmetric geometry that a function of current gives on a set of conductors
     and a set of sources of current. On the other hand, we may, knowing the
     group of transformations (B.23), by way of solving the functional equation

     $$
     I(R',\mathcal{E}',r') = I(R,\mathcal{E},r),
     $$ \\
     find the function of current $I=I(R,\mathcal{E},r)$ as its two-point
     invariant:

     $$
     I = \chi(\frac{\mathcal{E}}{R+r}),
     $$ \\
     where $\chi$ is a function of one variable. Thus, Ohm's law
     (В.19) is reconstructed with an accuracy up to a scaling transformation,
     that does not alter the physical meaning of the law, its phenomenological
     or group symmetries, and so it becomes possible to choose the scale
     of the ammeter.

     Summing up the above said we arrive at the conclusion that
     there exist not only physical but also mathematical prerequisites
     for introducing a geometry of two sets. The thing is that the
     metric function of such a geometry allows a group of motions
     that defines it uniquely. Thus, the "Erlangen programme" \ of 1872 of F.
     Klein is translated from the ordinary geometry on one set over onto
     the geometry of two sets, the group and phenomenological
     symmetries turning out to be equivalent in each.

\newpage

\begin{center}
{\bf \Large ГЛАВА I \\
\vspace{5mm}
 Геометрия как физическая структура \\
на одном множестве}
\end{center}

\vspace{25mm}

\begin{center}
{\bf \large \S1. Феноменологическая и групповая симметрии \\ в геометрии}
\end{center}

Пусть имеется множество $\mathfrak{M}$, являющееся $sn$-мерным многообразием,
где $s$ и $n$ -- натуральные числа, точки которого обозначим
строчными латинскими буквами, а также функция $f:\mathfrak{S}_{f} \rightarrow
R^{s}$, где $\mathfrak{S}_{f} \subseteq \mathfrak{M} \times \mathfrak{M}$,
сопоставляющая каждой паре $\langle  ij  \rangle$ $\in$ $\mathfrak{S}_{f}$ некоторую совокупность
$s$ вещественных чисел $f(ij) = (f^{1}(ij),\dotsc,f^{s}(ij)) \in R^{s}$.
Двухточечную функцию $f = (f^{1},\dotsc,f^{s})$ будем называть
{\it метрической},
не требуя, однако, положительной определенности ее  $s$ компонент и
выполнения для каждой из них аксиом обычной метрики. Заметим, что
в общем случае $\mathfrak{S}_{f} \subseteq \mathfrak{M} \times \mathfrak{M}$,
то есть, возможно, функция $f$ не всякой паре из
$\mathfrak{M} \times \mathfrak{M}$ сопоставляет $s$ чисел, но в
последующем изложении удобно в явной записи $f(ij)$
подразумевать, что $\langle ij \rangle$ $\in$ $\mathfrak{S}_{f}$. Обозначим через
$U(i)$ окрестность точки $i \in \mathfrak{M}$, через
$U(\langle ij \rangle)$ -- окрестность пары $\langle ij \rangle$ $\in$ $\mathfrak{M} \times
\mathfrak{M}$ и аналогично
окрестности кортежей из других прямых произведений множества
$\mathfrak{M}$ на себя.

Для некоторого кортежа $\langle k_{1}\dotsc k_{n} \rangle$ $\in$ $\mathfrak{M}^{n}$
длины $n$ введем функции \ $\bar{f^{n}} = \bar{f} [k_{1}\dotsc k_{n}]$ \ и
\ $\bar{\bar{f^{n}}} = \bar{\bar{f [}}k_{1}\dotsc k_{n}]$, \ сопоставляя
\ точке
\ $i \in \mathfrak{M}$ точки \ $(f(ik_{1}),\dotsc,$ $f(ik_{n})) \in
R^{sn}$ и $(f(k_{1}i),\dotsc,$ $f(k_{n}i)) \in R^{sn}$ соответственно,
если $\langle ik_{1} \rangle,\dotsc$, $\langle ik_{n} \rangle$ $\in$
$\mathfrak{S}_f$ и $\langle k_{1}i \rangle,\dotsc,\langle k_{n}i \rangle$ $\in$ $\mathfrak{S}_f$.
Заметим, что области определения введенных функций $\bar{f^{n}}$ и
$\bar{\bar{f^{n}}}$ могут не совпадать друг с другом и с самим
множеством $\mathfrak{M}$.

В отношении пространства $\mathfrak{M}$ с $s$-компонентной  метрической
функцией
$f = (f^{1},\dotsc,f^{s})$ будем предполагать выполнение следующих трех
аксиом:

\newpage

     \begin{center}
     {\bf \Large CHAPTER I \\
     \vspace{5mm}
     Geometry as a physical structure \\
     on one set }
     \end{center}

\vspace{25mm}

     \begin{center}
     {\bf \large \S1. Phenomenological and group symmetries \\ in geometry }
     \end{center}

     Let there be a set $\mathfrak{M}$ that is an $sn$-dimensional manifold,
     where $s$ and $n$ are natural numbers, whose points we shall designate with Latin
     lower-case letters, and a function  $f:\mathfrak{S}_{f} \rightarrow
     R^{s}$, where $\mathfrak{S}_{f} \subseteq \mathfrak{M} \times \mathfrak{M}$,
     that assigns to each pair $\langle  ij  \rangle$ $\in$ $\mathfrak{S}_{f}$ some real continuum
     of $s$ real numbers $f(ij) = (f^{1}(ij),\dotsc,f^{s}(ij)) \in R^{s}$.
     We shall call the two-point function $f = (f^{1},\dotsc,f^{s})$ a
     {\it metric} one, without demanding, however, that there should
     exist any positive definiteness of its  $s$ components or that the axioms
     of the ordinary metrics should be satisfied. We shall note that in most
     general case $\mathfrak{S}_{f} \subseteq \mathfrak{M} \times \mathfrak{M}$,
     i.e., the function $f$ does not possibly assign to each pair from
     $\mathfrak{M} \times \mathfrak{M}$ $s$ numbers, but in further discussion
     it will be convenient to understand in the explicit writing of $f(ij)$
     that $\langle ij \rangle$ $\in$ $\mathfrak{S}_{f}$. Let us designate by $U(i)$ the
     neighbourhood of the point $i \in \mathfrak{M}$, by $U(\langle ij \rangle)$ the
     neighbourhood of the pair $\langle ij \rangle$ $\in$ $\mathfrak{M} \times
     \mathfrak{M}$, and in the similar way the neighbourhoods of the
     corteges of other direct products of the set $\mathfrak{M}$ by itself.

     For some cortege $\langle k_{1}\dotsc k_{n} \rangle$ $\in$ $\mathfrak{M}^{n}$
     of length $n$, let us introduce functions \ $\bar{f^{n}} = \bar{f} [k_{1}\dotsc k_{n}]$ \
     and \ $\bar{\bar{f^{n}}} = \bar{\bar{f [}}k_{1}\dotsc k_{n}]$, \ by assigning to the
     point \ $i \in \mathfrak{M}$ the points \ $(f(ik_{1}),\dotsc,$ $f(ik_{n})) \in
     R^{sn}$ and $(f(k_{1}i),\dotsc,$ $f(k_{n}i)) \in R^{sn}$ respectively,
     if $\langle ik_{1} \rangle,\dotsc$, $\langle ik_{n} \rangle$ $\in$ $\mathfrak{S}_f$ and
     $\langle k_{1}i \rangle,\dotsc,\langle k_{n}i \rangle$ $\in$ $\mathfrak{S}_f$. We shall note that
     the domains of the functions $\bar{f^{n}}$ and $\bar{\bar{f^{n}}}$ introduced
     may not coincide with each other or with the set $\mathfrak{M}$ itself.

     We shall suppose in respect of the space $\mathfrak{M}$ with $s$-component
     metric function $f = (f^{1},\dotsc,f^{s})$ that three axioms hold as follows:

\newpage

{\bf I.} Область определения $\mathfrak{S}_{f}$ функции $f$
открыта и плотна в $\mathfrak{M} \times \mathfrak{M}$.

{\bf II.} Функция $f$  в области своего определения достаточно
гладкая.

{\bf III.} В $\mathfrak{M}^{n}$ плотно множество таких кортежей длины $n$,
для которых функция $\bar{f^{n}}(\bar{\bar{f^{n}}})$ имеет
максимальный ранг,
равный $sn$, в точках плотного в $\mathfrak{M}$ множества.

Достаточная гладкость означает, что в области ее определения
непрерывна как сама функция $f$, так и все ее производные
достаточно высокого порядка. Гладкую метрическую функцию
$f = (f^{1},\dotsc,f^{s})$,
для которой выполняется аксиома III, будем называть {\it невырожденной}.
Заметим, что ограничения в аксиомах I, II, III открытыми и плотными
множествами связано с тем, что исходные множества могут содержать
исключительные подмножества меньшей размерности, где эти аксиомы не
выполняются.

Пусть, далее, $m = n + 2$. На основе исходной метрической функции $f$ построим
функцию $F$, сопоставляя
кортежу   $\langle  ijk\ldots vw  \rangle$ длины $m$ из $\mathfrak{M}^{m}$ точку
$(f(ij),f(ik),\dotsc,f(vw)) \in R^{sm(m - 1)/2}$, координаты которой
в $R^{sm(m - 1)/2}$ определяются упорядоченной по исходному кортежу
последовательностью \ $sm(m - 1)/2$ расстояний для следующих пар его
точек: $\langle ij \rangle,\langle ik \rangle,\dotsc,\langle vw \rangle$, если все эти пары принадлежат
$\mathfrak{S}_{f}$. Область определения функции $F$ обозначим через
$\mathfrak{S}_{F}$. Очевидно, что область $\mathfrak{S}_{F}$
есть открытое и плотное в $\mathfrak{M}^{m}$ множество.

{\bf Определение 1.} Будем говорить, что функция $f = (f^{1},\dotsc,f^{s})$
задает на $sn$-мерном многообразии $\mathfrak{M}$
{\it феноменологически симметричную геометрию} (физическую структуру)
ранга $m = n + 2$, если, кроме аксиом I, II, III, дополнительно выполняется
следующая аксиома:

{\bf IV.} Существует плотное в $\mathfrak{S}_{F}$ множество, для
каждого кортежа \linebreak $\langle ijk\ldots vw \rangle$ длины $m = n + 2$
которого и некоторой его окрестности $U(\langle ijk\ldots vw \rangle)$ найдется
такая достаточно гладкая функция $\Phi: \mathcal{E} \rightarrow
R^{s}$, определенная в некоторой области $\mathcal{E} \subset
R^{sm(m - 1)/2}$, содержащей точку $F(\langle ijk\ldots vw \rangle)$, что в ней
$rang \ \Phi = s$ и множество $F(U(\langle ijk\ldots vw \rangle))$ является
подмножеством множества нулей функции $\Phi$, то есть

\newpage

     {\bf I.} The domain $\mathfrak{S}_{f}$ of the function $f$ is open and dense
     in $\mathfrak{M} \times \mathfrak{M}$.

     {\bf II.} The function $f$ is sufficiently smooth in its domain.

     {\bf III.} In $\mathfrak{M}^{n}$ a set is dense of such corteges of length $n$
     for which the rank of the function $\bar{f^{n}}(\bar{\bar{f^{n}}})$ is
     maximum, and is equal to $sn$, in the points of the set dense in
     $\mathfrak{M}$.

     Sufficient smoothness means that in the domain of the function $f$
     smooth are both the function $f$ and all the derivatives of it
     of sufficiently high order. We shall call the smooth metric function
     $f = (f^{1},\dotsc,f^{s})$ for which the axiom III is satisfied {\it nondegenerate}.
     We shall note that the constraint in the axioms I, II, and III by open and dense
     sets is introduced because the original sets may comprise exceptional
     subsets of smaller dimension where these axioms may fail to hold.

     Suppose, further, that $m = n + 2$. On the basis of the original metric
     function $f$, we shall construct a function $F$, by assigning to the cortege
     $\langle  ijk\ldots vw  \rangle$ of length $m$ from $\mathfrak{M}^{m}$ the point
     $(f(ij),f(ik),\dotsc,f(vw)) \in R^{sm(m - 1)/2}$ whose coordinates
     in $R^{sm(m - 1)/2}$ are determined by the sequence of \ $sm(m - 1)/2$
     distances for the pairs of points $\langle ij \rangle,\langle ik \rangle,\dotsc, \\ \langle vw \rangle$ of the original
     cortege ordered by that cortege, if these pairs all belong to
     $\mathfrak{S}_{f}$. We shall denote the domain of the function
     $F$ by $\mathfrak{S}_{F}$. The domain $\mathfrak{S}_{F}$ is obviously
     a set open and dense in $\mathfrak{M}^{m}$.

     {\bf Definition 1.} We shall say that the function $f = (f^{1},\dotsc,f^{s})$
     gives on an $sn$-dimensional manifold $\mathfrak{M}$ a {\it phenomenologically
     symmetric geometry} (physical structure) of rank $m = n + 2$, if,
     in addition to the axioms I, II, and III, the axiom is satisfied as follows:

     {\bf IV.} There exists a set dense in $\mathfrak{S}_{F}$ for each cortege \linebreak
     $\langle ijk\ldots vw \rangle$ of length $m = n + 2$ of which and some neighbourhood
     $U(\langle ijk\ldots \\ vw \rangle)$ of it if there exists a sufficiently smooth function
     $\Phi: \mathcal{E} \rightarrow R^{s}$ defined in some region
     $\mathcal{E} \subset R^{sm(m - 1)/2}$ that comprises the point
     $F(\langle ijk\ldots vw \rangle)$ such that $rang \ \Phi = s$ in it and the set
     $F(U(\langle ijk\ldots vw \rangle))$  is a subset of the set of zeros of the function
     $\Phi$, i.e.

\newpage

\begin{equation}
\label{a1}
\Phi(f(ij),f(ik),\dotsc,f(vw)) = 0
\tag{$1.1$}
\end{equation}
для всех кортежей из $U(\langle ijk\ldots vw \rangle)$.

Аксиома IV составляет содержание принципа феноменологической симметрии.
Эта аксиома выражает требование, чтобы \ $sm(m - 1)$/2 \ расстояний
между точками любого кортежа длины $m = n + 2$ из  $U(\langle ijk\ldots \\ vw \rangle)$
были функционально связаны, удовлетворяя системе $s$ уравнений
(1.1). Условие $rang \ \Phi = s$ означает, что уравнения
$\Phi = 0$ (то есть $\Phi_{1}=0,\dotsc,\Phi_{s}=0$) независимы.

Если $x = (x^{1},\dotsc,x^{sn})$ -- локальные координаты в
многообразии $\mathfrak{M}$, то для метрической функции
$f = (f^{1},\dotsc,f^{s})$
в некоторой окрестности $U(i) \times U(j)$ каждой пары
$\langle ij \rangle \ \in \mathfrak{S}_{f}$ можно выписать ее локальное
координатное представление
\begin{equation}
\label{a2}
f(ij)= f(x_i,x_j)=
f(x^{1}_i,\dotsc,x^{sn}_i,x^{1}_j,\dotsc,x^{sn}_j),
\tag{$1.2$}
\end{equation}
свойства которого определяются аксиомами II и III. Поскольку в
соответствии с аксиомой III ранги функций $\bar{f^{n}}$ и
$\bar{\bar{f^{n}}}$, равные $sn$, максимальны, координаты $x_i$ и
$x_j$ входят в представление (1.2) существенным образом.
Последнее означает, что никакая локально обратимая гладкая замена
координат не приведет к уменьшению их числа в этом представлении,
то есть не существует такой локальной системы
координат, в которой оно может быть записано в виде
\begin{displaymath}
f(ij) = f(x^{1}_i,\dotsc,x^{n'}_i,x^{1}_j,\dotsc,x^{n''}_j),
\end{displaymath}
где или $n' < sn$ или $n'' < sn$. Действительно, если,
например, $n' < sn$, то для любого кортежа $\langle j_{1}\ldots j_{n} \rangle$
$\in$ $(U(j))^n$ длины $n$ и для любой точки из $U(i)$ ранг функции
$\bar{f^{n}} = \bar{f}[j_{1}\ldots j_{n}]$ будет  заведомо меньше
$sn$, что противоречит аксиоме III. Заметим, однако, что существенная
зависимость представления (1.2) от локальных координат
$x_i$ и $x_j$ в общем случае не гарантирует выполнения аксиомы III.

Используя выражение (1.2), запишем локальное координатное
представление построенной выше функции $F$:

\newpage

     \begin{equation}
     \label{a1}
     \Phi(f(ij),f(ik),\dotsc,f(vw)) = 0
     \tag{$1.1$}
     \end{equation}
     for all the corteges of $U(\langle ijk\ldots vw \rangle)$.

     The Axiom IV gives the essence of the principle of phenomenological symmetry.
     That axiom expresses the requirement of \ $sm(m - 1)$/2 \ distances among
     the points of any cortege of length $m = n + 2$  from $U(\langle ijk\ldots vw \rangle)$ being
     functionally related, satisfying a system of $s$ equations (1.1). The condition
     $rang \ \Phi = s$ means that the equations $\Phi = 0$ (i.e.
     $\Phi_{1}=0,\dotsc,\Phi_{s}=0$) are independent.

     If $x = (x^{1},\dotsc,x^{sn})$ are local coordinates in the manifold $\mathfrak{M}$,
     then for the metric function $f = (f^{1},\dotsc,f^{s})$ in some neighbourhood
     $U(i) \times U(j)$ of each pair$\langle ij \rangle \ \in \mathfrak{S}_{f}$ it is possible to
     write its local coordinate representation
     \begin{equation}
     \label{a2}
     f(ij)= f(x_i,x_j)=
     f(x^{1}_i,\dotsc,x^{sn}_i,x^{1}_j,\dotsc,x^{sn}_j),
     \tag{$1.2$}
     \end{equation}
     whose properties are determined by the axioms II и III. Since, under the axiom
     III the ranks of the functions $\bar{f^{n}}$ and $\bar{\bar{f^{n}}}$, equal to $sn$,
     are maximum ones, the coordinates $x_i$ and $x_j$ are included in the
     representation (1.2) in an essential way. The latter implies that no locally
     invertible smooth change of coordinates will result in their number in the
     representation being decreased, i.e. there does not exist a local system of
     coordinates such in which it can be written as
     \begin{displaymath}
     f(ij) = f(x^{1}_i,\dotsc,x^{n'}_i,x^{1}_j,\dotsc,x^{n''}_j),
     \end{displaymath}
     where either $n' < sn$ or $n'' < sn$. Indeed, if for example $n' < sn$ then for
     any cortege $\langle j_{1}\ldots j_{n} \rangle$ \ $\in$ $(U(j))^n$ of length $n$ and for any point
     from $U(i)$ the rank of the function$\bar{f^{n}} = \bar{f}[j_{1}\ldots j_{n}]$ will a
     fortiori be less than$sn$, which is in contradiction with the axiom III. We shall
     note, however, that the essential dependence of the representation (1.2) on the
     local coordinates$x_i$ and $x_j$ in the general case does not guaranty that the
     axiom III be satisfied.

     Using the expression (1.2), we shall write a local coordinate
     representation of the function $F$ that we have constructed:

\newpage

\begin{equation}
\label{a3}
\left.\begin{array}{rcl}
f(ij) = f(x_i,x_j),\\
f(ik) = f(x_i,x_k),\\
\ldots \ldots \ldots \ldots \ldots \ldots \\
f(vw) = f(x_v,x_w)\;,
\end{array}\right\}
\tag{$1.3$}
\end{equation}
функциональная матрица для компонент которой
\begin{equation}
\label{a4}
\left|
\left|
\begin{array}[tcccccb]{cccccc}
\frac{\partial f(ij)}{\partial x_i} & \frac{\partial f(ij)}{\partial x_j}
& 0 & \ldots & 0 & 0\\
 & & & & & \\
\frac{\partial f(ik)}{\partial x_i} & 0 & \frac{\partial f(ik)}{\partial
x_k} & \ldots & 0 & 0\\
 & & & & & \\
 \ldots & \ldots & \ldots & \ldots & \ldots & \ldots \\
 & & & & & \\
0 & 0 & 0 & \ldots & \frac{\partial f(vw)}{\partial x_v} &
\frac{\partial f(vw)}{\partial x_w}\\
\end{array}
\right|
\right|
\tag{$1.4$}
\end{equation}
имеет $sm(m - 1)/2$ строк и $smn$ столбцов. Здесь через
$\partial f / \partial x$ кратко обозначена функциональная матрица
для $s$ компонент метрической функции $f = (f^{1},\dotsc,f^{s})$
по координатам $x = (x^{1},\dotsc,x^{sn})$:
\begin{equation}
\label{a5}
\frac{\partial f}{\partial x} =
\left|
\left|
\begin{array}{ccc}
\frac{\partial f^{1}}{\partial x^{1}} & \ldots &
\frac{\partial f^{1}}{\partial x^{sn}} \\
 & & \\
\ldots & \ldots & \ldots \\
 & & \\
\frac{\partial f^{s}}{\partial x^{1}} & \ldots &
\frac{\partial f^{s}}{\partial x^{sn}} \\
\end{array}
\right|
\right|.
\tag{$1.5$}
\end{equation}

Представление (1.3) задается системой $sm(m - 1)/2$ функций
$f(ij)$, $f(ik),\ldots,$ \ $f(vw)$, зависящих специальным образом
от $smn$ координат $x^{1}_i,\dotsc,x^{sn}_w$ всех точек кортежа
$\langle ijk\ldots vw \rangle$ длины $m = n + 2$. Поскольку общее число
компонент метрической функции $f$ в системе (1.3) не больше общего
числа координат, наличие связи (1.1) является нетривиальным
фактом, не имеющим места для произвольной системы функций (1.3).

Функция $F$, согласно ее локальному координатному представлению
(1.3), отображает окрестность $U(\langle ijk\ldots vw \rangle)$
$\subset \mathfrak{S}_{F}$ в $R^{sm(m - 1)/2}$.
Матрицей этого отображения
является функциональная матрица (1.4) системы функций (1.3),
а его рангом называется ранг этой матрицы.

\newpage

     \begin{equation}
     \label{a3}
     \left.\begin{array}{rcl}
     f(ij) = f(x_i,x_j),\\
     f(ik) = f(x_i,x_k),\\
     \ldots \ldots \ldots \ldots \ldots \ldots \\
     f(vw) = f(x_v,x_w)\;,
     \end{array}\right\}
     \tag{$1.3$}
     \end{equation}
     the functional matrix for the components of which
     \begin{equation}
     \label{a4}
     \left|
     \left|
     \begin{array}[tcccccb]{cccccc}
     \frac{\partial f(ij)}{\partial x_i} & \frac{\partial f(ij)}{\partial x_j}
     & 0 & \ldots & 0 & 0\\
      & & & & & \\
     \frac{\partial f(ik)}{\partial x_i} & 0 & \frac{\partial f(ik)}{\partial
     x_k} & \ldots & 0 & 0\\
      & & & & & \\
      \ldots & \ldots & \ldots & \ldots & \ldots & \ldots \\
      & & & & & \\
     0 & 0 & 0 & \ldots & \frac{\partial f(vw)}{\partial x_v} &
     \frac{\partial f(vw)}{\partial x_w}\\
     \end{array}
     \right|
     \right|
     \tag{$1.4$}
     \end{equation}
     has $sm(m - 1)/2$ rows and $smn$ columns. Here, it is the functional matrix
     for $s$ components of the metric function $f = (f^{1},\dotsc,f^{s})$
     in the coordinates $x = (x^{1},\dotsc,x^{sn})$ that is briefly designated by
     $\partial f / \partial x$:
     \begin{equation}
     \label{a5}
     \frac{\partial f}{\partial x} =
     \left|
     \left|
     \begin{array}{ccc}
     \frac{\partial f^{1}}{\partial x^{1}} & \ldots &
     \frac{\partial f^{1}}{\partial x^{sn}} \\
      & & \\
     \ldots & \ldots & \ldots \\
      & & \\
     \frac{\partial f^{s}}{\partial x^{1}} & \ldots &
     \frac{\partial f^{s}}{\partial x^{sn}} \\
     \end{array}
     \right|
     \right|.
     \tag{$1.5$}
     \end{equation}

     The representation (1.3) is defined by a system of $sm(m - 1)/2$
     functions $f(ij)$, $f(ik),\ldots,$ \ $f(vw)$ that depend in a special manner
     on $smn$ coordinates $x^{1}_i,\dotsc,x^{sn}_w$ of all the points of the
     cortege $\langle ijk\ldots vw \rangle$ of length $m = n + 2$. Since the total number of the
     components of the metric function $f$ in the system (1.3) is not bigger than
     the total number of the coordinates, the presence of the relation (1.1) is a
     nontrivial fact that will not take place in case of an arbitrary system of
     functions (1.3).

     The function $F$, according to the local coordinate representation
     (1.3), maps the neighbourhood $U(\langle ijk\ldots vw \rangle)$
     $\subset \mathfrak{S}_{F}$ в $R^{sm(m - 1)/2}$.
     The matrix of that mapping is the functional matrix (1.4) of the system
     of functions (1.3), and its rank is the rank of the matrix.

\newpage

\vspace{5mm} {\bf Теорема 1.} {\it Для того, чтобы метрическая
функция $f = (f^{1},\dotsc,f^{s})$ задавала на $sn$-мерном
многообразии $\mathfrak{M}$ феноменологически симметричную
геометрию (физическую структуру) ранга $m = n + 2$, необходимо и
достаточно, чтобы ранг отображения $F$ был равен $sm(m - 1)/2 - s$
на плотном в $\mathfrak{S}_{F}$ множестве.}

\vspace{5mm}
Полное доказательство теоремы 1 можно найти в монографии автора [10] и его
работе [11].

\vspace{5mm}

Рассмотрим теперь групповые свойства феноменологически
симметричной геометрии, введенной выше определением 1.

Пусть $U$ и $U'$ -- открытые области в многообразии $\mathfrak{M}$,
не обязательно связные. Гладкое инъективное отображение
\begin{equation}
\label{h1}
\lambda: U \rightarrow U'
\tag{$1.6$}
\end{equation}
называется локальным {\it движением}, если оно сохраняет метрическую функцию
$f = (f^{1},\dotsc,f^{s})$. Последнее означает, что для любой пары
$\langle ij \rangle  \ \in \mathfrak{S}_{f}$, такой что $i,j \in U$, и соответствующей
пары $\langle \lambda(i),\lambda(j) \rangle$, если она принадлежит $\mathfrak{S}_{f}$,
имеет место равенство
\begin{equation}
\label{h2}
f(\lambda(i),\lambda(j)) = f(ij),
\tag{$1.7$}
\end{equation}
выполняющееся  для каждой из компонент $f^{1},\dotsc,f^{s}$
метрической функции $f$.

Множество всех движений (1.6) есть локальная группа
преобразований, для которой метрическая функция согласно равенству
(1.7) является {\it двухточечным инвариантом}. Если метрическая
функция $f$ задана явно, например, в своем координатном
представлении (1.2), то равенство (1.7) является функциональным
уравнением, решая которое можно найти полную группу локальных
движений (1.6). Нам же о метрической функции известно только то,
что она невырождена и удовлетворяет некоторой системе $s$
уравнений (1.1). Но этого оказывается достаточно для того, чтобы
установить существование $sn(n + 1)/2$ -параметрической группы ее
движений.

Для большей ясности последующего изложения воспроизведемв наших
обозначениях определение локальной группы Ли преобразований,

\newpage

     \vspace{5mm}
     {\bf Theorem 1.}
     {\it For the metric function $f = (f^{1},\dotsc,f^{s})$ to give on an
     $sn$-dimensional manifold $\mathfrak{M}$ a phenomenologically symmetric
     geometry(physical structure) of rank $m = n + 2$, it is necessary and sufficient
     that the rank of the mapping $F$ be equal to $sm(m - 1)/2 - s$ on a set dense
     in $\mathfrak{S}_{F}$.}

     \vspace{5mm}
     The complete proof of Theorem 1 is in the author's monograph [10] and in his
     note [11].

     \vspace{5mm}

     Now we shall study the group properties of the
     phenomenologically \\
     symmetric geometry introduced by the definition 1 above.

     Let $U$ and $U'$ be open regions of the manifold $\mathfrak{M}$,
     that are not necessarily connected. A smooth injective mapping
     \vspace{5mm}
     \begin{equation}
     \label{h1}
     \lambda: U \rightarrow U'
     \tag{$1.6$}
     \end{equation} \\
     is called a local {\it motion} if it preserves the metric function
     $f = (f^{1},\dotsc,f^{s})$. The latter means that for any pair
     $\langle ij \rangle  \ \in \mathfrak{S}_{f}$, such that $i,j \in U$, and the corresponding
     pair $\langle \lambda(i),\lambda(j) \rangle$ if it belongs to $\mathfrak{S}_{f}$,
     the equality
    \vspace{5mm}
     \begin{equation}
     \label{h2}
     f(\lambda(i),\lambda(j)) = f(ij),
     \tag{$1.7$}
     \end{equation}
     \phantom{aaaaa} \\
     takes place that is satisfied for each of the components $f^{1},\dotsc,f^{s}$
     of the metric function $f$.

     The set of all the motions (1.6) is a local group of transformations for which the
     metric function, according to the equality (1.7) is a {\it two-point invariant}. If
     the metric function $f$ is defined explicitly, in its coordinate representation
     (1.2) for example, then the equality (1.7) is a functional equation, the solution
     of which gives the complete group of local motions (1.6). All we know about
     the metric function is that it is nondegenerate and satisfies some system of
     $s$ equations (1.1). But it turns out to be enough to establish the existence
     of the $sn(n + 1)/2$-parameter group of its motions.

      For the sake of making further discussion clearer, we shall reprouce in our
     designation the definition \ of \ the \ local Lie group \ of \ transformations,

\newpage

\noindent
 следуя монографии Л.С.Понтрягина "Непрерывные
группы" (см. [12], стр. 435). Пусть $G^{r}$ -- $r$-мерная
локальная группа Ли и $U$ -- некоторая область многообразия
$\mathfrak{M}$. Допустим, что каждому элементу $a \in G^{r}$
поставлено в соответствие непрерывно зависящее от $a$ инъективное
отображение $\lambda_{a}: U \rightarrow U'$ области $U$ в
некоторую область $U'$ многообразия $\mathfrak{M}$, относящее
каждой точке $i \in U$ некоторую точку $i' \in U'$, то есть $i' =
\lambda_{a}(i) = \lambda(i,a)$. Будем говорить, что $G^{r}$ есть
локальная группа Ли преобразований области $U$, если выполнены
следующие три условия:

1, Единице $e$ группы $G^{r}$ соответствует тождественное
преобразование $i' = \lambda(i,e)$ $ = i$ области $U$ на себя и
$\lambda(\lambda(i,a),b) = \lambda(i,ab)$, то есть произведению
$ab \in G^{r}$ соответствует композиция преобразований: сначала
$\lambda_{a}$ и затем $\lambda_{b}$ (возможен и другой порядок:
$\lambda(\lambda(i,a),b) = \lambda(i,ba))$.

2. Преобразование $\lambda_{a}$ является тождеством
лишь при условии, что $a$ есть единица $e$ группы $G^{r}$.

3. В координатной форме $\lambda(i,a)$ есть достаточное число раз
дифференцируемая функция точки $i \in U$ и элемента $a \in G^{r}$.

Определенная только что группа преобразований по условию 2 эффективна
и потому сами элементы группы $G^{r}$ могут считаться преобразованиями.
То есть можно говорить о $r$-мерной локальной группе преобразований
многообразия $\mathfrak{M}$, которую обозначим через $G^{r}(\lambda)$.
Таким образом, в области $U$ задано эффективное гладкое действие
группы $G^{r}$, причем условия 1, 2, 3 выполняются для некоторой ее
части, то есть некоторой, зависящей от $U$, окрестности единичного
элемента $e \in G^{r}$.

В последующем изложении удобно считать, что область
$U \subset \mathfrak{M}$ не обязательно связна, например, может
состоять из двух связных областей: $U = U_{1} \cup U_{2}$, причем
$U_{1} \cap U_{2} = \varnothing$.

{\bf Определение 2.} Будем говорить, что функция $f = (f^{1},\dotsc,f^{s})$
задает на $sn$-мерном многообразии $\mathfrak{M}$
{\it геометрию, наделенную групповой симметрией} степени $sn(n + 1)/2$,
если, кроме аксиом I, II, III, дополнительно выполняется следующая аксиома:

${\bf IV'.}$ Существует открытое и плотное в $\mathfrak{M}$
множество, для каждой точки $i$ которого задано эффективное
гладкое действие $sn(n + 1)/2$-мерной локальной группы Ли в
некоторой окрестности $U(i)$, такое, что действия ее в
окрестностях $U(i)$, $U(j)$ двух точек $i$, $j$ совпадают в пе-

\newpage

      \noindent
     following the monograph "Topological Groups"  \ by L.S. Pontriagin (see [12],
     P. 435). Let $G^{r}$ be an $r$-dimensional local Lie group and $U$ -- some
     region of the manifold $\mathfrak{M}$. Suppose, each element $a \in G^{r}$
     is assigned a continuous in $a$ injective mapping $\lambda_{a}: U \rightarrow U'$
     of the region $U$ into some region $U'$ of the manifold $\mathfrak{M}$,
     attributing to each point $i \in U$ some point $i' \in U'$,
     i.e. $i' = \lambda_{a}(i) = \lambda(i,a)$. We shall say that $G^{r}$ is the local Lie
     group of transformations of the region $U$ if three conditions hold as
     follows:

\vspace{1mm}

     1. A unit $e$ of the group $G^{r}$ is corresponded by an identical
     transformation $i' = \lambda(i,e)$ $ = i$ of the region $U$ on itself and
     $\lambda(\lambda(i,a),b) = \lambda(i,ab)$, i.e. the product
     $ab \in G^{r}$ is corresponded by a composition of transformations: first
     $\lambda_{a}$ and then $\lambda_{b}$ (another order is possible:
     $\lambda(\lambda(i,a),b) = \lambda(i,ba))$.

\vspace{1mm}

     2. The transformation $\lambda_{a}$ is only an identity on condition that
     $a$ is a unit $e$ of the group $G^{r}$.

\vspace{1mm}

     3. In the coordinate form $\lambda(i,a)$ there is a sufficiently differentiable
     function of the point $i \in U$ and the element $a \in G^{r}$.

\vspace{1mm}

     The group of transformations that we have just defined is, under condition 2, effective
     and so the elements of the group $G^{r}$ may themselves be considered transformations.
     That is, it is possible to speak of an $r$-dimensional local group of transformations
     of the manifold $\mathfrak{M}$ that we shall designate by $G^{r}(\lambda)$.
     Thus, there is an effective smooth action of the group $G^{r}$ defined in the
     region $U$, the conditions 1, 2, and 3 being satisfied for some part of it, i.e. for
     some, depending on $U$, neighbourhood of the unit element $e \in G^{r}$.

     In the further discussion it will be convenient to consider that the region
     $U \subset \mathfrak{M}$ is not necessarily connected, and may for
     example consist of two connected regions: $U = U_{1} \cup
     U_{2}$, \
     $U_{1} \cap U_{2}$ being equal to $\varnothing$.

     {\bf Definition 2.} We shall say that the function $f = (f^{1},\dotsc,f^{s})$
     gives on an $sn$-dimensional manifold $\mathfrak{M}$ a {\it geometry
     endowed with a group symmetry} of degree $sn(n + 1)/2$, if, in addition to
     the axioms I, II, and III, one more axiom holds as follows:

     ${\bf IV'.}$ There exists a set open and dense in $\mathfrak{M}$ for each
     point $i$ of which there is an effective smooth action of an
     $sn(n + 1)/2$-dimensional local Lie group defined in some neighbourhood
     $U(i)$, such that its actions in the neighbourhoods  $U(i)$, $U(j)$ of two points
     $i$, $j$ coincide \ in \ the \ intersection

\newpage

\noindent
 ресечении $U(i) \cap U(j)$ и что функция $f(ij)$ по каждой из
своих $s$ компонент является двухточечным инвариантом
соответствующей группы преобразований окрестности $U(i) \times
U(j)$.

Группа преобразований, о которой говорится в аксиоме $IV'$,
определяет локальную подвижность жестких фигур в $sn$-мерном
пространстве $\mathfrak{M}$, аналогичную подвижности твердых тел
в евклидовом пространстве. Заметим, что глобальной подвижности при
этом может и не быть, так как, хотя локальные действия группы
$G^{sn(n + 1)/2}$ определены согласно аксиоме $IV'$ в некоторой
окрестности каждой точки открытого и плотного в $\mathfrak{M}$
множества, может оказаться, что на всем этом множестве действует
только единичный элемент группы. Множество пар $\langle ij \rangle$, для которых
метрическая функция $f$ определена и является двухточечным инвариантом группы
локальных преобразований многообразия $\mathfrak{M}$,
очевидно, открыто и плотно в $\mathfrak{M} \times \mathfrak{M}$.
Будем также говорить, что метрическая функция {\it допускает}
$sn(n + 1)/2$-мерную локальную группу Ли локальных движений.

Из аксиомы $IV'$ следует также, что на открытом и плотном в
$\mathfrak{M}$ множестве задано $sn(n + 1)/2$-мерное линейное семейство
гладких векторных полей $X$, замкнутое относительно операции коммутирования,
то есть алгебра Ли преобразований (см. [12], \S60). В некоторой локальной
системе координат базисные векторные поля этого семейства запишем в
операторной форме:
\begin{equation}
\label{h3}
X _{\omega} = \lambda _{\omega} ^{\mu}(x) \partial / \partial x^{\mu},
\tag{$1.8$}
\end{equation}
где $\omega = 1,2,\dotsc,sn(n + 1)/2$, а по немому индексу $\mu$
производится суммирование в пределах от 1 до $sn$. Метрическая функция
$f = (f^{1},\dotsc,f^{s})$ будет двухточечным инвариантом
локальной группы Ли преобразований некоторых окрестностей
$U(i)$ и $U(j)$ точек $i$ и $j$ в том и только в том случае,
если она покомпонентно удовлетворяет системе $sn(n + 1)/2$ уравнений
\begin{equation}
\label{h4}
X_{\omega}(i)f(ij) + X_{\omega}(j)f(ij) = 0
\tag{$1.9$}
\end{equation}
с операторами (1.8):
\begin{displaymath}
\lambda^{\mu}_{\omega}(i) \partial f(ij)/ \partial x^{\mu}_i +
\lambda^{\mu}_{\omega}(j) \partial f(ij)/ \partial x^{\mu}_j = 0,
\end{displaymath}
где, например, $\lambda^{\mu}_{\omega}(i) = \lambda^{\mu}_{\omega}(x_i) =
\lambda^{\mu}_{\omega}(x^{1}_i,\dotsc,x^{sn}_i)$ (см. [13], стр. 229 и 237).

\newpage

      \noindent
     $U(i) \cap U(j)$ and the function $f(ij)$, with
     respect to each of its $s$ components is a two-point invariant of the
     corresponding group of transformations of the neighbourhood $U(i) \times U(j)$.

     The group of transformations mentioned in the axiom $IV'$
     defines the local mobility of rigid figures in an $sn$-dimensional
     space $\mathfrak{M}$, similar to the mobility of solid bodies
     in the Euclidean space. We shall note that that does not imply global mobility,
     for though the local actions of the group $G^{sn(n + 1)/2}$ are defined,
     according to axiom  $IV'$, in some neighbourhood of each point of a set
     open and dense in $\mathfrak{M}$, it may appear that there is only one
     single element of the group acting in the whole set. The set of pairs $\langle ij \rangle$
     for which the metric function $f$ is defined and is the two-point invariant of
     the group of local transformations of the manifold $\mathfrak{M}$ is,
     obviously, open and dense in $\mathfrak{M} \times \mathfrak{M}$.
     We shall also say that the metric function {\it allows} an $sn(n + 1)/2$-dimensional
     local Lie group of local motions. It also follows from the axiom $IV'$ that on a
     set open and dense in $\mathfrak{M}$ there is an $sn(n + 1)/2$-dimensional
     linear family of smooth vector fields $X$ defined that is commutation closed,
     i.e. an algebra of Lie transformations (see [12], \S60). In some local systems
     of coordinates, let us write basic vector fields in operator form:
    \begin{equation}
     \label{h3}
     X _{\omega} = \lambda _{\omega} ^{\mu}(x) \partial / \partial x^{\mu},
     \tag{$1.8$}
     \end{equation}
      where $\omega = 1,2,\dotsc,sn(n + 1)/2$, and by the dummy index $\mu$
     an operation of summation is performed within the range 1 to $sn$. The
     metric function $f = (f^{1},\dotsc,f^{s})$ will be the two-point invariant of
     the local Lie group of transformations of some neighbourhoods $U(i)$ and
     $U(j)$ of the points $i$ and $j$ if and only if it satisfies componentwise the
     system of $sn(n + 1)/2$ equations
  \vspace{5mm}
     \begin{equation}
     \label{h4}
     X_{\omega}(i)f(ij) + X_{\omega}(j)f(ij) = 0
     \tag{$1.9$}
     \end{equation}
     with the operators (1.8):
     \begin{displaymath}
     \lambda^{\mu}_{\omega}(i) \partial f(ij)/ \partial x^{\mu}_i +
     \lambda^{\mu}_{\omega}(j) \partial f(ij)/ \partial x^{\mu}_j = 0,
     \end{displaymath}
     where, for example, $\lambda^{\mu}_{\omega}(i) = \lambda^{\mu}_{\omega}(x_i) =
     \lambda^{\mu}_{\omega}(x^{1}_i,\dotsc,x^{sn}_i)$ (see [13], Pp. 229 and 237).

 \newpage

\vspace{5mm} {\bf Теорема 2.} {\it Для того, чтобы функция $f =
(f^{1},\dotsc,f^{s})$ задавала на $sn$-мерном многообразии
$\mathfrak{M}$ геометрию, наделенную групповой симметрией степени
$sn(n + 1)/2$, необходимо и достаточно, чтобы ранг отображения $F$
был равен $sm(m - 1)/2 - s$, где $m = n + 2$, на плотном в
$\mathfrak{S}_{F}$ множестве.} \vspace{5mm}

Полное доказательство теоремы 2, а также следующей ниже теоремы 4,
можно найти в монографии автора [10] и его работе [11].
\vspace {5mm}

Итоговым и очевидным результатом изложенного выше является вывод
об эквивалентности феноменологической и групповой симметрий
геометрии, задаваемой на $sn$-мерном многообразии $\mathfrak{M}$
функцией $f = (f^{1},\dotsc,f^{s})$. Эта эквивалентность является
следствием теорем 1 и 2 настоящего параграфа, необходимые и
достаточные условия которых о ранге отображения $F$ совпадают.

\vspace{5mm}
{\bf Теорема 3.}
{\it Для того, чтобы функция $f = (f^{1},\dotsc,f^{s})$ задавала на $sn$-мерном
многообразии $\mathfrak{M}$ феноменологически симметричную
геометрию (физическую структуру) ранга $m = n + 2$, необходимо и достаточно,
чтобы эта функция задавала на $\mathfrak{M}$
геометрию, наделенную групповой симметрией степени $sn(n + 1)/2$.}
\vspace{5mm}

Заметим, что условие о ранге отображения $F$ можно сформулировать
как четвертую аксиому в определении геометрии. Такая геометрия
будет, с одной стороны, феноменологически симметрична, а с другой
стороны -- наделена групповой симметрией, причем обе симметрии в
смысле теоремы 3 окажутся эквивалентными.

\vspace{5mm}
{\bf Теорема 4.}
{\it Размерность локальной группы локальных движений, допускаемых
\ метрической функцией $f = (f^{1},\dotsc,f^{s})$, \ задающей на \ $sn$-мерном
многообразии $\mathfrak{M}$ феноменологически симметричную геометрию
ранга $m = n + 2$, или геометрию, наделенную групповой симметрией
степени $sn(n + 1)/2$, не превышает этой степени.}
\vspace{5mm}

Таким образом, \ жесткие фигуры \ и \ твердые тела имеют не более

\newpage

    \vspace{5mm}
     {\bf Theorem 2.}
     {\it For the function $f = (f^{1},\dotsc,f^{s})$ to give on an $sn$- \\ dimensional
     manifold an $\mathfrak{M}$ geometry endowed with the group symmetry of degree
     $sn(n + 1)/2$ it is necessary and sufficient that the rank of the mapping $F$ be equal to
     $sm(m - 1)/2 - s$, where $m = n + 2$, on a set dense in $\mathfrak{S}_{F}$.}
     \vspace{5mm}

     The complete proof of theorem 2, as well as of theorem 4 below, is in the author's
     monograph [10], and in his note [11].
     \vspace {5mm}

     The final and obvious result of the above said is the conclusion of the
     phenomenological and group symmetries of the geometry defined on an
     $sn$-dimensional manifold $\mathfrak{M}$ by the function
     $f = (f^{1},\dotsc,f^{s})$ being equivalent. That equivalence is
     a corollary of Theorems 1 and 2 of this paragraph, the necessary and sufficient
     conditions of which concerning the rank of the mapping $F$ coincide.

     \vspace{5mm}
     {\bf Theorem 3.}
     {\it For the function $f = (f^{1},\dotsc,f^{s})$ to give on an $sn$- \\ dimensional
     manifold $\mathfrak{M}$ a phenomenologically symmetric
     geometry (physical structure) of rank $m = n + 2$ it is necessary and sufficient
     that that function give on $\mathfrak{M}$ a geometry endowed with a group
     symmetry of degree $sn(n + 1)/2$.}

     \vspace{5mm}

     We shall note that the condition concerning the rank of the mapping $F$ may be
     formulated as a fourth axiom of the definition of the geometry. Such a geometry
     will be phenomenologically symmetric, on the one hand, and will be endowed
     with a group symmetry, on the other, and both symmetries will be, in the sense
     of Theorem 3, equivalent.

     \vspace{5mm}
     {\bf Theorem 4.}
     {\it The dimension of the local group of motions allowed by the metric
     function $f = (f^{1},\dotsc,f^{s})$, giving on an $sn$-dimensional manifold
     $\mathfrak{M}$ a phenomenologically symmetric geometry of rank
     $m = n + 2$, or a geometry endowed with a group symmetry
     of degree $sn(n + 1)/2$, is not bigger than that degree.}
     \vspace{5mm}

     Thus, rigid \ figures and solid bodies have not more than \ $sn(n + 1)/2$

\newpage

\noindent
$sn(n + 1)/2$ степеней свободы при своем движении в пространстве.

\vspace{15mm}

\begin{center}
{\bf \large \S2. Классификация одномерных, двумерных \\ и трехмерных геометрий}
\end{center}

В настоящем параграфе будут приведены полные классификации
однометрических феноменологически симметричных геометрий, когда
однокомпонентная метрическая функция $f$ при $s=1$ паре точек сопоставляет
одно число. К настоящему времени такие классификации построены только
для одномерных, двумерных и трехмерных геометрий, то есть для $n=1,2,3$.
Классификация феноменологически симметричных геометрий более высокой
размерности в рамках разработанного метода наталкивается на серьезные
трудности чисто технического характера, которые еще не
преодолены. Возможно, что все эти трудности есть недостатки самого метода,
однако новые и более эффективные методы классификации еще не найдены.

Координатное представление метрической функции при переходе от
одной системы координат к другой меняется. Например, метрическая
функция плоскости Евклида в прямоугольной декартовой системе
координат $(x,y)$ задается выражением (В.1), а в полярной системе
координат $(r,\varphi)$ -- другим:
$$
f(ij)=r_i^2+r_j^2-2r_ir_j\cos(\varphi_i-\varphi_j).
$$

Дополнительное масштабное преобразование $\psi(f) \to f$, где
$\psi$ -- произвольная функция одной переменной, еще более изменит
исходное координатное представление (В.1), сделав его трудно
узнаваемым. Естественно выбрать такую систему координат в
многообразии $\mathfrak {M}$ и провести такое масштабное
преобразование самой метрической функции, при которых ее
координатное представление будет наиболее простым. Поэтому,
нижеследующие классификационные теоремы формулируются с точностью
до замены координат и масштабного преобразования.

\vspace {5mm}

Рассмотрим сначала по работе [5] простейшую одномерную $(s=1,\
n=1)$ \  феноменологически симметричную геометрию ранга 3. \ Та-

\newpage

\noindent
 degrees of freedom, in their motion in space.

     \vspace{15mm}

     \begin{center}
     {\bf \large \S2. Classification of one-, two- \\ and three-dimensional geometries}
     \end{center}

     In this paragraph, we shall give complete classifications of the unimetric
     phenomenologically symmetric geometries where a one-component metric
     function $f$ with $s=1$ assigns to a pair of points one number. By now, such
     classifications have only been built for the one-dimensional, two-dimensional,
     and three-dimensional geometries, i.e. for $n=1,2,3$. Using the method of
     classifying phenomenologically symmetric geometries that we have developed,
     when applying it to geometries of higher dimensionality, we encounter serious
     technical difficulties that we have not yet overcome. It is possible all those
     difficulties are the shortcomings of the method itself, but we have not yet
     discovered other and more effective methods.

     The coordinate representation of a metric function, while a transition from
     one system of coordinates to another performed, changes too. For instance,
     the metric function f the Euclidean plane in the Cartesian rectangu- \ lar coordinate
     system $(x,y)$ is defined by the expression (В.1), while in the polar coordinate
     system $(r,\varphi)$ by another:
     $$
     f(ij)=r_i^2+r_j^2-2r_ir_j\cos(\varphi_i-\varphi_j).
     $$

     An additional scaling transformation $\psi(f) \to f$, where  $\psi$ is an
     arbitrary function of one variable, will change the original coordinate
     representation (В.1) still more, to the point of it becoming hardly
     recognizable. It is most natural to choose such a system of coordinates in
     the manifold $\mathfrak {M}$ and perform such scaling transformation of
     the metric function itself that would make the coordinate representation of
     it as simple as possible.That is why the theorems of classification that follow
     are formulated with an accuracy up to change of coordinates and a scaling
     transformation.

     \vspace {5mm}

     Let us first consider, following note [5] the simplest $(s=1,n=1)$
     phenomenologically symmetric geometry of rank 3. Such geometry is
     defined

\newpage

\noindent
кая геометрия задается на одномерном многообразии $\mathfrak{M}$
метрической функцией $f:\mathfrak{S}_f\to R$, где
$\mathfrak{S}_f\subseteq\mathfrak{M}\times\mathfrak{M}$. Ее
координатное представление определяется выражением
\begin{equation}
\label{z1}
f(ij) = f(x_i,x_j),
\tag{$2.1$}
\end{equation}
где $x$ -- локальная координата в многообразии. Метрическая
функция (2.1) будет невырожденной, то есть удовлетворять аксиоме III из \S1,
при условии  отличия от нуля обеих производных по координатам $x_i$ и $x_j$
для плотного в $\mathfrak{M}\times\mathfrak{M}$
множества пар $\langle ij \rangle$. Уравнение, выражающее феноменологическую симметрию
рассматриваемой геометрии, согласно аксиоме IV из \S1 устанавливает
функциональную связь трех расстояний $f(ij), f(ik), f(jk)$ для
плотного в $\mathfrak{M}^3$ множества троек $\langle ijk \rangle$:

\begin{equation}
\label{z2}
\Phi(f(ij),f(ik),f(jk)) = 0.
\tag{$2.2$}
\end{equation}

\vspace{5mm}

{\bf Теорема 1.}  {\it С точностью до масштабного преобразования
$\psi(f)\to f$ и в надлежаще выбранной системе локальной координаты
$x$ невырожденная метрическая функция $(2.1)$, задающая на одномерном
многообразии феноменологически симметричную геометрию ранга три со связью
$(2.2)$, может быть представлена следующим
каноническим выражением:}

$$
f(ij)=x_i-x_j.
\eqno (2.3)
$$

\vspace {5mm}

Уравнение (2.2), выражающее феноменологическую симметрию этой
геометрии, легко находится: $f(ij)-f(ik)+f(jk) = 0$. Локальное
обратимое преобразование $x'=\lambda(x)$ одномерного многообразия
$\mathfrak {M}$ с отличной от нуля производной $\lambda'(x)$ будет
движением, если оно сохраняет метрическую функцию (2.3):
$\lambda(x_i)-\lambda(x_j)=x_i-x_j$. Полученное функциональное
уравнение на множество движений легко решается методом разделения
переменных: $\lambda(x)=x+a$, где $a$ -- произвольная постоянная.
Соответствующая однопараметрическая группа движений $x'=x+a$
определяет групповую симметрию степени 1 феноменологически
симметричной геометрии ранга 3, задаваемой на одномерном
многообразии

\newpage

     \noindent
     on a one-dimensional manifold $\mathfrak{M}$ by the
     metric function $f:\mathfrak{S}_f\to R$, where
     $\mathfrak{S}_f\subseteq\mathfrak{M}\times\mathfrak{M}$.
     Its coordinate representation is determined by the expression
     \begin{equation}
     \label{z1}
     f(ij) = f(x_i,x_j),
     \tag{$2.1$}
     \end{equation} \\
     where $x$ is a local coordinate in the manifold. The metric function (2.1)
     will be nondegenerate, i.e. satisfy the axiom III of \S1, on condition of
     both derivatives in the coordinates $x_i$ and $x_j$ being unequal to zero
     for the set of pairs $\langle ij \rangle$ dense in $\mathfrak{M}\times\mathfrak{M}$.
     The equation expressing the phenomenological symmetry  of the geometry
     in question, under the axiom IV of \S1, establishes a functional relation
     of the three distances $f(ij), f(ik)$, and $f(jk)$ for the set of triples
     $\langle ijk \rangle$ dense in $\mathfrak{M}^3$:

\vspace{5mm}

     \begin{equation}
     \label{z2}
     \Phi(f(ij),f(ik),f(jk)) = 0.
     \tag{$2.2$}
     \end{equation}

     \vspace{5mm}

     {\bf Theorem 1.}  {\it With an accuracy up to a scaling transformation
     $\psi(f)\to f$ and in a suitably chosen system of the local coordinate $x$ the
     nondegenerate metric function $(2.1)$ that defines on a one-dimensional
     manifold a \\ phenomenologically symmetric geometry of rank 3 with the relation
     $(2.2)$ may be represented with the following canonical expression:}

     $$
     f(ij)=x_i-x_j.
     \eqno (2.3)
     $$

\vspace{3mm}
     The equation (2.2) expressing the phenomenological symmetry of that
     geometry is readily found: $f(ij)-f(ik)+f(jk) = 0$. The local invertible
     transformation $x'=\lambda(x)$ of the one-dimensional manifold
     $\mathfrak {M}$ with the derivative $\lambda'(x)$ unequal to zero will
     be a motion if it preserves the metric function (2.3):
     $\lambda(x_i)-\lambda(x_j)=x_i-x_j$. The functional equation of the set of
     motions is easy to solve by the method of separating of variables:
     $\lambda(x)=x+a$, where $a$ is an arbitrary constant. The respective
     one-parameter group of motions \ $x'=x+a$ \ determines the group
     symmetry of degree 1 of the phenomenologically \ symmetric \
     geometry \
     of \ rank \ 3 \ defined \ on  \ a \ one- \\

\newpage

\noindent
 $\mathfrak{M}$ метрической функцией (2.3). В заключение отметим,
что двухточечный инвариант группы движений удовлетворяет
функциональному уравнению $f(x_i+a,x_j+a)=f(x_i,x_j)$. Это
уравнение решается методом его сведения к линейному однородному
дифференциальному уравнению в частных производных:
$f(ij)=\psi(x_i-x_j)$, где $\psi$ -- произвольная функция одной
переменной, откуда видно, что по группе движений метрическая
функция восстанавливается однозначно с точностью до масштабного
преобразования $\psi(f) \to f$.

\vspace{5mm}

Перейдем к рассмотрению двумерых $(s=1,n=2)$ феноменологически симметричных
геометрий, которые задаются на двумерном многообразии $\mathfrak{M}$
метрической функцией $f:\mathfrak{S}_f\to R$, где
$\mathfrak{S}_f\subseteq\mathfrak{M}\times\mathfrak{M}$.
Ее координатное представление определяется выражением
\begin{equation}
\label{z1}
f(ij) = f(x_i,y_i,x_j,y_j),
\tag{$2.4$}
\end{equation}
где $(x,y)$ -- локальные координаты в многообразии. Если эта
функция задает такую геометрию, то по аксиоме IV из \S1 шесть ее
значений для четверки $\langle  ijkl  \rangle$ функционально связаны:
\begin{equation}
\label{z2}
\Phi(f(ij),f(ik),f(il),f(jk),f(jl),f(kl)) = 0.
\tag{$2.5$}
\end{equation}

Невырожденная метрическая функция (2.4) по аксиоме III из \S1 должна,
очевидно, удовлетворять следующим двум условиям:
\begin{equation}
\label{z3}
\left.\begin{array}{rcl}
\partial(f(ik),f(il))/\partial(x_i,y_i) \neq 0,\\
\partial(f(kj),f(lj))/\partial(x_j,y_j)\neq 0\;
\end{array}\right\}
\tag{$2.6$}
\end{equation}
для открытого и плотного в $\mathfrak{M}^{3}$ множества троек
$\langle  ikl  \rangle$ и $\langle  klj  \rangle$.

Плоскость Евклида с метрической функцией (В.1) и функциональной связью (В.2),
выражающей ее феноменологическую симметрию, которая
была рассмотрена во Введении в качестве
примера, является одной из таких геометрий. Но сколько их может
быть? На этот вопрос отвечает следующая теорема (см. [6]):

\vspace{5mm} {\bf Теорема 2.}  {\it С точностью до масштабного
преобразования $\psi(f)\to f$ и в надлежаще выбранной системе
локальных координат $(x,y)$ невырожденная метрическая функция
$(2.4)$, задающая на двумерном мно-}

\newpage

\noindent
dimensional manifold
$\mathfrak{M}$ by
     the metric function (2.3). In conclusion, we shall note that the two-point
     invariant of the group of motions satisfies the functional equation
     $f(x_i+a,x_j+a)=f(x_i,x_j)$. That equation is solved by reducing it to a
     linear homogeneous differential  equation in partial derivatives:
     $f(ij)=\psi(x_i-x_j)$, where $\psi$ is an arbitrary function of one variable,
     wherefrom it can be seen that by the group of motions the metric function
     is reconstructed uniquely with an accuracy up to a scaling  \\ transformation
     $ \psi(f) \to f$.

     \vspace{5mm}

     Let us now proceed to the two-dimensional $(s=1,n=2)$ \\ phenomenologically
     symmetric geometries that are defined on a two- \\ dimensional manifold $\mathfrak{M}$
     by the metric function $f:\mathfrak{S}_f\to R$, where
     $\mathfrak{S}_f\subseteq\mathfrak{M}\times\mathfrak{M}$. Its coordinate
     representation is determined by the expression
     \begin{equation}
     \label{z1}
     f(ij) = f(x_i,y_i,x_j,y_j),
     \tag{$2.4$}
     \end{equation}
     where $(x,y)$ are local coordinates in the manifold. If that function really gives
     such a geometry then, under the axiom IV of \S1 the six values of it for the
     quadruple $\langle  ijkl  \rangle$ are functionally related:
     \begin{equation}
     \label{z2}
     \Phi(f(ij),f(ik),f(il),f(jk),f(jl),f(kl)) = 0.
     \tag{$2.5$}
     \end{equation}

     Obviously, the nondegenerate metric function (2.4) must, under the axiom
     III of \S1, satisfy the following two conditions:
     \begin{equation}
     \label{z3}
     \left.\begin{array}{rcl}
     \partial(f(ik),f(il))/\partial(x_i,y_i) \neq 0,\\
     \partial(f(kj),f(lj))/\partial(x_j,y_j)\neq 0\;
     \end{array}\right\}
     \tag{$2.6$}
     \end{equation}
     for the open and dense in $\mathfrak{M}^{3}$ set of triples
     $\langle  ikl  \rangle$ and $\langle  klj  \rangle$.

     The Euclidean plane with the metric function (В.1) and the functional relation (В.2)
     that expresses its phenomenological symmetry, that was \\ discussed in the
     Introduction by way of an example, is such a geometry. But how many of them
     are there? That question is answered by the theorem as follows (see [6]):

     \vspace{5mm}

     {\bf Theorem 2.}  {\it With an accuracy up to a scaling transformation $\psi(f)\to f$
     and in a suitably chosen system of local coordinates $(x,y)$ the nondegenerate metric
     function $(2.4)$ defining on a two-dimensional manifold a
     phenomeno-}

\newpage

\noindent {\it гообразии феноменологически симметричную геометрию
ранга четыре со связью $(2.5)$, может быть представлена одним из
следующих одиннадцати канонических выражений:}

\begin{equation}
\label{z4}
f(ij) = (x_i - x_j)^{2} + (y_i - y_j)^{2},
\tag{$2.7$}
\end{equation}

\begin{equation}
\label{z5}
f(ij) = \sin y_i \sin y_j\cos(x_i - x_j) + \cos y_i \cos y_j,
\tag{$2.8$}
\end{equation}

\begin{equation}
f(ij) = \textsl{sh}y_i \textsl{sh} y_j \cos(x_i - x_j) -
\textsl{ch} y_i \textsl{ch} y_j,
\tag{$2.9$}
\end{equation}

\begin{equation}
\label{z7}
f(ij) = (x_i - x_j)^{2} - (y_i - y_j)^{2},
\tag{$2.10$}
\end{equation}

\begin{equation}
\label{z8}
f(ij) = \textsl{ch}y_i\textsl{ch}y_j\cos(x_i-x_j)-\textsl{sh}y_i\textsl{sh}y_j,
\tag{$2.11$}
\end{equation}

\begin{equation}
\label{z9}
f(ij) = x_i y_j-x_jy_i,
\tag{$2.12$}
\end{equation}

\begin{equation}
\label{z10}
f(ij) = \frac{y_i - y_j}{x_i - x_j},
\tag{$2.13$}
\end{equation}

\begin{equation}
\label{z11}
\begin{array}{rcl}
f(ij) = ((x_i - x_j)^{2} - (y_i - y_j)^{2})
\exp \left(2 \beta \textsl{ar(c)th} \displaystyle
\frac{y_i - y_j}{x_i - x_j}\right),\
\end{array}
\tag{$2.14$}
\end{equation}

\begin{equation}
\label{z12}
f(ij) = (x_i - x_j)^{2}\exp\left(2\frac{y_i - y_j}{x_i - x_j}\right),
\tag{$2.15$}
\end{equation}

\begin{equation}
\label{z13}
f(ij) = ((x_i - x_j)^{2} + (y_i - y_j)^{2})
\exp \left(2 \gamma \textsl{arctg} \displaystyle
\frac{y_i - y_j}{x_i - x_j}\right),
\tag{$2.16$}
\end{equation}

\begin{equation}
\label{z14}
f(ij) = \frac{(x_i - x_j)^{2} + \varepsilon_{i}y^{2}_i
+ \varepsilon_{j}y^{2}_j}{y_iy_j},
\tag{$2.17$}
\end{equation}
{ \it где $\beta >0$ и $\beta \neq 1;$ $\gamma > 0;$
$\varepsilon_{i} = 0, \pm1;$
$\varepsilon_{j} = 0, \pm1$,
причем не обязательно
$\varepsilon_{i} = \varepsilon_{j}$.}
\vspace{5mm}

Щсть выражений (2.7)--(2.12) определяют метрические функции хорошо
известных двумерных геометрий: {\bf (2.7)} -- {\it плоскости
Евклида}; {\bf (2.8)} -- {\it двумерной \ сферы} \ в \ трехмерном
\ евклидовом \ пространстве;

 \newpage

     \noindent
      {\it  logically
     symmetric geometry of rank 4 with the relation$(2.5)$, may be represen- \ ted by one
     of the following eleven canonical expressions:}

     \begin{equation}
     \label{z4}
     f(ij) = (x_i - x_j)^{2} + (y_i - y_j)^{2},
     \tag{$2.7$}
     \end{equation}

     \begin{equation}
     \label{z5}
     f(ij) = \sin y_i \sin y_j \cos(x_i - x_j) + \cos y_i \cos y_j,
     \tag{$2.8$}
     \end{equation}

     \begin{equation}
     f(ij) = \sinh y_i\sinh y_j \cos(x_i - x_j) -
     \cosh y_i \cosh y_j,
     \tag{$2.9$}
     \end{equation}

     \begin{equation}
     \label{z7}
     f(ij) = (x_i - x_j)^{2} - (y_i - y_j)^{2},
     \tag{$2.10$}
     \end{equation}

     \begin{equation}
     \label{z8}
     f(ij) = \cosh y_i \cosh y_j \cos(x_i-x_j)-\sinh y_i \sinh y_j,
     \tag{$2.11$}
     \end{equation}

     \begin{equation}
     \label{z9}
     f(ij) = x_i y_j-x_jy_i,
     \tag{$2.12$}
     \end{equation}

     \begin{equation}
     \label{z10}
     f(ij) = \frac{y_i - y_j}{x_i - x_j},
     \tag{$2.13$}
     \end{equation}

     \begin{equation}
     \label{z11}
     \begin{array}{rcl}
     f(ij) = ((x_i - x_j)^{2} - (y_i - y_j)^{2})
     \exp \left(2 \beta \text{arc(cot)tanh} \displaystyle
     \frac{y_i - y_j}{x_i - x_j}\right)\
     \end{array}
     \tag{$2.14$}
     \end{equation}

     \begin{equation}
     \label{z12}
     f(ij) = (x_i - x_j)^{2}\exp\left(2\frac{y_i - y_j}{x_i - x_j}\right),
     \tag{$2.15$}
     \end{equation}

     \begin{equation}
     \label{z13}
     f(ij) = ((x_i - x_j)^{2} + (y_i - y_j)^{2})
     \exp \left(2 \gamma \arctan \displaystyle
     \frac{y_i - y_j}{x_i - x_j}\right),
     \tag{$2.16$}
     \end{equation}

     \begin{equation}
     \label{z14}
     f(ij) = \frac{(x_i - x_j)^{2} + \varepsilon_{i}y^{2}_i
     + \varepsilon_{j}y^{2}_j}{y_iy_j},
     \tag{$2.17$}
     \end{equation}
     { \it where $\beta >0$ and $\beta \neq 1;$ $\gamma > 0;$
     $\varepsilon_{i} = 0, \pm1;$ $\varepsilon_{j} = 0, \pm1$,$\varepsilon_{i}$
     not necessarily being equal to $\varepsilon_{j}$.}
     \vspace{5mm}

     Six of the expressions, (2.7)--(2.12), define metric functions of the two-dimensional
     geometries that are well-known: {\bf (2.7)} -- {\it of the Euclidean plane}; {\bf (2.8)}
     -- {\it of the two-dimensional sphere} in the three-dimensional Euclidean

\newpage

\noindent
 {\bf (2.9)} -- {\it плоскости Лобачевского} как
двумерного двухполостного гиперболоида в трехмерном
псевдоевклидовом пространстве; {\bf (2.10)} -- {\it плоскости
Минковского}; {\bf (2.11)} -- {\it двумерного однополостного
гиперболоида} в трехмерном псевдоевклидовом пространстве; {\bf
(2.12)} -- {\it симплектической плоскости}.

Существование четырех метрических функций (2.13)--(2.16), задающих
двумерные феноменологически симметричные геометрии ранга четыре,
впервые было установлено автором [6]. Профессор А.М.Широков
(кафедра геометрии Казанского госуниверситета) обратил внимание
автора на то, что три метрические функции: (2.14), (2.15) и (2.16)
можно записать единообразно, используя три типа комплексных чисел:
\begin{displaymath}
f(ij) = (z_i - z_j)\overline{(z_i - z_j)}\exp\ 2\gamma \text{arg}(z_i-z_j),
\end{displaymath}
где $z=x+ey$,  $\bar{z} = x - ey$, причем $e^{2} = + 1, \
\gamma>0$ и дополнительно $\gamma\neq1$ для выражения (2.14);
$e^{2} = 0$ и $\gamma=1$ для выражения (2.15);  $e^{2} = - 1$ и
$\gamma>0$ для выражения (2.16). Таким образом, все три возможные
типа комплексных чисел на плоскости, а именно: двойные $(e^2=+1)$,
дуальные $(e^2=0)$ и обычные $(e^2=-1)$, естественно вписались в
полную классификацию двумерных феноменологически симметричных
геометрий ранга четыре. По-видимому, соответствующие геометрии
никогда геометрами не изучались и потому специального
общепринятого названия не имеют. Двумерную геометрию с метрической
функцией {\bf (2.16)} автор назвал {\it плоскостью Гельмгольца},
так как окружностью в ней является логарифмическая спираль, о чем
кратко сообщает Гельмгольц в своей работе [4], считая это
отрицательной характеристикой такой геометрии. Соответственно
метрическая функция {\bf (2.14)} задает {\it псевдогельмгольцеву
плоскость}, а метрическая функция {\bf (2.15)} -- {\it
дуальногельмгольцеву} плоскость. Метрическая функция {\bf (2.13)}
задает так называемую {\it симплициальную плоскость}, название
которой было подсказано автору известным геометром Р.Пименовым, в
исследованиях которого такое выражение встречалось. Последнее
выражение {\bf (2.17)} определяет метрическую функцию, задающую
двумерную феноменологически симметричную геометрию {\it на
несвязном двумерном многообразии}, на связных компонентах которого
будет \ либо \ симплектическая

\newpage

     \noindent
     space;
     {\bf (2.9)} -- {\it of the Lobachevski plane} as a two-dimensional two-sheet
     hyperboloid in the three-dimensional pseudo-Euclidean space; {\bf (2.10)} --
     {\it of the Minkowski plane}; {\bf (2.11)} -- {\it of the two-dimensional two-sheet
     hyperboloid} in the three-dimensional pseudo-Euclidean space; {\bf (2.12)} -- {\it of
     the simplectic plane}.

     The existence of four metric functions, (2.13)--(2.16) defining two-dimen- \ sional
     phenomenologically symmetric geometries of rank 4 was established for the first
     time by the author [6]. Professor A.V. Shirokov (the Chair of Geometry of the Kazan
     State University) drew the author's attention to the possibility to write three of the
     metric functions, (2.14), (2.15) and (2.16), uniformly, using the three types of
     complex numbers:
     \begin{displaymath}
     f(ij) = (z_i - z_j)\overline{(z_i - z_j)}\exp\ 2\gamma \text{arg}(z_i-z_j),
     \end{displaymath}
     where $z=x+ey$,  $\bar{z} = x - ey$, $e^{2} = + 1, \ and
     \gamma>0$, and, additionally, $\gamma\neq1$ for the expression (2.14); $e^{2} = 0$
     and $\gamma=1$ for the expression (2.15);  $e^{2} = - 1$ and $\gamma>0$ for the
     expression (2.16). Thus, all the three possible types of complex numbers on the plane,
     namely double $(e^2=+1)$, dual $(e^2=0)$, and common $(e^2=-1)$, fit quite
     naturally into the complete classification of the two-dimensional phenomenologically
     symmetric geometries of rank 4. Apparently, the respective geometries have not
     ever got under scrutiny by the geometricians, and so have no universally accepted
     conventional names. The two-dimensional geometry with the metric function {\bf (2.16)}
     was given by the author the name of the {\it Helmholtz plane}, for the circle in it is
     the logarithmic spiral, which was stated in a few words in his work [4] by Helmholtz,
     who thought it to be a negative feature of such a geometry. Correspondingly, the
     metric function {\bf (2.14)} defines a {\it pseudo-Helmholtz plane}, and the metric
     function {\bf (2.15)} {\it a dual-Helmholtz } plane. The metric function {\bf (2.13)}
     defines the so called {\it simplicial plane}, the name suggested to the author by the
     well-known geometrician R. Pimenov, in whose papers one comes across it. The last
     of the expressions, {\bf (2.17)}, defines a metric function that gives a two-dimensional
     phenomenologically symmetric geometry {\it on a disconnected two-dimensional
     manifold}, on the connected components of which there will be present either \ the
     \ simplectic \ plane \ $(\varepsilon_{i} =\varepsilon_{j} = 0)$ \ or

\newpage

\noindent
плоскость $(\varepsilon_{i} = \varepsilon_{j} =
0)$, либо плоскость Лобачевского в модели Пуанкаре
$(\varepsilon_{i} = \varepsilon_{j} = +1)$, либо двумерный
однополостной гиперболоид $(\varepsilon_{i}=\varepsilon_{j}= -
1)$.

Феноменологическая симметрия ранга 4, выражаемая уравнением (2.5),
для всех двумерных геометрий, перечисленных в теореме 2, легко
устанавливается по рангу функциональной матрицы для шести функций
$f(ij),f(ik),f(il),f(jk),f(jl)$ ,$f(kl)$, специальным образом
зависящих от восьми переменных -- координат $x_i,y_i,x_j$,
$y_j,x_k,y_k,x_l,y_l$ четырех точек кортежа $\langle ijkl \rangle$. Ранг этой
матрицы, как показывает компьютерная проверка, равен 5, что
свидетельствует о наличии функциональной связи, задаваемой
уравнением (2.5) и выражающей феноменологическую симметрию всех
одиннадцати геометрий (2.7) -- (2.17). Что же касается самого
уравнения (2.5), то в явном виде оно найдено для всех двумерных
геометрий, кроме гельмгольцевых, задаваемых метрическими функциями
(2.14), (2.15) и (2.16). Например, для плоскости Евклида -- (2.7)
и псевдоевклидовой плоскости Минковского -- (2.10) таким
уравнением будет общее для них уравнение (В.2) из Введения,
обращающее в нуль определитель Кэли-Менгера пятого порядка для
четверки $\langle ijkl \rangle$ точек этих плоскостей. Для следующих пяти
двумерных геометрий, а именно: двумерной сферы -- (2.8), плоскости
Лобачевского -- (2.9), однополостного гиперболоида -- (2.11),
симплектической плоскости -- (2.12) и геометрии на несвязном
двумерном многообразии -- (2.17), в уравнении (2.5) слева стоит
определитель Грама четвертого порядка для четверки $\langle ijkl \rangle$,
причем диагональными элементами определителя являются значения
метрической функции для диагональных пар $\langle ii \rangle,\langle jj \rangle,\langle kk \rangle,\langle ll \rangle$.
Например, для двумерной сферы -- (2.8) и однополостного
гиперболоида -- (2.11) уравнение (2.5) будет таким:
$$
\left|
\begin{array}[tccb]{cccc}
 1 & f(ij) & f(ik) & f(il) \\
 f(ij) & 1 & f(jk) & f(jl) \\
 f(ik) & f(jk) & 1 & f(kl) \\
 f(il) & f(jl) & f(kl) & 1 \\
\end{array}
\right| = 0.
$$

Найдено в явном виде уравнение (2.5) также и для симплициальной

\newpage

      \noindent
      the Lobachevski plane
     on the Poincare model $(\varepsilon_{i} = \varepsilon_{j} = +1)$, or the two-dimensional
     one-sheet hyperboloid $(\varepsilon_{i}=\varepsilon_{j}= - 1)$.

     The phenomenological symmetry of rank 4 expressed by the expression (2.5),
     for all the two-dimensional geometries mentioned in Theorem 2, is easily established
     by the rank of the functional matrix for the six functions, $f(ij),f(ik),f(il),f(jk),f(jl)$,
     $f(kl)$, that depend in a special manner on the eight variables, the coordinates
     $x_i,y_i,x_j$, $y_j,x_k,y_k,x_l,y_l$ of four points of the cortege $\langle ijkl \rangle$. The rank
     of that matrix, as is checkout by computer says, is equal to 5, which indicates the
     presence of the functional relation defined by the equation (2.5) and expressing
     the phenomenological symmetry of all the eleven geometries (2.7) -- (2.17). As
     to the equation (2.5) itself, in the explicit form it has been found for all the
     two-dimensional geometries defined by the metric functions (2.14), (2.15), and (2.16),
     except for the Helmholtz geometry. For example, for the Euclidean plane -- (2.7)
     and the pseudo-Euclidean Minkowski plane -- (2.10) that will be the equation (В.2)
     from the Introduction, that turns into zero the Cayly-Menger determinant of the
     fifth order for the quadruple $\langle ijkl \rangle$ of the points of those planes. For five other
     two-dimensional geometries, i.e. the two-dimensional sphere -- (2.8), the Lobachevski
     plane -- (2.9), one-sheet hyperboloid -- (2.11), the simplectic plane -- (2.12), and the
     geometry on an unconnected two-dimensional manifold -- (2.17), in the equation
     (2.5) on the left there stands a gramian of the fourth order for the quadruple$\langle ijkl \rangle$,
     the diagonal elements of the determinant being the values of the metric function for
     the diagonal pairs $\langle ii \rangle,\langle jj \rangle,\langle kk \rangle,\langle ll \rangle$.
     For example, for the two-dimensional sphere -- (2.8) and the one-sheet
     hyperboloid -- (2.11), the equation (2.5) will be as follows:

     $$
      \left|
     \begin{array}[tccb]{cccc}
      1 & f(ij) & f(ik) & f(il) \\
      f(ij) & 1 & f(jk) & f(jl) \\
      f(ik) & f(jk) & 1 & f(kl) \\
      f(il) & f(jl) & f(kl) & 1 \\
     \end{array}
     \right| = 0.
     $$

\vspace{5mm}
     There has
     also been found the explicit form of the equation (2.5) for the

\newpage

\noindent
плоскости, задаваемой метрической функцией (2.13):
$$
\left|\begin{array}{ccc}
f(ij)-f(jk) & f(jk)-f(ik) & 0 \\
f(ij)-f(jl) & 0 & f(il)-f(jl) \\
0 & f(ik)-f(kl) & f(il)-f(kl)
\end{array}\right|=0.
$$

Как отмечалось выше, для гельмгольцевых плоскостей -- (2.14), \
(2.15) и \ (2.16) \ уравнение \ (2.15) \ не найдено. Есть
предположение, что его нельзя записать через известные
элементарные функции.

Групповая симметрия двумерных геометрий является естественным следствием
феноменологической симметрии согласно теореме 3 из \S1. Локально обратимое
преобразование

$$
x'=\lambda(x,y), \ \ y'=\sigma(x,y),
$$
удовлетворяющее условию $\partial(\lambda,\sigma)/\partial(x,y)\neq0$, будет
движением, если оно сохраняет метрическую функцию (2.4):

$$
f(\lambda(i),\sigma(i),\lambda(j),\sigma(j))=f(x_i,y_i,x_j,y_j),
\eqno(2.18)
$$
где, например, $\lambda(i)=\lambda(x_i,y_i)$. Решением функционального
уравнения (2.18) для каждой двумерной геометрии (2.7) -- (2.17) находится
полная локальная трехпараметрическая группа движений, которая и
определяет ее групповую симметрию степени 3. Метрическая функция (2.4)
является также решением дифференциального уравнения

$$
X(i)f(ij)+X(j)f(ij)=0,
\eqno(2.19)
$$
с операторами $X=\lambda(x,y)\partial/\partial x+
\sigma(x,y)\partial/\partial y$ соответствующей трехмерной алгебры
Ли. Уравнение же (2.19) можно рассматривать как функциональное на
коэффициенты $\lambda$ и $\sigma$ оператора $X$. Оказывается, что
так трактуемое при известной метрической функции (2.4) уравнение
(2.19) решается более простыми методами (см. [14]), чем исходное
функциональное уравнение (2.18). Напомним, что согласно известным
теоремам Ли между группой Ли и соответствующей ей алгеброй Ли
имеется взаимно однозначное соответствие.

\newpage

      \noindent
     simplicial plane
     defined by the metric function (2.13):
     $$
     \left|\begin{array}{ccc}
     f(ij)-f(jk) & f(jk)-f(ik) & 0 \\
     f(ij)-f(jl) & 0 & f(il)-f(jl) \\
     0 & f(ik)-f(kl) & f(il)-f(kl)
     \end{array}\right|=0.
     $$

     As has been said above, for the Helmholtz planes -- (2.14), (2.15) and (2.16) the
     equation (2.15) has not been discovered. There is a supposition that it cannot be
     written through the known elementary functions.

     The group symmetry of two-dimensional geometries is a natural corollary of the
     phenomenological symmetry, under Theorem 3 of \S1. The locally invertible
     transformation
     $$
     x'=\lambda(x,y), \ \ y'=\sigma(x,y),
     $$
     satisfying the condition of $\partial(\lambda,\sigma)/\partial(x,y)\neq0$, will be a
     motion if it preserves the metric function  (2.4):

     $$
     f(\lambda(i),\sigma(i),\lambda(j),\sigma(j))=f(x_i,y_i,x_j,y_j),
     \eqno(2.18)
     $$
     \phantom{aaaaa} \\
     where, for example, $\lambda(i)=\lambda(x_i,y_i)$. By way of solving the functional
     equation (2.18) of each of the two-dimensional geometries (2.7) -- (2.17) the complete
     local three-parameter group of motions can be found that defines the group
     symmetry of it of degree 3. The metric function (2.4) is also a solution of the differential
     equation
     $$
     X(i)f(ij)+X(j)f(ij)=0,
     \eqno(2.19)
     $$
     with the operators $X=\lambda(x,y)\partial/\partial x+\sigma(x,y)\partial/\partial y$
     of the corresponding three-dimensional Lie algebra. As to the equation (2.19), it can
     be considered a functional one for the coefficients $\lambda$ and $\sigma$ of the
     operator $X$. It turns out that so interpreted and with the known metric function (2.4)
     the equation (2.19) can be solved by employing simpler methods (see [14]), than the
     original functional equation (2.18). We shall remind that under the known Lie
     theorems, there is between a Lie group and the corresponding Lie algebra a one-to-one
     correspondence.

\newpage

Рассмотрим еще трехмерные $(s=1,n=3)$  феноменологически
симметричные геометрии, которые задаются на трехмерном
многообразии $\mathfrak{M}$ метрической функцией
$f:\mathfrak{S}_f\to R$, где
$\mathfrak{S}_f\subseteq\mathfrak{M}\times\mathfrak{M}$. Ее
координатное представление определяется выражением
\begin{equation}
\label{z1}
f(ij) = f(x_i,y_i,z_i,x_j,y_j,z_j),
\tag{$2.20$}
\end{equation}
где $(x,y,z)$ -- локальные координаты в многообразии. Если эта
функция задает такую геометрию, то по аксиоме IV из \S1 десять ее
значений для пятерки $\langle  ijklm  \rangle$ функционально связаны:
\begin{equation}
\begin{split}
\Phi(f(ij), &f(ik), f(il),f(im),f(jk),f(jl),f(jm),
\\  &f(kl),f(km),f(lm)) =0.
\end{split}
\tag{$2.21$}
\end{equation}

Невырожденная метрическая функция (2.20) по аксиоме III из \S1 должна,
очевидно, удовлетворять следующим двум условиям:
\begin{equation}
\label{z3}
\left.\begin{array}{rcl}
\partial(f(ik),f(il),f(im))/\partial(x_i,y_i,z_i) \neq 0,\\
\partial(f(kj),f(lj),f(mj))/\partial(x_j,y_j,z_j)\neq 0\;
\end{array}\right\}
\tag{$2.22$}
\end{equation}
для открытого и плотного в $\mathfrak{M}^{4}$ множества четверок
$\langle  iklm  \rangle$ и $\langle  klmj  \rangle$.

Примером трехмерной феноменологически симметричной геометрии является
трехмерное пространство Евклида. Для метрической функции $f(ij)$,
сопоставляющей паре точек $\langle  ij  \rangle$ квадрат обычного расстояния,
в декартовой прямоугольной системе координат $(x,y,z)$
представление будет следующим:
$$
f(ij)=(x_i-x_j)^2+(y_i-y_j)^2+(z_i-z_j)^2.
$$

Хорошо известно, что в пространстве Евклида десять взаимных
расстояний для пятерки $\langle  ijklm  \rangle$ точек функционально между
собой связаны, обращая в ноль определитель Кэли-Менгера шестого
порядка, строение которого аналогично строению определителя (В.2):

$$
\left|\begin{array}{cccccc}
0 & 1 & 1 & 1 & 1 & 1 \\
1 & 0 & f(ij) & f(ik) & f(il) & f(im) \\
1 & f(ij) & 0 & f(jk) & f(jl) & f(jm) \\
1 & f(ik) & f(jk) & 0 & f(kl) & f(km) \\
1 & f(il) & f(jl) & f(kl) & 0 & f(lm) \\
1 & f(im) & f(jm) & f(km) & f(lm) & 0
\end{array}\right|=0.
$$

\newpage

     Let us also consider the three-dimensional $(s=1,n=3)$  phenomenologi- \ cally symmetric
     geometries defined on a three-dimensional manifold $\mathfrak{M}$ by the metric
     function $f:\mathfrak{S}_f\to R$, where
     $\mathfrak{S}_f\subseteq\mathfrak{M}\times\mathfrak{M}$. The coordinate
     representation for it is determined by the expression
     \begin{equation}
     \label{z1}
     f(ij) = f(x_i,y_i,z_i,x_j,y_j,z_j),
     \tag{$2.20$}
     \end{equation}
     where $(x,y,z)$ are local coordinates in the manifold. If the function gives such a
     geometry, then, under the axiom IV of \S1 the ten values of it for the quintuple
     $\langle  ijklm  \rangle$ are functionally related:
     \begin{equation}
     \begin{split}
     \Phi(f(ij), &f(ik),f(il),f(im),f(jk),f(jl),f(jm), \\ &f(kl),f(km),f(lm)) = 0.
     \end{split}
      \tag{$2.21$}
     \end{equation}

     The nondegenerate metric function (2.20) must, under the axiom III of \S1,
     obviously satisfy the two conditions as follows:
     \begin{equation}
     \label{z3}
     \left.\begin{array}{rcl}
     \partial(f(ik),f(il),f(im))/\partial(x_i,y_i,z_i) \neq 0,\\
     \partial(f(kj),f(lj),f(mj))/\partial(x_j,y_j,z_j)\neq 0\;
     \end{array}\right\}
     \tag{$2.22$}
     \end{equation}
     for the open and dense in $\mathfrak{M}^{4}$ set of quadruples
     $\langle  iklm  \rangle$ and $\langle  klmj  \rangle$.

     An example of a three-dimensional phenomenologically symmetric geomet- \ ry is
     the three-dimensional Euclidean space. For the metric function $f(ij)$ assigning to
     the pair of points $\langle  ij  \rangle$ the squared ordinary distance, in the Cartesian coordinate
     system  $(x,y,z)$ the representation will be as follows:
     $$
     f(ij)=(x_i-x_j)^2+(y_i-y_j)^2+(z_i-z_j)^2.
     $$

     It is well known that in the Euclidean space the ten reciprocal distances for the
     quintuple $\langle  ijklm  \rangle$ of points are among themselves functionally related, turning
     into zero the Cayly-Menger determinant of the sixth order the structure of which is
     similar to that of the determinant (В.2):

     $$
     \left|\begin{array}{cccccc}
     0 & 1 & 1 & 1 & 1 & 1 \\
     1 & 0 & f(ij) & f(ik) & f(il) & f(im) \\
     1 & f(ij) & 0 & f(jk) & f(jl) & f(jm) \\
     1 & f(ik) & f(jk) & 0 & f(kl) & f(km) \\
     1 & f(il) & f(jl) & f(kl) & 0 & f(lm) \\
     1 & f(im) & f(jm) & f(km) & f(lm) & 0
     \end{array}\right|=0.
     $$

\newpage

 Полная классификация трехмерных феноменологически
симметричных геометрий ранга пять была построена В.Х. Левом.
Приведем ее по его работе [7]:

\vspace{5mm}

{\bf Теорема 3.}  {\it С точностью до масштабного преобразования
$\psi(f)\to f$ и в надлежаще выбранной системе локальных координат
$(x,y,z)$ невырожденная метрическая функция $(2.20)$, задающая на трехмерном
многообразии феноменологически симметричную геометрию ранга пять со связью
$(2.21)$, может быть представлена одним из следующих пятнадцати
канонических выражений:}

\begin{equation}
f(ij) = (x_i - x_j)^2 + (y_i - y_j)^2 + (z_i - z_j)^2,
\tag{$2.23$}
\end{equation}

\begin{equation}
\begin{split}
f(ij) = &\sin z_i\sin z_j[\sin y_i\sin y_j\cos(x_i - x_j) + \\
&+\cos y_i\cos y_j] + \cos z_i\cos z_j,
\end{split}
\tag{$2.24$}
\end{equation}

\begin{equation}
f(ij) = \text{sh}z_i\text{sh}z_j[\sin y_i\sin y_j\cos(x_i - x_j) +
\cos y_i\cos y_j] - \text{ch}z_i\text{ch}z_j,
\tag{$2.25$}
\end{equation}

\begin{equation}
f(ij) = (x_i - x_j)^2 + (y_i - y_j)^2 - (z_i - z_j)^2,
\tag{$2.26$}
\end{equation}

\begin{equation}
f(ij) = \text{ch}z_i\text{ch}z_j[\sin y_i\sin y_j\cos(x_i - x_j) +
\cos y_i\cos y_j] -\text{sh}z_i\text{sh}z_j,
\tag{$2.27$}
\end{equation}

\begin{equation}
f(ij) = \text{ch}z_i\text{ch}z_j[\text{ch}y_i\text{ch}y_j\cos(x_i - x_j) -
\text{sh}y_i\text{sh}y_j] - \text{sh}z_i\text{sh}z_j,
\tag{$2.28$}
\end{equation}

\begin{equation}
f(ij) = x_iy_j - x_jy_i + z_i - z_j,
\tag{$2.29$}
\end{equation}

\begin{equation}
f(ij) = \displaystyle \frac{y_i - y_j}{x_i - x_j} +z_i + z_j,
\tag{$2.30$}
\end{equation}

\begin{equation}
f(ij) = \displaystyle \frac{y_i - y_j}{x_i - x_j}
\exp(z_i + z_j),
\tag{$2.31$}
\end{equation}

\newpage

     The complete classification of the three-dimensional phenomenologically symmetric
     geometries of rank 5 was built by V.H. Lev. We shall reproduce it after his note [7]:

     \vspace{5mm}

     {\bf Theorem 3.}  {\it With an accuracy up to a scaling transformation $\psi(f)\to f$
     and in a suitably chosen system of local coordinates $(x,y,z)$ the nondegene- \ rate metric
     function $(2.20)$ that gives on a three-dimensional manifold a \\ phenomenologically
     symmetric geometry of rank 5 with the relation $(2.21)$ may be represented by one of
     the following fifteen canonical expressions:}

     \vspace{5mm}

     \begin{equation}
     f(ij) = (x_i - x_j)^2 + (y_i - y_j)^2 + (z_i - z_j)^2,
     \tag{$2.23$}
     \end{equation}

     \begin{equation}
     \begin{split}
     f(ij) = &\sin z_i\sin z_j[\sin y_i\sin y_j\cos(x_i - x_j) + \\
    &+\cos y_i\cos y_j] + \cos z_i\cos z_j,
     \end{split}
     \tag{$2.24$}
     \end{equation}

     \begin{equation}
     \begin{split}
     f(ij) = &\sinh z_i\sinh z_j[\sin y_i\sin y_j\cos(x_i - x_j) + \\
    &+\cos y_i\cos y_j] - \cosh z_i\cosh z_j,
     \end{split}
     \tag{$2.25$}
     \end{equation}

     \begin{equation}
     f(ij) = (x_i - x_j)^2 + (y_i - y_j)^2 - (z_i - z_j)^2,
     \tag{$2.26$}
     \end{equation}

     \begin{equation}
     \begin{split}
     f(ij) = &\cosh z_i\cosh z_j[\sin y_i\sin y_j\cos(x_i - x_j) + \\
    &+ \cos y_i\cos y_j] -\sinh z_i\sinh z_j,
     \end{split}
     \tag{$2.27$}
     \end{equation}

     \begin{equation}
      \begin{split}
      f(ij) = &\cosh z_i\cosh z_j[\cosh y_i\cosh y_j\cos(x_i - x_j) - \\
     &-\sinh y_i\sinh y_j] - \sinh z_i\sinh z_j,
      \end{split}
     \tag{$2.28$}
     \end{equation}

     \begin{equation}
     f(ij) = x_iy_j - x_jy_i + z_i - z_j,
     \tag{$2.29$}
     \end{equation}

     \begin{equation}
     f(ij) = \displaystyle \frac{y_i - y_j}{x_i - x_j} +z_i + z_j,
     \tag{$2.30$}
     \end{equation}

     \begin{equation}
     f(ij) = \displaystyle \frac{y_i - y_j}{x_i - x_j}
     \exp(z_i + z_j),
     \tag{$2.31$}
     \end{equation}

\newpage

\begin{equation}
f(ij) = [(x_i - x_j)^2 - (y_i - y_j)^2]\exp2(z_i + z_j),
\tag{$2.32$}
\end{equation}

\begin{equation}
f(ij) = [(x_i - x_j)^2 + (y_i - y_j)^2]\exp2(z_i + z_j),
\tag{$2.33$}
\end{equation}

\begin{equation}
f(ij) = [(x_i - x_j)^2 - (y_i - y_j)^2]\exp[\displaystyle 2(\beta
\text{ar(c)th}\frac{y_i - y_j}{x_i - x_j} + z_i + z_j)],
\tag{$2.34$}
\end{equation}

\begin{equation}
f(ij) = (x_i - x_j)^2\exp[\displaystyle
2(\frac{y_i - y_j}{x_i - x_j} + z_i + z_j)],
\tag{$2.35$}
\end{equation}

\begin{equation}
f(ij) = [(x_i - x_j)^2 + (y_i - y_j)^2]\exp[\displaystyle 2(\gamma
\text{arctg}\frac{y_i - y_j}{x_i - x_j} + z_i + z_j)],
\tag{$2.36$}
\end{equation}

\begin{equation}
\label{z14}
f(ij) = \frac{(x_i - x_j)^{2} \pm (y_i-y_j)^2+ \varepsilon_{i}z^{2}_i
+ \varepsilon_{j}z^{2}_j}{z_iz_j},
\tag{$2.37$}
\end{equation}
{ \it где $\beta >0$ и $\beta \neq 1;$ $\gamma > 0;$
$\varepsilon_{i} = 0, \pm1;$
$\varepsilon_{j} = 0, \pm1$,
причем не обязательно
$\varepsilon_{i} = \varepsilon_{j}$.}
\vspace{5mm}

\vspace{5mm}

Семь выражений (2.23)--(2.29) определяют метрические функции
хорошо известных трехмерных геометрий: {\bf (2.23)} -- {\it
пространства Евклида} как естественного трехмерного расширения
плоскости Евклида с метрической функцией (2.7); {\bf (2.24)} --
{\it трехмерной сферы} в четырехмерном евклидовом пространстве;
{\bf (2.25)} -- {\it пространства Лобачевского} как трехмерного
двухполостного гиперболоида в четырехмерном псевдоевклидовом
пространстве сигнатуры $\langle  +++-  \rangle$; {\bf (2.26)} -- {\it
пространства Минковского} как естественного трехмерного расширения
плоскости Минковского с метрической функцией (2.10); {\bf (2.27)}
-- {\it трехмерного однополостного гиперболоида} I в четырехмерном
псевдоевклидовом пространстве той же сигнатуры $\langle  +++-  \rangle$; {\bf
(2.28)} -- {\it трехмерного однополостного гиперболоида} II в
четырехмерном псевдоевклидовом пространстве, но другой сигнатуры
$\langle  ++--  \rangle$; {\bf (2.29)} -- {\it симплектического пространства}
как естественного расширения симплектической плоскости с
метрической функцией (2.12) на нечетную размерность, равную трем.

\newpage

     \begin{equation}
     f(ij) = [(x_i - x_j)^2 - (y_i - y_j)^2]\exp2(z_i + z_j),
     \tag{$2.32$}
     \end{equation}

     \begin{equation}
     f(ij) = [(x_i - x_j)^2 + (y_i - y_j)^2]\exp2(z_i + z_j),
     \tag{$2.33$}
     \end{equation}

     \begin{equation}
     f(ij) = [(x_i - x_j)^2 - (y_i - y_j)^2]\exp[\displaystyle 2(\beta
     \text{arc(cot)tanh}\frac{y_i - y_j}{x_i - x_j} + z_i + z_j)],
     \tag{$2.34$}
     \end{equation}

     \begin{equation}
     f(ij) = (x_i - x_j)^2\exp[\displaystyle
     2(\frac{y_i - y_j}{x_i - x_j} + z_i + z_j)],
     \tag{$2.35$}
     \end{equation}

     \begin{equation}
     f(ij) = [(x_i - x_j)^2 + (y_i - y_j)^2]\exp[\displaystyle 2(\gamma
     \arctan\frac{y_i - y_j}{x_i - x_j} + z_i + z_j)],
     \tag{$2.36$}
     \end{equation}

     \begin{equation}
     \label{z14}
     f(ij) = \frac{(x_i - x_j)^{2} \pm (y_i-y_j)^2+ \varepsilon_{i}z^{2}_i
     + \varepsilon_{j}z^{2}_j}{z_iz_j},
     \tag{$2.37$}
     \end{equation}
     { \it where $\beta >0$ and $\beta \neq 1;$ $\gamma > 0;$ $\varepsilon_{i} = 0, \pm1;$
     $\varepsilon_{j} = 0, \pm1$,$\varepsilon_{i}$ not necessarily equal to $\varepsilon_{j}$.}
     \vspace{5mm}

     \vspace{5mm}
     The seven expressions (2.23)--(2.29) determine the metric functions of well-known
     three-dimensional geometries: {\bf (2.23)} -- {\it the Euclidean space} as a natural
     three-dimensional extension of the Euclidean plane with the metric function (2.7);
     {\bf (2.24)} -- {\it the three-dimensional sphere} in the four-dimensional Euclidean space;
     {\bf (2.25)} -- {\it The Lobachevski space} as of a three-dimensional two-sheet hyperboloid
     in the four-dimensional pseudo-Euclidean space of the signature $\langle  +++-  \rangle$; {\bf (2.26)}
     -- {\it the Minkowski space} as of a natural three-dimensional extension of the Minkowski
     plane with the metric function (2.10); {\bf (2.27)} -- {\it three-dimensional one-sheet
     hyperboloid} I in the four- \\ dimensional pseudo-Euclidean space of the same signature
     $\langle  +++-  \rangle$; {\bf (2.28)} -- {\it three-dimensional one-sheet hyperboloid } II in the four-
     dimensional pseudo-Euclidean space, but of another signature $\langle  ++--  \rangle$; {\bf (2.29)}
     --{\it the simplectic space} as a natural extension of the simplectic plane with the metric
     function (2.12) for the odd dimensionality equal to three.

\newpage

Следующие семь выражений (2.30)--(2.36) определяют метрические
функции таких трехмерных геометрий, которые впервые были
обнаружены В.Х.Левом [15], никогда геометрами не изучались и
потому специального названия не имеют: {\bf (2.30)} -- {\it
симплициального пространства} I как аддитивного трехмерного
феноменологически симметричного расширения симплициальной
плоскости с метрической функцией (2.13); {\bf (2.31)} -- {\it
симплициального пространства} II как мультипликативного
трехмерного феноменологически симметричного расширения
симплициальной плоскости с метрической функцией (2.13); {\bf
(2.32)} -- {\it особого трехмерного феноменологически
симметричного расширения плоскости Минковского} с метрической
функцией (2.10); {\bf (2.33)} -- {\it особого трехмерного
феноменологически симметричного расширения плоскости Евклида} с
метрической функцией (2.7); {\bf (2.34)} -- {\it
псевдогельмгольцева пространства} как трехмерного
феноменологически симметричного расширения псевдогельмгольцевой
плоскости с метрической функцией (2.14); {\bf (2.35)} -- {\it
дуальногельмгольцева пространства} как трехмерного
феноменологически симметричного расширения  дуальногельмгольцевой
плоскости с метрической функцией (2.15); {\bf (2.36)} -- {\it
пространства Гельмгольца} как трехмерного феноменологически
симметричного расширения плоскости Гельмгольца с функцией (2.16).

\vspace{5mm}

Последнее выражение {\bf (2.37)} определяет метрическую функцию трехмерной
геометрии {\it на несвязном трехмерном многообразии}, на связных компонентах
которого она задает либо расширения (2.32), (2.33), либо сферы (2.25), (2.27),
(2.28) в четырехмерных псевдоевклидовых пространствах.

\vspace {5mm}

Феноменологическая симметрия каждой из пятнадцати перечисленных
выше трехмерных геометрий, задаваемых метрическими функциями
(2.23)--(2.37), подтверждается рангом соответствующей
функциональной матрицы для десяти функций
$f(ij),f(ik),f(il),f(im),f(jk),$
\linebreak$f(jl),f(jm),f(kl),f(km),f(lm)$, \ специальным образом
зависящих от пятнадцати переменных -- координат
$x_i,y_i,z_i,x_j,y_j,z_j,x_k,y_k,z_k,x_l,y_l,z_l,$
\linebreak$x_m,y_m,z_m$ всех точек пятерки \ $\langle ijklm
\rangle$, \ который оказывается равным

\newpage

      The next seven expressions, (2.30)--(2.36), define the metric
      functions for the
     three-dimensional geometries discovered by V.H.Lev [15], which have never been
     explored, and so have no well-established and unanimously accepted names:
     {\bf (2.30)} -- {\it simplicial space} I as the additive three-dimensional
     phenomenologically symmetric extension of the simplicial plane with the metric
     function (2.13); {\bf (2.31)} -- {\it simplicial space} II as the multiplicative
     three-dimensional phenomenologically symmetric extension of the
     simplicial plane with the metric function (2.13); {\bf (2.32)} -- {\it the special
     phenomenologically symmetric extension of the Minkowski plane} with the
     metric function (2.10); {\bf (2.33)} -- {\it the special three-dimensional
     phenomenologically symmetric exten- \ sion of the Euclidean plane} with the
     metric function (2.7); {\bf (2.34)} -- {\it pseudo-Helmholtz space} as the
     three-dimensional phenomenologically symmetric extension of the pseudo-Helmholtz
     plane with the metric function (2.14); {\bf (2.35)} -- {\it dual-Helmholtz space} as the
     three-dimensional phenomenologically symmetric extension of the dual-Helmholtz
     plane with the metric function (2.15); {\bf (2.36)} -- {\it Helmholtz space}
     as the three-dimensional phenomenologically symmetric extension
     of the Helmholtz plane with the metric function (2.16).

    \vspace{20mm}

     And the last one, {\bf (2.37)}, defines the metric function of the three-dimensio- \ nal
     geometry {\it on an unconnected three-dimensional manifold}, on the connec- \ ted
     components of which it gives either the extensions (2.32), (2.33), or the spheres (2.25), (2.27),
     (2.28) in four-dimensional pseudo-Euclidean spaces.

     \vspace {20mm}

     The phenomenological symmetry of each of the fifteen above said three-dimensional
     geometries given by the metric functions (2.23)--(2.37) is proved by the rank of the
     respective functional matrix for the ten functions  $f(ij), \\
     f(ik),f(il),f(im),f(jk),f(jl),f(jm),f(kl),f(km),f(lm)$,
     that  depend in a special manner on the fifteen variables -- the
     coordinates \
     $x_i,y_i,z_i,x_j,y_j,z_j, \\ x_k,y_k,z_k,x_l,y_l,z_l,x_m,y_m,z_m$
     of all the points of  \ the  \ quintuple \ $\langle ijklm \rangle$,

\newpage

\noindent
девяти. Тем самым доказывается, что для каждой трехмерной
геометрии существует некоторое уравнение (2.21), выражающее ее
феноменологическую симметрию. Явный вид этого уравнения найден для
всех трехмерных геометрий, кроме симплициальных и гельмгольцевых
пространств: (2.30), (2.31) и (2.34), (2.35), (2.36). Для всех
других геометрий уравнение (2.21) записывается в виде обращения в
ноль или определителя Кэли-Менгера шестого порядка или
определителя Грама пятого порядка, диагональными элементами
которых являются значения метрической функции для диагональных пар
$\langle ii \rangle,\langle jj \rangle,\langle kk \rangle,$ \linebreak$\langle ll \rangle,\langle mm \rangle$.

Групповая симметрия трехмерных геометрий, как и двумерных,
эквивалентна феноменологической симметрии согласно теореме 3 из
\S1. Локально обратимое преобразование
$$
x'=\lambda(x,y,z), \ \  y'=\sigma(x,y,z), \ \ z'=\tau(x,y,z),
$$
удовлетворяющее условию $\partial(\lambda,\sigma,\tau)/\partial(x,y,z)\neq0$,
будет локальным движением, если оно сохраняет метрическую функцию (2.20):
$$
f(\lambda(i),\sigma(i),\tau(i),\lambda(j),\sigma(j),\tau(j))=
f(x_i,y_i,z_i,x_j,y_j,z_j),
$$
где, например, $\lambda(i)=\lambda(x_i,y_i,z_i)$. Решением этого
функционального уравнения для каждой трехмерной геометрии (2.23)--(2.37)
находится полная локальная шестипараметрическая группа движений, которая и
определяет ее групповую симметрию степени шесть. Метрическая функция (2.20)
является также решением дифференциального уравнения (2.19)
с операторами
$$X=\lambda(x,y,z)\partial/\partial x+
\sigma(x,y,z)\partial/\partial y+\tau(x,y,z)\partial/\partial z
$$
соответствующей шестимерной алгебры Ли. Кроме того, это уравнение
можно, в свою очередь, рассматривать как функциональное на
коэффициенты $\lambda$, $\sigma$, $\tau$ оператора $X$. Так
трактуемое при известной метрической функции (2.20) уравнение
(2.19) решается достаточно простыми методами (см. [14]). По
известной же группе движений трехмерной геометрии или ее
шестимерной алгебре Ли исходная метрическая функция как
двухточечный инвариант восстанавливается однозначно с точностью до
масштабного преобразования $\psi(f)\to f$.

\newpage

     \noindent
     that turns out to be equal to nine. Thereby it is
     proved that for each three-dimensional geometry there exists some equation (2.21)
     that expresses its phenomenological symmetry. The explicit form of that
     equation is found for all the three-dimensional geometries, except for the simplicial and
     Helmholtz spaces: (2.30), (2.31) and (2.34), (2.35), (2.36). For all the other
     geometries the equation (2.21) is written in the form of the vanishing to zero of either
     the Cayly-Menger determinant of the sixth order or of the gramian of the fifth
     order whose diagonal elements are the values of the metric function
     for the diagonal pairs $\langle ii \rangle,\langle jj \rangle,\langle kk \rangle,\langle ll \rangle,\langle mm \rangle$.

     The group symmetry of the three-dimensional geometries, just as in case of the two-dimensional
     ones, is equivalent to the phenomenological symmetry, according Theorem 3 of
     \S1. The locally invertible transformation

     $$
     x'=\lambda(x,y,z), \ \  y'=\sigma(x,y,z), \ \ z'=\tau(x,y,z)
     $$
     satisfying the condition $\partial(\lambda,\sigma,\tau)/\partial(x,y,z)\neq0$,
     will be a local motion if it preserves the metric function (2.20):
     $$
     f(\lambda(i),\sigma(i),\tau(i),\lambda(j),\sigma(j),\tau(j))=
     f(x_i,y_i,z_i,x_j,y_j,z_j),
     $$
     where, for example, $\lambda(i)=\lambda(x_i,y_i,z_i)$. The solution of that
     functional equation for each of the three-dimensional geometries (2.23)--(2.37)
     yields the complete local six-parameter group of motions, and it is that that
     defines its group symmetry of degree six. The metric function (2.20)
     is also a solution of the differential equation (2.19) with the operators
     $$X=\lambda(x,y,z)\partial/\partial x+
     \sigma(x,y,z)\partial/\partial y+\tau(x,y,z)\partial/\partial z
     $$
     of the respective six-dimensional Lie algebra. Besides, it is possible to consider that
     equation as a functional one for the coefficients $\lambda$, $\sigma$, $\tau$ of the
     operator $X$. Interpreted this way, and with the known metric function (2.20),
     the equation (2.19) is solved by easy enough methods (see [14]). And by the known
     group of motions of the three-dimensional geometry or its six-dimensional Lie algebra
     the original metric function is, as a two-point invariant, reconstructed uniquely with an
     accuracy up to a scaling transformation $\psi(f)\to f$.

\newpage

В заключение заметим, что феноменологически симметричные
геометрии более высокой размерности $(s=1,n>3)$ задаются на $n$-мерном многообразии невырожденной
однокомпонентной функцией с координатным представлением
$$
f(ij)=f(x^1_i,x^2_i,\ldots,x^n_i,x^1_j,x^2_j,\ldots,x^n_j),
$$
причем для кортежа $\langle  ijk\ldots vw  \rangle$ длины $n+2$ и некоторой
его окрестности, таких, что множество пар
$\langle  ij  \rangle,\langle  ik  \rangle,\ldots,\langle  vw  \rangle$ принадлежит области ее
определения, соответствующее множество значений
$f(ij),f(ik),\ldots, \\ f(vw)$ функционально связано некоторым
уравнением (1.1). Классификации таких геометрий еще не построены,
однако можно выписать некоторые выражения для метрической функции
$n$-мерной феноменологически симметричной геометрии ранга $n+2$
как естественные и особые расширения отдельных выражений из
классификации геометрий меньшей размерности. Например, для
четырехмерной феноменологически симметричной геометрии $(s=1,
n=4)$ ранга 6 предварительная, но пока не окончательная,
классификация с точностью до замены локальных координат $x,y,z,t$
в многообразии и масштабного преобразования $\psi(f)\to f$ будет
(см. [15]) следующей:

$$
f(ij)=(x_i-x_j)^2+(y_i-y_j)^2+(z_i-z_j)^2+(t_i-t_j)^2,
\eqno (2.38)
$$

$$
f(ij)=(x_i-x_j)^2+(y_i-y_j)^2+(z_i-z_j)^2-(t_i-t_j)^2,
\eqno (2.39)
$$

$$
f(ij)=(x_i-x_j)^2+(y_i-y_j)^2-(z_i-z_j)^2-(t_i-t_j)^2,
\eqno (2.40)
$$

$$
f(ij)=[(x_i-x_j)^2+(y_i-y_j)^2+(z_i-z_j)^2]\exp2(t_i+t_j),
\eqno (2.41)
$$

$$
f(ij)=[(x_i-x_j)^2+(y_i-y_j)^2-(z_i-z_j)^2]\exp2(t_i+t_j),
\eqno (2.42)
$$

\newpage

     We shall note  in conclusion that phenomenologically symmetric
     geometries of higher dimensionality $(s=1,n>3)$ are defined on an $n$-dimensional
     manifold by the nondegenerate one-component function with the coordinate
     representation

     $$
     f(ij)=f(x^1_i,x^2_i,\ldots,x^n_i,x^1_j,x^2_j,\ldots,x^n_j),
     $$
     the respective set of values $f(ij),f(ik),\ldots,f(vw)$ for the cortege $\langle  ijk\ldots \\ vw  \rangle$
     of length $n+2$ and some neighbourhood of it, such that the set of pairs $\langle  ij  \rangle,\langle  ik  \rangle,
     \ldots,\langle  vw  \rangle$ belongs to its domain, being functionally related by some equation (1.1).
     Classifications of such geometries have not been built yet, however we can write some
     expressions for the metric function of the $n$-dimensional phenomenologically symmetric
     geometry of rank  $n+2$ as natural and special extensions of some certain expressions
     of a classification of geometries of smaller dimensionality. For example, for the
     four-dimensional phenomenologically symmetric geometry $(s=1, n=4)$ of rank 6, the
     classification, only preliminary and far from being complete, will be, with an accuracy up
     to a change of the local coordinates $x,y,z,t$ in the manifold and a scaling transformation
     $\psi(f)\to f$, (see [15]) as follows:

     $$
     f(ij)=(x_i-x_j)^2+(y_i-y_j)^2+(z_i-z_j)^2+(t_i-t_j)^2,
     \eqno (2.38)
     $$

     $$
     f(ij)=(x_i-x_j)^2+(y_i-y_j)^2+(z_i-z_j)^2-(t_i-t_j)^2,
     \eqno (2.39)
     $$

     $$
     f(ij)=(x_i-x_j)^2+(y_i-y_j)^2-(z_i-z_j)^2-(t_i-t_j)^2,
     \eqno (2.40)
     $$

     $$
     f(ij)=[(x_i-x_j)^2+(y_i-y_j)^2+(z_i-z_j)^2]\exp2(t_i+t_j),
     \eqno (2.41)
     $$

     $$
     f(ij)=[(x_i-x_j)^2+(y_i-y_j)^2-(z_i-z_j)^2]\exp2(t_i+t_j),
     \eqno (2.42)
     $$

\newpage

\begin{equation}
\begin{split}
f(ij)=&\sin t_i\sin t_j[\sin z_i\sin z_j(\sin y_i\sin y_j\cos(x_i-x_j)+ \\
      &+\cos y_i\cos y_j)+\cos z_i\cos z_j]+\cos t_i\cos t_j,
\end{split}
\tag{2.43}
\end{equation}

\begin{equation}
\begin{split}
f(ij)=&\text{ch}t_i\text{ch}t_j[\sin z_i\sin z_j(\sin y_i\sin y_j\cos(x_i-x_j)+ \\
      &+\cos y_i\cos y_j)+\cos z_i\cos z_j]-\text{sh}t_i\text{sh}t_j,
\end{split}
\tag{2.44}
\end{equation}

\begin{equation}
\begin{split}
f(ij)=&\text{sh}t_i\text{sh}t_j[\sin z_i\sin z_j(\sin y_i\sin y_j\cos(x_i-x_j)+ \\
      &+\cos y_i\cos y_j)+\cos z_i\cos z_j]-\text{ch}t_i\text{ch}t_j,
\end{split}
\tag{2.45}
\end{equation}

\begin{equation}
\begin{split}
f(ij)=&\text{ch}t_i\text{ch}t_j[\text{ch}z_i\text{ch}z_j(\sin y_i\sin y_j\cos(x_i-x_j)+ \\
      &+\cos y_i\cos y_j)-\text{sh}z_i\text{sh}z_j]-\text{sh}t_i\text{sh}t_j,
\end{split}
\tag{2.46}
\end{equation}

\begin{equation}
\begin{split}
f(ij)=&\text{sh}t_i\text{sh}t_j[\text{ch}z_i\text{ch}z_j(\sin y_i\sin y_j\cos(x_i-x_j)+ \\
      &+\cos y_i\cos y_j)-\text{sh}z_i\text{sh}z_j]-\text{ch}t_i\text{ch}t_j,
\end{split}
\tag{2.47}
\end{equation}

$$
f(ij)=x_iy_j-x_jy_i+z_it_j-z_jt_i,
\eqno (2.48)
$$

\begin{equation}
\label{z14}
f(ij) = \frac{(x_i - x_j)^{2} \pm (y_i-y_j)^2 \pm (z_i-z_j)^2 +
\varepsilon_{i}t^{2}_i + \varepsilon_{j}t^{2}_j}{t_it_j},
\tag{$2.49$}
\end{equation}
{ \it где
$\varepsilon_{i} = 0, \pm1;$
$\varepsilon_{j} = 0, \pm1$,
причем не обязательно
$\varepsilon_{i} = \varepsilon_{j}$.}

\vspace {5mm}

Заметим, что в приведенном списке (2.38)--(2.49) отсутствуют
четырехмерные симплициальные и гельмгольцевы пространства.
По-види- мому, их среди феноменологически симметричных
четырехмерных геометрий ранга шесть просто нет. Для приведенных же
четырехмерных геометрий легко находятся уравнения, выражающие их
феноменологическую симметрию. Соответствующие определители
Кэли-Менгера седьмого порядка и определители Грама шестого порядка
обращаются в ноль. Групповая же симметрия степени 10 определяется
согласно теореме 3 из \S1 десятипараметрической группой движений,
сохраняющих метрическую функцию.

\newpage

     \begin{equation}
     \begin{split}
     f(ij)=\sin t_i\sin t_j[\sin z_i\sin z_j(\sin y_i\sin y_j\cos(x_i-x_j)+ \\
           +\cos y_i\cos y_j)+\cos z_i\cos z_j]+\cos t_i\cos t_j,
     \end{split}
     \tag{2.43}
     \end{equation}

     \begin{equation}
     \begin{split}
     f(ij)=&\cosh t_i\cosh t_j[\sin z_i\sin z_j(\sin y_i\sin y_j\cos(x_i-x_j)+ \\
           &+\cos y_i\cos y_j)+\cos z_i\cos z_j]-\sinh t_i\sinh t_j,
     \end{split}
     \tag{2.44}
     \end{equation}

     \begin{equation}
     \begin{split}
     f(ij)=&\sinh t_i\sinh t_j[\sin z_i\sin z_j(\sin y_i\sin y_j\cos(x_i-x_j)+ \\
           &+\cos y_i\cos y_j)+\cos z_i\cos z_j]-\cosh t_i\cosh t_j,
     \end{split}
     \tag{2.45}
     \end{equation}

     \begin{equation}
     \begin{split}
     f(ij)=&\cosh t_i\cosh t_j[\cosh z_i\cosh z_j(\sin y_i\sin y_j\cos(x_i-x_j)+ \\
           &+\cos y_i\cos y_j)-\sinh z_i\sinh z_j]-\sinh t_i\sinh t_j,
     \end{split}
     \tag{2.46}
     \end{equation}

     \begin{equation}
     \begin{split}
     f(ij)=&\sinh t_i\sinh t_j[\cosh z_i\cosh z_j(\sin y_i\sin y_j\cos(x_i-x_j)+ \\
           &+\cos y_i\cos y_j)-\sinh z_i\sinh z_j]-\cosh t_i\cosh t_j,
     \end{split}
     \tag{2.47}
     \end{equation}

     $$
     f(ij)=x_iy_j-x_jy_i+z_it_j-z_jt_i,
     \eqno (2.48)
     $$

     \begin{equation}
     \label{z14}
     f(ij) = \frac{(x_i - x_j)^{2} \pm (y_i-y_j)^2 \pm (z_i-z_j)^2 +
     \varepsilon_{i}t^{2}_i + \varepsilon_{j}t^{2}_j}{t_it_j},
     \tag{$2.49$}
     \end{equation}
     { \it where $\varepsilon_{i} = 0, \pm1;$ $\varepsilon_{j} = 0, \pm1$, $\varepsilon_{i}$
     being not necessarily equal to $\varepsilon_{j}$.}

     \vspace {5mm}

     We shall note that neither the 4-dimensional simplicial nor Helmholtz spaces  are on
     the list. Apparently, they are not among the phenomenologically
     symmetric \
     4-dimensional geometries of rank 6 at all. As to the  \ 4-dimensional geometries that we
     did manage to come by, the equations expressing their phenomenological symmetry
     are found quite easily. The respective Cayly-Menger determinants of the seventh
     order and the  sextic Gramians vanish to zero. The group symmetry of degree 10 is
     determined according to Theorem 3 of \S1 by the 10-parameter group of motions
     that preserve the metric function.

\newpage

\begin{center}
{\bf \large \S3. Двуметрические геометрии на плоскости и триметричекие
геометрии в пространстве}
\end{center}

Согласно общим определениям \S1 двуметрическая двумерная геометрия
$(s=2, n=1)$ задается на двумерном многообразии
$\mathfrak{M}$ двухкомпонентной метрической функцией $f=(f^1,f^2)$,
сопоставляющей любой паре $\langle  ij  \rangle$ из области ее определения
$\mathfrak{S}_f\subseteq\mathfrak{M}\times\mathfrak{M}$ два
действительных числа $f(ij)=(f^1(ij),f^2(ij))\in R^2$.
Предполагается, что область определения $\mathfrak{S}_f$ метрической функции
$f$ есть открытое и плотное в $\mathfrak{M}\times\mathfrak{M}$ множество.

Если $x,y$ -- локальные координаты в $\mathfrak{M}$, то ее
координатное представление
$$
f(ij)=f(x_i,y_i,x_j,y_j),
\eqno(3.1)
$$
или в более подробной по компонентам записи:
$$
\left.\begin{array}{rcl}
f^{1}(ij) = f^{1}(x_i,y_i,x_j,y_j),\\
f^{2}(ij) = f^{2}(x_i,y_i,x_j,y_j),\;
\end{array}\right\}
$$
является гладкой невырожденной функцией координат $x_i,y_i$ и $x_j,y_j$,
которые должны входить в нее существенным образом. Условие
невырожденности метрической функции $f=(f^1,f^2)$ математически выражается
двумя неравенствами:
\begin{equation}
\left.\begin{array}{rcl}
\partial(f^{1}(ij),f^{2}(ij))/\partial(x_i,y_i) \neq 0,\\
\partial(f^{1}(ij),f^{2}(ij))/\partial(x_j,y_j)\neq 0\;
\end{array}\right\}
\tag{$3.2$}
\end{equation}
для открытого и плотного в $\mathfrak{M} \times \mathfrak{M}$
множества пар $\langle  ij  \rangle$.

Если двухкомпонентная метрическая функция (3.1), удовлетворяющая
перечисленным выше трем условиям, задает на двумерном многообразии
$\mathfrak{M}$ феноменологически симметричную геометрию ранга три,
то для любой тройки $\langle  ijk  \rangle$ из плотного и открытого в $\mathfrak{M}^{3}$
множества, такой что пары $\langle  ij  \rangle, \langle  ik  \rangle, \langle  jk  \rangle$ принадлежат
$\mathfrak{S}_{f}$, шесть взаимных расстояний $f(ij), \ f(ik), \ f(jk)$
функционально связаны двумя независимыми уравнениями:
\begin{equation}
\label{t3}
\Phi(f(ij),f(ik),f(jk)) = 0,
\tag{$3.3$}
\end{equation}
где \ $\Phi=(\Phi_1,\Phi_2)$ -- двухкомпонентная функция шести
переменных. В

\newpage

     \begin{center}
     {\bf \large \S3. Dimetric geometries on a plane and \\ trimetric
     geometries in space}
     \end{center}

     Under the general definitions of \S1 the 2-dimensional dimetric geometry
     $(s=2, n=1)$ is defined on a two-dimensional manifold $\mathfrak{M}$ by a
     two-component metric function $f=(f^1,f^2)$ that assigns to each pair $\langle  ij  \rangle$
     from its domain $\mathfrak{S}_f\subseteq\mathfrak{M}\times\mathfrak{M}$ two
     real numbers $f(ij)=(f^1(ij),f^2(ij))\in R^2$. The domain $\mathfrak{S}_f$ of the metric
     function $f$ is supposed to be a set open and dense in $\mathfrak{M}\times\mathfrak{M}$.

     If $x,y$ are local coordinates  in $\mathfrak{M}$ its coordinate representation

     $$
     f(ij)=f(x_i,y_i,x_j,y_j),
     \eqno(3.1)
     $$ \\
     or in the more detailed, with respect to the components, writing:
     $$
     \left.\begin{array}{rcl}
     f^{1}(ij) = f^{1}(x_i,y_i,x_j,y_j),\\
     f^{2}(ij) = f^{2}(x_i,y_i,x_j,y_j),\;
     \end{array}\right\}
     $$
     is a smooth nondegenerate function of the coordinates $x_i,y_i$ and $x_j,y_j$, that
     must be included into it in an essential manner. The condition of the metric function
     $f=(f^1,f^2)$ being nondegenerate is expressed, mathematically, by two inequalities:
     \begin{equation}
     \left.\begin{array}{rcl}
     \partial(f^{1}(ij),f^{2}(ij))/\partial(x_i,y_i) \neq 0,\\
     \partial(f^{1}(ij),f^{2}(ij))/\partial(x_j,y_j)\neq 0\;
     \end{array}\right\}
     \tag{$3.2$}
     \end{equation}
     for the set of pairs $\langle  ij  \rangle$ open and dense in $\mathfrak{M} \times \mathfrak{M}$.

     If the two-component metric function (3.1) that satisfies the above three conditions
     gives on a two-dimensional manifold $\mathfrak{M}$ a phenomenologically symmetric
     geometry of rank 3, then for any triple $\langle  ijk  \rangle$ of a set dense and open in
     $\mathfrak{M}^{3}$, such that the pairs $\langle  ij  \rangle, \langle  ik  \rangle, \langle  jk  \rangle$ belong to
     $\mathfrak{S}_{f}$, the six reciprocal distances $f(ij), \ f(ik), \ f(jk)$ are functionally related
     by two independent equations:
     \begin{equation}
     \label{t3}
     \Phi(f(ij),f(ik),f(jk)) = 0,
     \tag{$3.3$}
     \end{equation}
     where  $\Phi=(\Phi_1,\Phi_2)$ is a two-component function of six variables.
     In  a more

\newpage

\noindent
более подробной записи по компонентам имеем:
$$
\left.\begin{array}{rcl}
\Phi_{1}(f^{1}(ij),f^{2}(ij),f^{1}(ik),f^{2}(ik),f^{1}(jk),f^{2}(jk)) = 0,\\
\Phi_{2}(f^{1}(ij),f^{2}(ij),f^{1}(ik),f^{2}(ik),f^{1}(jk),f^{2}(jk)) = 0,\;
\end{array}\right\}
$$
причем независимость этих уравнений означает, что $\text{rang} \ \Phi = 2$.

Мы хорошо уже знаем, что феноменологическая симметрия
геометрии тесно связана с ее
групповой симметрией. В частности, плоскость термодинамических состояний,
рассмотренная во Введении, с одной стороны, феноменологически
симметрична с рангом 3, а с другой -- наделена групповой симметрией степени
2. Такое соотношение ранга одной симметрии и степени другой не
случайно и является следствием их эквивалентности, устанавливаемой теоремой 3 из
\S1. Действительно,
в рассматриваемом случае трехточечная жесткая фигура
должна свободно двигаться с двумя
степенями свободы и не более, так как ее положение задается шестью
координатами $x_i,y_i,x_j,y_j,x_k,y_k$, на которые наложены четыре связи, происходящие от
сохранения шести расстояний, удовлетворяющих двум соотношениям $(3.3)$.
Эквивалентность феноменологической и групповой симметрий двуметрических
двумерных геометрий выразим для большей ясности отдельной теоремой:

\vspace{5mm}

{\bf Теорема 1.} {\it Для того, чтобы невырожденная двухкомпонентная
метрическая функция $f=(f^1,f^2)$ задавала на двумерном многообразии
$\mathfrak{M}$ двуметрическую
феноменологически симметричную двумерную геометрию
ранга три, необходимо и достаточно,
чтобы она задавала на том же многообразии двуметрическую двумерную
геометрию, наделенную групповой симметрией степени два. }

\vspace{5mm}

Таким образом, двухкомпонентная метрическая функция (3.1) феноменологически
симметричной ранга три двумерной геометрии
допускает двухпараметрическую группу движений,
то есть таких эффективных гладких локальных действий
в двумерном многообразии некоторой локальной группы Ли $G^{2}$:
\begin{equation}
x' = \lambda(x,y;a^{1},a^{2}),\,\,\
y' = \sigma(x,y;a^{1},a^{2}),
\tag{$3.4$}
\end{equation}

\newpage

     \noindent
     detailed, in the components, writing, we have:
      $$
     \left.\begin{array}{rcl}
     \Phi_{1}(f^{1}(ij),f^{2}(ij),f^{1}(ik),f^{2}(ik),f^{1}(jk),f^{2}(jk)) = 0,\\
     \Phi_{2}(f^{1}(ij),f^{2}(ij),f^{1}(ik),f^{2}(ik),f^{1}(jk),f^{2}(jk)) = 0,\;
     \end{array}\right\}
     $$
     the independence of these equations meaning that $\text{rang} \ \Phi = 2$.

     We know very well already that the phenomenological symmetry of a
     geometry is closely connected with the group symmetry of it. In particular, the plane
     of thermodynamical states, that we discussed in the Introduction, is, on the one
     hand, phenomenologically symmetric with rank 3, and on the other - endowed with
     a group symmetry of degree2. Such a relation of the rank of one symmetry and of the
     degree of the other is not occasional and is a corollary of their being equivalent,
     which is established by Theorem 3 of \S1. Indeed, in the case in question, a three-point
     rigid figure must move freely with two degrees of freedom and not more, because its
     position is defined by the six coordinates $x_i,y_i,x_j,y_j,x_k,y_k$, with the four
     relations that spring up from the six relations being preserved satisfying the relations $(3.3)$.
     For the sake of clarity,we shall express the equivalence of the phenomenological and group
     symmetries of the two-dimensional dimetric geometries with a special theorem:

     \vspace{10mm}

     {\bf Theorem 1.} {\it For the nondegenerate two-component metric function
     $f=(f^1,f^2)$ to give on a two-dimensional manifold $\mathfrak{M}$ a dimetric
     \\ phenomenologically symmetric two-dimensional geometry  of rank 3 it is necessary
     and sufficient that it should give on the same manifold a dimetric geometry endowed
     with group symmetry of degree 2. }

     \vspace{10mm}

     Thus, the two-component metric function (3.1) of a phenomenologically
     symmetric two-dimensional geometry of rank 3 allows a 2-parameter group of motions,
     i.e. of such effective smooth local actions in a two-dimensional manifold of some local
     Lie group $G^{2}$:

     \begin{equation}
     x' = \lambda(x,y;a^{1},a^{2}),\,\,\
     y' = \sigma(x,y;a^{1},a^{2}),
     \tag{$3.4$}
     \end{equation}

\newpage

\noindent
что каждая компонента метрической функции сохраняется:
\begin{equation}
f(\lambda(i),\sigma(i),\lambda(j),\sigma(j)) = f(x_i,y_i,x_j,y_j),
\tag{$3.5$}
\end{equation}
где $(a^{1},a^{2}) \in G^{2}$ и, например,
$\lambda(i)=\lambda(x_i,y_i;a^1,a^2)$.

Запишем базисные векторные поля $X_{1}$, $X_{2}$ двумерной алгебры
Ли локальных преобразований (3.4) двумерного многообразия
$\mathfrak{M}$ в операторной форме:
\begin{equation}
\left.\begin{array}{rcl}
X_{1} = \lambda_{1}(x,y) \partial _{x} + \sigma_{1}(x,y) \partial _{y},\\
X_{2} = \lambda_{2}(x,y) \partial _{x} + \sigma_{2}(x,y) \partial _{y},\;
\end{array}\right\}
\tag{$3.6$}
\end{equation}
где \ $\partial _{x} = \partial /\partial x$, \  $\partial _{y} = \partial
/\partial y$ \ и, \ например, \ $\lambda_{1}(x,y)=\partial \lambda
(x,y;a^{1},a^{2})/ \\ /\partial a^{1}|_{a^{1} = a^{2} = 0}$, предполагая, что
при $a^{1} = a^{2} = 0$ имеем тождественное преобразование в
группе (3.4). Метрическая функция (3.1), будучи по равенству
(3.5) двухточечным инвариантом группы преобразований
(3.4), необходимо является решением следующей системы двух дифференциальных
уравнений:
\begin{equation}
\left.\begin{array}{rcl}
X_{1}(i) f(ij) + X_{1}(j) f(ij) = 0,\\
X_{2}(i) f(ij) + X_{2}(j) f(ij) = 0.\;
\end{array}\right\}
\tag{$3.7$}
\end{equation}
с операторами (3.6).

В свое время (1893) Софус Ли дал полную классификацию конечномерных
локальных групп преобразований двумерного многообразия [16]. Из этой классификации можно
выделить и записать в надлежаще выбранной системе локальных координат
$(x,y)$ базисные операторы (3.6) четырех соответствующих
двумерных алгебр Ли:

\begin{equation}
X_{1} = \partial _{x},\,\,\,\
X_{2} = y \partial _{x};
\tag{$3.8$}
\end{equation}
\begin{equation}
X_{1} = \partial _{x},\,\,\,\
X_{2} = \partial _{y};
\tag{$3.9$}
\end{equation}
\begin{equation}
X_{1} = \partial _{x},\,\,\,\
X_{2} = x \partial _{x};
\tag{$3.10$}
\end{equation}
\begin{equation}
X_{1} = \partial _{x},\,\,\,\
X_{2} = x \partial _{x} + \partial _{y}.
\tag{$3.11$}
\end{equation}

\vspace{5mm}

{\bf Теорема 2.} {\it Существуют две и только две не сводимые друг
к другу двухкомпонентные метрические функции $f = (f^{1},f^{2})$,
задающие на двумерном многообразии $\mathfrak{M}$
феноменологически симметричные геометрии ранга три. С \ точностью
\ до \ масштабного \ преобразования}

\newpage

\noindent
that each component of the
metric function is preserved:
     \begin{equation}
     f(\lambda(i),\sigma(i),\lambda(j),\sigma(j)) = f(x_i,y_i,x_j,y_j),
     \tag{$3.5$}
     \end{equation}
     where $(a^{1},a^{2}) \in G^{2}$ and, for example, $\lambda(i)=\lambda(x_i,y_i;a^1,a^2)$.

     We shall write the basic vector fields $X_{1}$, $X_{2}$ of the two-dimensional Lie algebra
     of the local transformations (3.4) of the two-dimensional manifold $\mathfrak{M}$ in the
     operator from:
     \begin{equation}
     \left.\begin{array}{rcl}
     X_{1} = \lambda_{1}(x,y) \partial _{x} + \sigma_{1}(x,y) \partial _{y},\\
     X_{2} = \lambda_{2}(x,y) \partial _{x} + \sigma_{2}(x,y) \partial _{y},\;
     \end{array}\right\}
     \tag{$3.6$}
     \end{equation}
     where $\partial _{x} = \partial /\partial x$, $\partial _{y} = \partial/\partial y$ and, for
     example, $\lambda_{1}(x,y)=\partial \lambda(x,y;a^{1},a^{2})/ \\ \partial a^{1}|_{a^{1} = a^{2} = 0}$,
     supposing that with $a^{1} = a^{2} = 0$ we have an identity substitution in the group (3.4). The
     metric function (3.1), which is, in force of the equality (3.5), the two-point invariant of the group
     of transformations (3.4), is necessarily a solution of the system of two differential equations as
     follows:
     \begin{equation}
     \left.\begin{array}{rcl}
     X_{1}(i) f(ij) + X_{1}(j) f(ij) = 0,\\
     X_{2}(i) f(ij) + X_{2}(j) f(ij) = 0.\;
     \end{array}\right\}
     \tag{$3.7$}
     \end{equation}
     with the operators (3.6).

     In 1893, Sophus Lie gave a complete classification of the finite dimensional
     local groups of transformations of the two-dimensional manifold [16]. Out of
     his classification, it is possible to single out and write, in a suitably chosen
     system of local coordinates $(x,y)$, the basic operators (3.6) of the four
     respective two-dimensional Lie algebras:

     \begin{equation}
     X_{1} = \partial _{x},\,\,\,\
     X_{2} = y \partial _{x};
     \tag{$3.8$}
     \end{equation}
     \begin{equation}
     X_{1} = \partial _{x},\,\,\,\
     X_{2} = \partial _{y};
     \tag{$3.9$}
     \end{equation}
     \begin{equation}
     X_{1} = \partial _{x},\,\,\,\
     X_{2} = x \partial _{x};
     \tag{$3.10$}
     \end{equation}
     \begin{equation}
     X_{1} = \partial _{x},\,\,\,\
     X_{2} = x \partial _{x} + \partial _{y}.
     \tag{$3.11$}
     \end{equation}

     \vspace{5mm}

     {\bf Theorem 2.} {\it There are two and only two irreducible
     to each other two-component metric functions $f =
     (f^{1},f^{2})$
     that give on a two-dimensional manifold $\mathfrak{M}$
     a phenomenologically symmetric geometry of rank 3. With an}

\newpage

\noindent $\psi(f)\to f$, {\it  где $\psi=(\psi_1,\psi_2)$, и в
надлежаще выбранной системе локальных координат $(x, y)$ эти
метрические функции могут быть представлены следующими двумя
каноническими выражениями:}
\begin{equation}
f^{1}(ij) = x_i - x_j,\,\,\
f^{2}(ij) = y_i - y_j;
\tag{$3.12$}
\end{equation}
\begin{equation}
f^{1}(ij) = (x_i - x_j) y_i,\,\,\
f^{2}(ij) = (x_i - x_j) y_j.
\tag{$3.13$}
\end{equation}

\vspace{5mm}

Компоненты метрической функции $f = (f^{1},f^{2})$ являются
независимыми решениями системы уравнений (3.7). Поскольку в этих
уравнениях с операторами (3.8) и (3.10) отсутствует оператор
дифференцирования $\partial / \partial y_i$, их независимыми
решениями будут функции $y_i$ и $\varphi(ij)$. То есть для
метрической функции получаем выражение $f(ij) =
\psi(y_i,\varphi(ij))$, где $\psi: R^2 \to R^2$. Но для нее не
выполняется второе из условий (3.2), и она оказывается
вырожденной. Решения системы (3.7) с операторами (3.9) легко
находятся методом характеристик, и совпадают в своем явном
координатном представлении с компонентами метрической функции
(3.12). Решения системы (3.7) с операторами (3.11)  также легко
находятся, но совпадают в своем явном координатном представлении с
компонентами метрической функции (3.13) только при дополнительной
замене координат $x\to x, \ \exp(-y)\to y$. Обе метрические
функции невырождены, так как каждый из якобианов условия (3.2) для
них отличен от нуля.

Метрическую функцию (3.12) можно интерпретировать, например,
проекциями вектора $\vec{ji}$ на координатные оси. Соответствующая
функциональная связь (3.3) для нее задается двумя независимыми уравнениями

\begin{displaymath}
\left.\begin{array}{rcl}
f^{1}(ij) - f^{1}(ik) + f^{1}(jk) = 0,\\
f^{2}(ij) - f^{2}(ik) + f^{2}(jk) = 0.\;
\end{array}\right\}
\end{displaymath}

\vspace{5mm}

 Метрическая функция (3.13) допускает содержательную
физическую интерпретацию в термодинамике, подробно рассмотренную
во Введении. Соответствующая функциональная связь (3.3) для нее
задается двумя
независимыми уравнениями

     \newpage

      \noindent
     {\it accuracy up to
     a scaling transformation  $\psi(f)\to f$, where
     $\psi=(\psi_1,\psi_2)$, and in a suitably chosen system of local coordinates
     $(x, y)$ these metric
     functions may be represented
     by the two canonical expressions as follows:}
     \begin{equation}
     f^{1}(ij) = x_i - x_j,\,\,\
     f^{2}(ij) = y_i - y_j;
     \tag{$3.12$}
     \end{equation}
     \begin{equation}
     f^{1}(ij) = (x_i - x_j) y_i,\,\,\
     f^{2}(ij) = (x_i - x_j) y_j.
     \tag{$3.13$}
     \end{equation}

     \vspace{5mm}

     The components of the metric function $f = (f^{1},f^{2})$ are independent
     solutions of the system of equations (3.7). Since in these equations with the
     operators (3.8) and (3.10) there is missing an operator of differentiation
     $\partial / \partial y_i$, their independent solutions will be the functions $y_i$
     and $\varphi(ij)$. I.e. we get for the metric function the expression
     $f(ij) = \psi(y_i,\varphi(ij))$, where $\psi: R^2 \to R^2$. But it does not satisfy the
     second of the conditions (3.2), so it turns out to be a degenerate one. The solutions
     of the system (3.7) with the operators (3.9) are found easily by the method of
     characteristics, and they coincide in their explicit coordinate representation with the
     components of the metric function (3.12). The solutions of the system (3.7) with the
     operators (3.11)  are found as easily, but they only coincide in their explicit coordinate
     representation with the components of the metric function (3.13) on condition that
     the additional change of coordinates $x\to x,\exp(-y)\to y$ is introduced. Both
     metric functions are nondegenerate ones, because each of the two Jacobians of
     theirs in the condition (3.2) is unequal to zero.

     The metric function (3.12) may be interpreted, for example, by the projections of the
     vector $\vec{ji}$ on the coordinate axes. The corresponding functional relation (3.3) for
     it is defined by two independent equations

     \begin{displaymath}
     \left.\begin{array}{rcl}
     f^{1}(ij) - f^{1}(ik) + f^{1}(jk) = 0,\\
     f^{2}(ij) - f^{2}(ik) + f^{2}(jk) = 0.\;
     \end{array}\right\}
     \end{displaymath}

 \vspace{5mm}

     The metric function (3.13) allows the essential physical interpretation in
     thermodynamics discussed in detail in the Introduction. The corresponding
     functional relation (3.3) for it is defined by the two independent equations

\newpage

$$
\left.\begin{array}{cc}
\left|\begin{array}{ccc}
0 & -f^{2}(ij) & -f^{2}(ik) \\
f^{1}(ij) & 0 & -f^{2}(jk) \\
f^{1}(ik) & f^{1}(jk) & 0 \\
\end{array}
\right| = 0,\\
\phantom{aaaaabbbbbccccc} \\
\left|
\begin{array}{ccc}
f^{1}(ij) & f^{1}(jk) & -f^{2}(ik) \\
f^{1}(ik) & 0 & -f^{2}(ik) \\
f^{1}(ik) & -f^{2}(ij) & -f^{2}(jk) \\
\end{array}
\right| = 0.
\end{array}\right\}
$$

\vspace{5mm}

Далее рассмотрим триметрические
феноменологически симметричные геометрии ранга три, задаваемые на трехмерном
многообразии $\mathfrak{M}$ трехкомпонентной метрической функцией
$f=(f^1,f^2,f^3)$, которая каждой паре $\langle  ij  \rangle$ из области
ее определения $\mathfrak{S}_f\subseteq\mathfrak{M} \times \mathfrak{M}$
сопоставляет три числа
$f(ij)=(f^1(ij),f^2(ij),f^3(ij))\in R^3$.

Пусть $(x,y,z)$ -- локальные координаты в $\mathfrak{M}$.
Для метрической функции $f$ в некоторой окрестности
пары $\langle  ij  \rangle$ $\in\mathfrak{S}_{f}$ можно записать ее гладкое
координатное представление:
\begin{equation}
f(ij) = f(x_i,y_i,z_i,x_j,y_j,z_j).
\tag{$3.14$}
\end{equation}

Невырожденность метрической функции (3.14), в частности, ее
существенная зависимость от координат $x_i,y_i,z_i$ и
$x_j,y_j,z_j$ точек $i$ и $j$, означает необращение в нуль двух
якобианов третьего порядка:
\begin{equation}
\left.\begin{array}{rcl}
\partial(f^{1}(ij),f^{2}(ij),f^{3}(ij))/\partial(x_i,y_i,z_i) \neq 0,\\
\partial(f^{1}(ij),f^{2}(ij),f^{3}(ij))/\partial(x_j,y_j,z_j) \neq 0\;
\end{array}\right\}
\tag{$3.15$}
\end{equation}
для открытого и плотного в $\mathfrak{M} \times \mathfrak{M}$
множества пар $\langle ij \rangle$.

Если трехкомпонентная метрическая функция (3.14) задает на трехмерном многообразии
$\mathfrak{M}$ феноменологически симметричную геометрию ранга три,
то найдется такая трехкомпонентная функция
$\Phi = (\Phi_{1},\Phi_{2},\Phi_{3})$ от девяти переменных,
что девять взаимных расстояний между точками открытого и плотного в
$\mathfrak{M}^{3}$ множества троек $\langle  ijk  \rangle$
функционально связаны тремя независимыми уравнениями
\begin{equation}
\Phi(f(ij),f(ik),f(jk)) = 0.
\tag{$3.16$}
\end{equation}

\newpage

     $$
     \left.\begin{array}{cc}
     \left|\begin{array}{ccc}
     0 & -f^{2}(ij) & -f^{2}(ik) \\
     f^{1}(ij) & 0 & -f^{2}(jk) \\
     f^{1}(ik) & f^{1}(jk) & 0 \\
     \end{array}
     \right| = 0,\\
     \phantom{aaaaabbbbbccccc} \\
     \left|

     \begin{array}{ccc}
     f^{1}(ij) & f^{1}(jk) & -f^{2}(ik) \\
     f^{1}(ik) & 0 & -f^{2}(ik) \\
     f^{1}(ik) & -f^{2}(ij) & -f^{2}(jk) \\
     \end{array}
     \right| = 0.
     \end{array}\right\}
     $$

     \vspace{5mm}

     Further we shall discuss trimetric phenomenologically symmetric geomet- \ ries of
     rank 3 defined on a three-dimensional manifold $\mathfrak{M}$ by the three-
     component metric function $f=(f^1,f^2,f^3)$ which assigns to each pair $\langle  ij  \rangle$
     of its domain $\mathfrak{S}_f\subseteq\mathfrak{M} \times \mathfrak{M}$ three
     numbers $f(ij)=(f^1(ij),f^2(ij),f^3(ij))\in \ \in R^3$.

     Let $(x,y,z)$ be local coordinates in $\mathfrak{M}$. For the metric function $f$, in
     some neighbourhood of the pair $\langle  ij  \rangle$ $\in\mathfrak{S}_{f}$, it is possible to
     write down its smooth coordinate representation:

     \begin{equation}
     f(ij) = f(x_i,y_i,z_i,x_j,y_j,z_j).
     \tag{$3.14$}
     \end{equation}

     The nondegeneracy of the metric function (3.14), in particular its essential
     dependence on the coordinates $x_i,y_i,z_i$ and $x_j,y_j,z_j$ of the points $i$ and $j$,
     means nonvanishing to zero of two Jacobians of third order:
     \begin{equation}
     \left.\begin{array}{rcl}
     \partial(f^{1}(ij),f^{2}(ij),f^{3}(ij))/\partial(x_i,y_i,z_i) \neq 0,\\
     \partial(f^{1}(ij),f^{2}(ij),f^{3}(ij))/\partial(x_j,y_j,z_j) \neq 0\;
     \end{array}\right\}
     \tag{$3.15$}
     \end{equation}
     for the set of pairs $\langle ij \rangle$ open and dense in $\mathfrak{M} \times \mathfrak{M}$.

     If the three-component metric function (3.14) gives on a three-dimensional manifold
     $\mathfrak{M}$ a phenomenologically symmetric geometry of rank 3, then there exists
     a three-component function $\Phi = (\Phi_{1},\Phi_{2},\Phi_{3})$ of nine variables, such that
     the nine reciprocal distances among the points of the set of triples $\langle  ijk  \rangle$ open and dense
     in $\mathfrak{M}^{3}$ are functionally related by three independent equations
     \begin{equation}
     \Phi(f(ij),f(ik),f(jk)) = 0.
     \tag{$3.16$}
     \end{equation}

\newpage

Метрическая функция (3.14) допускает трехпараметрическую группу
движений:
\begin{equation}
\left.\begin{array}{rcl}
x' = \lambda(x,y,z;a^{1},a^{2},a^{3}),\\
y' = \sigma(x,y,z;a^{1},a^{2},a^{3}),\\
z' = \tau(x,y,z;a^{1},a^{2},a^{3}),\
\end{array}\right\}
\tag{$3.17$}
\end{equation}
относительно которой она является невырожденным двухточечным инвариантом,
удовлетворяя следующему функциональному уравнению:
$$
f(\lambda(i),\sigma(i),\tau(i),\lambda(j),\sigma(j),\tau(j)) =
f(x_i,y_i,z_i,x_j,y_j,z_j),
\eqno(3.18)
$$
где, например, $\lambda(i)=\lambda(x_i,y_i,z_i;a^1,a^2,a^3)$.
Вследствие локальной обратимости преобразований (3.17) должно выполняться условие:
$$
\partial(\lambda,\sigma,\tau) / \partial(x,y,z) \neq 0.
$$

Обозначим через
\begin{equation}
\left.\begin{array}{rcl}
X_{1} = \lambda_{1}(x,y,z) \partial _{x} + \sigma_{1}(x,y,z) \partial _{y}
+\tau_{1}(x,y,z) \partial_{z},\\
X_{2} = \lambda_{2}(x,y,z) \partial _{x} + \sigma_{2}(x,y,z) \partial _{y}
+ \tau_{2}(x,y,z) \partial_{z},\\
X_{3} = \lambda_{3}(x,y,z) \partial _{x} + \sigma_{3}(x,y,z) \partial _{y}
+ \tau_{3}(x,y,z) \partial_{z}\;
\end{array}\right\}
\tag{$3.19$}
\end{equation}
базисные операторы трехмерной алгебры Ли группы (3.17). Тогда для
метрической функции (3.14) как двухточечного инварианта получаем
из функционального уравнения (3.18) систему трех линейных
однородных дифференциальных уравнений в частных производных:
\begin{equation}
\left.\begin{array}{rcl}
X_{1}(i) f(ij) + X_{1}(j) f(ij) = 0,\\
X_{2}(i) f(ij) + X_{2}(j) f(ij) = 0,\\
X_{3}(i) f(ij) + X_{3}(j) f(ij) = 0\;
\end{array}\right\}
\tag{$3.20$}
\end{equation}
с операторами (3.19).

Таким образом, задача классификации метрических функций (3.14) сводится к
классификации трехмерных алгебр Ли преобразований трехмерного многообразия
с базисными операторами (3.19) и к интегрированию
соответствующих систем уравнений (3.20). Легко убедиться в том, что решение
этой системы определит невырожденную метрическую функцию только в том случае,
если группа преобразований (3.17) транзитивна, для чего, как известно,
необходимо и достаточно, чтобы ранг матрицы коэффициентов операторов (3.19)
был равен трем.

\newpage

     The metric function (3.14) allows a three-parameter group of motions:
     \begin{equation}
     \left.\begin{array}{rcl}
     x' = \lambda(x,y,z;a^{1},a^{2},a^{3}),\\
     y' = \sigma(x,y,z;a^{1},a^{2},a^{3}),\\
     z' = \tau(x,y,z;a^{1},a^{2},a^{3}),\
     \end{array}\right\}
     \tag{$3.17$}
     \end{equation}
     with respect to which it is a nondegenerate two-point invariant and satisfies the
     functional equation as follows:
     $$
     f(\lambda(i),\sigma(i),\tau(i),\lambda(j),\sigma(j),\tau(j)) = f(x_i,y_i,z_i,x_j,y_j,z_j),
     \eqno(3.18)
     $$
     where, for example, $\lambda(i)=\lambda(x_i,y_i,z_i;a^1,a^2,a^3)$. As a corollary of the
     local invertibility of the transformations (3.17), the condition must hold as follows:
     $$
     \partial(\lambda,\sigma,\tau) / \partial(x,y,z) \neq 0.
     $$

     We shall designate by
     \begin{equation}
     \left.\begin{array}{rcl}
     X_{1} = \lambda_{1}(x,y,z) \partial _{x} + \sigma_{1}(x,y,z) \partial _{y}
     +\tau_{1}(x,y,z) \partial_{z},\\
     X_{2} = \lambda_{2}(x,y,z) \partial _{x} + \sigma_{2}(x,y,z) \partial _{y}
     + \tau_{2}(x,y,z) \partial_{z},\\
     X_{3} = \lambda_{3}(x,y,z) \partial _{x} + \sigma_{3}(x,y,z) \partial _{y}
     + \tau_{3}(x,y,z) \partial_{z}\;
     \end{array}\right\}
     \tag{$3.19$}
     \end{equation}
     the basic operators of the three-dimensional Lie algebra of the group (3.17). Then, for
     the metric function (3.14), as a two-point invariant, we have, form the functional equation
     (3.18), a system of three linear homogeneous differential equations in partial derivatives:
     \begin{equation}
     \left.\begin{array}{rcl}
     X_{1}(i) f(ij) + X_{1}(j) f(ij) = 0,\\
     X_{2}(i) f(ij) + X_{2}(j) f(ij) = 0,\\
     X_{3}(i) f(ij) + X_{3}(j) f(ij) = 0\;
     \end{array}\right\}
     \tag{$3.20$}
     \end{equation}
     with the operators (3.19).

     Thus, the task of the classification of the metric functions (3.14) narrows down to that of
     classification of the three-dimensional Lie algebras of the transformations of the three-dimensional
     manifold with the basic operators (3.19) and to that of integrating the respective systems of
     equations (3.20). It is quite easy to make sure that the solution of the system will only give a
     nondegenerate metric function if the group of transformations (3.17) is transitive, for which, as
     is known, it is necessary and sufficient that the rank of the matrix of the coefficients of the
     operators (3.19) be equal to 3.

     \newpage

{\bf Теорема 3.} {\it Базисные операторы $(3.19)$ трехмерной
алгебры Ли локальной группы Ли локально транзитивных
преобразований трехмерного многообразия с точностью до изоморфизма
и в надлежаще выбранной системе локальных координат $(x,y,z)$
задаются следующими выражениями:}

\begin{equation}
X_{1} = \partial _{x},\,\,\,\
X_{2} = \partial _{y},\,\,\,\
X_{3} = \partial_{z};
\tag{$3.21$}
\end{equation}

\begin{equation}
X_{1} = \partial _{x},\,\,\,\
X_{2} = \partial _{y},\,\,\,\
X_{3} = y \partial_{x} + \partial_{z};
\tag{$3.22$}
\end{equation}

\begin{equation}
X_{1} = \partial _{x},\,\,\,\
X_{2} = \partial _{y},\,\,\,\
X_{3} = (x + y) \partial_{x} + y \partial_{y} + \partial_{z};
\tag{$3.23$}
\end{equation}

\begin{equation}
X_{1} = \partial _{x},\,\,\,\
X_{2} = \partial _{y},\,\,\,\
X_{3} = x \partial_{x} + py \partial_{y} + \partial_{z};
\tag{$3.24$}
\end{equation}

\begin{equation}
X_{1} = \partial _{x},\,\,\,\
X_{2} = \partial _{y},\,\,\,\
X_{3} = - y \partial_{x} + (x + qy) \partial_{y} + \partial_{z};
\tag{$3.25$}
\end{equation}

\begin{equation}
\left.\begin{array}{rcl}
X_{1} = \partial _{x},\\
X_{2} = \text{tg} \ y \ \sin x \ \partial_{x} + \cos x \ \partial_{y} +
\sec y \ \sin x \ \partial_{z},\\
X_{3} = \text{tg} \ y \ \cos x \ \partial_{x} - \sin x \ \partial_{y} +
\sec y \ \cos x \ \partial_{z};\;
\end{array}\right\}
\tag{$3.26$}
\end{equation}

\begin{equation}
\left.\begin{array}{rcl}
X_{1} = \partial _{x},\\
X_{2} = \sin x \ \partial_{x} + \cos x \ \partial_{y} +
\exp y \ \sin x \ \partial_{z},\\
X_{3} = \cos x \ \partial_{x} - \sin x \ \partial_{y} +
\exp y \ \cos x \ \partial_{z},\;
\end{array}\right\}
\tag{$3.27$}
\end{equation}
{\it где $- 1 \leqslant p \leqslant 1$, $0 \leqslant q < 2$.}

\vspace{5mm}

Приведенная в теореме 3 классификация построена автором и приведена по его
работе [17].

\vspace{5mm}

{\bf Теорема 4.}
{\it С точностью до масштабного преобразования $\psi(f)\to f$, где
$\psi=(\psi_1,\psi_2,\psi_3)$, и в надлежаще выбранной системе
локальных координат $(x,y,z)$ метрическая функция $f = (f^{1},$ $f^{2},f^{3})$,
задающая на трехмерном многообразии $\mathfrak{M}$
феноменологически симметричную геометрию ранга три, может быть
представлена одним из следующих одиннадцати канонических выражений:}

\newpage

     {\bf Theorem  3.}
     {\it The basic operators $(3.19)$ of the three-dimensional Lie algebra of the local Lie group of
     the locally transitive transformations of a three-dimensional manifold with an accuracy up to
     isomorphism and in a suitably chosen system of local coordinates $(x,y,z)$ are defined by the
     expressions as follows:}

     \begin{equation}
     X_{1} = \partial _{x},\,\,\,\
     X_{2} = \partial _{y},\,\,\,\
     X_{3} = \partial_{z};
     \tag{$3.21$}
     \end{equation}

     \begin{equation}
     X_{1} = \partial _{x},\,\,\,\
     X_{2} = \partial _{y},\,\,\,\
     X_{3} = y \partial_{x} + \partial_{z};
     \tag{$3.22$}
     \end{equation}

     \begin{equation}
     X_{1} = \partial _{x},\,\,\,\
     X_{2} = \partial _{y},\,\,\,\
     X_{3} = (x + y) \partial_{x} + y \partial_{y} + \partial_{z};
     \tag{$3.23$}
     \end{equation}

     \begin{equation}
     X_{1} = \partial _{x},\,\,\,\
     X_{2} = \partial _{y},\,\,\,\
     X_{3} = x \partial_{x} + py \partial_{y} + \partial_{z};
     \tag{$3.24$}
     \end{equation}

     \begin{equation}
     X_{1} = \partial _{x},\,\,\,\
     X_{2} = \partial _{y},\,\,\,\
     X_{3} = - y \partial_{x} + (x + qy) \partial_{y} + \partial_{z};
     \tag{$3.25$}
     \end{equation}

     \begin{equation}
     \left.\begin{array}{rcl}
     X_{1} = \partial _{x},\\
     X_{2} = \tan y \ \sin x \ \partial_{x} + \cos x \ \partial_{y} +
     \sec y \ \sin x \ \partial_{z},\\
     X_{3} = \tan y \ \cos x \ \partial_{x} - \sin x \ \partial_{y} +
     \sec y \ \cos x \ \partial_{z};\;
     \end{array}\right\}
     \tag{$3.26$}
     \end{equation}

     \begin{equation}
     \left.\begin{array}{rcl}
     X_{1} = \partial _{x},\\
     X_{2} = \sin x \ \partial_{x} + \cos x \ \partial_{y} +
     \exp y \ \sin x \ \partial_{z},\\
     X_{3} = \cos x \ \partial_{x} - \sin x \ \partial_{y} +
     \exp y \ \cos x \ \partial_{z},\;
     \end{array}\right\}
     \tag{$3.27$}
     \end{equation}
     {\it where $- 1 \leqslant p \leqslant 1$, $0 \leqslant q < 2$.}

     \vspace{5mm}

     The classification in this Theorem 3 was built up by the author and can be found in his
     note [17].

     \vspace{5mm}

     {\bf Theorem 4.}
     {\it With an accuracy up to a scaling transformation $\psi(f)\to f$, where
     $\psi=(\psi_1,\psi_2,\psi_3)$, and in a suitably chosen system of local coordinates
     $(x,y,z)$ the metric function $f = (f^{1},$ $f^{2},f^{3})$ that defines on a three-dimensional
     manifold $\mathfrak{M}$ a phenomenologically symmetric geometry of rank 3 may be
     represented by one of the following eleven canonical expressions:}

   \newpage

\begin{equation}
f^{1}(ij) = x_i - x_j, \ f^{2}(ij) = y_i - y_j, \ f^{3}(ij) = z_i - z_j;\;
\tag{3.28}
\end{equation}

\begin{equation}
\left.\begin{array}{r}
f^{1}(ij) = y_i - y_j,\\
f^{2}(ij) = (x_i - x_j) y_i + z_i - z_j,\\
f^{3}(ij) = (x_i - x_j) y_j + z_i - z_j;\;
\end{array}\right\}
\tag{$3.29$}
\end{equation}

\begin{equation}
\left.\begin{array}{rcl}
f^{1}(ij) = (x_i - x_j)^{2}\
\exp\left(2\displaystyle\frac{y_i - y_j}{x_i -x_j}\right),\\
f^{2}(ij) = (x_i - x_j) z_i, \ f^{3}(ij) = (x_i - x_j) z_j;\;
\end{array}\right\}
\tag{$3.30$}
\end{equation}

\begin{equation}
\left.\begin{array}{rcl}
f^{1}(ij) = \displaystyle\frac{y_i - y_j}{x_i - x_j},\\
f^{2}(ij) = (x_i - x_j) z_i, \\ f^{3}(ij) = (x_i - x_j) z_j;\;
\end{array}\right\}
\tag{$3.31$}
\end{equation}

\begin{equation}
\left.\begin{array}{rcl}
f^{1}(ij) = (x_i - x_j)(y_i - y_j),\\
f^{2}(ij) = (x_i - x_j) z_i, \\ f^{3}(ij) = (x_i - x_j) z_j;\;
\end{array}\right\}
\tag{$3.32$}
\end{equation}

\begin{equation}
\left.\begin{array}{rcl}
f^{1}(ij) = y_i - y_j,\\
f^{2}(ij) = (x_i - x_j) z_i, \\ f^{3}(ij) = (x_i - x_j) z_j;\;
\end{array}\right\}
\tag{$3.33$}
\end{equation}

\begin{equation}
\left.\begin{array}{rcl}
f^{1}(ij) = \displaystyle\frac{(x_i - x_j)^p}{y_i - y_j},\\
f^{2}(ij) = (x_i - x_j) z_i, \\ f^{3}(ij) = (x_i - x_j) z_j;\;
\end{array}\right\}
\tag{$3.34$}
\end{equation}

\begin{equation}
\left.\begin{array}{rcl}
f^{1}(ij) = (x_i - x_j)^{2} + (y_i - y_j)^{2}, \\
f^{2}(ij) = z_i +
\text{arctg}\left(\displaystyle\frac{y_i - y_j}{x_i -x_j}\right),\\
f^{3}(ij) = z_j +
\text{arctg}\left(\displaystyle\frac{y_i - y_j}{x_i -x_j}\right);\;
\end{array}\right\}
\tag{$3.35$}
\end{equation}

\begin{equation}
\left.\begin{array}{rcl}
f^{1}(ij) = ((x_i - x_j)^{2} + (y_i - y_j)^{2})\times \\
\times\exp\left(2\gamma\text{arctg}\displaystyle\frac{y_i-y_j}{x_i-x_j}\right), \\
f^{2}(ij) = z_i +
\text{arctg}\left(\displaystyle\frac{y_i - y_j}{x_i -x_j}\right),\\
f^{3}(ij) = z_j +
\text{arctg}\left(\displaystyle\frac{y_i - y_j}{x_i -x_j}\right);\;
\end{array}\right\}
\tag{$3.36$}
\end{equation}

     \begin{equation}
     f^{1}(ij) = x_i - x_j, \ f^{2}(ij) = y_i - y_j, \ f^{3}(ij) = z_i - z_j;\;
     \tag{3.28}
     \end{equation}

     \begin{equation}
     \left.\begin{array}{r}
     f^{1}(ij) = y_i - y_j,\\
     f^{2}(ij) = (x_i - x_j) y_i + z_i - z_j,\\
     f^{3}(ij) = (x_i - x_j) y_j + z_i - z_j;\;
     \end{array}\right\}
     \tag{$3.29$}
     \end{equation}

     \begin{equation}
     \left.\begin{array}{rcl}
     f^{1}(ij) = (x_i - x_j)^{2}\
     \exp\left(2\displaystyle\frac{y_i - y_j}{x_i -x_j}\right),\\
     f^{2}(ij) = (x_i - x_j) z_i, \ f^{3}(ij) = (x_i - x_j) z_j;\;
     \end{array}\right\}
     \tag{$3.30$}
     \end{equation}

     \begin{equation}
     \left.\begin{array}{rcl}
     f^{1}(ij) = \displaystyle\frac{y_i - y_j}{x_i - x_j},\\
     f^{2}(ij) = (x_i - x_j) z_i, \\ f^{3}(ij) = (x_i - x_j) z_j;\;
     \end{array}\right\}
     \tag{$3.31$}
     \end{equation}

     \begin{equation}
     \left.\begin{array}{rcl}
     f^{1}(ij) = (x_i - x_j)(y_i - y_j),\\
     f^{2}(ij) = (x_i - x_j) z_i, \\ f^{3}(ij) = (x_i - x_j) z_j;\;
     \end{array}\right\}
     \tag{$3.32$}
     \end{equation}

     \begin{equation}
     \left.\begin{array}{rcl}
     f^{1}(ij) = y_i - y_j,\\
     f^{2}(ij) = (x_i - x_j) z_i, \\ f^{3}(ij) = (x_i - x_j) z_j;\;
     \end{array}\right\}
     \tag{$3.33$}
     \end{equation}

     \begin{equation}
     \left.\begin{array}{rcl}
     f^{1}(ij) = \displaystyle\frac{(x_i - x_j)^p}{y_i - y_j},\\
     f^{2}(ij) = (x_i - x_j) z_i, \\ f^{3}(ij) = (x_i - x_j) z_j;\;
     \end{array}\right\}
     \tag{$3.34$}
     \end{equation}

     \begin{equation}
     \left.\begin{array}{rcl}
     f^{1}(ij) = (x_i - x_j)^{2} + (y_i - y_j)^{2}, \\
     f^{2}(ij) = z_i + \arctan \left(\displaystyle\frac{y_i - y_j}{x_i -x_j}\right),\\
     f^{3}(ij) = z_j + \arctan \left(\displaystyle\frac{y_i - y_j}{x_i -x_j}\right);\;
     \end{array}\right\}
     \tag{$3.35$}
     \end{equation}

     \begin{equation}
     \left.\begin{array}{rcl}
     f^{1}(ij) = ((x_i - x_j)^{2} + (y_i - y_j)^{2})\times \\
     \times\exp\left(2\gamma\arctan \displaystyle\frac{y_i-y_j}{x_i-x_j}\right), \\
     f^{2}(ij) = z_i + \arctan \left(\displaystyle\frac{y_i - y_j}{x_i -x_j}\right),\\
     f^{3}(ij) = z_j + \arctan \left(\displaystyle\frac{y_i - y_j}{x_i -x_j}\right);\;
     \end{array}\right\}
     \tag{$3.36$}
     \end{equation}

 \newpage

\begin{equation}
\left.\begin{array}{rcl}
f^{1}(ij) = \sin y_i \sin y_j\cos(x_i - x_j) + \cos y_i \cos y_j,\\
f^{2}(ij) = z_i - \text{arcsin} \left( \displaystyle
\frac{\sin (x_i - x_j) \sin y_j}{\sqrt{1
- (f^{1}(ij))^{2}}}\right),\\
f^{3}(ij) = z_j +
\text{arcsin}\left( \displaystyle
\frac{\sin (x_i - x_j) \sin y_i}{\sqrt{1 - (f^{1}(ij))^{2}}}\right);\;
\end{array}\right\}
\tag{$3.37$}
\end{equation}

\begin{equation}
\left.\begin{array}{rcl}
f^{1}(ij) = (x_i - x_j) y_i y_j,\\
\phantom{aaaaabbbbbccccc} \\
f^{2}(ij) = z_i + \displaystyle\frac{1}{(x_i - x_j) y^2_i},\\
\phantom{aaaaabbbbbccccc}  \\
f^{3}(ij) = z_j - \displaystyle\frac{1}{(x_i - x_j) y^2_j},\;
\end{array}\right\}
\tag{$3.38$}
\end{equation}
{\it причем здесь $0<|p|<1$ и \ $0<\gamma < \infty$, где
$\gamma = q/\sqrt{4 - q^{2}}$ при \  $0<q < 2$.}
\vspace{5mm}

Подробное доказательство теоремы 4 можно найти в \S5 монографии автора [10].
Оно состоит в последовательном решении всех систем уравнений (3.20) с
операторами (3.21)--(3.27), что в техническом отношении
особых трудностей не представляет. При окончательной записи приведенных выше
канонических выражений произведена в некоторых случаях удобная замена
координат и выделены случаи $p=0,\pm1$; $\gamma = 0$.

\vspace{5mm}

Феноменологическая симметрия всех одиннадцати геометрий, задаваемых
метрическими функциями (3.28)--(3.38), устанавливается по рангу функциональной
матрицы для девяти функций $f(ij),f(ik),f(jk)$,  специальным образом
зависящих от девяти переменных -- координат точек тройки $\langle ijk \rangle$, который
оказывается равен шести. Соответствующие функциональные уравнения (3.16),
выражающие эту симметрию, могут быть найдены в явном виде, что утверждается
следующей теоремой, доказанной Р.М. Мурадовым
в работе, отправленной в печать:

\vspace {5mm}

\textbf{Теорема 5.} \emph{Если трехкомпонентная метрическая
функция $(3.14)$} \emph{задает на трехмерном многообразии}
$\mathfrak{M}$ \emph{феноменологически симметричную геометрию
ранга 3, \ то с точностью \ до \ замены локальных}

\newpage

     \begin{equation}
     \left.\begin{array}{rcl}
     f^{1}(ij) = \sin y_i \sin y_j\cos(x_i - x_j) + \cos y_i \cos y_j,\\
     f^{2}(ij) = z_i - \text{arcsin} \left( \displaystyle
     \frac{\sin (x_i - x_j) \sin y_j}{\sqrt{1
     - (f^{1}(ij))^{2}}}\right),\\
     f^{3}(ij) = z_j + \text{arcsin}\left( \displaystyle
     \frac{\sin (x_i - x_j) \sin y_i}{\sqrt{1 - (f^{1}(ij))^{2}}}\right);\;
     \end{array}\right\}
     \tag{$3.37$}
     \end{equation}

     \begin{equation}
     \left.\begin{array}{rcl}
     f^{1}(ij) = (x_i - x_j) y_i y_j,\\
     \phantom{aaaaabbbbbccccc} \\
     f^{2}(ij) = z_i + \displaystyle\frac{1}{(x_i - x_j) y^2_i},\\
     \phantom{aaaaabbbbbccccc}  \\
     f^{3}(ij) = z_j - \displaystyle\frac{1}{(x_i - x_j) y^2_j},\;
     \end{array}\right\}
     \tag{$3.38$}
     \end{equation}
     {\it herein $0<|p|<1$ и \ $0<\gamma < \infty$, where $\gamma = q/\sqrt{4 - q^{2}}$ with \  $0<q < 2$.}
     \vspace{5mm}

     The detailed proof of Theorem 4 is given in \S5 of the author's monograph [10]. Essentially,
     it is the solution, one after another, of all the systems of equations (3.20) with the operators
     (3.21)--(3.27), which does not pose any technical difficulties. In the final form of the canonical
     expressions given above, in some places a suitable change of coordinates was performed and
     the cases of $p=0,\pm1$; $\gamma = 0$ were singled out.

     \vspace{5mm}

     The phenomenological symmetry of all the eleven geometries defined by the metric
     functions (3.28)--(3.38) is established by the rank of the functional matrix for the nine
     functions $f(ij),f(ik),f(jk)$ that depend in special manner on nine variables - the coordinates
     of the points of the triple $\langle ijk \rangle$, and that rank turns out to be equal to 6. The respective
     functional equations (3.16) that express that symmetry may be found in the explicit form,
     which is the essence of the following theorem, that was proved by R.M. Muradov in his
     note (in print).
     \vspace {5mm}

     \textbf{Theorem 5.} \emph{If a three-component metric function $(3.14)$} \emph{gives
     on a three-dimensional manifold} $\mathfrak{M}$\emph {a phenomenologically
     symmetric geometry of rank 3, then with an accuracy up to a change of local coordinates
     in the}

\newpage

\noindent
\emph{координат в многообразии и масштабного преобразования
$\psi(f)\to f$, где $\psi=(\psi_1,\psi_2,\psi_3)$, она определяет
в} $R^3$ \emph{такую локальную квазигрупповую операцию с правой
единицей, что правый обратный элемент совпадает с исходным и
уравнение, выражающее феноменологическую симметрию, имеет подобный
самой метрической функции вид:}

$$
f(ij)=f(f^1(ik), f^2(ik), f^3(ik), f^1(jk), f^2(jk), f^3(jk)).
\eqno (3.39)
$$

\vspace{5mm}

В указанной работе для каждой из метрических функций (3.28)--(3.38) найдены
такие, соответствующие теореме 5, ее координатные представления, которые позволяют
записать уравнение (3.16) по формуле (3.39) в явном виде.

\vspace{5mm}

Компоненты метрической функции (3.28) можно интерпретировать
проекциями вектора $\vec{ji}$ на координатные оси.
Соответствующая функциональная связь (3.16) задается
системой трех  независимых уравнений:
\begin{displaymath}
\left.\begin{array}{rcl}
f^{1}(ij) - f^{1}(ik) + f^{1}(jk)=0,\\
f^{2}(ij) - f^{2}(ik) + f^{2}(jk)=0,\\
f^{3}(ij) - f^{3}(ik) + f^{3}(jk)=0.\;
\end{array}\right\}
\end{displaymath}

Метрическая функция (3.29) допускает содержательную физическую интерпретацию
в термодинамике. Первую ее компоненту представим как разность температур
$T_i$ и $T_j$ термодинамической системы в состояниях $i$ и $j$,
а вторую и третью -- как работы $A^{TS}(ij)$ и $A^{ST}(ij)$
внешних тел над ней при ее переходе из состояния $i$ в состояние
$j$ по двум путям, составленным из равновесных
изотермического ($T = \text{const}$) и адиабатического ($S = \text{const}$)
процессов:
\begin{displaymath}
\left.\begin{array}{rcl}
f^{1} (ij) = T_i - T_j,\\
f^{2} (ij) = A^{TS}(ij) = (S_i - S_j)T_i - U_i + U_j,\\
f^{3} (ij) = A^{ST}(ij) = (S_i - S_j)T_j - U_i + U_j,\;
\end{array}\right\}
\end{displaymath}
где $S$, $T$ и $U$ - энтропия, температура и внутренняя энергия
системы. Соответствующая функциональная связь (3.16) задается
тремя независимыми уравнениями:

\newpage

      \noindent
     \emph{ manifold and a scaling
     transformation $\psi(f)\to f$, where
     $\psi=(\psi_1,\psi_2,\psi_3)$, it defines in}$R^3$ \emph{a local quazigroup operation
     with a right identity such that the right inverse coincides with the original one and
     the equation that expresses the phenomenological symmetry has a form similar to that
     of the metric function itself:}

     $$
     f(ij)=f(f^1(ik), f^2(ik), f^3(ik), f^1(jk), f^2(jk), f^3(jk)).
     \eqno (3.39)
     $$

     \vspace{5mm}

     In the note pointed out, for each of the metric functions (3.28)--(3.38) there are
     coordinate representations corresponding Theorem 5 discovered such that make it
     possible to write the equation (3.16) by the formula (3.39) in the explicit form.

     \vspace{5mm}

     The components of the metric function (3.28) may be interpreted by way of the
     projections of the vector $\vec{ji}$ onto the coordinate axes. The corresponding functional
     relation (3.16) is defined by the system of three independent equations:
     \begin{displaymath}
     \left.\begin{array}{rcl}
     f^{1}(ij) - f^{1}(ik) + f^{1}(jk)=0,\\
     f^{2}(ij) - f^{2}(ik) + f^{2}(jk)=0,\\
     f^{3}(ij) - f^{3}(ik) + f^{3}(jk)=0.\;
     \end{array}\right\}
     \end{displaymath}

     The metric function (3.29) allows an essential physical interpretation in thermodynamics.
     Let us consider the first component of it the difference between the temperatures $T_i$
     and $T_j$ of a thermodynamic system in the states $i$ and $j$, and the second and third
     - works $A^{TS}(ij)$ and $A^{ST}(ij)$ done by outward bodies when transferring the
     system from the state $i$ to the state $j$, along the two-way process, i.e. comprised of
     an equilibrium isothermic ($T = \text{const}$) and adiabatic ($S = \text{const}$)
     processes:
     \begin{displaymath}
     \left.\begin{array}{rcl}
     f^{1} (ij) = T_i - T_j,\\
     f^{2} (ij) = A^{TS}(ij) = (S_i - S_j)T_i - U_i + U_j,\\
     f^{3} (ij) = A^{ST}(ij) = (S_i - S_j)T_j - U_i + U_j,\;
     \end{array}\right\}
     \end{displaymath}
     where $S$, $T$ and $U$ are respectively the entropy, temperature, and internal
     energy of the system. The corresponding functional relation (3.16) is defined
     by three independent equations:

\newpage

\begin{displaymath}
\left.\begin{array}{rcl}
f^{1} (ij) - f^{1} (ik) + f^{1} (jk) = 0,\\
\phantom{aaaaabbbbbccccc} \\
\displaystyle\frac{f^{2} (ij) - f^{3} (ij)}{f^{1} (ij)} -
\displaystyle\frac{f^{2} (ik) - f^{3} (ik)}{f^{1} (ik)} +
\displaystyle\frac{f^{2} (jk) - f^{3} (jk)}{f^{1} (jk)} = 0, \\
\phantom{aaaaabbbbbccccc} \\
\displaystyle\frac{f^{3} (ij) - f^{3} (ik) + f^{2} (jk)}{f^{1} (jk)} -
\displaystyle\frac{f^{2} (ik) - f^{3} (ik)}{f^{1} (ik)} = 0.\;
\end{array}\right\}
\end{displaymath}

В термодинамике же можно интерпретировать еще и компоненты метрической функции
(3.33) разностью температур и работами среды над системой
при ее переходе из состояния $i$ в состояние $j$
по путям $PV$ и $VP$, где $P$ и $V$ -- давление и объем системы.
Вопрос об интерпретации остальных
триметрических геометрий остается пока открытым. Их нетривиальные симметрии,
групповая и феноменологическая, обуславливающие друг друга,
дают основание надеяться, что такие интерпретации будут найдены
и для других метрических функций
классификационного списка (3.28)--(3.38).

\vspace{5mm}

К настоящему времени В.А.Кыровым построена классификация
четыреметрических (s=4,\ n=1) феноменологически симметричных геометрий
ранга три, которую приведем по его работе [18]:

\vspace{5mm}

{\bf Теорема 6.}
{\it С точностью до масштабного преобразования $\psi(f)\to f$, где
$\psi=(\psi_1,\psi_2,\psi_3,\psi_4)$, и в надлежаще выбранной системе
локальных координат $(x,y,z,t)$ метрическая функция
$f = (f^{1},f^{2},f^{3},f^4)$,
задающая на четырехмерном многообразии $\mathfrak{M}$
феноменологически симметричную геометрию ранга три, может быть
представлена {\bf явно} одним из следующих двенадцати канонических выражений:}

$$
\left. {\begin{array}{ccc}
f^1(ij)=(x_i-x_j)^2\exp[\varepsilon(t_i+t_j)],\,\,
f^2(ij)=(y_i-y_j)^2\exp[k(t_i+t_j)], \\
f^3(ij)=(z_i-z_j)^2\exp[l(t_i+t_j)],\,\, f^4(ij)=t_i-t_j;
\end{array}}
\right\}
$$

 \newpage

     \begin{displaymath}
     \left.\begin{array}{rcl}
     f^{1} (ij) - f^{1} (ik) + f^{1} (jk) = 0,\\
     \phantom{aaaaabbbbbccccc} \\
     \displaystyle\frac{f^{2} (ij) - f^{3} (ij)}{f^{1} (ij)} -
     \displaystyle\frac{f^{2} (ik) - f^{3} (ik)}{f^{1} (ik)} +
     \displaystyle\frac{f^{2} (jk) - f^{3} (jk)}{f^{1} (jk)} = 0, \\
     \phantom{aaaaabbbbbccccc} \\
     \displaystyle\frac{f^{3} (ij) - f^{3} (ik) + f^{2} (jk)}{f^{1} (jk)} -
     \displaystyle\frac{f^{2} (ik) - f^{3} (ik)}{f^{1} (ik)} = 0.\;
     \end{array}\right\}
     \end{displaymath}

     Moreover, it is also possible to attach essential thermodynamic interpreta- \ tions
     to the components of the metric function (3.33), by way of the tempera- \ ture difference
     and the works of the mediums done to the system in turning it from the state $i$ to the
     state $j$ along the ways $PV$ and $VP$, where $P$ is the pressure and $V$ the volume of
     the system. The question of interpretation of other trimetric geometries is still open. Their
     nontrivial symmetries, the group and the phenomenological ones, that condition each
     other, give some grounds for hopes that such interpretations will be found for other
     metric functions of the classification list (3.28)--(3.38).

     \vspace{5mm}

     By now, V.A. Kyrov has built a classification of four-metric (s=4,\ n=1) phenomenologically
     symmetric geometries of rank 3 that we shall give after his note [18]:

     \vspace{5mm}

     {\bf Theorem 6.}
     {\it With an accuracy up to a scaling transformation $\psi(f)\to f$ where
     $\psi=(\psi_1,\psi_2,\psi_3,\psi_4)$ in a suitably chosen system of local coordinates
     $(x,y,z,t)$, the metric function $f = (f^{1},f^{2},f^{3},f^4)$ that defines on a four-
     dimensional manifold $\mathfrak{M}$ a phenomenologically symmetric geometry
     of rank 3 may be represented {\bf explicitly} by one of the following twelve canonical
     expressions:}

     $$
     \left. {\begin{array}{ccc}
     f^1(ij)=(x_i-x_j)^2\exp[\varepsilon(t_i+t_j)],\,\,
     f^2(ij)=(y_i-y_j)^2\exp[k(t_i+t_j)], \\
     f^3(ij)=(z_i-z_j)^2\exp[l(t_i+t_j)],\,\, f^4(ij)=t_i-t_j;
     \end{array}}
     \right\}
     $$

\newpage

$$
\left. {\begin{array}{ccc} f^1(ij)=[(x_i-x_j)^2+
(y_i-y_j)^2]\exp\left(-2k\
\text{arctg}\dfrac{y_i-y_j}{x_i-x_j}\right),
\\
f^2(ij)=2\text{arctg}\dfrac{y_i-y_j}{x_i-x_j}+t_i+t_j, \\
f^3(ij)=(z_i-z_j)^2\exp[l(t_i+t_j)],\,\, f^4(ij)=t_i-t_j;
\end{array}}
\right\}
$$

$$
\left. {\begin{array}{ccc}
f^1(ij)=(x_i-x_j)^2\exp\left(-2k\dfrac{y_i-y_j}{x_i-x_j}\right),\,\,
f^2(ij)=2\dfrac{y_i-y_j}{x_i-x_j}+t_i+t_j, \\
f^3(ij)=(z_i-z_j)^2\exp[\varepsilon(t_i+t_j)],\,\,
f^4(ij)=t_i-t_j;
\end{array}}
\right\}
$$

$$
\left. {\begin{array}{ccc}
f^1(ij)=x_i-x_j,\,\,
f^2(ij)=2\dfrac{y_i-y_j}{x_i-x_j}-(t_i+t_j), \\
f^3(ij)=z_i-z_j-\dfrac{(y_i-y_j)^2}{2(x_i-x_j)},\,\,
f^4(ij)=t_i-t_j;
\end{array}}
\right\}
$$

$$
\left. {\begin{array}{ccc} f^1(ij)=x_i-x_j, \ \
f^2(ij)=2\dfrac{y_i-y_j}{x_i-x_j}-(t_i+t_j), \\
f^3(ij)=(x_i-x_j)\ln(z_i-z_j+y_i-y_j+x_i-x_j)-y_i+y_j, \\
f^4(ij)=t_i-t_j;
\end{array}}
\right\}
$$

$$
\left. {\begin{array}{ccc}
f^1(ij)=(x_i-x_j)^2\exp\left(-2k\dfrac{y_i-y_j}{x_i-x_j}\right),\,\,
f^2(ij)=2\dfrac{y_i-y_j}{x_i-x_j}-(t_i+t_j), \\
f^3(ij)=k(y_i-y_j)-(x_i-x_j)-k^2(z_i-z_j),\,\, f^4(ij)=t_i-t_j;
\end{array}}
\right\}
$$

$$
\left. {\begin{array}{ccc}
f^1(ij)=(x_i-x_j)^2\exp\left(-2k\dfrac{y_i-y_j}{x_i-x_j}\right),\,\,
f^2(ij)=2\dfrac{y_i-y_j}{x_i-x_j}-(t_i+t_j), \\
f^3(ij)=2\dfrac{z_i-z_j}{x_i-x_j}-k\left(\dfrac{y_i-y_j}{x_i-x_j}\right)^2,\,\,
f^4(ij)=t_i-t_j;
\end{array}}
\right\}
$$

$$
\left. {\begin{array}{ccc}
f^1(ij)=(x_i-x_j-z_i(y_i-y_j))^2\exp[c(t_i+t_j)], \\
f^2(ij)=(x_i-x_j-z_j(y_i-y_j))^2\exp[c(t_i+t_j)], \\
f^3(ij)=(y_i-y_j)^2\exp(t_i+t_j),\,\, f^4(ij)=t_i-t_j;
\end{array}}
\right\}
$$

$$
\left. {\begin{array}{ccc} f^1(ij)=(x_i-x_j)e^{z_i},\,\,
f^2(ij)=(x_i-x_j)e^{z_j}, \\
f^3(ij)=(y_i-y_j)e^{t_i},\,\, f^4(ij)=(y_i-y_j)e^{t_j};
\end{array}}
\right\}
$$

$$
\left. {\begin{array}{ccc}
f^1(ij)=[(x_i-x_j)^2+(y_i-y_j)^2]\exp(z_i+z_j), \\
f^2(ij)=2\text{arctg}\dfrac{y_i-y_j}{x_i-x_j}+t_i+t_j,\,\,
f^3(ij)=z_i-z_j,\,\, f^4(ij)=t_i-t_j;
\end{array}}
\right\}
$$

\newpage

     $$
     \left. {\begin{array}{ccc}
     f^1(ij)=[(x_i-x_j)^2+
     (y_i-y_j)^2]\exp\left(-2k\ \arctan \dfrac{y_i-y_j}{x_i-x_j}\right), \\
     f^2(ij)=2\arctan \dfrac{y_i-y_j}{x_i-x_j}+t_i+t_j, \\
     f^3(ij)=(z_i-z_j)^2\exp[l(t_i+t_j)],\,\, f^4(ij)=t_i-t_j;
     \end{array}}
     \right\}
     $$

     $$
     \left. {\begin{array}{ccc}
     f^1(ij)=(x_i-x_j)^2\exp\left(-2k\dfrac{y_i-y_j}{x_i-x_j}\right),\,\,
     f^2(ij)=2\dfrac{y_i-y_j}{x_i-x_j}+t_i+t_j, \\
     f^3(ij)=(z_i-z_j)^2\exp[\varepsilon(t_i+t_j)],\,\,
     f^4(ij)=t_i-t_j;
     \end{array}}
     \right\}
     $$

     $$
     \left. {\begin{array}{ccc}
     f^1(ij)=x_i-x_j,\,\,
     f^2(ij)=2\dfrac{y_i-y_j}{x_i-x_j}-(t_i+t_j), \\
     f^3(ij)=z_i-z_j-\dfrac{(y_i-y_j)^2}{2(x_i-x_j)},\,\,
     f^4(ij)=t_i-t_j;
     \end{array}}
     \right\}
     $$

     $$
     \left. {\begin{array}{ccc}
     f^1(ij)=x_i-x_j,\,\,
     f^2(ij)=2\dfrac{y_i-y_j}{x_i-x_j}-(t_i+t_j), \\
     f^3(ij)=(x_i-x_j)\ln(z_i-z_j+y_i-y_j+x_i-x_j)-y_i+y_j,\,\, \\
     f^4(ij)=t_i-t_j;
     \end{array}}
     \right\}
     $$

     $$
     \left. {\begin{array}{ccc}
     f^1(ij)=(x_i-x_j)^2\exp\left(-2k\dfrac{y_i-y_j}{x_i-x_j}\right),\,\,
     f^2(ij)=2\dfrac{y_i-y_j}{x_i-x_j}-(t_i+t_j), \\
     f^3(ij)=k(y_i-y_j)-(x_i-x_j)-k^2(z_i-z_j),\,\, f^4(ij)=t_i-t_j;
     \end{array}}
     \right\}
     $$

     $$
     \left. {\begin{array}{ccc}
     f^1(ij)=(x_i-x_j)^2\exp\left(-2k\dfrac{y_i-y_j}{x_i-x_j}\right),\,\,
     f^2(ij)=2\dfrac{y_i-y_j}{x_i-x_j}-(t_i+t_j), \\
     f^3(ij)=2\dfrac{z_i-z_j}{x_i-x_j}-k\left(\dfrac{y_i-y_j}{x_i-x_j}\right)^2,\,\,
     f^4(ij)=t_i-t_j;
     \end{array}}
     \right\}
     $$

     $$
     \left. {\begin{array}{ccc}
     f^1(ij)=(x_i-x_j-z_i(y_i-y_j))^2\exp[c(t_i+t_j)], \\
     f^2(ij)=(x_i-x_j-z_j(y_i-y_j))^2\exp[c(t_i+t_j)], \\
     f^3(ij)=(y_i-y_j)^2\exp(t_i+t_j),\,\, f^4(ij)=t_i-t_j;
     \end{array}}
     \right\}
     $$

     $$
     \left. {\begin{array}{ccc} f^1(ij)=(x_i-x_j)e^{z_i},\,\,
     f^2(ij)=(x_i-x_j)e^{z_j}, \\
     f^3(ij)=(y_i-y_j)e^{t_i},\,\, f^4(ij)=(y_i-y_j)e^{t_j};
     \end{array}}
     \right\}
     $$

     $$
     \left. {\begin{array}{ccc}
     f^1(ij)=[(x_i-x_j)^2+(y_i-y_j)^2]\exp(z_i+z_j), \\
     f^2(ij)=2\arctan \dfrac{y_i-y_j}{x_i-x_j}+t_i+t_j,\,\,
     f^3(ij)=z_i-z_j,\,\, f^4(ij)=t_i-t_j;
     \end{array}}
     \right\}
     $$

\newpage

$$
\left. {\begin{array}{ccc}
f^1(ij)=\sin y_i\sin y_j\cos(x_i-x_j)+\cos y_i\cos y_j, \\
f^2(ij)=z_i-\arcsin\dfrac{\sin(x_i-x_j)\sin
y_j}{\sqrt{1-(f^1(ij))^2)}}, \\
f^3(ij)=z_j+\arcsin\dfrac{\sin(x_i-x_j)\sin
y_i}{\sqrt{1-(f^1(ij))^2)}}, \\
 f^4(ij)=t_i-t_j;
\end{array}}
\right\}
$$

$$
\left. {\begin{array}{ccc} f^1(ij)=(x_i-x_j)y_iy_j,\,\,
f^2(ij)=z_i+\dfrac{1}{(x_i-x_j)y_i^2}, \\
f^3(ij)=z_j-\dfrac{1}{(x_i-x_j)y_j^2},\,\, f^4(ij)=t_i-t_j;
\end{array}}
\right\}
$$
\emph{а также  {\bf неявно} еще двумя выражениями:}

$$
\left. {\begin{array}{ccc}
f^1(ij)=f^1(x_i-x_j-z_i(y_i-y_j),x_i-x_j-z_j(y_i-y_j),y_i-y_j,t_i,t_j), \\
f^2(ij)=f^2(x_i-x_j-z_i(y_i-y_j),x_i-x_j-z_j(y_i-y_j),y_i-y_j,t_i,t_j), \\
f^3(ij)=f^3(x_i-x_j-z_i(y_i-y_j),x_i-x_j-z_j(y_i-y_j),y_i-y_j,t_i,t_j), \\
f^4(ij)=t_i-t_j,
\end{array}}
\right\}
$$
\emph{в которых четыре компоненты функции $f=f(u,v,w,t_i,t_j)$
являются независимыми интегралами либо уравнения}

$$
\left(qu-\dfrac{1}{2}v^2+\dfrac{1}{2}\left(\dfrac{v-u}{w}\right)^2\right)
\dfrac{\partial f}{\partial u}
+\left(qu+\dfrac{1}{2}v^2-\dfrac{1}{2}\left(\dfrac{v-u}{w}\right)^2\right)
\dfrac{\partial f}{\partial v}
-
$$

$$
-\dfrac{v-u}{w}\dfrac{\partial f}{\partial w}+
\dfrac{\partial f}{\partial t_i}+\dfrac{\partial f}{\partial t_j}=0,
$$
\emph{ либо уравнения}

$$
\left(2u-\dfrac{1}{2}\left(\dfrac{v-u}{w}\right)^2\right)
\dfrac{\partial f}{\partial u}
+\left(2v+\dfrac{1}{2}\left(\dfrac{v-u}{w}\right)^2\right)
\dfrac{\partial f}{\partial v}
+\left(2v+\dfrac{v-u}{w}\right)\dfrac{\partial f}{\partial w}
+$$

$$+\dfrac{\partial f}{\partial t_i}+\dfrac{\partial f}{\partial t_j}=0,
$$
\emph{где} $k,l,c,q$ -- {\it произвольные числа}, $\varepsilon=0,1.$

\vspace{5mm}

В отношении этой классификации в общих чертах можно сказать тоже
самое, что и в отношении классификации, содержащейся в теореме 4.
\  В частности,  феноменологическая \ симметрия \
четыреметрических

\newpage

$$
     \left. {\begin{array}{ccc}
     f^1(ij)=\sin y_i\sin y_j\cos(x_i-x_j)+\cos y_i\cos y_j, \\
     f^2(ij)=z_i-\arcsin\dfrac{\sin(x_i-x_j)\sin
     y_j}{\sqrt{1-(f^1(ij))^2)}}, \\
     f^3(ij)=z_j+\arcsin\dfrac{\sin(x_i-x_j)\sin
     y_i}{\sqrt{1-(f^1(ij))^2)}}, \\

      f^4(ij)=t_i-t_j;
     \end{array}}
     \right\}
     $$

     $$
     \left. {\begin{array}{ccc} f^1(ij)=(x_i-x_j)y_iy_j,\,\,
     f^2(ij)=z_i+\dfrac{1}{(x_i-x_j)y_i^2}, \\
     f^3(ij)=z_j-\dfrac{1}{(x_i-x_j)y_j^2},\,\, f^4(ij)=t_i-t_j;
     \end{array}}
     \right\}
     $$
     \emph{and, {\bf implicitly}, by two more expressions:}

     $$
     \left. {\begin{array}{ccc}
     f^1(ij)=f^1(x_i-x_j-z_i(y_i-y_j),x_i-x_j-z_j(y_i-y_j),y_i-y_j,t_i,t_j), \\
     f^2(ij)=f^2(x_i-x_j-z_i(y_i-y_j),x_i-x_j-z_j(y_i-y_j),y_i-y_j,t_i,t_j), \\
     f^3(ij)=f^3(x_i-x_j-z_i(y_i-y_j),x_i-x_j-z_j(y_i-y_j),y_i-y_j,t_i,t_j), \\
     f^4(ij)=t_i-t_j,
     \end{array}}
     \right\}
     $$
     \emph{wherein the four components of the function $f=f(u,v,w,t_i,t_j)$
     are inde- \ pendent integrals of either the equation}
     $$
     \left(qu-\dfrac{1}{2}v^2+\dfrac{1}{2}\left(\dfrac{v-u}{w}\right)^2\right)
     \dfrac{\partial f}{\partial u}
     +\left(qu+\dfrac{1}{2}v^2-\dfrac{1}{2}\left(\dfrac{v-u}{w}\right)^2\right)
     \dfrac{\partial f}{\partial v}
     -
     $$
     $$
     -\dfrac{v-u}{w}\dfrac{\partial f}{\partial w}+
     \dfrac{\partial f}{\partial t_i}+\dfrac{\partial f}{\partial t_j}=0,
     $$
     \emph{ or of the equation}
     $$
     \left(2u-\dfrac{1}{2}\left(\dfrac{v-u}{w}\right)^2\right)
     \dfrac{\partial f}{\partial u}
     +\left(2v+\dfrac{1}{2}\left(\dfrac{v-u}{w}\right)^2\right)
     \dfrac{\partial f}{\partial v}
     +\left(2v+\dfrac{v-u}{w}\right)\dfrac{\partial f}{\partial w}
     +$$
     $$
     +\dfrac{\partial f}{\partial t_i}+\dfrac{\partial f}{\partial t_j}=0,
     $$
     \emph{where} $k,l,c,q$ are {\it arbitrary numbers}, and $\varepsilon=0,1.$

     \vspace{5mm}

     The same, generally, may be said about that classification as about that given by
     Theorem 4. In particular, the phenomenological symmetry of four-

\newpage

\noindent
геометрий устанавливается по рангу функциональной матрицы для 12
компонент $f(ij),f(ik),f(jk)$ метрической функции $f$, специальным
образом зависящих от 12 переменных -- координат тройки $\langle ijk \rangle$,
который оказывается равен 8. Некоторые четырехкомпонентные
метрические функции имеют содержательную физическую интерпретацию.

Классификация $s$-метрических феноменологически симметричных
\linebreak геометрий ранга три для $s>4$ к настоящему моменту
никем не проводилась вследствие возникающих трудностей чисто
технического характера. Они большей частью связаны с особенностью
применяемого метода, в котором предварительно проводится
классификация $s$-мерных групп Ли преобразований пространства
$R^s$, а затем только находятся метрические функции как
невырожденные двухточечные инварианты. Возможно, что эти трудности
удастся преодолеть, если при проведении классификации групп
преобразований сразу ввести условие существования у них
невырожденных двухточечных инвариантов.

\vspace{15mm}

\begin{center}
{\bf \large \S4. К вопросу о симметрии расстояния в геометрии}
\end{center}

Обычно расстояние между точками пространства $\mathfrak{M}$
определяется с помощью функции $\rho:\mathfrak{M} \times
\mathfrak{M} \to R$, сопоставляющей каждой паре точек $\langle ij \rangle$
некоторое число $\rho(ij)$ и удовлетворяющей известной системе
аксиом: {\bf 1.} $\rho(ij) \geqslant 0$, причем $\rho(ij) = 0$
тогда и только тогда, когда $i = j$; {\bf 2.} $\rho(ij) =
\rho(ji)$; {\bf 3.} $\rho(ik) + \rho(jk) \geqslant \rho(ij)$.
Согласно аксиоме {\bf 2.} расстояние симметрично. Однако в
геометрии нельзя исключить из рассмотрения так называемые
симплектические пространства, в которых определяемое между двумя
точками расстояние антисимметрично. С другой стороны, симметричный
интервал между двумя событиями в псевдоевклидовом
пространстве-времени Минковского, не удовлетворяющий аксиомам {\bf
1.} и {\bf 3.}, тоже можно рассматривать как расстояние.
Естественно возникает вопрос: почему в геометрии допускаются
только симметричные или антисимметричные расстояния? Оказывается,
если предположить существование функциональной связи между
расстояниями $\rho(ij)$ и $\rho(ji)$, то будут возможны только
такие два типа симметрии.

\newpage

     \noindent
     dimensional geometries is established by the rank of the functional  matrix
     for the twelve
     components $f(ij),f(ik),f(jk)$ of the metric function $f$ that depend in a special manner
     on 12 variables, the coordinates of the triple $\langle ijk \rangle$, which rank equals 8. Some four-
     component metric functions have an essential physical interpretation.

     Classifying of $s$-metric phenomenologically symmetric geometries of rank three
      for $s>4$ has not been attempted by anyone, because of the difficulties of sheerly
     technical nature. They are, for the most part, consequence of the method employed,
     the essence of which is in classifying first the $s$-dimensional Lie groups of the
     transformations of the space$R^s$, only then followed by the finding of the metric
     functions as nondegenerate two-point invariants. It is possible these difficulties will
     be overcome if, in classifying groups of transformations, an outright condition be
     introduced of the exis- \ tence of nondegenerate two-point invariants for them.

     \vspace{30mm}

     \begin{center}
     {\bf \large \S4. The symmetry of distance in geometry}
     \end{center}

     Ordinarily, the distance between points of a space $\mathfrak{M}$ is determined
     by the function $\rho:\mathfrak{M} \times \mathfrak{M} \to R$ that assigns to each
     pair of points $\langle ij \rangle$ some number $\rho(ij)$ and satisfies the axioms as follows:
     {\bf 1.} $\rho(ij) \geqslant 0$, $\rho(ij)$ being equal to $0$ if and only if $i = j$;
     {\bf 2.} $\rho(ij) = \rho(ji)$; {\bf 3.} $\rho(ik) + \rho(jk) \geqslant \rho(ij)$. Under the
     axiom {\bf 2.} distance is symmetric. However, the so called simplectic spaces, where
     the distance between two points is antisymmetric  are not to be wiped off the slate of
     geometry either. On the other hand, the symmetric interval between events in the
     pseudo-Euclidean space-time of Minkowski, that does not satisfy the axioms {\bf 1.}
     and {\bf 3.}, can also be considered as a distance. The question that naturally and
     logically suggests itself is why only symmetric and antisymmetric distances are allowed.
     It appears that if the existence of a functional relation between the distances $\rho(ij)$
     and $\rho(ji)$ is assumed, then only these two types of symmetry will be possible. Let
     us

\newpage

\noindent
Перейдем к точным формулировкам.

Пусть имеются множество $\mathfrak{M} = \{i,j,k,\ldots \}$
произвольной природы и функция $f: G_{f} \to R$, где
$G_{f} \subseteq \mathfrak{M} \times \mathfrak{M}$, сопоставляющая
упорядоченной паре $\langle ij \rangle \ \in G_{f}$ вещественное число
$f(ij) \in R$, рассматриваемое как расстояние в некотором обобщенном смысле.
Двухточечную функцию $f$ будем называть метрической, не требуя от нее
выполнения аксиом обычной метрики. В общем случае область определения $G_{f}$
функции $f$ не обязательно совпадает со всем прямым произведением
$\mathfrak{M} \times \mathfrak{M}$. Однако естественно предположить,
что если $\langle ij \rangle \ \in G_{f}$, то и $\langle ji \rangle \ \in G_{f}$, то есть
расстояния $f(ij)$ и $f(ji)$ определены или не определены одновременно.

{\bf Определение.} Будем говорить, что метрические функции $f$
и $g$ эквивалентны, если совпадают их области определения
$G_{f}$ и $G_{g}$ в  $\mathfrak{M} \times \mathfrak{M}$ и
существует строго монотонная функция $\psi: f(G_{f}) \to R$
 такая, что для любой пары $\langle ij \rangle \ \in G_{f}$ имеет место равенство
$g(ij) = \psi(f(ij))$.

Исходя из замечания в работе [2], аксиому симметрии
будем формулировать следующим образом [19]:

{\bf A.S.} Для любых точек $i,j \in \mathfrak{M}$ таких, что
пары $\langle ij \rangle$ и $\langle ji \rangle$ принадлежат $G_{f}$, расстояния $f(ij)$ и
$f(ji)$ связаны соотношением
\begin{equation}
\label{po1} f(ij) = \varTheta(f(ji)), \tag{$4.1$}
\end{equation}
где $\varTheta$ -- некоторая строго \ монотонная \ функция
\ одной переменной, область определения и область значений
которой совпадают с областью значений $f(G_{f}))$ исходной
метрической функции.

\vspace{5mm}

{\bf Теорема.}
{\it Если расстояние между \ точками пространства \ $\mathfrak{M}$,
определяемое метрической функцией $f: G_{f} \to R$, где
$G_{f} \subseteq \mathfrak{M} \times \mathfrak{M}$, удовлетворяет
аксиоме симметрии {\bf A.S.}, то это расстояние может быть либо
симметричным, либо, с точностью до эквивалентности,
антисимметричным.}

\vspace{5mm}

Из соотношения (4,1) для любой пары $\langle ij \rangle \ \in G_{f}$ получаем
тождество $\varTheta(\varTheta(f(ij))) = f(ij)$, означающее,
что функция $\varTheta$ является решением функционального
уравнения
\begin{equation}
\label{po2}
\varTheta(\varTheta(x)) = x,
\tag{$4.2$}
\end{equation}

\newpage

     \noindent
     now move on to exact formulations.

     Let $\mathfrak{M} = \{i,j,k,\ldots \}$ be a set of arbitrary nature and
     $f: G_{f} \to R$ a function where $G_{f} \subseteq \mathfrak{M} \times \mathfrak{M}$
     that assigns to an ordered pair $\langle ij \rangle \ \in G_{f}$ a real number $f(ij) \in R$ considered
     as a distance, in some generalized sense. We shall call the two-point function $f$ a
     metric one and not require it should satisfy ordinary metric axioms. In a general case
     the domain $G_{f}$ of the function $f$ does not necessarily coincide with the whole of
     the direct product $\mathfrak{M} \times \mathfrak{M}$. But it is natural to suppose
     that if $\langle ij \rangle \ \in G_{f}$, then $\langle ji \rangle \ \in G_{f}$ too, i.e. the distances $f(ij)$ and $f(ji)$
     are simultaneously either defined or undefined.

     {\bf Definition.} We shall say that the metric functions $f$ and$g$ are equivalent if
     their domains $G_{f}$ and $G_{g}$ coincide in $\mathfrak{M} \times \mathfrak{M}$
     and there exists a strictly monotone function $\psi: f(G_{f}) \to R$ such that for any
     pair $\langle ij \rangle \ \in G_{f}$ the equality$g(ij) = \psi(f(ij))$ takes place.

     Basing on the remark in the note [2], we shall formulate the symmetry axiom
     as follows [19]:

     {\bf A.S.} For any points $i,j \in \mathfrak{M}$ such that the pairs $\langle ij \rangle$ and $\langle ji \rangle$
     belong to $G_{f}$, the distances $f(ij)$ and$f(ji)$ are tied by the relation

     \begin{equation}
     \label{po1}
     f(ij) = \varTheta(f(ji)),
     \tag{$4.1$}
     \end{equation}    \\
     where $\varTheta$ is some strictly monotone function of one variable whose domain
     and range of values coincide with the domain $f(G_{f}))$ of the original metric function.

     \vspace{5mm}

     {\bf Theorem.}
     {\it If the distance between \ points of a space \ $\mathfrak{M}$ determined by the metric
     function $f: G_{f} \to R$, where $G_{f} \subseteq \mathfrak{M} \times \mathfrak{M}$,
     satisfies the symmetry axiom {\bf A.S.}, then that distance may only be either symmetric
     or, with an accuracy up to equivalence, antisymmetric.}

     \vspace{5mm}

     Out of the relation (4.1) for any pair $\langle ij \rangle \ \in G_{f}$ we get the identity
     $\varTheta(\varTheta(f(ij))) = f(ij)$ that means that the function $\varTheta$ is a
     solution of the functional equation
     \begin{equation}
     \label{po2}
     \varTheta(\varTheta(x)) = x,
     \tag{$4.2$}
     \end{equation}

\newpage

\noindent
где $x \in f(G_{f}) \subseteq R$. По предположению функция
$\varTheta$ строго монотонная и поэтому имеет к себе обратную.
Если функция $\varTheta$ монотонно возрастает, то $\varTheta(x) =
x$ и расстояние оказывается симметричным. Если же функция
$\varTheta$ монотонно убывает, то, перейдя к эквивалентной
метрической функции $g = \psi(f)$, где $\psi(f) = f -
\varTheta(f)$, имеем в силу (4.1): $g(ij) = f(ij) -
\varTheta(f(ij)) = f(ij) - f(ji) = - g(ji)$, то есть
антисимметричное расстояние. Теорема доказана.

\vspace{5mm}

Симметрия или антисимметрия расстояния в геометрии при
наличии связи (4.1) ранее были установлены
автором в работе "Некоторые следствия гипотезы о бинарной
структуре пространства" [20] для того случая, когда
это расстояние определялось с помощью функции
$F: G_{F} \to R$, где $G_{F} \subseteq \mathfrak{M} \times \mathfrak{N}$,
задающей на $n$-мерных многообразиях $\mathfrak{M} = \{i,j,k,\ldots \}$
и $\mathfrak{N} = \{\alpha,\beta,\gamma,\ldots \}$
феноменологически симметричную геометрию двух множеств
(физическую структуру) ранга $(n + 1,n + 1)$, и некоторого локального
диффеоморфизма $\varphi: \mathfrak{M} \to \mathfrak{N}$:
\begin{equation}
\label{po3}
f(ij) = F(i,\varphi(j)).
\tag{$4.3$}
\end{equation}

Для расстояния (4.3) соотношение (4.1) при известной функции $F$
становится функциональным уравнением относительно функции
$\varTheta$ и диффеоморфизма $\varphi$:
\begin{displaymath}
F(i,\varphi(j)) = \varTheta(F(j,\varphi(i))).
\end{displaymath}

Решая это уравнение для функций
\begin{gather*}
F(i\alpha) = x^{1}_i\xi^{1}_\alpha + \cdots + x^{n}_i\xi^{n}_\alpha,\\
F(i\alpha) = x^{1}_i\xi^{1}_\alpha + \cdots +
x^{n-1}_i\xi^{n-1}_\alpha + x^{n}_i + \xi^{n}_\alpha,
\end{gather*}
где $x^{1},\ldots,x^{n}$ и $\xi^{1},\ldots,\xi^{n}$ локальнве
координаты в многообразиях $\mathfrak{M}$ и $\mathfrak{N}$,
можно найти одновременно и диффеоморфизм $\varphi$ и функцию
$\varTheta$, определяющую тип симметрии расстояния (4.3).
В надлежаще выбранной в многообразии $\mathfrak{M}$ системе
локальных координат выражения для расстояния $f(ij)$ с
точностью до локальной эквивалентности можно записать
в следующем виде:
\begin{equation}
\label{po4}
\left.\begin{array}{rcl}
f(ij) = g_{\lambda\sigma}x^{\lambda}_ix^{\sigma}_j,\\
f(ij) = h_{\mu\nu}x^{\mu}_ix^{\nu}_j + x^{n}_i + ax^{n}_j,\;
\end{array}\right\}
\tag{$4.4$}
\end{equation}

\newpage

     \noindent
     where $x \in f(G_{f}) \subseteq R$. By assumption, the function $\varTheta$ is a
     strictly monotone one and so must have its inverse. If the function $\varTheta$
     monotone increasing, then $\varTheta(x) = x$ and the distance turns out to be
     symmetric. And if the function $\varTheta$ monotone decreasing, then, switching
     over to the equivalent metric function $g = \psi(f)$, where $\psi(f) = f - \varTheta(f)$,
     we have, under (4.1), $g(ij) = f(ij) - \varTheta(f(ij)) = f(ij) - f(ji) = - g(ji)$, i.e. an
     antisymmetric distance. The theorem has been proved.

     \vspace{5mm}

     Symmetry or antisymmetry of a distance in geometry, with the relation (4.1)
     present were established by the author in his note "Some
     consequences of the hypothesis of binary structure of space"
     \    [20] for the case where that distance is
     determined by the function $F: G_{F} \to R$, where $G_{F} \subseteq \mathfrak{M} \times
     \mathfrak{N}$, defining on $n$-dimensional manifolds $\mathfrak{M} = \{i,j,k,\ldots \}$
     and $\mathfrak{N} = \{\alpha,\beta,\gamma,\ldots \}$ a phenomenologically symmetric
     geometry of two sets (physical structure) of rank $(n + 1,n + 1)$ and of some local
     diffeomorphism $\varphi: \mathfrak{M} \to \mathfrak{N}$:
     \begin{equation}
     \label{po3}
     f(ij) = F(i,\varphi(j)).
     \tag{$4.3$}
     \end{equation}

\vspace{5mm}

     For the distance (4.3), the relation (4.1), with the known function $F$ becomes the
     functional equation with respect to the function$\varTheta$ and the diffeomorphism
     $\varphi$:
     \begin{displaymath}
     F(i,\varphi(j)) = \varTheta(F(j,\varphi(i))).
     \end{displaymath}

     By solving that equation for the functions
     \begin{gather*}
     F(i\alpha) = x^{1}_i\xi^{1}_\alpha + \cdots + x^{n}_i\xi^{n}_\alpha,\\
     F(i\alpha) = x^{1}_i\xi^{1}_\alpha + \cdots +
     x^{n-1}_i\xi^{n-1}_\alpha + x^{n}_i + \xi^{n}_\alpha,
     \end{gather*}
     where $x^{1},\ldots,x^{n}$ and $\xi^{1},\ldots,\xi^{n}$ are local coordinates in
     the manifolds $\mathfrak{M}$ and $\mathfrak{N}$, it is possible to find both the
     diffeomorphism $\varphi$ and the function $\varTheta$ determining the type of
     the distance symmetry (4.3). In a suitably chosen in the manifold $\mathfrak{M}$
     system of local coordinates, the expressions for the distance $f(ij)$ may be written
     with an accuracy up to a local equivalence as follows:
     \begin{equation}
     \label{po4}
     \left.\begin{array}{rcl}
     f(ij) = g_{\lambda\sigma}x^{\lambda}_ix^{\sigma}_j,\\
     f(ij) = h_{\mu\nu}x^{\mu}_ix^{\nu}_j + x^{n}_i + ax^{n}_j,\;
     \end{array}\right\}
     \tag{$4.4$}
     \end{equation}

    \newpage

\noindent
где $a = + 1, - 1$; $g_{\lambda\sigma} = ag_{\sigma\lambda}$,
$\lambda,\sigma = 1,\ldots,n$; $h_{\mu\nu} = ah_{\nu\mu}$,
$\mu,\nu = 1,\ldots,n-1$, причем для $a = -1$ размерность $n$
многообразия $\mathfrak{M}$ четна в первом из выражений (4.4) и
нечетна во втором.

Из выражений (4.4) при некоторых естественных дополнительных условиях
в случае $a = + 1$ можно получить симметричные метрические функции римановых
и псевдоримановых пространств постоянной кривизны. В случае же
$a = - 1$ выражения (4.4) определяют антисимметричные метрические
функции симплектических пространств {\it четной} и, обратим внимание,
{\it нечетной} размерности.

\vspace{15mm}

\begin{center}
{\bf \large \S5. Бинарные и тернарные геометрии}
\end{center}

Бинарные феноменологически симметричные геометрии определяются на
одном множестве. Двухточечная функция, задающая такую геометрию,
допускает нетривиальную группу движений с конечным числом
непрерывных параметров, которое было названо степенью групповой
симметрии. При определенных соотношениях между рангом
феноменологической симметрии, числом существенных параметров
группы движений и размерностью многообразия групповая и
феноменологическая симметрии оказываются эквивалентными. Эти
соотношения были заложены в определение геометрии, ее
феноменологической и групповой симметрий. Естественно возникает
вопрос об их происхождении и обосновании. Кроме того, имеется
много возможностей обобщения и развития понятия геометрии, одна из
которых была реализована в \S1, когда двум точкам сопоставлялось
несколько действительных чисел. Другая возможность обобщения
реализуется в определении, например, тернарных геометрий, когда
метрическая функция сопоставляет число не двум точкам, а трем.
Однако уже предварительное исследование показало, что тернарные
геометрии, в отличие от бинарных, не могут быть наделены групповой
симметрией, то есть трехточечная метрическая функция не допускает
нетривиальную группу движений. Поэтому возникает еще вопрос о
причинах такого различия между бинарными и тернарными геометриями.

\newpage

  \noindent
  where $a = + 1, - 1$; $g_{\lambda\sigma} = ag_{\sigma\lambda}$,$\lambda,\sigma = 1,\ldots,n$;
     $h_{\mu\nu} = ah_{\nu\mu}$, $\mu,\nu = 1,\ldots,n-1$, the dimensionality $n$ of the manifold
     $\mathfrak{M}$ being even for $a = -1$ in the former expression (4.4) and odd in the latter.

     Out of the expressions (4.4), by introducing some natural additional conditions,
     in the case of $a = + 1$, we can get symmetric metric functions of Riemannian and
     pseudo-Riemannian spaces of constant curvature. And in case$a = - 1$ the expressions
     (4.4) define antisymmetric metric functions of simplectic spaces of {\it even} and, also,
     we shall point it out, {\it odd} dimensionality.

     \vspace{25mm}

     \begin{center}
     {\bf \large \S5. Binary and ternary geometries}
     \end{center}

     Binary phenomenologically symmetric geometries are defined on one set. A
     two-point function that gives such a geometry allows a nontrivial group of motions
     with a finite number of continuous parameters that has been called the degree of a
     group symmetry. With certain relations among the rank of the phenomenological
     symmetry, the number of the essential parameters of the group of motions, and the
     dimensionality of the manifold, the group and phenomenological symmetry turn out to
     be equivalent. Those relations were incorporated into the definition of the geometry,
     and its phenomenological and group symmetries. Naturally, the question arises of their
     origin and interpretation. Besides, there are a lot of opportunities of generalization and
     development of the notion of geometry, one of which is realized in \S1 when two points
     are assigned more than one real numbers. Another opportunity for generalization is
     realized in defining of, for example, ternary geometries, where a metric function
     assigns a number not to two but to three points. However, as early as on the stage of
     preliminary exploration it turned out that the ternary geometries, in contrast to binary,
     may not have a group symmetry, i.e. a three-point metric function does not allow a nontrivial
     group of motions. So additionally there arises the question of the causes of such difference
     between the binary and ternary geometries.

\newpage

Для ответа на эти вопросы необходимо исходить из более общего
определения полиарных геометрий. Тогда можно будет установить, при
каких соотношениях между основными характеристиками геометрии она
может быть наделена групповой симметрией, а при каких -- не может.
Естественно предположить, что только те геометрии содержательны в
физическом и математическом смыслах, группы движений которых
нетривиальны.

Пусть имеется множество $\mathfrak{M}$
произвольной природы, которое в математическом смысле представляет
собой гладкое многообразие размерности $m$. Пусть также имеется функция
$$
f:\mathfrak{S}_f\to R^s,
\eqno(5.1)
$$
где $\mathfrak{S}_f\subseteq\mathfrak{M}^{q}$, сопоставляющая каждому
кортежу длины $q$ из $\mathfrak{S}_f$ некоторую точку
из $R^s$, то есть $s$ действительных
чисел. Предполагается, что область определения $\mathfrak{S}_f$ функции $f$
открыта и плотна в $q$-арном прямом произведении $\mathfrak{M}^{q}$
множества $\mathfrak{M}$ на себя, а ее координатное представление
достаточно гладкое.
Число $q$ назовем арностью, а $q$-арную и $s$-компонентную
функцию (5.1) -- метрической.

Пусть, далее, $M>q$ -- произвольное целое число. Построим отображение
$$
F:\mathfrak{S}_F\to R^{sC^{q}_{M}},
\eqno(5.2)
$$
где $\mathfrak{S}_F\subseteq\mathfrak{M}^{M}$, сопоставляя
каждому кортежу длины $M$
из $\mathfrak{S}_F$ упорядоченную по нему совокупность $sC^{q}_{M}$ чисел,
соответствующих всем кортежам длины
$q$, которые являются проекциями исходного кортежа на область
$\mathfrak{S}_f$. Область определения $\mathfrak{S}_F$ функции (5.2) будет,
очевидно, открытой и плотной в прямом произведении $\mathfrak{M}^{M}$.
Аналогично построим второе отображение
$$
F':\mathfrak{S}_{F'}\to R^{sC^{q}_{M'}},
\eqno(5.2')
$$
где $\mathfrak{S}_{F'}\subseteq\mathfrak{M}^{M'}$ и $M'\geq M$. Проекцию
отображения $F'$ получим, опуская из области
его определения $\mathfrak{S}_{F'}$
некоторую совокупность кортежей длины $q$, а из области его значений
соответствующие по функции (5.1) числа.

{\bf Определение 1.} Будем говорить, что функция (5.1) задает на
$m$-мерном многообразии $\mathfrak{M}$ {\it $q$-арную
$s$-метрическую феноменологически}

\newpage

     To answer these questions it is necessary to proceed from a more general definition of
     polyary geometries. Then it will be possible to find out at what relations among the basic
     characteristics of a geometry it may be endowed with a group symmetry, and at what
     it may not. It is natural to suppose that only those geometries are meaningful, physically
     and mathematically, whose groups of motions are nontrivial.

     Let there be a set $\mathfrak{M}$ of arbitrary nature that is, mathematically, a smooth
     manifold of dimension $m$. Let also there be a function
     $$
     f:\mathfrak{S}_f\to R^s,
     \eqno(5.1)
     $$
     where $\mathfrak{S}_f\subseteq\mathfrak{M}^{q}$, that assigns to each cortege of
     length $q$ from $\mathfrak{S}_f$ some point from $R^s$, i.e. $s$ real numbers. It is
     assumed that the domain $\mathfrak{S}_f$ of the function $f$ is open and dense in
     the $q$-ary direct product $\mathfrak{M}^{q}$ of the set $\mathfrak{M}$ by itself, and
     that its coordinate representation is sufficiently smooth. We shall call the number $q$ arity,
     and a $q$-ary and $s$-component function (5.1) - a metric one.

     Let, further, $M>q$ be an arbitrary integer number. Let us construct the mapping
     $$
     F:\mathfrak{S}_F\to R^{sC^{q}_{M}},
     \eqno(5.2)
     $$  \\
     where $\mathfrak{S}_F\subseteq\mathfrak{M}^{M}$, by assigning to each cortege of
     length $M$ from $\mathfrak{S}_F$ a collection of numbers $sC^{q}_{M}$ ordered with
     respect to that cortege, the numbers of which collection correspond to all the corteges of
     length $q$, that are projections of the original cortege onto the domain $\mathfrak{S}_f$.
     The domain $\mathfrak{S}_F$ of the function (5.2) is, obviously, open and dense in the direct
     product $\mathfrak{M}^{M}$. Similarly, we shall build another mapping
     $$
     F':\mathfrak{S}_{F'}\to R^{sC^{q}_{M'}},
     \eqno(5.2')
     $$
     where $\mathfrak{S}_{F'}\subseteq\mathfrak{M}^{M'}$ and $M'\geq M$. The projection of the
     mapping $F'$ is obtained from the domain $\mathfrak{S}_{F'}$ of it by way of dropping some
     collection of corteges of length $q$, along with the corresponding numbers from the region of
     its values with respect to the function (5.1).

     {\bf Definition 1.} We shall say that the function (5.1) gives on an $m$-dimensional
     manifold $\mathfrak{M}$ a {\it $q$-ary $s$-metric phenomenologically
     symmetric }

    \newpage

\noindent
{\it симметричную геометрию ранга $M$}, если на плотном в
$\mathfrak{S}_F$ множестве ранг отображения $F$ равен
$s(C^{q}_{M}-1)$, а ранг любой проекции отображения $F'$, не
включающей в себя область отображения $F$, максимален на плотном в
$\mathfrak{S}_{F'}$ множестве.

Другими словами, локально множество значений
отображения $F$ в $R^{sC^{q}_{M}}$
принадлежит множеству нулей системы $s$ независимых функций $\Phi=
(\Phi_1,\ldots,\Phi_s)$ от $sC^{q}_{M}$
переменных, причем $s$ функциональных связей
$$
\Phi=(\Phi_1,\ldots,\Phi_s)=0
\eqno(5.3)
$$
являются порождающими в том смысле, что любые другие нетривиальные связи будут
только их следствием.

{\bf Определение 2.} Будем говорить, что определенная выше феноменологически
симметричная геометрия {\it наделена групповой симметрией
конечной степени $r$}, если задано
такое эффективное гладкое локальное действие некоторой
$r$-мерной локальной группы
Ли $G^r$ в многообразии $\mathfrak{M}$, что компоненты метрической
функции (5.1), задающей эту геометрию, являются $q$-точечными инвариантами.

Поскольку преобразуемое многообразие конечномерно,
естественно в определении 2 условие, что максимальное число
существенных параметров группы локальных движений конечно.

Запишем систему $sC^{q}_{M'}$ уравнений, следующих из условия
сохранения компонент метрической функции (5.1):
$$
Df|_{F'}=0
\eqno(5.4)
$$
относительно $M'm$ дифференциалов координат точек кортежа
из $\mathfrak{S}_{F'}$. Если введенная определением 1 феноменологически
симметричная геометрия наделена групповой
симметрией конечной степени, то однородная система (5.4),
с одной стороны, должна иметь хотя
бы одно ненулевое решение, а с другой, число ее линейно независимых
ненулевых решений для любого числа $M'$ не должно превышать
некоторого конечного значения, равного степени групповой симметрии. Число
таких решений равно, как известно, числу неизвестных в системе минус ранг ее
матрицы. Но матрица системы уравнений (5.4) есть функциональная
матрица для системы
функций $f$, соответствующих всем упорядоченным проекциям области определения
$\mathfrak{S}_{F'}$ отображения $(5.2')$ на область

\newpage

      \noindent
     {\it geometry of rank $M$} if on a set dense in $\mathfrak{S}_F$ the rank of the mapping $F$ is
     equal to $s(C^{q}_{M}-1)$, and the rank of any projection of the mapping $F'$ that does not
     include the range of the mapping $F$ is maximal in the set dense in $\mathfrak{S}_{F'}$.

     In other words, locally the range of the mapping $F$ in $R^{sC^{q}_{M}}$ belongs to the
     set of zeros of the system of $s$ independent functions $\Phi=(\Phi_1,\ldots,\Phi_s)$ of
     $sC^{q}_{M}$ variables, $s$ functional relations
     $$
     \Phi=(\Phi_1,\ldots,\Phi_s)=0
     \eqno(5.3)
     $$
     being generating ones in that particular sense that any other nontrivial relations will be
     just their consequence.

     {\bf Definition 2.} We shall say that the phenomenologically symmetric geo- \ metry that we
     have defined above is {\it endowed with the group symmetry of finite degree $r$}, if an
     effective smooth local action of some $r$-dimensional local Lie group $G^r$ in the manifold
     $\mathfrak{M}$ is defined, such that the components of the metric function (5.1) that gives
     that geometry are $q$-point invariants.

     Since the manifold transformed is finite-dimensional, Definition 2 naturally implies the condition
     that the maximum number of the essential parameters of the group of local motions is finite.

     We shall write down a system of $sC^{q}_{M'}$ equations that are result of the condition
     of preservation of the components of the metric function (5.1):
     $$
     Df|_{F'}=0
     \eqno(5.4)
     $$
     with respect to $M'm$ differentials of the points of the cortege of $\mathfrak{S}_{F'}$. If the
     phenomenologically symmetric geometry introduced by Definition 1 is endowed with a group
     symmetry of finite degree, then, on the one hand, the homogeneous system (5.4) must have
     at least one nonzero solution, and, on the other hand, the number of its independent nonzero
     solutions for any number $M'$ may not be bigger than some finite value equal to the degree of
     the group symmetry. The number of such solutions is, as is known, equal to the number of
     unknowns in the system minus the rank of matrix of it. But the matrix of the system of equations
     (5.4) is the functional matrix for the system of functions $f$ corresponding to all the ordered
     projections of the domain $\mathfrak{S}_{F'}$ of the mapping $(5.2')$ onto the domain
     $\mathfrak{S}_f$ of the original

     \newpage

\noindent
определения $\mathfrak{S}_f$ исходной функции (5.1). Ранг матрицы
системы уравнений (5.4), очевидно, не изменится, если из системы
функций $f|_{F'}$ исключить зависимые по связи (5.3). Исключив их,
получим максимальную проекцию отображения $(5.2')$, не содержащую
в себе отображения (5.2). Обозначим число функций $f$ в этой
максимальной проекции через $N(M')$. Тогда по определению 1 ранг
матрицы системы уравнений (5.4) будет равен
$$
\min(M'm; \ N(M')).
\eqno(5.5)
$$

Если \ найдется \ такое \ значение \ числа \ $M'$, \ для \
которого \ $M'm\leq N(M')$, то ранг матрицы системы уравнений
(5.4) для него будет равен $M'm$, то есть числу неизвестных в ней.
Но тогда система (5.4) будет иметь только нулевое решение, что
означает отсутствие нетривиальной группы движений в
рассматриваемой феноменологически симметричной геометрии. Если же
для любого значения $M'$ выполняется строгое неравенство
$N(M')<M'm$, то ранг матрицы системы уравнений (5.4) будет равен
$N(M')$, а число ее линейно независимых ненулевых решений окажется
равным
$$
r'=M'm-N(M')>0.
\eqno(5.6)
$$

Число $r'$, как было отмечено выше, в случае наделения
феноменологически симметричной геометрии групповой симметрией
конечной степени не должно превышать некоторого конечного
значения. Из этого условия установим, при каком соотношении между
размерностью множества и рангом феноменологической симметрии
задаваемая по определению 1 метрической функцией (5.1) геометрия
может быть наделена групповой симметрией и определим степень $r$
этой симметрии.

Рассмотрим сначала бинарные (q=2) геометрии.

Бинарная $s$-метрическая феноменологически симметричная геометрия
ранга $M\geq3$ на одном множестве $\mathfrak{M}$, которое является
$m$-мерным многообразием, задается метрической функцией (5.1), где
$\mathfrak{S}_f\subseteq\mathfrak{M\times M}$, причем по
определению 1 ранг отображения $F:\mathfrak{S}_F\to
R^{sM(M-1)/2}$, где $\mathfrak{S}_F\subseteq\mathfrak{M}^M$, равен
$sM(M-1)/2-s$. Найдем, сколько в системе $sM'(M'-1)/2$ функций
отображения $F':\mathfrak{S}_{F'}\to R^{sM'(M'-1)/2}$, где
$\mathfrak{S}_{F'}\subseteq \mathfrak{M}^{M'}$ и $M'\geq M$, при
этом будет зависимых. На матрицу пар

\newpage

       \noindent
      function (5.1). Obviously, the rank of the matrix of the system of
     equations (5.4) will not change if we eliminate from the system of functions $f|_{F'}$  the
     dependents with respect to the relation (5.3). Eliminating them yields the maximal projection of
     the mapping $(5.2')$ that does not contain in itself the mapping (5.2). We shall designate by
     $N(M')$ the number of functions $f$ in that maximal projection. Then, under Definition 1, the
     rank of the matrix of the system of equations (5.4) will be
     $$
     \min(M'm; \ N(M')).
     \eqno(5.5)
     $$

     If there exists such a value of the number $M'$, for which \ $M'm\leq N(M')$, the rank of the
     matrix of the system of equations (5.4) for it will be equal to $M'm$, i.e. to the number of
     unknowns in it. But then the system (5.4) will only have a zero solution, which means the
     absence of a nontrivial group of motions in the phenomenologically symmetric geometry in
     question. But if the strict inequality $N(M')<M'm$ holds for any value of $M'$, then the rank
     of the matrix of the system of equations (5.4) will be equal to $N(M')$, and the number of the
     linearly independent nonzero solutions of it will be equal to
     $$
     r'=M'm-N(M')>0.
     \eqno(5.6)
     $$

     The number $r'$, as has already been stated, in case of a phenomenologically symmetric
     geometry having a group symmetry of finite degree may not exceed some finite value. Out
     of that condition, let us establish at what relation between the dimensionality of the set
     and the rank of the \\ phenomenological symmetry the geometry defined under Definition 1
     by the metric function (5.1) may have a group symmetry and shall define the degree $r$ of
     that symmetry.

     Let us take first binary (q=2) geometries.

     A binary $s$-metric phenomenologically symmetric geometry of rank $M\geq3$ on one
     set $\mathfrak{M}$ that is an $m$-dimensional manifold is defined by the metric function
     (5.1), where $\mathfrak{S}_f\subseteq\mathfrak{M\times M}$, and, under Definition 1,
     the rank of the mapping $F:\mathfrak{S}_F\to R^{sM(M-1)/2}$, where
     $\mathfrak{S}_F\subseteq\mathfrak{M}^M$, is equal to $sM(M-1)/2-s$. We shall find
     the number of the dependents in the system of $sM'(M'-1)/2$ functions of the mapping
     $F':\mathfrak{S}_{F'}\to R^{sM'(M'-1)/2}$, where $\mathfrak{S}_{F'}\subseteq \mathfrak{M}^{M'}$
     and $M'\geq M$. \  We shall sequentially superpose the matrix of

     \newpage

\noindent
для кортежа длины $M'$ из $\mathfrak{S}_{F'}$ будем
последовательно налагать матрицу пар для кортежа длины $M$ из
$\mathfrak{S}_F$. При каждом полном наложении вычеркнем одну пару,
например, последнюю. Эта процедура повторяется до тех пор, пока
возможно наложение без пропуска. Число зависимых функций будет
равно, очевидно, числу состоявшихся наложений, умноженному на
число $s$. Нетрудно установить, что искомое число равно
$s(M'-M+1)(M'-M+2)/2$, и потому ранг функциональной матрицы всей
системы функций $f|_{F'}$ по определению 1 будет равным
$$
\min(M'm; \ sM'(M'-2)/2-s(M'-M+1)(M'-M+2)/2).
$$

Если $m<s(M-2)$, то для достаточно больших значений $M'$ ранг матрицы системы
уравнений (5.4) в рассматриваемом случае равен $M'm$,
то есть числу неизвестных в
ней, и потому она имеет для этих значений $M'$ только нулевое решение.
Следовательно при $m<s(M-2)$ бинарная $s$-метрическая феноменологически
симметричная геометрия ранга $M$ не может быть наделена
групповой симметрией. Если же $m\geq s(M-2)$, то для всех $M'\geq M$ ранг
матрицы системы (5.4) меньше $M'm$ и она по формуле (5.6) имеет
$$
r'=M'm-sM'(M-2)+s(M-1)(M-2)/2
$$
линейно независимых ненулевых решений. При $m>s(M-2)$ с ростом $M'$ число
решений $r'$ может стать сколь угодно большим, что противоречит условию
конечности степени групповой симметрии согласно определению 2.
Поэтому, если рассматриваемая бинарная геометрия
наделена групповой симметрией конечной степени,
то размерность $m$ многообразия $\mathfrak{M}$ и ее ранг
должны быть связаны соотношением
$$
{\bf m=s(M-2).}
\eqno(5.7)
$$

При соотношении (5.7) число линейно независимых ненулевых решений $r'$
системы уравнений (5.4) равно числу существенных и независимых параметров
группы движений, то есть степени $r$ групповой симметрии:
$$
{\bf r=s(M-1)(M-2)/2=m(m+s)/2s.}
\eqno(5.8)
$$

\newpage

     \noindent
     the pairs of the cortege of length
     $M$ from $\mathfrak{S}_F$ onto the matrix of the pairs for the cortege of length $M'$ from
     $\mathfrak{S}_{F'}$. At each complete superposition, we shall cross out one pair, for example,
     the last one. The procedure is repeated so long as superposition without blanks  is possible.
     The number of the dependent functions will obviously be equal to the number of successful
     superpositions multiplied by $s$. It is readily established that the number is equal to
     $s(M'-M+1)(M'-M+2)/2$, and hence the rank of the functional matrix of the system of functions
     $f|_{F'}$, in accordance with Definition 1, will be equal to
     $$
     \min(M'm; \ sM'(M'-2)/2-s(M'-M+1)(M'-M+2)/2).
     $$

     If $m<s(M-2)$, then for sufficiently large values of $M'$ the rank of the matrix of the
     system of equations (5.4), in the case in question, is equal to $M'm$, i.e. to the number
     of the unknowns in it, and so it only has for those values of $M'$ a zero solution.
     Therefore, with $m<s(M-2)$ a binary $s$-metric phenomenologically symmetric
     geometry of rank $M$ may not have a group symmetry. But if $m\geq s(M-2)$, then, for
     every $M'\geq M$, the rank of the matrix of the system (5.4) is less than $M'm$,
     and it has, according to the  expression (5.6),
     $$
     r'=M'm-sM'(M-2)+s(M-1)(M-2)/2
     $$
     linearly independent nonnull solutions. At $m>s(M-2)$, with $M'$ increasing, the number
     of solutions $r'$ may become arbitrarily large, which is in \\ contradiction with the condition,
     under Definition 2, of the degree of the group symmetry being finite. So, if the binary geometry
     in question is endowed with a group symmetry of finite degree the dimension $m$ of the
     manifold $\mathfrak{M}$ and its rank must be tied by the relation
     $$
     {\bf m=s(M-2).}
     \eqno(5.7)
     $$

     Under the relation (5.7), the number of the linearly independent nonzero solutions $r'$ of
     the system of equations (5.4) is equal to the number of the essential and independent
     parameters of the group of motions, i.e. equal to the degree $r$ of the group symmetry:
     $$
     {\bf r=s(M-1)(M-2)/2=m(m+s)/2s.}
     \eqno(5.8)
     $$

\newpage

Полученные соотношения (5.7) и (5.8) между размерностью $m$
многообразия $\mathfrak{M}$, рангом $M$ феноменологической
симметрии задаваемой  на нем функцией (5.1) бинарной $(q=2)$
$s$-метрической геометрии и степенью $r$  ее групповой симметрии
были использованы в определениях \S1 настоящей монографии, а также
в монографии автора [10] и в его работах [11], [21], [22]. При
сопоставлении соотношений (5.7) и (5.8) с соответствующими
соотношениями \S1 и других указанных выше источников необходимо,
очевидно, сделать следующие замены: $m\to sn, M\to m$.

Перейдем к рассмотрению тернарных (q=3) геометрий.

Для тернарной феноменологически
симметричной геометрии ранга $M$, где $M\geq4$,
задаваемой на одном множестве $\mathfrak{M}$, представляющем собой $m$-мерное
многообразие, метрической функцией (5.1), где $\mathfrak{S}_f\subseteq
\mathfrak{M}^3$, ранг отображения $F:\mathfrak{S}_F\to R^{sM(M-1)(M-2)/6}$,
где $\mathfrak{S}_F\subseteq\mathfrak{M}^M$, по определению 1 равен
$sM(M-1)(M-2)/6-s$. Найдем ранг отображения $F':\mathfrak{S}_{F'}\to
R^{sM'(M'-1)(M'-2)/6}$, где $\mathfrak{S}_{F'}\subseteq\mathfrak{M}^{M'}$ и
$M'\geq M$. Среди всех $sM'(M'-1)(M'-2)/6$ функций $f|_{F'}$ этого
отображения число зависимых определяется методом наложения матрицы троек для
кортежа длины $M$ из $\mathfrak{S}_F$ на матрицу троек для кортежа длины
$M'$ из $\mathfrak{S}_{F'}$. Этот метод был описан выше при рассмотрении
бинарных геометрий на одном множестве. Для числа зависимых функций отображения
$F'$ аналогично получаем значение $s(M'-M+1)(M'-M+2)(M'-M+3)/6$. По
определению 1 ранг матрицы системы уравнений (5.4) будет
равен
\begin{eqnarray*}
\min(M'm; \ sM'(M'-1)(M'-2)/6- \phantom{aaaaa} \\
\phantom{aaa} \mbox{}-s(M'-M+1)(M'-M+2)(M'-M+3)/6).
\end{eqnarray*}

Поскольку $M>3$, для достаточно больших значений $M'$ этот ранг
равен $M'm$, то есть числу неизвестных в системе уравнений (5.4),
и она поэтому для таких значений $M'$ имеет только нулевое
решение. Таким образом, тернарные феноменологически симметричные
геометрии на одном множестве не могут быть наделены групповой
симметрией. Аналогичный результат и аналогичным методом может быть
получен для любой $q$-арной феноменологически симметричной
геометрии, арность которой больше трех.

\newpage

    The relations (5.7) and (5.8) among the dimensionality $m$ of the manifold $\mathfrak{M}$,
     the rank $M$ of the phenomenological symmetry defined on it by the
     function (5.1) of the
     binary $(q=2)$ $s$-metric geometry, and the degree $r$ of its group symmetry are used in
     the definitions of \S1 of this monograph, as well as in the author's monograph [10] and in his
     notes [11], [21], and [22]. When comparing the relations (5.7) and (5.8) with the respective
     relations in \S1, and those in the other works mentioned, it is obviously necessary to keep in
     mind the following replacements: $m\to sn, M\to m$.

     Now let us take the ternary (q=3) geometries.

     For a ternary phenomenologically symmetric geometry of rank $M$, where $M\geq4$,
     defined on one set $\mathfrak{M}$ that is an $m$-dimensional manifold by the metric
     function (5.1), where $\mathfrak{S}_f\subseteq\mathfrak{M}^3$, the rank of the mapping
     $F:\mathfrak{S}_F\to R^{sM(M-1)(M-2)/6}$, where $\mathfrak{S}_F\subseteq\mathfrak{M}^M$,
     is, under Definition 1, equal to$sM(M-1)(M-2)/6-s$. We shall find the rank of the mapping
     $F':\mathfrak{S}_{F'}\to R^{sM'(M'-1)(M'-2)/6}$, where $\mathfrak{S}_{F'}\subseteq\mathfrak{M}^{M'}$
     and $M'\geq M$. Among all the $sM'(M'-1)(M'-2)/6$ functions $f|_{F'}$ of that mapping
     the number of the dependents is found by the method of superposition of the matrix of
     the triples for the cortege of length $M$ from $\mathfrak{S}_F$ on the matrix of the
     triples for the cortege of length $M'$ from $\mathfrak{S}_{F'}$. The method is described
     above, where the binary geometries on one set are discussed. For the number of the
     dependent functions of the mapping $F'$, we similarly get the value $s(M'-M+1)(M'-M+2)(M'-
     M+3)/6$. Under Definition 1, the rank of the matrix of the system of equations (5.4) will be
     equal to
     \begin{eqnarray*}
     \min(M'm; \ sM'(M'-1)(M'-2)/6- \phantom{aaaaa} \\
     \phantom{aaa} \mbox{}-s(M'-M+1)(M'-M+2)(M'-M+3)/6).
     \end{eqnarray*}

     Since $M>3$, for sufficiently great values of $M'$ that rank is equal to $M'm$, i.e. to the number
     of the unknowns in the system of equations (5.4), and it only has, for that reason, for such values
     of $M'$ the zero solution. Thus, the ternary phenomenologically symmetric geometries on one
     set may not be endowed with a group symmetry. The similar result, and by the similar method,
     may be obtained for any $q$-ary phenomenologically symmetric geometry whose arity is more
     than three.

\newpage

Окончательный вывод по результатам проведенного выше исследования
выражает следующая

\vspace{5mm}

{\bf Теорема.}  {\it Групповой симметрией конечной степени
могут быть наделены только бинарные
$(q=2)$ феноменологически симметричные геометрии, в то время как для
$q$-арных феноменологически симметричных геометрий с $q\geq3$ задающая их
метрическая функция $(5.1)$
не допускает никаких нетривиальных локальных движений.}

\vspace{5mm}

Только что сформулированной теореме об отсутствии групповой симметрии у всех
$q$-арных феноменологически симметричных геометрий с $q\geq3$ как будто
бы противоречит следующий простой пример.
Сопоставим каждой тройке точек $\langle  ijk  \rangle$
координатной плоскости $(x,y)$ ориентированную площадь
треугольника с вершинами в этих точках:
$$
S(ijk)=\frac 12\left|
\begin{array}{ccc}
x_i & x_j & x_k \\
y_i & y_j & y_k \\
1 & 1 & 1
\end{array}
\right| .
\eqno(5.9)
$$

Для произвольной четверки $\langle ijkl \rangle$ точек этой плоскости четыре площади
треугольников $\langle ijk \rangle$, $\langle ijl \rangle$, $\langle ikl \rangle$, $\langle jkl \rangle$ функционально
связаны следующим очевидным соотношением:
$$
S(ijk)-S(ijl)+S(ikl)-S(jkl)=0.
\eqno(5.10)
$$

Кроме того, функция (5.9) допускает пятипараметрическую группу
движений:
$$
x'=ax+by+c, \ \
y'=gx+hy+d,
\eqno(5.11)
$$
где $ah-bg=1$.

Таким образом, вроде бы приходим к выводу, что трехточечная
функция площади $(5.9)$ все-таки задает на координатной плоскости
$(x,y)$ тернарную геометрию, которая, с одной стороны,
феноменологически симметрична, а с другой  наделена групповой
симметрией степени 5.

Однако в соответствии с определением 1 наличие функциональной
связи (5.10) не является достаточным условием того, \ чтобы
функция

\newpage

     The final conclusion on the result of the above investigation is expressed in the following
     theorem.

     \vspace{5mm}

     {\bf Theorem.}  {\it Only binary $(q=2)$ phenomenologically symmetric geometries
     may be endowed with a group symmetry of finite degree, while with the $q$-ary
     phenomenologically symmetric geometries with $q\geq3$ the metric function $(5.1)$
     defining them does not allow any nontrivial motions.}

     \vspace{5mm}

     This Theorem of the $q$-ary phenomenologically symmetric geometries with $q\geq3$
     having no group symmetry seems to be refuted by a very simple example. Let us assign
     to each triple of the points $\langle  ijk  \rangle$ of the coordinate plane $(x,y)$ the oriented area of
     the triangle with the apeces in these points:

     $$
     S(ijk)=\frac 12\left|
     \begin{array}{ccc}
     x_i & x_j & x_k \\
     y_i & y_j & y_k \\
     1 & 1 & 1
     \end{array}
     \right| .
     \eqno(5.9)
     $$   \\

     For an arbitrary quadruple $\langle ijkl \rangle$ of points of the plane, the four areas of the triangles $\langle ijk \rangle$,
     $\langle ijl \rangle$, $\langle ikl \rangle$, and $\langle jkl \rangle$ are functionally related by the following obvious relation:
     $$
     S(ijk)-S(ijl)+S(ikl)-S(jkl)=0.
     \eqno(5.10)
     $$

     Besides, the function (5.9) allows a 5-parameter group of motions:

     $$
     x'=ax+by+c, \ \
     y'=gx+hy+d,
     \eqno(5.11)
     $$
     where $ah-bg=1$.

     Thus, we seemingly come to the conclusion that the 3-point function of area $(5.9)$ does
     give on the coordinate plane $(x,y)$ a ternary geometry that is, on the one hand,
     phenomenologically symmetric, and on the other is endowed with a group symmetry of
     degree 5.

     However, under Definition 1, the presence of the functional relation (5.10) is not
     sufficient condition for the function \ (5.9) \ to give a \ planar
     ternary

    \newpage

\noindent
 (5.9) задавала на плоскости тернарную феноменологически
симметричную геометрию ранга 4. Необходимо еще по нему, чтобы
связь (5.10) была порождающей в следующем смысле: {\it всякая
другая нетривиальная связь должна быть ее следствием.}

Покажем, что именно этому условию уравнение (5.10) и не удовлетворяет. Возьмем на
плоскости семерку точек $\langle ijklpmn \rangle$, которой по функции (5.9) сопоставим все
тридцать пять площадей $S(ijk),S(ijl),\ldots,$ $S(imn); \ S(jkl),S(jkp),\ldots,$\
$S(pmn)$ соответствующих треугольников $\langle ijk \rangle,\ldots,\langle pmn \rangle$. Из этих тридцати
пяти площадей по связи (5.10) можно исключить двадцать: $S(jkl),\ldots,
S(pmn)$. Между \ оставшимися \ пятнадцатью \ площадями
\ $S(ijk),S(ijl),\ldots,$
$S(imn)$ имеется одна тривиальная связь, так как они зависят только от четырнадцати
координат $x_i,y_i,x_j,y_j,\ldots,x_n,y_n$. Если связь (5.10) порождающая, то,
исключая, например, площадь $S(ijk)$, получаем четырнадцать площадей
$$
S(ijl),S(ijp),\ldots,S(imn),
\eqno(5.12)
$$
которые должны быть независимыми. Тривиальной связи между ними
быть не может, так как они являются функциями четырнадцати
координат точек семерки $\langle ijklpmn \rangle$, но нетривиальные связи между
ними, в действительности, есть, в чем можно убедиться из следующих
простых соображений. Если бы таких связей не было, то семиточечная
фигура $\langle ijklpmn \rangle$ не имела бы для своего движения ни одной
степени свободы, так как на четырнадцать координат ее точек было
бы наложено столько же независимых соотношений, обусловленных
инвариантностью четырнадцати площадей (5.12). В действительности
же, семиточечная фигура $\langle ijklpmn \rangle$, сохраняя площади всех
тридцати пяти треугольников, может перемещаться по плоскости с
пятью степенями свободы: одно вращение, два параллельных переноса
и два сдвига. Соответствующая пятипараметрическая группа движений
задается уравнениями (5.11).

Установленное противоречие и доказывает, что между четырнадцатью
площадями (5.12) должны существовать дополнительные нетривиальные
функциональные связи, не являющиеся следствием основной связи
(5.10), которая потому не является порождающей.

\newpage

      \noindent
     phenomenologically
     symmetric geometry of rank 4. It is also necessary, under the same Definition, that the
     relation (5.10) be a generating one in the sense as follows: {\it any other nontrivial
     relation must be a corollary of that relation (5.10).}

     We shall demonstrate that that very condition the equation (5.10) does not satisfy. We
     shall take on a plane a heptuple of points $\langle ijklpmn \rangle$ and assign to it, according the
     function (5.9), all the thirty-five areas $S(ijk),S(ijl), \\ \ldots,S(imn); \ S(jkl),S(jkp),\ldots,$\
     $S(pmn)$ of the respective triangles $\langle ijk \rangle,\ldots, \\ \langle pmn \rangle$. Out of these thirty-five areas,
     it is possible to eliminate, under the relation (5.10), with respect to the relation, twenty:
     $S(jkl),\ldots, S(pmn)$. Among the remaining fifteen areas, \ $S(ijk),S(ijl),\ldots,$ $S(imn)$
     there is one trivial relation, for they depend on fourteen coordinates $x_i,y_i,x_j,y_j, \\ \ldots,x_n,y_n$
     only. If the relation (5.10) is an originating one, then, eliminating for example the area $S(ijk)$
     yields fourteen areas

     $$
     S(ijl),S(ijp),\ldots,S(imn),
     \eqno(5.12)
     $$  \\
     that must be independent. There may not exist a trivial relation among them, because
     they are functions of the fourteen coordinates of the septuple $\langle ijklpmn \rangle$, but there are
     nontrivial relations among them, which is easily made sure. If there did not exist such
     relations then the seven-point figure $\langle ijklpmn \rangle$ would not have had any single degree
     of freedom of motion, because on the fourteen coordinates of the points of it there
     would have been imposed as many independent correlations conditioned by the invariance
     of the fourteen areas (5.12). But in reality, the seven-point figure$\langle ijklpmn \rangle$ may move
     on the plane, while preserving the areas of all the thirty-five triangles, with five degrees
     of freedom: one rotation, two parallel translations, and two shifts. The corresponding
     5-parameter group of motions is defined by the equations (5.11).

      The contradiction
     established is that what proves that among the fourteen areas (5.12) there must exist
     additional nontrivial functional relations that are not corollaries of the relation (5.10), and
     thus that last relation is not an originating  one.

     \newpage

Таким образом, функция (5.9) не задает на плоскости
$\mathfrak{M}=R^2$ тернарную феноменологически симметричную
геометрию ранга 4 в смысле определения 1, несмотря на наличие
функциональной связи (5.10), так как ранг проекции отображения
$F':\mathfrak{M}^7\to R^{35}$, задаваемой четырнадцатью площадями
(5.12) и не содержащей отображения $F:\mathfrak{M}^4\to R^4$,
меньше четырнадцати, то есть не максимален.

Групповая симметрия бинарных феноменологически симметричных геометрий,
которым было уделено основное
внимание в монографии автора "Полиметрические геометрии" [10] и в
первой главе настоящей его
монографии, является определяющей. То есть функция $f:\mathfrak{S}_f\to R^s$,
где $\mathfrak{S}_f\subseteq\mathfrak{M}^2$,
будет задавать феноменологически симметричную геометрию
в том и только в
том случае, если она допускает нетривиальную конечномерную группу движений.
Условие наделения феноменологически симметричной геометрии
групповой симметрией конечной степени определяет эту
степень, устанавливая ее связь с размерностью многообразия и рангом
феноменологической симметрии соотношениями (5.7) и (5.8). С другой стороны,
без предположения о групповой симметрии в смысле
определения 2 даже соотношение (5.7),
устанавливающее связь размерности многообразия и ранга феноменологической
симметрии и не содержащее
степень групповой симметрии, должно оговариваться
дополнительно без достаточно убедительного обоснования этой связи.

\vspace{15mm}

\begin{center}
{\bf \large \S6. Функциональные уравнения в геометрии }
\end{center}

Рассмотрим одномерную геометрию, которая задается
невырожденной метрической функцией
$$
f(ij)=f(x_i,x_j).
\eqno(6.1)
$$
Согласно определению 1 из \S1 ее феноменологическая симметрия
ранга 3 выражается уравнением
\begin{equation}
\label{pr1}
\Phi(f(ij),f(ik),f(jk)) = 0.
\tag{$6.2$}
\end{equation}

Уравнение (6.2) должно обращаться в тождество по координатам
$x_i,x_j,x_k$ точек тройки $\langle ijk \rangle$ при подстановке в него функции
(6.1).

\newpage

     So the function (5.9) does not give on the
     plane $\mathfrak{M}=R^2$ a ternary phenomenologically symmetric geometry of rank 4 in
     the sense of Definition 1, despite the presence of the functional relation (5.10), because the
     rank of the projection of the mapping $F':\mathfrak{M}^7\to R^{35}$ defined by the fourteen
     areas (5.12) and not including the mapping $F:\mathfrak{M}^4\to R^4$ is less than fourteen,
     i.e. it is not maximal.

     The group symmetry of binary phenomenologically symmetric geometries, which are the
     main subject in the author's monograph "Polymetric geomet- \ ries" \ [10], as well as in
     Chapter 1 of this monograph, is a determining one. That is, the function $f:\mathfrak{S}_f\to R^s$,
     where $\mathfrak{S}_f\subseteq\mathfrak{M}^2$, will define a phenomenologically symmetric
     geometry if and only if it allows a nontrivial finite-dimensional group of motions. The condition of a
     phenomenologically symmetric geometry being endowed with a group symmetry of finite degree
     determines that degree, by establishing the relation of it with the dimensio- \\ nality of the manifold
     and the rank of the phenomenological symmetry, by the relations (5.7) and (5.8). On the other
     hand, without the assumption of the group symmetry in the sense of Definition 2 even the relation
     (5.7), that establishes the connection between the dimensionality of the manifold and the rank
     of the phenomenological symmetry, and having no degree of the group symmetry, must be
     stipulated in a complementary way without any sufficient grounding of such connection.

     \vspace{10mm}

     \begin{center}
     {\bf \large \S6. Functional equations in geometry  }

     \end{center}

     Let us consider the one-dimensional geometry that is defined by the nondegenerate
     metric function
     $$
     f(ij)=f(x_i,x_j).
     \eqno(6.1)
     $$
     Under Definition 1 of \S1, its phenomenological symmetry of rank 3 is expressed by
     the equation
     \begin{equation}
     \label{pr1}
     \Phi(f(ij),f(ik),f(jk)) = 0.
     \tag{$6.2$}
     \end{equation}

     The equation (6.2) must turn into an identity in the coordinates $x_i,x_j,x_k$ of the
     triple of points $\langle ijk \rangle$ when the function (6.1) is substituted into
     it.

\newpage

Таким образом, уравнение (6.2) фактически является функциональным
уравнением как относительно метрической функции (6.1), так и
относительно функции $\Phi$, с помощью которой выражается
феноменологическая симметрия одномерной геометрии.

Ниже кратко опишем метод решения функционального уравнения (6.2).
Сначала представим его в виде, разрешенном
относительно одного из аргументов:
$$
f(ij) = \varphi(f(ik),f(jk)),
\eqno(6.3)
$$
где $\varphi(u,v)$ -- гладкая функция двух переменных с отличными
от нуля производными $\varphi_{u}$ и $\varphi_{v}$.

Возьмем, далее, упорядоченную четверку точек $\langle ijkl \rangle$ и запишем
уравнение (6.3) для троек $\langle ijk \rangle, \ \langle ijl \rangle, \ \langle ikl \rangle, \ \langle jkl \rangle$:
$$
\left.\begin{array}{rcl}
f(ij) = \varphi(f(ik),f(jk)),\\
f(ij) = \varphi(f(il),f(jl)),\\
f(ik) = \varphi(f(il),f(kl)),\\
f(jk) = \varphi(f(jl),f(kl)),\;
\end{array}\right\}
$$
откуда легко получаем равенство
\begin{displaymath}
\varphi[\varphi(f(il),f(kl)),\varphi(f(jl),f(kl))] =
\varphi(f(il),f(jl)),
\end{displaymath}
в котором, очевидно, независимы переменные $f(il), f(jl),
f(kl)$. Если для них ввести обозначения $x = f(il), y = f(jl),
z = f(kl)$, то приходим к функциональному уравнению
\begin{equation}
\label{pr2}
\varphi(\varphi(x,z),\varphi(y,z)) = \varphi(x,y).
\tag{$6.4$}
\end{equation}
имеющему следующее нетривиальное решение:
\begin{equation}
\label{pr3}
\varphi(u,v) = \psi(\psi^{-1}(u) - \psi^{-1}(v)),
\tag{$6.5$}
\end{equation}
где $\psi$ -- произвольная гладкая функция одной переменной с
$\psi' \neq 0$, \ $\psi^{-1}$ -- обратная к ней функция.

С помощью решения (6.5) от уравнения (6.3) приходим к уравнению
(6.2):
$$
\psi^{-1}(f(ij))-\psi^{-1}(f(ik))+\psi^{-1}(f(jk))=0.
\eqno(6.6)
$$

Явный вид самой метрической функции (6.1) можно
найти из того же уравнения (6.3) с решением (6.5), если в нем зафиксировать
координату

\newpage

       Thus, the equation
     (6.2) is in fact the functional equation both with respect to the metric function (6.1), and
     as concerns the function$\Phi$ that expresses the phenomenological symmetry of the
     one-dimensional geometry.

     Further we shall describe in brief a method of solution of the functional
     equation (6.2). First, we shall set it down in the form solved with respect
     to one of the arguments:
     $$
     f(ij) = \varphi(f(ik),f(jk)),
     \eqno(6.3)
     $$
     where $\varphi(u,v)$ is a smooth function of 2 variables with unequal to zero
     derivatives $\varphi_{u}$ and $\varphi_{v}$.

     We take, then, an ordered quadruple of points $\langle ijkl \rangle$ and write the equation
     (6.3) for the triples $\langle ijk \rangle, \ \langle ijl \rangle, \ \langle ikl \rangle, \ \langle jkl \rangle$:
     $$
     \left.\begin{array}{rcl}
     f(ij) = \varphi(f(ik),f(jk)),\\
     f(ij) = \varphi(f(il),f(jl)),\\
     f(ik) = \varphi(f(il),f(kl)),\\
     f(jk) = \varphi(f(jl),f(kl)),\;
     \end{array}\right\}
     $$
     wherefrom we readily get the equality
     \begin{displaymath}
     \varphi[\varphi(f(il),f(kl)),\varphi(f(jl),f(kl))] =
     \varphi(f(il),f(jl)),
     \end{displaymath}
     where, obviously, the variables $f(il), f(jl)$, and $f(kl)$ are independent. If we
     introduce for them designation $x = f(il), y = f(jl), z = f(kl)$, we arrive at the functional
     equation
     \begin{equation}
     \label{pr2}
     \varphi(\varphi(x,z),\varphi(y,z)) = \varphi(x,y).
     \tag{$6.4$}
     \end{equation}
     that has a nontrivial solution as follows:
     \begin{equation}
     \label{pr3}
     \varphi(u,v) = \psi(\psi^{-1}(u) - \psi^{-1}(v)),
     \tag{$6.5$}
     \end{equation}
     where $\psi$ is an arbitrary smooth function of one variable with $\psi' \neq 0$, \
     and $\psi^{-1}$ is the function that is its inverse.

     By using the solution (6.5) of the equation (6.3), we arrive at the equation (6.2):
     $$
     \psi^{-1}(f(ij))-\psi^{-1}(f(ik))+\psi^{-1}(f(jk))=0.
     \eqno(6.6)
     $$

     The explicit form of the metric function (6.1) itself can again be found from the equation
     (6.3) with the solution (6.5), if we fix in it the coordinate

     \newpage

\noindent $x_k$ точки $k$:
$$
f(ij)=\psi(\varphi(x_i)-\varphi(x_j)),
\eqno(6.7)
$$
где $\varphi(x)=\psi^{-1}(f(x,x_k))|_{x_k=const}$.

В совокупности уравнение (6.6) и функция (6.7) являются общим решением
функционального уравнения (6.2). С точностью до замены координаты
$\varphi(x)\to x$ в одномерном многообразии и масштабного
преобразования $\psi^{-1}(f)\to f$ метрической функции это решение может быть
записано в следующей канонической форме:
$$
\left.\begin{array}{cccl}
f(ij) = x_i-x_j,\\
f(ij)-f(ik)+f(jk)=0.
\end{array}\right\}
\eqno(6.8)
$$

Гладкое обратимое преобразование одномерного многообразия
$$
x'=\lambda(x)
\eqno(6.9)
$$
называется движением, если оно сохраняет метрическую функцию: \linebreak
$f(i'j')=f(ij)$. Отсюда при известной метрической функции (6.1) получаем
функциональное уравнение на множество движений:
$$
f(\lambda(x_i),\lambda(x_j))=f(x_i,x_j),
\eqno(6.10)
$$
решением которого для одномерной феноменологически симметричной геометрии
является однопараметрическая группа
$$
x'=\lambda(x;a).
\eqno(6.11)
$$

Обратно, если известна однопараметрическая группа преобразований (6.11), то
метрическая функция одномерной геометрии, для
которой эта группа будет группой движений, найдется как ее двухточечный
инвариант решением следующего функционального уравнения
$$
f(\lambda(x_i;a),\lambda(x_j;a))=f(x_i,x_j). \eqno(6.12)
$$

Пусть инфинитезимальный оператор
$$
X=\lambda(x)\partial/\partial x
\eqno(6.13)
$$
принадлежит алгебре Ли группы движений (6.11). Тогда метрическая функция
одновременно является решением дифференциального уравнения
$$
X(i)f(ij)+X(j)f(ij)=0,
\eqno(6.14)
$$

\newpage

      \noindent
     $x_k$ of the point $k$:
     $$
     f(ij)=\psi(\varphi(x_i)-\varphi(x_j)),
     \eqno(6.7)
     $$
     where $\varphi(x)=\psi^{-1}(f(x,x_k))|_{x_k=const}$.

     Together, the equation (6.6) and the function (6.7) are a general solution of the
     functional equation (6.2). With an accuracy up to a change of coordina- \ tes $\varphi(x)\to x$
     in the one-dimensional manifold and a scaling transformation $\psi^{-1}(f)\to f$ of the metric
     function, that solution may be written in the following canonical form:
     $$
     \left.\begin{array}{cccl}
     f(ij) = x_i-x_j,\\
     f(ij)-f(ik)+f(jk)=0.
     \end{array}\right\}
     \eqno(6.8)
     $$

     A smooth invertible transformation of the one-dimensional manifold
     $$
     x'=\lambda(x)
     \eqno(6.9)
     $$
     is called a motion if it preserves the metric function: $f(i'j')=f(ij)$. Hence, with the
     known metric function (6.1), we get the functional equation for the set of motions:
     $$
     f(\lambda(x_i),\lambda(x_j))=f(x_i,x_j),
     \eqno(6.10)
     $$
     whose solution for a one-dimensional phenomenologically symmetric geometry is
     the 1-parameter group
     $$
     x'=\lambda(x;a).
     \eqno(6.11)
     $$

     And vice versa, if the one-parameter group of transformations (6.11) is known the
     metric function of the one-dimensional geometry for which that group is the group
     of motions will be found as its two-point invariant, through the solution of the functional
     equation as follows
     $$
     f(\lambda(x_i;a),\lambda(x_j;a))=f(x_i,x_j).
     \eqno(6.12)
     $$

     Let the infinitesimal operator
     $$
     X=\lambda(x)\partial/\partial x
     \eqno(6.13)
     $$
      belongs to the Lie algebra of the group of motions (6.11). Then the metric
     function is simultaneously the solution of the differential equation
     $$
     X(i)f(ij)+X(j)f(ij)=0,
     \eqno(6.14)
     $$

    \newpage

\noindent
которое при известной метрической функции (6.1)
является функциональным уравнением на коэффициент $\lambda(x)$
оператора (6.13).

Двумерная геометрия задается невырожденной метрической функцией
$$
f(ij)=f(x_i,y_i,x_j,y_j),
\eqno(6.15)
$$
а ее феноменологическая симметрия выражается уравнением
$$
\Phi(f(ij),f(ik),f(il),f(jk),f(jl),f(kl))=0.
\eqno(6.16)
$$

При подстановке метрической функции (6.15) в уравнение (6.16) оно
по восьми координатам $x_i,y_i,x_j,y_j,x_k,y_k,x_l,y_l$ точек
четверки $\langle ijkl \rangle$ должно обратиться в тождество. Таким образом,
это уравнение в действительности является особого рода
функциональным уравнением относительно как самой метрической
функции $f$, так и относительно функции $\Phi$, с помощью которой
выражается феноменологическая симметрия двумерной геометрии. С
точностью до замены координат в двумерном многообразии и
масштабного преобразования $\psi(f)\to f$ все возможные решения
уравнения (6.16) относительно метрической функции (6.15) могут
быть записаны в одиннадцати канонических формах (2.7) -- (2.17).
Что же касается функции $\Phi$, то не всегда ее можно записать в
явном виде, о чем более подробно было сказано в \S2 после
указанной классификации.

Множество обратимых движений
$$
x'=\lambda(x,y), \ y'=\sigma(x,y),
\eqno(6.17)
$$
сохраняющих метрическую функцию (6.15), является совокупностью
решений функционального уравнения
$$
f(\lambda(i),\sigma(i),\lambda(j),\sigma(j))=f(x_i,y_i,x_j,y_j).
\eqno(6.18)
$$
Согласно теореме 3 из \S1 это множество есть трехпараметрическая
группа реобразований многообразия:
$$
x'=\lambda(x,y;a^1,a^2,a^3), \ y'=\sigma(x,y;a^1,a^2,a^3),
\eqno(6.19)
$$
определяющая групповую симметрию соответствующей двумерной
геометрии. Если же группа преобразований \ (6.19) \  известна, то
с точно-

\newpage

     \noindent
     which, with the metric function (6.1) known, is the functional equation
     for the coefficient $\lambda(x)$ of the operator (6.13).

     The two-dimensional geometry is defined by the nondegenerate metric function
     $$
     f(ij)=f(x_i,y_i,x_j,y_j),
     \eqno(6.15)
     $$
     and its phenomenological symmetry is expressed by the equation
     $$
     \Phi(f(ij),f(ik),f(il),f(jk),f(jl),f(kl))=0.
     \eqno(6.16)
     $$

     If the metric function (6.15) is substituted into the equation (6.16) then, with respect to
     the eight coordinates $x_i,y_i,x_j,y_j,x_k,y_k,x_l,y_l$ of the points of the quadruple $\langle ijkl \rangle$,
     it must turn into an identity. Thus, that equation is really a special kind of a functional
     equation both with respect of the metric function $f$ and with respect of the function
     $\Phi$, which is part of the phenomenological symmetry of the two-dimensional geometry.
     With an accuracy up to a change of coordinates in a two-dimensional manifold and a scaling
     transformation $\psi(f)\to f$, all the possible solutions of the equation (6.16), with respect
     of the metric function (6.15), may be written in the eleven canonical forms (2.7) -- (2.17).
     As to the function $\Phi$, it cannot always be written in the explicit form, of which we
     spoke in more detail in \S2, after the said classification had been given.

     The set of invertible motions
     $$
     x'=\lambda(x,y), \ y'=\sigma(x,y),
     \eqno(6.17)
     $$
     that preserve the metric function (6.15) is the totality of the solutions of the functional
     equation
     $$
     f(\lambda(i),\sigma(i),\lambda(j),\sigma(j))=f(x_i,y_i,x_j,y_j).
     \eqno(6.18)
     $$
     Under Theorem 3 of \S1, that totality is the three-parameter group of \\ transformations
     of the manifold:
     $$
     x'=\lambda(x,y;a^1,a^2,a^3), \ y'=\sigma(x,y;a^1,a^2,a^3),
     \eqno(6.19)
     $$
     and defines the group symmetry of the corresponding two-dimensional geo- \ metry. And
     if the group of transformations (6.19) is known, then, with
     an

     \newpage

\noindent
 стью до масштабного преобразования
метрическая функция (6.15) восстанавливается как ее невырожденный
двухточечный инвариант решением другого функционального уравнения:
$$
f(\lambda(i;a),\sigma(i;a),\lambda(j;a),\sigma(j;a))=f(x_i,y_i,x_j,y_j),
\eqno(6.20)
$$
где, например, $\lambda(i;a)=\lambda(x_i,y_i;a^1,a^2,a^3)$.

Пусть
$$
X=\lambda(x,y)\partial/\partial x+\sigma(x,y)\partial/\partial y
\eqno(6.21)
$$
есть инфинитезимальный оператор трехмерной алгебры Ли группы движений (6.19).
Тогда метрическая функция (6.15) является также решением
дифференциального уравнения
$$
X(i)f(ij)+X(j)f(ij)=0.
\eqno(6.22)
$$
Однако при известной метрической функции (6.15) оно уже будет функциональным
уравнением на коэффициенты $\lambda$ и $\sigma$ оператора (6.21).

 {\it Циклом двумерной геометрии} назовем такую гладкую невырожденную кривую
$$
x=x(t), \ y=y(t),
\eqno(6.23)
$$
по которой свободно может катиться жесткий треугольник $\langle ijk \rangle$. Ясно, что на
множестве точек этой кривой метрическая функция (6.15) должна задавать
феноменологически симметричную одномерную геометрию, наделенную групповой
симметрией степени 1. В результате получаем следующее функциональное
уравнение на цикл [23, \S12]:
$$
f(x(t_i),y(t_i),x(t_j),y(t_j))=\psi(t_i-t_j).
\eqno(6.24)
$$

Например, для плоскости Евклида функциональное уравнение
$$
(x(t_i)-x(t_j))^2+(y(t_i)-y(t_j))^2=\psi(t_i-t_j)
$$
имеет два решения [23, \S15]:
$$
{\bf 1)} \ \ x=at+b, \ y=ct+d;  \ \ \
{\bf 2)} \ \ x=R\cos t + x_0, \ y=R\sin t + y_0,
$$
где $a^2+c^2\neq 0, \ R>0$, которые задают на ней множество прямых
и множество окружностей.

\newpage

     \noindent
     accuracy up to a scaling
     transformation, the metric function (6.15) is reconst- \ ructed as its nondegenerate two-point
     invariant by way of the solution of another functional equation:
     $$
     f(\lambda(i;a),\sigma(i;a),\lambda(j;a),\sigma(j;a))=f(x_i,y_i,x_j,y_j),
     \eqno(6.20)
     $$
     where, for example, $\lambda(i;a)=\lambda(x_i,y_i;a^1,a^2,a^3)$.

     Let
     $$
     X=\lambda(x,y)\partial/\partial x+\sigma(x,y)\partial/\partial y
     \eqno(6.21)
     $$
     be an infinitesimal operator of the three-dimensional Lie algebra of the group of
     motions (6.19). Then the metric function (6.15) is also a solution of the differential
     equation
     $$
     X(i)f(ij)+X(j)f(ij)=0.
     \eqno(6.22)
     $$
     However, with the metric function (6.15) known, it is already the functional
     equation for the coefficients $\lambda$ and $\sigma$ of the operator (6.21).

     We shall call {\it a cycle of a two-dimensional geometry} such a smooth
     nondegenerate curve
     $$
     x=x(t), \ y=y(t),
     \eqno(6.23)
     $$
     along which a rigid triangle $\langle ijk \rangle$ may roll freely. It is obvious that on the set of
     the points of that curve the metric function (6.15) must give a phenomenologically
     symmetric one-dimensional geometry endowed with a group symmetry of degree 1.
     As result, we have the functional equation for
     the cycle [23, \S12] as follows:
     $$
     f(x(t_i),y(t_i),x(t_j),y(t_j))=\psi(t_i-t_j).
     \eqno(6.24)
     $$

     For example, for the Euclidean plane the functional equation
     $$
     (x(t_i)-x(t_j))^2+(y(t_i)-y(t_j))^2=\psi(t_i-t_j)
     $$
     has two solutions [23, \S15]:
     $$
     {\bf 1)} \ \ x=at+b, \ y=ct+d;  \ \ \
     {\bf 2)} \ \ x=R\cos t + x_0, \ y=R\sin t + y_0,
     $$
     where $a^2+c^2\neq 0, \ R>0$ giving on the plane a set of straight lines and \\ a
     set of circles.

\newpage

Заметим, что функциональные уравнения, аналогичные уравнениям
(6.18), (6.20), (6.22), (6.24), могут быть записаны для любой
геометрии, задаваемой метрической функцией (1.2). Например, для
трехмерной геометрии, задаваемой метрической функцией (2.20),
функциональное уравнение на цикл будет следующим:
$$
f(x(t_i),y(t_i),z(t_i),x(t_j),y(t_j),z(t_j))=\psi(t_i-t_j),
$$
а для трехмерной геометрии, задаваемой трехкомпонентной метрической функцией
(3.14), система функциональных уравнений на цикл запишется точно также, но в
ней уже $f=(f^1,f^2,f^3)$ \ и  \ $\psi = (\psi^1,\psi^2,\psi^3)$. Метод решения
большинства геометрических функциональных уравнений состоит в сведении их к
дифференциальным и разделении переменных.

\vspace{15mm}

\begin{center}
{\bf \large \S7. Вопросы классификации феноменологически симметричных
геометрий}
\end{center}

В теории физических структур классификация феноменологически симметричных
геометрий является одной из наиболее важных задач. Дело в том, что и сами
метрические функции (1.2), задающие такие геометрии, и уравнения (1.1),
выражающие их феноменологическую симметрию, могут иметь содержательную
физическую интерпретацию.

В \S2 и \S3 приведены построенные к настоящему времени полные классификации
одномерных, двумерных и трехмерных феноменологически симметричных
геометрий соответствующих рангов 3, 4 и 5,
а также двуметрических, триметрических и
четыреметрических феноменологически симметричных геометрий
минимального ранга, равного 3. Другие классификации пока не
построены, так как еще не найдены новые методы решения подобных задач.

Ниже будут перечислены те классификационные задачи, которые, с одной стороны,
являются естественным продолжением уже решенных, а с другой -- могут
привлечь тех читателей, которым удастся найти более эффективные
методы их решения.

\newpage

     We shall note that functional equations similar to the equations (6.18), (6.20), (6.22),
     (6.24) may be written for any geometry defined by the metric function (1.2). For example,
     for the three-dimensional geometry defined by the metric function (2.20), the functional
     equation for the cycle will be as follows:
     $$
     f(x(t_i),y(t_i),z(t_i),x(t_j),y(t_j),z(t_j))=\psi(t_i-t_j),
     $$
     and for the three-dimensional geometry defined by the metric function (3.14), the system of
     functional equations for the cycle will be written in the same way, but in it $f=(f^1,f^2,f^3)$ \
     and \ $\psi = (\psi^1,\psi^2,\psi^3)$. The method of solution for most geometrical functional
     equations is that of reduction to differential ones and of separating the variables.

     \vspace{25mm}

     \begin{center}

     {\bf \large \S7. Problems of classification of phenomenologically \\ symmetric geometries }
     \end{center}

     In the theory of physical structures, classification of phenomenologically symmetric
     geometries is one of the most important problems. The thing is, both the metric functions
     (1.2) themselves, defining such geometries, and the equations (1.1), expressing their
     phenomenological symmetry, may have an essential physical interpretation.

     In \S2 and \S3 the complete classifications are given that have been built of one-dimensional,
     two-dimensional, and three-dimensional phenomenologically symmetric geometries of
     respective ranks 3, 4, and 5, as well as of the dimetric, trimetric and four-metric
     phenomenologically symmetric geometries of the minimal rank, i.e. that equal to 3. Any
     other classifications have not yet been built, because up till now we have not found any
     new methods of solving problems of that kind.

     Now, we shall give the classification problems that are, on the one hand, a natural
     extension of those already solved, and, on the other, may interest those readers who will
     be able to find more effective methods of their solution.

   \newpage

{\bf 1.} {\it Четырехмерные геометрии}

\vspace{5mm}

Четырехмерная геометрия задается на четырехмерном многообразии
$\mathfrak{M}$ невырожденной метрической функцией
$$
f(ij)=f(x_i,y_i,z_i,t_i,x_j,y_j,z_j,t_j),
\eqno(7.1)
$$
а ее феноменологическая симметрия выражается уравнением
$$
\Phi(f(ij),f(ik),f(il),f(im),f(in),f(jk),...,f(jn),...,f(mn))=0,
\eqno(7.2)
$$
устанавливающим связь 15 взаимных расстояний между шестью точками
кортежа $\langle ijklmn \rangle$ из некоторого открытого и плотного в $\mathfrak{M}^6$
множества. Наделена же такая геометрия групповой симметрией степени 10.

Предварительная классификация четырехмерных геометрий
была \linebreak приведена в конце \S2. Но эту классификацию нельзя считать
окончательной, так как используемые методы не позволили преодолеть
встретившиеся трудности технического характера. В настоящее время В.А. Кыровым
разрабатывается новый метод их классификации, опирающийся на гипотезу о
вложении. Суть этой гипотезы
иллюстрирует классификация (2.23) -- (2.37) трехмерных феноменологически
симметричных геометрий, из которой
видно, что каждая ее метрическая функция содержит внутри себя как целое
метрическую функцию, задающую двумерную феноменологически
симметричную геометрию:
$$
f(ij)=f(g(ij),z_i,z_j),
$$
где выражение для $g(ij)=g(x_i,y_i,x_j,y_j)$ берется из классификации
(2.7) -- (2.17). К сожалению, строгого доказательства гипотезы о
вложении пока нет, однако уже имеющимися полными классификациями она
подтверждается.

\vspace{5mm}

{\bf 2.} {\it Двуметрические и триметрические геометрии}

\vspace{5mm}

Двуметрические и триметрические феноменологически симметричные
геометрии минимального ранга 3 были рассмотрены в \S3, \ и \ для

\newpage

{\bf 1.} {\it Four-dimensional geometries}

     \vspace{5mm}

     A four-dimensional geometry is defined on a four-dimensional manifold $\mathfrak{M}$
     by the nondegenerate metric function

     $$
     f(ij)=f(x_i,y_i,z_i,t_i,x_j,y_j,z_j,t_j),
     \eqno(7.1)
     $$ \\
     and its phenomenological symmetry is expressed by the equation
    $$
     \Phi(f(ij),f(ik),f(il),f(im),f(in),f(jk),..., f(jn),...,f(mn))=0,
     \eqno(7.2)
     $$
     that establishes the relation of 15 reciprocal distances among the six points
     of a cortege $\langle ijklmn \rangle$ of some set open and dense in $\mathfrak{M}^6$.
     The group symmetry such a geometry is endowed with is of degree 10.

     A preliminary classification of four-dimensional geometries was given in the
     end of \S2. That classification cannot be considered complete, for the methods
     employed in building it cannot help overcome technical difficulties that have
     been encountered. Recently, V.A. Kyrov has been developing a new method of
     their classification, one based on the hypothesis of enclosure. The essence of the
     hypotheses is illustrated by the classification (2.23) -- (2.37) of the three-dimensional
     phenomenologically symmetric geometries where it can be seen that each metric
     function of it contains in itself as a whole a metric function defining a
     two-dimensional phenomenologically symmetric geometry:
     $$
     f(ij)=f(g(ij),z_i,z_j),
     $$
     where the expression for $g(ij)=g(x_i,y_i,x_j,y_j)$ is borrowed from the classification
     (2.7) -- (2.17). Unfortunately, we have not any rigorous proof of the hypothesis
     of enclosure, but it seems to be confirmed by those complete classifications that
     are already available.

     \vspace{5mm}

     {\bf 2.} {\it Dimetric and trimetric geometries}

     \vspace{5mm}

     The dimetric and trimetric phenomenologically symmetric geometries of
     minimal rank 3 were discussed \ in \S3, and complete classifications (3.12) -

    \newpage

\noindent
них построены полные классификации (3.12) - (3.13) и (3.28) -
(3.38). Естественно поэтому перейти к классификации двуметрических
и триметрических феноменологически симметричных геометрий более
высокого ранга равного четырем. Например двуметрическая
феноменологически симметричная геометрия ранга 4 задается на
четырехмерном многообразии двухкомпонентной метрической функцией
$$
f(ij)=(f^1(ij),f^2(ij))=f(x_i,y_i,z_i,t_i,x_j,y_j,z_j,t_j),
\eqno(7.3),
$$
а ее феноменологическая симметрия выражается уравнением
$$
\Phi(f(ij),f(ik),f(il),f(jk),f(jl),f(kl))=0,
\eqno(7.4)
$$
где $\Phi$ -- двухкомпонентная функция 12 переменных. Степень ее групповой
симметрии равна 6.

В полную классификацию таких геометрий, очевидно, входят все
пары метрических функций из классификации (2.7) -- (2.17), причем
$f^1(ij)=f^1(x_i,y_i,x_j,y_j)$, \ $f^2(ij)=f^2(z_i,t_i,z_j,t_j)$. Например,
сочетание функций (2.7) и (2.12), задающих плоскость Евклида и
симплектическую плоскость:
$$
f^1(ij)=(x_i-x_j)^2+(y_i-y_j)^2, \ f^2(ij)=z_it_j-z_jt_i
$$
есть двухкомпонентная метрическая функция, задающая одну из
двуметрических четырехмерных феноменологически симметричных
геометрий ранга 4. Общее число сочетаний, включая диагональные,
равно 65. Но, возможно, существуют и такие двуметрические
геометрии, метрические функции (7.3) которых подобными сочетаниями
не являются.

\vspace{5mm}

{\bf 3.} {\it Полиметрические геометрии}

\vspace{5mm}

Классификации некоторых феноменологически симметричных полиметрических
геометрий минимального ранга, равного 3, были представлены в \S3. Такие
геометрии задаются в пространстве $R^s$ $s$-компонентной метрической функцией
$$
f(ij)=(f^1(ij),...,f^s(ij))=f(x^1_i,...,x^s_i,x^1_j,...,x^s_j),
$$
а их феноменологическая симметрия выражается уравнением
$$
\Phi(f(ij),f(ik),f(jk))=0,
$$

\newpage

      \noindent
     (3.13)
     and (3.28) - (3.38) have been built for them. So it is
     natural to undertake next
     classifying dimetric and trimetric phenomenologically symmet- \ ric geometries
     of higher rank, that is of rank 4. For example, the dimetric phenomenologically
     symmetric geometry of rank 4 is defined on a four-dimensional manifold by the
     two-component metric function
     $$
     f(ij)=(f^1(ij),f^2(ij))=f(x_i,y_i,z_i,t_i,x_j,y_j,z_j,t_j),
     \eqno(7.3),
     $$
     and its phenomenological symmetry is expressed by the equation
     $$
     \Phi(f(ij),f(ik),f(il),f(jk),f(jl),f(kl))=0,
     \eqno(7.4)
     $$
     where $\Phi$ is a two-component function of 12 variables. The degree of its
     group symmetry is 6.

     The complete classification of such geometries naturally includes all
     the pairs of the metric functions of the classification (2.7) -- (2.17), with
     $f^1(ij)=f^1(x_i,y_i,x_j,y_j)$, \ $f^2(ij)=f^2(z_i,t_i,z_j,t_j)$. For example,
     the combination of the functions (2.7) and (2.12), giving Euclidean plane
     and the simplectic plane:
     $$
     f^1(ij)=(x_i-x_j)^2+(y_i-y_j)^2, \ f^2(ij)=z_it_j-z_jt_i
     $$
     is a two-component metric function that defines one of the
     dimetric four-dimensional phenomenologically symmetric geometries
     of rank 4. The comp- \ lete number of combinations, including the diagonal
     ones, is 65. But it is possible there exist such dimetric
     geometries whose metric functions (7.3) are not such combinations.

     \vspace{8mm}

     {\bf 3.} {\it Polymetric geometries}

          \vspace{8mm}

     The classifications of some phenomenologically symmetric polymetric
     geometries of minimal rank, equal to 3, were given in \S3. Such geometries
     are defined in the space $R^s$ by an $s$-component metric function
     $$
     f(ij)=(f^1(ij),...,f^s(ij))=f(x^1_i,...,x^s_i,x^1_j,...,x^s_j),
     $$
     and their phenomenological symmetry is expressed by the equation
     $$
     \Phi(f(ij),f(ik),f(jk))=0,
     $$

    \newpage

    \noindent
 где $\Phi$ -- $s$-компонентная функция $3s$
переменных. Полная классификация построена только для $s=1,2,3,4$
и отсутствует для $s\geq5$.

Классификация феноменологически симметричных
геометрий еще не завершена. Поэтому имеет смысл представить в
обозримом виде перечень задач, как решенных,
так и не решенных. Тогда каждый, у кого появится желание испытать
свои силы и способности, сможет выбрать
для этого любую из них (в том числе и решенную -- для
разработки новых методов классификации  и проверки уже полученных
результатов). Приведем этот перечень в виде следующей таблицы:
\begin{center}
\begin{tabular}{|c|c|c|c|c|c|c|c|}\hline
\multicolumn{8}{|c|}{\bf Феноменологически симметричные геометрии
}
\\ \hline № & $s$ & $n$ & $sn$ & $m = n+2$ & $sn(n + 1)/2$ & реш.
& ист.\\ \hline
1 & 1 & 1 & 1 & 3 & 1& $+$ & \S2 \\
2 & 1 & 2 & 2 & 4 & 3 & $+$ & \S2 \\
3 & 1 & 3 & 3 & 5 & 6 & $+$ &  \S2\\
4 & 1 & 4 & 4 & 6 & 10 & $-$  & $-$ \\
5 & 1 & $\geqslant 5$ & $5,6,\ldots$ & $7,8,\ldots$ &
$15,21,\ldots$ & $-$ & $-$ \\ \hline
6 & 2 & 1 & 2 & 3 & 2 & + & \S3 \\
7 & 2 & $\geqslant 2$ & $4,6,\ldots$ & $4,5,\ldots$ &
$6,12,\ldots$ & $-$ & $-$ \\ \hline
8 & 3 & 1 & 3 & 3 & 3 & $+$ & \S3 \\
9 & 3 & $\geqslant 2$ & $6,9,\ldots$ & $4,5,\ldots$ &
$9,18,\ldots$ & $-$ & $-$ \\ \hline
10 & 4 & 1 & 4 & 3 & 4 & $+$ & \S3 \\
11 & 4 & $\geqslant 2$ & $8,12,\ldots$ & $4,5,\ldots$ &
$12,24,\ldots$ & $-$ & $-$ \\ \hline 12 & $\geqslant 5$ &
$\geqslant 1$ & $\geqslant 5$ & $\geqslant 3$ & $\geqslant 5$ &
$-$ & $-$ \\ \hline
\end{tabular}
\end{center}

\vspace{5mm}

Напомним, что $s$ -- число компонент метрической функции
\linebreak $f = (f^{1},\ldots,f^{s})$, которая на $sn$-мерном
многообразии задает феноменологически симметричную геометрию ранга
$m = n + 2$, группа движений которой зависит от $sn(n + 1)/2$
непрерывных параметров. В последних двух столбцах таблицы знаком
плюс сделана отметка о решении данной задачи и указан параграф, в
котором находится соответствующая полная классификация.

Метрические функции, задающие на многообразии феноменологически
симметричные геометрии, являются невырожденными двухточеч-

\newpage

      \noindent
      where $\Phi$ is an $s$-component function of $3s$ variables. The
      complete
     classifica- \ tion has only been built for $s=1,2,3$, and $4$ and is not available for
     $s\geq5$.

     Thus, the classification of the phenomenologically symmetric geometries is not
     finished yet. So it makes sense to present in a visually graspable form the
     problems both solved and not yet solved. Then, anyone who would like to
     try one's abilities can choose any of them (the solved included, for developing
     new methods of classification and inspecting the results obtained). We shall
     represent the list of problems in a table format as follows:

      \vspace{2mm}

     \begin{center}
     \begin{tabular}{|c|c|c|c|c|c|c|c|}\hline
     \multicolumn{8}{|c|}{\bf The phenomenologically symmetric
     geometries } \\ \hline
     № & $s$ & $n$ & $sn$ & $m = n+2$ & $sn(n + 1)/2$ & solved & source\\ \hline
     1 & 1 & 1 & 1 & 3 & 1& $+$ & \S2 \\
     2 & 1 & 2 & 2 & 4 & 3 & $+$ & \S2 \\
     3 & 1 & 3 & 3 & 5 & 6 & $+$ &  \S2\\
     4 & 1 & 4 & 4 & 6 & 10 & $-$  & $-$ \\
     5 & 1 & $\geqslant 5$ & $5,6,\ldots$ & $7,8,\ldots$ & $15,21,\ldots$
     & $-$ & $-$ \\ \hline
     6 & 2 & 1 & 2 & 3 & 2 & $+$ & \S3 \\
     7 & 2 & $\geqslant 2$ & $4,6,\ldots$ & $4,5,\ldots$ & $6,12,\ldots$
     & $-$ & $-$ \\ \hline
     8 & 3 & 1 & 3 & 3 & 3 & $+$ & \S3 \\
     9 & 3 & $\geqslant 2$ & $6,9,\ldots$ & $4,5,\ldots$ & $9,18,\ldots$
     & $-$ & $-$ \\ \hline
     10 & 4 & 1 & 4 & 3 & 4 & $+$ & \S3 \\
     11 & 4 & $\geqslant 2$ & $8,12,\ldots$ & $4,5,\ldots$ & $12,24,\ldots$
     & $-$ & $-$ \\ \hline
     12 & $\geqslant 5$ & $\geqslant 1$ & $\geqslant 5$ & $\geqslant 3$ &
     $\geqslant 5$ & $-$ & $-$ \\ \hline
     \end{tabular}
     \end{center}

     \vspace{2mm}

     We shall remind that $s$ is the number of the components of the metric
     function $f = (f^{1},\ldots,f^{s})$ that defines on an $sn$-dimensional manifold
     a phenomenologically symmetric geometry of rank $m = n + 2$ whose group
     of motions depends on $sn(n + 1)/2$ continuous parameters. In the two
     right-hand columns the plus sign means that the problem has been solved
     and the number of the paragraph is given where the corresponding complete
     classification is to be found.

     The metric functions defining on the manifold phenomenologically
     sym- \ metric geometries are nondegenerate two-point invariants of some
     groups

    \newpage

\noindent
ными инвариантами некоторых групп преобразований этого
многообразия. Задача их классификации предполагала предварительную
классификацию групп преобразований с определенным числом
непрерывных параметров. Однако с ростом числа компонент
метрической функции и ранга феноменологической симметрии
задаваемой ею геометрии проведение классификации групп
преобразований становится технически очень сложной задачей.
Поэтому естественно возникает вопрос: является ли предварительная
классификация групп преобразований многообразия действительно
необходимой. Ведь для многих из них двухточечные инварианты
оказываются вырожденными. Следовательно, имеет смысл сначала
установить, какому условию должна удовлетворять группа
преобразований, чтобы ее двухточечный инвариант был невырожденным.
Например, при построении классификации (3.28)--(3.38) трехмерных
триметрических феноменологически симметричных геометрий ранга 3
таким условием была транзитивность группы преобразований. В целом
же вопрос о дополнительных ограничениях на группы преобразований,
следующих из невырожденности их двухточечных инвариантов, пока
остается открытым. Отметим еще, что все определения и результаты
главы 1 локальны. Их глобализация требует качественно нового шага
в развитии методов исследования и является новой проблемой,
значимость которой обусловливается содержательной интерпретацией
феноменологически симметричных геометрий не только в самой
геометрии, но и в физике.

      \newpage

      \noindent
      of transformations of the manifold. The problem of their classification
     implied building
     up a preliminary classification of the groups of transformations
     with the certain number of the continuous parameters. But, with the number
     of the components of the metric function and the rank of the phenomenologi- \ cal
     symmetry of the geometry it defines increasing, classifying groups of motions
     becomes rather a strenuous task, in the technical sense. So, quite naturally a
     question suggests itself: Is a preliminary classification of the groups of
     transformations of a manifold really necessary. After all, for many of them the
     two-point invariants turn out to be degenerate. Therefore, it seems to make
     sense first to establish what condition must a group of transformations satisfy
     in order that the two-point invariant of it be nondegenerate. For example, in
     building the classification (3.28)--(3.38) of the three-dimensional trimetric
     phenomenologically symmetric geometries of rank 3 such condition was that of
     transitivity of the group of transforma- \ tions. But at large the problem of
     additional constraints on groups of transformations following from their
     two-point invariants being nondegene- \ rate is still open. We shall also note that
     all the definitions and results of Chapter 1 are local. Their globalization requires
     a new qualitative approach in the development of the methods of research and
     presents a new problem, whose significance is conditioned by the possibility of
     essential interpreta- \ tions of the phenomenologically symmetric geometries, not
     only in geometry, but in physics too.

    \newpage

\begin{center}
{\bf
\Large ГЛАВА II \\  Физическая структура как геометрия  \\ двух множеств}
\end{center}

\vspace{20mm}

\begin{center}
{\bf \large \S8. Феноменологическая и групповая симметрии \\ физических структур}
\end{center}

Пусть имеются два множества $\mathfrak{M}$ и $\mathfrak{N}$,
являющиеся $sm$-мерным и $sn$-мерным многообразиями, где $s,m$ и
$n$ -- натуральные числа, точки которых будем обозначать строчными
латинскими и греческими буквами соответственно, а также функция
$f:\mathfrak{M} \times \mathfrak{N} \rightarrow R^s$,
сопоставляющая паре $\langle  i\alpha  \rangle$ из области ее определения
$\mathfrak{S}_f \subseteq \mathfrak{M} \times \mathfrak{N}$
некоторую совокупность $s$ вещественных чисел
$f(i\alpha)=(f^1(i\alpha),\ldots,f^s(i\alpha))\in R^s$. Заметим,
что в общем случае $\mathfrak{S}_f \neq \mathfrak{M} \times
\mathfrak{N}$, то есть функция $f$ не любой паре из $\mathfrak{M}
\times \mathfrak{N}$ сопоставляет $s$ чисел, но в последующем
изложении удобно в явной записи значения $f(i\alpha)$ этой функции
для пары $\langle  i\alpha  \rangle$ подразумевать, что $\langle  i\alpha  \rangle \ \in
\mathfrak{S}_f$. Обозначим через $U(i)$ и $U(\alpha)$ окрестности
точек $i \in \mathfrak{M}$ и $\alpha \in \mathfrak{N}$, через
$U(\langle  i\alpha  \rangle)$ -- окрестность пары $\langle  i\alpha  \rangle \ \in
\mathfrak{M} \times \mathfrak{N}$ и аналогично окрестности
кортежей из других прямых произведений множеств $\mathfrak{M}$ и
$\mathfrak{N}$ на себя или друг на друга.

Для некоторых кортежей $\langle  \alpha_1 \ldots \alpha_m  \rangle \ \in \mathfrak{N}^m$ и
$\langle  i_1 \ldots i_n  \rangle \ \in \mathfrak{M}^n$ введем функции
$f^m=f[\alpha_1 \ldots \alpha_m]$ и $f^n=f[i_1 \ldots i_n]$, сопоставляя
точкам $i \in \mathfrak{M}$ и $\alpha \in \mathfrak{N}$ точки
$(f(i\alpha_1), \ldots ,f(i\alpha_m))\in R^{sm}$ и
$(f(i_1\alpha), \ldots ,$ $f(i_n\alpha)) \in R^{sn}$, если пары
$\langle  i\alpha_1  \rangle, \ldots ,\langle  i\alpha_m  \rangle$ и $\langle  i_1\alpha  \rangle, \ldots ,
\langle  i_n\alpha  \rangle$
принадлежат $\mathfrak{S}_f$. Заметим, что функции $f^m$ и $f^n$ не
обязательно определены всюду на множествах $\mathfrak{M}$ и $\mathfrak{N}$.
Будем предполагать выполнение следующих трех аксиом:

{\bf I}. Область определения $\mathfrak{S}_f$ функции $f$ есть открытое и
плотное в $\mathfrak{M} \times \mathfrak{N}$ множество.

{\bf II}. Функция $f$ в области своего определения достаточно гладкая.

{\bf III}. В $\mathfrak{N}^m$ и $\mathfrak{M}^n$ плотны множества таких кортежей
длины $m$ и $n$ для которых функции $f^m$ и $f^n$ имеют максимальные ранги,
равные $sm$ и $sn$, в точках плотных в $\mathfrak{M}$ и $\mathfrak{N}$
множеств соответственно.

\newpage

\begin{center}
     {\bf
     \Large CHAPTER II \\
     A physical structure as a geometry of  \\ two sets}
     \end{center}

     \vspace{15mm}

     \begin{center}
     {\bf \large \S8. The phenomenological and group symmetry of \\ physical structures}
     \end{center}

     Let there be two sets $\mathfrak{M}$ and $\mathfrak{N}$ that are an
     $sm$-dimensional and an $sn$-dimensional manifolds, where  $s,m$ and
     $n$ are natural numbers, whose points we shall designate by Latin and Greek
     lower-case letters respectively, and a function $f:\mathfrak{M} \times \mathfrak{N} \rightarrow R^s$
     that assigns to a pair $\langle  i\alpha  \rangle$ from its  domain $\mathfrak{S}_f \subseteq \mathfrak{M} \times \mathfrak{N}$
     some collection of $s$ real numbers $f(i\alpha)=(f^1(i\alpha),\ldots,f^s(i\alpha))\in R^s$.
     We shall note that in the general case $\mathfrak{S}_f \neq \mathfrak{M} \times \mathfrak{N}$,
     i.e. the function $f$ does not assign $s$ numbers to {\it every} pair from
     $\mathfrak{M} \times \mathfrak{N}$ but in the further discussion it will
     be convenient in the explicit writing of the value $f(i\alpha)$ of the function
     for a pair $\langle  i\alpha  \rangle$ to consider that $\langle  i\alpha  \rangle \ \in \mathfrak{S}_f$.
     We shall designate by $U(i)$ and $U(\alpha)$ the neighbourhoods of the points
     $i \in \mathfrak{M}$ and $\alpha \in \mathfrak{N}$ and by $U(\langle  i\alpha  \rangle)$ the
     neighbourhood of the pair $\langle  i\alpha  \rangle \ \in \mathfrak{M} \times \mathfrak{N}$,
     and similarly for the corteges from the other direct products
     of the sets $\mathfrak{M}$ and $\mathfrak{N}$ each by itself of by each other.

     For some corteges $\langle  \alpha_1 \ldots \alpha_m  \rangle \ \in \mathfrak{N}^m$ and
     $\langle  i_1 \ldots i_n  \rangle \ \in \mathfrak{M}^n$, we shall introduce a function
     $f^m=f[\alpha_1 \ldots \alpha_m]$ and $f^n=f[i_1 \ldots i_n]$, by assigning to
     the points $i \in \mathfrak{M}$ and $\alpha \in \mathfrak{N}$ points
     $(f(i\alpha_1), \ldots ,f(i\alpha_m))\in R^{sm}$ and
     $(f(i_1\alpha), \ldots ,$ $f(i_n\alpha)) \in R^{sn}$ if the pairs
     $\langle  i\alpha_1  \rangle, \ldots ,\langle  i\alpha_m  \rangle$ and $\langle  i_1\alpha  \rangle, \ldots ,
     \langle  i_n\alpha  \rangle$ belong to $\mathfrak{S}_f$. We shall note that the functions
     $f^m$ and $f^n$ are not necessarily defined everywhere on the sets $\mathfrak{M}$
     and $\mathfrak{N}$. We shall assume that three axioms hold as follows:

     {\bf I}. The domain $\mathfrak{S}_f$ of the function $f$ is a set that is
     open and dense in $\mathfrak{M} \times \mathfrak{N}$.

     {\bf II}. The function $f$ in its domain is sufficiently smooth.

     {\bf III}. In $\mathfrak{N}^m$ and $\mathfrak{M}^n$ the sets of corteges of
     lengths $m$ and $n$ are dense such for which the functions $f^m$ and $f^n$
     have maximal ranks equal to $sm$ and $sn$ in the points of sets dense in $\mathfrak{M}$
     and $\mathfrak{N}$respectively.

\newpage

Достаточная гладкость означает, что в области своего определения
непрерывна как сама функция $f$, так и все ее производные
достаточно высокого порядка. Гладкую функцию $f$, для которой
выполняется условие III, будем называть {\it невырожденной}.
Заметим также, что ограничения в аксиомах I, II, III открытыми и
плотными подмножествами связано с тем, что исходные множества
могут содержать исключительные подмножества меньшей размерности,
где эти аксиомы не выполняются.

Введем еще функцию $F$, сопоставляя кортежу
$\langle  ijk \ldots v, \alpha \beta \gamma \ldots \tau  \rangle$ длины $m+n+2$ из
$\mathfrak{M}^{n+1} \times \mathfrak{N}^{m+1}$ точку
$(f(i\alpha),f(i\beta), \ldots ,f(v\tau)) \in R^{s(m+1)(n+1)}$, координаты
которой в $R^{s(m+1)(n+1)}$ определяются упорядоченной по исходному кортежу
последовательностью значений функции $f$ для всех пар его элементов
$(\langle  i\alpha  \rangle$, $\langle  i\beta  \rangle, \ldots ,\langle  v\tau  \rangle)$, если эти пары принадлежат
$\mathfrak{S}_f$. Область определения введенной функции есть, очевидно,
открытое и плотное в $\mathfrak{M}^{n+1} \times \mathfrak{N}^{m+1}$ множество,
которое обозначим через $\mathfrak{S}_F$.

{\bf Определение 1.} Будем говорить, что функция $f=(f^1, \ldots ,f^s)$
задает на $sm$-мерном и $sn$-мерном
многообразиях $\mathfrak{M}$ и $\mathfrak{N}$ \ {\it физическую
структуру (феноменологически симметричную геометрию двух множеств)
ранга $(n+1,m+1)$}, если, кроме аксиом I, II, III, дополнительно
выполняется следующая аксиома:

{\bf IY}. Существует плотное в $\mathfrak{S}_F$ множество, для каждого кортежа \linebreak
$\langle  ijk \ldots v,\alpha\beta\gamma$ $\ldots \tau  \rangle$ длины $m+n+2$ которого и
некоторой его окрестности $U(\langle  i \ldots \tau  \rangle)$ найдется такая достаточно
гладкая $s$-компонентная функция $\Phi:\mathcal{E} \rightarrow R^s$,
определенная в некоторой области $\mathcal{E} \subset R^{s(m+1)(n+1)}$,
содержащей точку $F(\langle i \ldots \tau \rangle)$, что \ в \ ней \ rang $\Phi =s$ \ и \ множество \\
$F(U(\langle i \ldots \tau \rangle))$ является подмножеством множества нулей функции $\Phi$,
то есть
$$
\Phi(f(i\alpha),f(i\beta), \ldots ,f(v\tau))=0
\eqno(8.1)
$$
для всех кортежей из $U(\langle ijk \ldots v,\alpha\beta\gamma \ldots \tau \rangle)$.

Аксиома IY составляет содержание принципа феноменологической симметрии.
Уравнения (8.1) задают $s$ функциональных
связей между $s(m+1)(n+1)$ измеряемыми в опыте значениями $s$ физических величин

\newpage

      Sufficient smoothness means that both the function $f$ and all its derivatives
     of sufficiently high order are continuous in the domain of the function $f$. We
     shall call a smooth function $f$ that satisfies Condition III a {\it nondegenerate}
     one. We shall also note that the restriction in Axioms I, II, and III by open and
     dense subsets is due to the possibility that the original sets may contain
     exceptional subsets of smaller dimensionality where those axioms do not hold.

     We shall also introduce a function $F$, by assigning to a cortege
     $\langle  ijk \ldots v, \\ \alpha \beta \gamma \ldots \tau  \rangle$ of length $m+n+2$ from
     $\mathfrak{M}^{n+1} \times \mathfrak{N}^{m+1}$ the point
     $(f(i\alpha),f(i\beta), \ldots , \\ f(v\tau)) \in R^{s(m+1)(n+1)}$ whose coordinates
     in $R^{s(m+1)(n+1)}$ are determined by the series of values of the function $f$
     for all the pairs of the elements of the original cortege
     $(\langle  i\alpha  \rangle$, $\langle  i\beta  \rangle, \ldots ,\langle  v\tau  \rangle)$ ordered by that original cortege
     if all those pairs belong to $\mathfrak{S}_f$. The domain of the function introduced
     is, obviously, a set open and dense in $\mathfrak{M}^{n+1} \times \mathfrak{N}^{m+1}$.
     We shall designate it by $\mathfrak{S}_F$.

     {\bf Definition 1.} We shall say that the function $f=(f^1, \ldots ,f^s)$
     gives on an $sm$-dimensional and an $sn$-dimensional
     manifolds $\mathfrak{M}$ and $\mathfrak{N}$ \ a {\it physical
     structure (a phenomenologically symmetric geometry of two sets)
     of rank $(n+1,m+1)$} if, in addition to Axioms I, II, and III, one more
     axiom holds as follows:

     {\bf IY}. There exists a set dense in $\mathfrak{S}_F$ for whose every cortege
     $\langle  ijk \ldots v,\alpha\beta\gamma$ $\ldots \tau  \rangle$ of length $m+n+2$ and
     some neighbourhood $U(\langle  i \ldots \tau  \rangle)$ of it a sufficiently smooth
     $s$-component function $\Phi:\mathcal{E} \rightarrow R^s$ may be found
     defined in some region $\mathcal{E} \subset R^{s(m+1)(n+1)}$ that contains
     the point $F(\langle  i \ldots \tau  \rangle)$, such that rang $\Phi =s$ and the set
     $F(U(\langle  i \ldots \tau  \rangle))$ is a subset of the set of zeros of the function $\Phi$,
     that is
     $$
     \Phi(f(i\alpha),f(i\beta), \ldots ,f(v\tau))=0
     \eqno(8.1)
     $$ \\
     for all the corteges from $U(\langle  ijk \ldots v,\alpha\beta\gamma \ldots \tau  \rangle)$.

     Axiom IY gives the essence of the principle of phenomenological symmetry.
     The equations (8.1) define $s$ functional relations among $s(m+1)(n+1)$ values
     of \ $s$ \ physical quantities \ $f=(f^1, \ldots ,f^s)$ \ measured by experiment

     \newpage

\noindent
$f=(f^1, \ldots ,f^s)$ и являются аналитическим выражением
физического закона, записанного в феноменологически симметричной
форме. Условие rang $\Phi =s$ означает, что уравнения $\Phi=0$ (то
есть $\Phi_1=0, \ldots ,\Phi_s=0$) независимы.

Пусть $x=(x^1, \ldots ,x^{sm})$ и $\xi=(\xi^1, \ldots ,\xi^{sn})$ -- локальные
координаты в многообразиях $\mathfrak{M}$ и $\mathfrak{N}$. Для исходной
функции $f$ в некоторой окрестности $U(i) \times U(\alpha)$ каждой пары
$\langle  i\alpha  \rangle \ \in \mathfrak{S}_f$ получаем тогда локальное координатное
представление

$$
f(i\alpha)=f(x_i,\xi_\alpha)=
f(x^1_i, \ldots ,x^{sm}_i,\xi^1_\alpha, \ldots ,\xi^{sn}_\alpha),
\eqno(8.2)
$$ \\
свойства которого определяются аксиомами II и III. Поскольку по
аксиоме III ранги функций $f^m$ и $f^n$ максимальны, координаты
$x$ и $\xi$ входят в представление (8.2) существенным образом.
Последнее означает, что никакая гладкая локально обратимая замена
координат не приведет к уменьшению их числа в представлении (8.2),
то есть ни для какой локальной системы координат его невозможно
записать в виде

$$
f(i\alpha)=f(x^1_i, \ldots ,x^{m'}_i,\xi^1_\alpha, \ldots ,\xi^{n'}_\alpha),
$$
где или $m'< sm$, или $n'< sn$. Действительно, если, например,
$m'< sm$, то для любого кортежа $\langle  \alpha_1 \ldots \alpha_m  \rangle \
\in (U(\alpha))^m$ длины $m$ и для любой точки из $U(i)$ ранг
функции $f^m=f[\alpha_1 \ldots \alpha_m]$ будет заведомо меньше
$sm$, что противоречит аксиоме III. Заметим, однако, что
существенная зависимость представления (8.2) от локальных
координат $x_i$ и $\xi_\alpha$ еще не гарантирует выполнения
аксиомы III. То есть при наличии всех координат в любом таком
представлении функция $f$ может оказаться вырожденной.

Функцию $f=(f^1, \ldots ,f^s)$ будем рассматривать как метрическую в геометрии
двух множеств. Но поскольку $s$ расстояний $f(i\alpha)$ определены для точек
разных множеств, обычные аксиомы метрики здесь не имеют смысла.

Используя представление (8.2), запишем локальное координатное
задание для введенной выше функции $F$:

\newpage

    \noindent
    and
     are an analytical expression of a law of physics written in the  \\ phenomenologically
     symmetric form. The condition of rang $\Phi =s$ means that the
      equations $\Phi=0$ (i.e. $\Phi_1=0, \ldots ,\Phi_s=0$) are independent.

     Let $x=(x^1, \ldots ,x^{sm})$ and $\xi=(\xi^1, \ldots ,\xi^{sn})$ be local
     coordinates in the manifolds $\mathfrak{M}$ and $\mathfrak{N}$. Then for the original
     function $f$, we have, in some neighbourhood $U(i) \times U(\alpha)$ of every pair
     $\langle  i\alpha  \rangle \ \in \mathfrak{S}_f$, the local coordinate representation

     $$
     f(i\alpha)=f(x_i,\xi_\alpha)=
     f(x^1_i, \ldots ,x^{sm}_i,\xi^1_\alpha, \ldots ,\xi^{sn}_\alpha),
     \eqno(8.2)
     $$ \\
     whose properties are determined by Axioms II and III. Since under Axiom III
     the ranks of the functions $f^m$ and $f^n$ are maximal, the coordinates
     $x$ and $\xi$ are included in the representation (8.2) in an essential manner.
     The latter implies that no smooth local invertible change of coordinates may result
     in their number in the representation (8.2) being decreased, i.e. it may not, in any
     local system of coordinates, be written as

     $$
     f(i\alpha)=f(x^1_i, \ldots ,x^{m'}_i,\xi^1_\alpha, \ldots ,\xi^{n'}_\alpha),
     $$
     where either $m'< sm$, or $n'< sn$. Indeed, if for example$m'< sm$, then for
     any cortege $\langle  \alpha_1 \ldots \alpha_m  \rangle \ \in (U(\alpha))^m$ of length $m$
     and for any point of $U(i)$ the rank of the function $f^m=f[\alpha_1 \ldots \alpha_m]$
     will a fortiori be less than $sm$, which is in contradiction with Axiom III.
     We shall note, however, that the essential dependence of the representation(8.2)
     on the local coordinates $x_i$ and $\xi_\alpha$ is not a sufficient guaranty of Axiom
     III being satisfied. That is, the function $f$ may turn out to be degenerate, with all
     the coordinates present in any such representation.

     We shall consider the function $f=(f^1, \ldots ,f^s)$ as a metric one in a geometry
     of two sets. But since $s$ distances $f(i\alpha)$ are determined for the points
     of more than one set, the ordinary metric axioms are of no sense here.

     Using the representation (8.2), let us write the local coordinate definition for the
     function $F$ that we have introduced:

\newpage

$$
\left.\begin{array}{c}
f(i\alpha )=f(x_i,\xi_\alpha), \\
f(i\beta )=f(x_i,\xi_\beta), \\
\cdots \cdots \cdots \cdots \cdots \cdots \\
f(v\tau )=f(x_v,\xi_\tau),
\end{array}\right\}
\eqno(8.3) \\
$$ \\
функциональная матрица которого

$$
\left\|
\begin{array}{cccccccccc}
\displaystyle\frac{\partial f(i\alpha )}{\partial x_i}  & 0 & ... & 0 & 0 &
\displaystyle\frac{\partial f(i\alpha )}{\partial \xi_\alpha} & 0 & ... & 0 & 0 \\
... & ... & ... & ... & ... & ... & ... & ... & ... & ... \\
0 & 0 & ... & 0 & \displaystyle\frac{\partial f(v\tau )}{\partial x_v} & 0 & 0 & ... & 0
& \displaystyle\frac{\partial f(v\tau )}{\partial \xi_\tau}
\end{array}
\right\|
\eqno(8.4)
$$ \\
имеет $s(m+1)(n+1)$ строк и $s(2mn+m+n)$ столбцов. Здесь через
$\partial f/\partial x$ и $\partial f/\partial \xi$ кратко обозначены
соответствующие функциональные матрицы для компонент функции
$f=(f^1, \ldots , f^s)$
по координатам $x=(x^1, \ldots ,x^{sm})$ и $\xi=(\xi^1, \ldots ,\xi^{sn})$
соответственно:

$$
\partial f/\partial x=\left\|
\begin{array}{ccc}
\partial f^1/\partial x^1 & \cdots & \partial f^1/\partial x^{sm} \\
\cdots & \cdots & \cdots \\
\partial f^s/\partial x^1 & \cdots & \partial f^s/x^{sm}
\end{array}
\right\| ,
$$

$$
\partial f/\partial \xi =\left\|
\begin{array}{ccc}
\partial f^1/\partial \xi ^1 & \cdots & \partial f^1/\partial \xi ^{sn} \\
\cdots & \cdots & \cdots \\
\partial f^s/\partial \xi ^1 & \cdots & \partial f^s/\partial \xi ^{sn}
\end{array}
\right\| .
$$ \\

Задание (8.3) для функции $F$ представляет собой систему $s(m+1)(n+1)$ функций
$f^1(i\alpha), \ldots ,f^s(i\alpha), \ldots ,f^1(v\tau), \ldots ,f^s(v\tau)$,
специальным образом зависящих от $s(2mn+m+n)$ переменных $x^1_i, \ldots ,
x^{sm}_i,$
$\ldots ,\xi^1_\tau, \ldots ,$
$\xi^{sn}_\tau$ -- координат всех
точек кортежа $\langle  ijk ... v,\alpha\beta\gamma ... \tau  \rangle$ длины $m+n+2$.
Поскольку число функций в системе (8.3) не больше общего числа переменных,
наличие связи (8.1) является нетривиальным фактом, не имеющим места для
произвольных функций в этой системе.

\newpage

     $$
     \left.\begin{array}{c}
     f(i\alpha )=f(x_i,\xi_\alpha), \\
     f(i\beta )=f(x_i,\xi_\beta), \\
     \cdots \cdots \cdots \cdots \cdots \cdots \\
     f(v\tau )=f(x_v,\xi_\tau),
     \end{array}\right\}
     \eqno(8.3)
     $$ \\
     whose functional matrix

     $$
     \left\|
     \begin{array}{cccccccccc}
     \displaystyle\frac{\partial f(i\alpha )}{\partial x_i}  & 0 & ... & 0 & 0 &
     \displaystyle\frac{\partial f(i\alpha )}{\partial \xi_\alpha} & 0 & ... & 0 & 0 \\

     ... & ... & ... & ... & ... & ... & ... & ... & ... & ... \\
     0 & 0 & ... & 0 & \displaystyle\frac{\partial f(v\tau )}{\partial x_v} & 0 & 0 & ... & 0
     & \displaystyle\frac{\partial f(v\tau )}{\partial \xi_\tau}

     \end{array}
     \right\|
     \eqno(8.4)
     $$ \\
     has $s(m+1)(n+1)$ rows and $s(2mn+m+n)$ columns. Here, by $\partial f/\partial x$ and $\partial f/\partial \xi$
     the respective functional matrices are briefly designated for the components
     of the function $f=(f^1, \ldots , f^s)$ with respect to the coordinates
     $x=(x^1, \ldots ,x^{sm})$ and $\xi=(\xi^1, \ldots ,\xi^{sn})$ respectively:

     $$
     \partial f/\partial x=\left\|
     \begin{array}{ccc}
     \partial f^1/\partial x^1 & \cdots & \partial f^1/\partial x^{sm} \\
     \cdots & \cdots & \cdots \\
     \partial f^s/\partial x^1 & \cdots & \partial f^s/x^{sm}
     \end{array}
     \right\| ,
     $$

     $$
     \partial f/\partial \xi =\left\|
     \begin{array}{ccc}
     \partial f^1/\partial \xi ^1 & \cdots & \partial f^1/\partial \xi ^{sn} \\
     \cdots & \cdots & \cdots \\
     \partial f^s/\partial \xi ^1 & \cdots & \partial f^s/\partial \xi ^{sn}
     \end{array}
     \right\| .
     $$ \\

     The definition (8.3) for the function $F$ is a system of $s(m+1)(n+1)$ functions
     $f^1(i\alpha), \ldots ,f^s(i\alpha), \ldots ,f^1(v\tau), \ldots ,f^s(v\tau)$ that
     depend in a special manner on $s(2mn+m+n)$ variables $x^1_i, \ldots , x^{sm}_i,$
     $\ldots ,\xi^1_\tau, \ldots ,$ $\xi^{sn}_\tau$ -- the coordinates of all the
     points of the cortege $\langle  ijk ... v,\alpha\beta\gamma ... \tau  \rangle$ of length $m+n+2$.
     Since the number of functions in the system (8.3) is not more than the total number
     of the variables, the presence of the relation (8.1) is a nontrivial fact, not taking place
     for arbitrary functions in that system.

    \newpage

Рассмотренные в первой главе геометрии одного множества
показывают, что их феноменологическая и групповая симметрии
взаимно обуславливают друг друга. Так, связь между шестью
расстояниями для любых четырех точек в двумерной геометрии, не
обязательно евклидовой, приводит к существованию в ней
трехпараметрической группы движений. Но движение в геометрии двух
множеств имеет свою специфику, отличную от привычных свойств
движения в геометрии одного множества. Поэтому необходимо дать
точные определения.

Под локальным движением в геометрии двух множеств $\mathfrak{M}$ и
$\mathfrak{N}$ будем понимать такую пару взаимно однозначных гладких
отображений (преобразований)
$$
\lambda:U \rightarrow U' \ \ \text{и} \ \ \sigma:V \rightarrow V',
\eqno(8.5)
$$
где $U,U'\subset \mathfrak{M}$ и $V,V'\subset \mathfrak{N}$ -- открытые
области, при которых функция $f$ сохраняется. Последнее означает, что для
каждой пары $\langle  i\alpha  \rangle \ \in \mathfrak{S}_f$, такой что $i \in U, \ \alpha \in
V$ и $\langle  i'\alpha'  \rangle \ \in \mathfrak{S}_f$, где $i'=\lambda(i)\in U'$,
$\alpha'=\sigma (\alpha)\in V'$, имеет место равенство
$$
f(\lambda(i),\sigma(\alpha))=f(i\alpha),
\eqno(8.6)
$$
выполняющееся для каждой из компонент $f^1, \ldots ,f^s$ функции $f$.

Множество всех движений (8.5) есть локальная группа преобразований, для которой
функция $f$, согласно равенству (8.6), является {\it двухточечным инвариантом}.
Преобразования $\lambda$ и $\sigma$ в движениях (8.5) сами составляют две
отдельные группы, а группа движений есть их взаимное расширение. Если функция
$f$ известна, например, в своем локальном координатном представлении (8.2),
то равенство (8.6) представляет собой функциональное уравнение относительно
преобразований $\lambda$ и $\sigma$. Нам же о функции $f$ известно только, что
она невырождена и удовлетворяет
некоторой системе $s$ независимых уравнений (8.1). Но этого оказывается
достаточно для установления факта существования группы ее движений, зависящей
от $smn$ параметров.

{\bf Определение 2.} Будем говорить, что функция $f=(f^1, \ldots
,f^s)$ задает на $sm$-мерном и $sn$-мерном многообразиях
$\mathfrak{M}$ и $\mathfrak{N}$ {\it геометрию двух множеств,
наделенную групповой симметрией степени $smn$}, если, кроме аксиом
I, II, III,  выполняется  ещё и следующая аксиома:

\newpage

     The geometries of one set discussed in Chapter I demonstrate that their
     phenomenological and group symmetry mutually condition each other. Thus,
     the relation among the six distances for any four points of a two-dimensional
     geometry, not only the Euclidean one, results in the appearing of a three-parameter
     group of motions in it. But motion in a geometry of two sets is specific, and
     very different from that in a geometry of one set. That is why it is pertinent that
     exact definitions should be given.

     Under the name of a local motion in a geometry of two sets $\mathfrak{M}$ and
     $\mathfrak{N}$ we shall understand such a pair of biunique (one-to-one) smooth
     mappings (transformations)
     $$
     \lambda:U \rightarrow U' \ \ \text{и} \ \ \sigma:V \rightarrow V',
     \eqno(8.5)
     $$
     where $U,U'\subset \mathfrak{M}$ and $V,V'\subset \mathfrak{N}$ are open
     regions, at which the function $f$ is preserved. The latter means that for every
     pair $\langle  i\alpha  \rangle \ \in \mathfrak{S}_f$, such that $i \in U, \ \alpha \in
     V$ and $\langle  i'\alpha'  \rangle \ \in \mathfrak{S}_f$, where $i'=\lambda(i)\in U'$,
     $\alpha'=\sigma (\alpha)\in V'$, the equality

  $$
     f(\lambda(i),\sigma(\alpha))=f(i\alpha),
     \eqno(8.6)
     $$ \\
     takes place for each component $f^1, \ldots ,f^s$ of the function $f$.

     The set of all motions (8.5) is a local group of transformations for which the function
     $f$, under the equality (8.6), is a {\it two-point invariant}. The transformations $\lambda$
     and $\sigma$ in the motions (8.5) are themselves two separate groups, and the
     group of motions is their mutual extension. If the function$f$ is known, in its local
     coordinate representation (8.2), for example, then the equality (8.6) is the functional
     equation with respect to the transformations $\lambda$ and $\sigma$. However, we
     only know about the function $f$ that it is nondegenerate and satisfies some system
     of $s$ independent equations (8.1). But that turns out to be enough to establish the
     fact of existence of the group of motions of it depending on $smn$ parameters.

     {\bf Definition 2.} We shall say that the function $f=(f^1, \ldots ,f^s)$ gives on an
     $sm$-dimensional and an $sn$-dimensional manifolds $\mathfrak{M}$ and
     $\mathfrak{N}$ a {\it geometry of two sets endowed with a group symmetry of degree
     $smn$} if in addition to Axioms I, II, and III, one more axiom holds as follows:

     \newpage

{\bf IY$'$}. Существуют открытые и плотные в $\mathfrak{M}$ и
$\mathfrak{N}$ множества, для всех точек $i$ и $\alpha$ которых
определены эффективные гладкие  действия $smn$-мерной локальной
группы Ли в некоторых окрестностях $U(i)$ и $U(\alpha)$, такие что
действия ее в окрестностях $U(i),U(j)$ и $U(\alpha),U(\beta)$
точек $i,j$ и $\alpha,\beta$ совпадают в пересечениях $U(i) \cap
U(j)$ и $U(\alpha) \cap U(\beta)$, и функция $f$ является
двухточечным инвариантом по каждой из своих $s$ компонент.

Напомним, что группы Ли преобразований гладких
многообразий были описаны в \S1 перед формулировкой аналогичной аксиомы IV$'$
в геометрии одного множества. Группы преобразований, о которых говорится
в аксиоме IY$'$ настоящего параграфа, определяют
своеобразную локальную подвижность жестких фигур ("твердых тел") в
пространстве $\mathfrak{M} \times \mathfrak{N}$ с $smn$ степенями свободы.
Заметим, что глобальной подвижности при этом может и не быть. Множество пар
$\langle  i\alpha  \rangle$, для которых функция $f$ определена и одновременно является
двухточечным инвариантом, очевидно, открыто и плотно в
$\mathfrak{M} \times \mathfrak{N}$.

Согласно аксиоме IY$'$, на открытых и плотных в $\mathfrak{M}$ и $\mathfrak{N}$
множествах заданы $smn$-мерные линейные семейства гладких векторных полей $X$
и $\Xi$, замкнутые относительно операции коммутирования, то есть алгебры Ли
преобразований. В некоторых локальных системах координат в
многообразиях $\mathfrak{M}$ и $\mathfrak{N}$ базисные векторные поля этих
семейств запишем в операторной форме:
$$
\left.\begin{array}{c}
X_{\omega}=\lambda^{\mu}_{\omega}(x)\partial/\partial x^{\mu}, \\
\Xi_{\omega}=\sigma^{\nu}_{\omega}(\xi)\partial/\partial \xi^{\nu},
\end{array}\right\}
\eqno(8.7)
$$
где $\omega=1,\ldots ,smn$, а по "немым" индексам $\mu$ и $\nu$ производится
суммирование от 1 до $sm$ и от 1 до $sn$ соответственно.
По критерию инвариантности функция
$f(i\alpha)$ будет инвариантом локальной группы преобразований некоторой
окрестности $U(i) \times U(\alpha)$, то есть двухточечным инвариантом, в том и
только в том случае, если она покомпонентно удовлетворяет системе $smn$
уравнений
$$
X_{\omega}(i)f(i\alpha)+\Xi_{\omega}(\alpha)f(i\alpha)=0
\eqno(8.8)
$$
с операторами (8.7):
$$
\lambda^{\mu}_{\omega}(i)\frac{\partial f(i\alpha)}{\partial x^{\mu}_i} +
\sigma^{\nu}_{\omega}(\alpha)\frac {\partial f(i\alpha)}
{\partial \xi^{\nu}_\alpha} =0,
\eqno(8.9)
$$

\newpage

     {\bf IY$'$}. There exist sets open and dense in $\mathfrak{M}$ and $\mathfrak{N}$
     for all the points $i$ and $\alpha$ of which effective smooth actions of an
     $smn$-dimensional local Lie group are defined in some neighbourhoods $U(i)$
     and $U(\alpha)$, such that its actions in the neighbourhoods $U(i),U(j)$ and
     $U(\alpha),U(\beta)$ of the points $i,j$ and $\alpha,\beta$ coincide in the intersections
     $U(i) \cap U(j)$ and $U(\alpha) \cap U(\beta)$ and the function $f$ is a two-point
     invariant in each of its $s$ components.

     We shall remind that groups of Lie transformations of smooth manifolds were
     described in \S1 when a similar axiom, Axiom IV$'$ was formulated of a geometry
     of one set. The groups of transformations in question in Axiom IY$'$ of this
     paragraph define a peculiar local mobility of rigid figures ("solid bodies") in the
     space $\mathfrak{M} \times \mathfrak{N}$ with $smn$ degrees of freedom.
     We shall note that global mobility is not necessarily implied. The set of pairs
     $\langle  i\alpha  \rangle$ for which the function $f$ is defined and is simultaneously the
     two-point invariant is, obviously, open and dense in $\mathfrak{M} \times \mathfrak{N}$.

     Under Axiom IY$'$, there are $smn$-dimensional linear families of smooth vector fields
     $X$ and $\Xi$ defined on sets open and dense in $\mathfrak{M}$ and $\mathfrak{N}$
     that are commutation closed, i.e. algebras of Lie transformations. We shall write the
     basic vector fields of these families, in some local systems of coordinates  in the manifolds
     $\mathfrak{M}$ and $\mathfrak{N}$, in the operator form:
     $$
     \left.\begin{array}{c}
     X_{\omega}=\lambda^{\mu}_{\omega}(x)\partial/\partial x^{\mu}, \\
     \Xi_{\omega}=\sigma^{\nu}_{\omega}(\xi)\partial/\partial \xi^{\nu},
     \end{array}\right\}
     \eqno(8.7)
     $$
     where $\omega=1,\ldots ,smn$, and with respect to "mute" \ indexes $\mu$ and $\nu$
     summation is performed from 1 to $sm$ and from 1 to $sn$ respectively. By the
     criterion of invariance, the function $f(i\alpha)$ will be the invariant of the local group
     of transformations of some neighbourhood $U(i) \times U(\alpha)$, i.e. a two-point
     invariant, if and only if it satisfies component-wise a system of $smn$ equations
     $$
     X_{\omega}(i)f(i\alpha)+\Xi_{\omega}(\alpha)f(i\alpha)=0
     \eqno(8.8)
     $$ \\
     with the operators (8.7):

     $$
     \lambda^{\mu}_{\omega}(i)\frac{\partial f(i\alpha)}{\partial x^{\mu}_i} +
     \sigma^{\nu}_{\omega}(\alpha)\frac {\partial f(i\alpha)}
     {\partial \xi^{\nu}_\alpha} =0,
     \eqno(8.9)
     $$

     \newpage

\noindent
где $\lambda^{\mu}_{\omega}(i)=\lambda^{\mu}_{\omega}(x_i)=
\lambda^{\mu}_{\omega}(x^1_i, \ldots ,x^{sm}_i)$ и
$\sigma^{\nu}_{\omega}(\alpha)=\sigma^{\nu}_{\omega}(\xi_\alpha)=
\sigma^{\nu}_{\omega}(\xi^1_\alpha, \ldots ,\xi^{sn}_\alpha)$.

\vspace{5mm}

{\bf Теорема 1.} {\it Если функция $f=(f^1, \ldots ,f^s)$
задает на $sm$-мерном и $sn$-мерном многообразиях
$\mathfrak{M}$ и $\mathfrak{N}$ геометрию двух множеств,
наделенную групповой симметрией степени $smn$, то она на тех же многообразиях
задает физическую структуру (феноменологически симметричную
геометрию двух множеств) ранга $(n+1,m+1)$.}

\vspace{5mm}

{\bf Теорема 2.}  {\it Если функция $f=(f^1, \ldots ,f^s)$
задает на $sm$-мерном и $sn$-мерном многообразиях
$\mathfrak{M}$ и $\mathfrak{N}$ физическую структуру
(феноменологически симметричную геометрию двух множеств) ранга $(n+1,m+1)$,
то она на тех же многообразиях задает геометрию двух множеств,
наделенную групповой симметрией степени $smn$.}

\vspace{5mm}

Полные доказательства сформулированных выше теорем 1 и 2, каждая
из которых является обратной по отношению к другой, можно найти в
\S1 монографии автора [24]. Их следствием является вывод об
эквивалентности феноменологической и групповой симметрий геометрии
двух множеств, задаваемой на $sm$-мерном и $sn$-мерном
многообразиях $\mathfrak{M}$ и $\mathfrak{N}$ $s$-компонентной
метрической функцией $f=(f^1,\ldots ,f^s)$.

\vspace{5mm}

{\bf Теорема 3.}  {\it Для того, чтобы функция $f=(f^1,\ldots ,f^s)$
задавала на $sm$-мерном и $sn$-мерном
многообразиях $\mathfrak{M}$ и $\mathfrak{N}$ геометрию
двух множеств,
наделенную групповой симметрией степени $smn$, необходимо и достаточно, чтобы
она на тех же многообразиях задавала физическую структуру
(феноменологически симметричную геометрию двух множеств)
ранга $(n+1,m+1)$.}

\vspace{5mm}

В следующем \S9 приведена полная классификация однометрических
$(s=1)$ физических структур произвольного ранга $(n+1,m+1)$.
Заметим, что для $s\geq2$ полная классификация полиметрических
физических структур ранга $(n+1,m+1)$ еще не построена. Однако и
по ним получены некоторые предварительные результаты. В частности,
устанавливаемая теоремой 3 эквивалентность феноменологической и
груп-

\newpage

    \noindent
    where $\lambda^{\mu}_{\omega}(i)=\lambda^{\mu}_{\omega}(x_i)=
     \lambda^{\mu}_{\omega}(x^1_i, \ldots ,x^{sm}_i)$, \  $\sigma^{\nu}_{\omega}(\alpha)=\sigma^{\nu}_{\omega}(\xi_\alpha)=
     \sigma^{\nu}_{\omega}(\xi^1_\alpha, \ldots ,\xi^{sn}_\alpha)$.

     \vspace{9mm}

     {\bf Theorem 1.} {\it If a function $f=(f^1, \ldots ,f^s)$ gives on an $sm$-dimensional
     and an $sn$-dimensional manifolds $\mathfrak{M}$ and $\mathfrak{N}$ a geometry
     of two sets endowed with a group symmetry of degree $smn$ then, on the same
     manifolds, it gives a physical structure (a phenomenologically symmetric geometry of
     two sets) of rank $(n+1,m+1)$.}

     \vspace{8mm}

     {\bf Theorem 2.}  {\it If a function $f=(f^1, \ldots ,f^s)$ gives on an $sm$-dimensional
     and an $sn$-dimensional manifolds $\mathfrak{M}$ and $\mathfrak{N}$ a physical
     structure \\ (a phenomenologically symmetric geometry of two sets) of rank $(n+1,m+1)$
     then, on the same manifolds, it gives a geometry of two sets endowed with a group
     symmetry of degree $smn$.}

     \vspace{7mm}

     The complete proofs of these theorems, that are each the inverse of the other, may
     be found in \S1 of the author's monograph [24]. Their corollary is the conclusion
     about the phenomenological and the group symmetries of a geometry of two sets
     defined on an $sm$-dimensional and an $sn$-dimensional manifolds $\mathfrak{M}$
     and $\mathfrak{N}$ by the $s$-component metric function $f=(f^1,\ldots ,f^s)$ being
     equivalent.

     \vspace{7mm}

     {\bf Theorem 3.}  {\it For a function $f=(f^1,\ldots ,f^s)$ to define on an $sm$-dimensional
     and an $sn$-dimensional manifolds $\mathfrak{M}$ and $\mathfrak{N}$ a geometry of two
     sets endowed with a group symmetry of degree $smn$ it is necessary and sufficient that it
     should give, on the same manifolds, a physical structure (a phenomenologically symmetric
     geometry of two sets) of rank $(n+1,m+1)$.}

     \vspace{7mm}

     In \S9 a complete classification of unimetric $(s=1)$ physical structures of arbitrary
     rank $(n+1,m+1)$ will be given. We shall note that for $s\geq2$ a complete classification
     of polymetric physical structures of rank $(n+1,m+1)$ has not been built yet. However,
     with respect to those functions too some preliminary results have been obtained. In
     particular, the equivalence of the phenomenological and group symmetries established
     by Theorem \ 3 \  was

     \newpage

\noindent
повой симметрий была использована автором [25], [26] при
построении классификации двуметрических физических структур ранга
$(n+1,2)$, то есть для случая $s=2$ \ и $\ m=1, \ n\geq1$. Эта
классификация приведена в \S10 настоящей монографии. Кроме того,
поскольку триметрические физические структуры ранга (2,2)
допускают трехмерные группы движений, оказалось возможным по
имеющейся классификации трехмерных алгебр Ли транзитивных
преобразований пространства [17] построить в монографии [24] и их
полную классификацию, которая приведена в том же \S10.

Особый интерес представляют комплексные физические
структуры. Например, Ю.С. Владимиров [27] комплексную структуру
ранга (3,3) использовал для обоснования размерности и
сигнатуры  классического  пространства-времени.
Комплексные физические
структуры более высокого ранга были применены им для построения
единой теории физических взаимодействий.

С математической точки зрения комплексные физические структуры есть частный
случай вещественных двуметрических физических структур. Поэтому, если бы была
построена полная
классификация последних, то по ней можно было бы воспроизвести соответствующую
классификацию первых.
Комплексные физические структуры могут быть также получены из
вещественных однометрических с помощью их
комплексификации, состоящей в замене вещественных координат и функций
комплексными. Однако при этом нет никакой гарантии того,
что получающаяся классификация комплексных физических структур
окажется полной.

\vspace{15mm}

\begin{center}
{\bf \large \S9. Классификация однометрических \\ физических структур}
\end{center}

Согласно \S8 однометрическая феноменологически симметричная
геометрия двух множеств (физическая структура) ранга $(n+1,m+1)$ в
общих чертах и кратко может быть определена следующим образом.
Пусть множества $\mathfrak{M}$ и $\mathfrak{N}$ есть
соответственно \ $m$-мерное \ и \ $n$-мерное

\newpage

      \noindent
     used  by the author [25], [26] in building the classification of the dimetric
     physical structures of rank$(n+1,2)$, i.e.for the case of $s=2$ \ and $\ m=1, \ n\geq1$.
     That classification is given in \S10 of this monograph. Besides, since trimetric physical
     structures of rank (2,2) allow three- dimensional groups of motions, it turned out to be
     possible to use the classification available in [17] of three-dimensional Lie algebras of
     transitive transformations of the space  to build the classification of the physical structures
     of rank (2,2) too (given in \S10 of this monograph an in the monograph [24]).

     Of special interest are complex physical structures. For example, Yu.S. Vladimirov [27]
     used a complex structure of rank (3,3) in a theoretical justification of the dimensionality
     and signature of the classical space-time. He used complex physical structures of higher
     rank for building a unified theory of physical interactions.

     Mathematically, the complex physical structures are a special case of real dimetric
     physical structures. So, if a complete classification of the latter should be built one of
     the former could be reproduced after it. Also, complex physical structures may be
     derived from real unimetric ones, by way of their complexification which consists in
     replacing of the real coordinates and functions by complex ones. But that method does
     not carry with itself any guaranty of a classification of the complex physical structures
     expected to be quite complete.

     \vspace{35mm}

     \begin{center}
     {\bf \large \S9. A classification of unimetric \\ physical structures}
     \end{center}

     Under \S8, a unimetric phenomenologically symmetric geometry of two sets
     (a physical structure) of rank $(n+1,m+1)$ may, grosso modo, be defined as follows.
     Let the sets $\mathfrak{M}$ and $\mathfrak{N}$ be respectively an $m$-dimensional
     and an $n$-dimensional smooth manifolds. We shall designate the local
     coordinates

    \newpage

\noindent
гладкие многообразия. Обозначим локальные координаты этих
многообразий через $x=(x^1,\ldots,x^m)$ и
$\xi=(\xi^1,\ldots,\xi^n)$, считая для определенности, что $m\leq
n$. Пусть, далее, имеется функция
$f:\mathfrak{M}\times\mathfrak{N}\to R$ с открытой и плотной в
$\mathfrak{M}\times\mathfrak{N}$ областью определения,
сопоставляющая каждой паре из нее некоторое число. Функцию $f$
будем называть метрической, не требуя от нее выполнения обычных
аксиом метрики, тем более, что расстояния для двух точек только из
$\mathfrak{M}$ или двух точек только из $\mathfrak{N}$ не
определены. Предполагается, что ее локальное координатное
представление задается достаточно гладкой функцией (8.2), которую
удобно записать, не конкретизируя обозначения точек $i$ и
$\alpha$:

$$
f=f(x,\xi)=f(x^1,\ldots,x^m,\xi^1,\ldots,\xi^n).
\eqno(9.1)
$$

Вследствие невырожденности метрической функции $f$, в
представление (9.1) координаты $x$ и $\xi$ входят существенным
образом. Последнее означает, что никакая гладкая локально
обратимая замена координат не приведет к уменьшению их числа в
представлении (9.1).

Построим функцию
$F:\mathfrak{M}^{n+1}\times\mathfrak{N}^{m+1}\to R^{(n+1)(m+1)}$
с естественной в
$\mathfrak{M}^{n+1}\times\mathfrak{N}^{m+1}$ областью определения,
сопоставляя каждому кортежу длины $m+n+2$ из нее
все $(m+1)(n+1)$ возможные по метрической функции $f$ и упорядоченные
по нему расстояния. Будем
говорить, что функция $f$ задает на $m$-мерном и $n$-мерном многообразиях
$\mathfrak{M}$ и $\mathfrak{N}$ феноменологически симметричную геометрию
двух множеств
(физическую структуру) ранга $(n+1,m+1)$, если локально множество значений
построенной функции $F$ в $R^{(m+1)(n+1)}$ является
подмножеством множества нулей
некоторой достаточно гладкой функции $\Phi$ от $(m+1)(n+1)$ переменных с
$grad\Phi\neq0$ на плотном в области определения функции $F$ подмножестве,
удовлетворяя уравнению (8.1).

В работе автора [28] приведена полная классификация
однометрических физических структур произвольного ранга
$(n+1,m+1)$ в естественном предположении, что $n\geq m\geq 1$, так
как обратный случай $m\geq n\geq 1$ легко воспроизводится по
симметрии, а в его работах [29], [30] и монографии [31] показаны
математические методы, которыми она получена.

\newpage

     \noindent
     of these manifolds by $x=(x^1,\ldots,x^m)$ and $\xi=(\xi^1,\ldots,\xi^n)$ considering,
     for the sake of definiteness, that $m\leq n$. Let, further, there exist a function
     $f:\mathfrak{M}\times\mathfrak{N}\to R$ with the domain open and dense in
     $\mathfrak{M}\times\mathfrak{N}$ that assigns some number to every pair from it.
     We shall call the function $f$ a metric one, and we shall not require that it should
     satisfy the usual metric axioms, especially as because the distances for two points
     from only $\mathfrak{M}$, or for two points from only $\mathfrak{N}$, are not
     defined. It is assumed that its local coordinate representation is defined by a
     sufficiently smooth function (8.2), which is more convenient to write down
     without specifying the designations for the points $i$ and $\alpha$:
     $$
     f=f(x,\xi)=f(x^1,\ldots,x^m,\xi^1,\ldots,\xi^n).
     \eqno(9.1)
     $$

     In consequence of the metric function $f$ being nondegenerate, the coordi- \ nates $x$
     and $\xi$ are included into the representation (9.1) in an essential manner. The latter
     means that no smooth locally invertible change of coordinates will result in their number
     in the representation (9.1) being decreased.

     We shall construct a function $F:\mathfrak{M}^{n+1}\times\mathfrak{N}^{m+1}\to R^{(n+1)(m+1)}$
     with the domain natural in $\mathfrak{M}^{n+1}\times\mathfrak{N}^{m+1}$, by
     assigning to every cortege of length $m+n+2$ of it all $(m+1)(n+1)$ distances possible
     with respect to the metric function $f$ and ordered with respect to the aforesaid
     manifold. We shall say that the function $f$ defines on an $m$-dimensional and an
     $n$-dimensional manifolds $\mathfrak{M}$ and $\mathfrak{N}$ a phenomenologically
     symmetric geometry of two sets (a physical structure) of rank $(n+1,m+1)$, if locally
     the set of values of the constructed function $F$ in $R^{(m+1)(n+1)}$ is a subset of the
     set of zeros of some sufficiently smooth function $\Phi$ of $(m+1)(n+1)$ variables with
     $grad\Phi\neq0$ on a subset dense in the domain of the function $F$, satisfying the
     equation (8.1).

     In the author's note [28] the complete classification of unimetric physical structures of
     arbitrary rank$(n+1,m+1)$ is given, with the natural supposition that $n\geq m\geq 1$,
     as the opposite case, that of $m\geq n\geq 1$ is easily reproduced by the symmetry, and
     in his notes [29] and [30] and his monograph [31] mathematical methods are given that
     were used to build it.

     \newpage

{\it Запишем ниже соответствующие классификационные результаты с
точностью до локально обратимой замены координат в многообразиях
$\mathfrak{M},\mathfrak{N}$ и масштабного преобразования
$\chi(f)\rightarrow f$, где $\chi$ -- произвольная гладкая функция
одной переменной с отличной от нуля производной:

\vspace{5mm}

$m=1,n=1$:
$$
f=x+\xi;
\eqno(9.2)
$$

$m=1,n=2$:
$$
f=x\xi+\eta;
\eqno(9.3)
$$

$m=1,n=3$:
$$
f=(x\xi+\eta)/(x+\vartheta);
\eqno(9.4)
$$

$m=n\geq2$:
$$
f=x^1\xi^1+\ldots+x^{m-1}\xi^{m-1}+x^m\xi^m,
\eqno(9.5)
$$
$$
 f=x^1\xi^1+\ldots+x^{m-1}\xi^{m-1}+x^m+\xi^m; \eqno(9.6)
$$

$m=n-1\geq2$:
$$
f=x^1\xi^1+\ldots+x^m\xi^m+\xi^{m+1}.
\eqno(9.7)
$$

Для всех остальных пар значений натуральных чисел $m$ и $n$ при оговоренном
выше условии $n\geq m\geq 1$ физические структуры ранга $(n+1,m+1)$
не существуют.}

\vspace{5mm}

Феноменологическая симметрия геометрий двух множеств (физических структур),
задаваемых вышеперечисленными метрическими функциями (9.2)--(9.7),
выражается, соответственно, следующими уравнениями:

$$
f(i\alpha)-f(i\beta)-f(j\alpha)+f(j\beta)=0;
\eqno(9.2')
$$
$$
$$
$$
\left|
\begin{array}{ccc}
f(i\alpha ) & f(i\beta ) & 1 \\
f(j\alpha ) & f(j\beta ) & 1 \\
f(k\alpha ) & f(k\beta ) & 1
\end{array}
\right| =0;
\eqno(9.3')
$$

\newpage

     {\it Further we shall write the corresponding classification results with an accuracy up
     to a locally invertible change of coordinates in the manifolds $\mathfrak{M},\mathfrak{N}$
     and to the scaling transformation $\chi(f)\rightarrow f$, where $\chi$ is an arbitrary
     smooth function of one variable with the derivative unequal to zero.

     \vspace{5mm}

     $m=1,n=1$:
     $$
     f=x+\xi;
     \eqno(9.2)
     $$

     $m=1,n=2$:
     $$
     f=x\xi+\eta;
     \eqno(9.3)
     $$

     $m=1,n=3$:
     $$
     f=(x\xi+\eta)/(x+\vartheta);
     \eqno(9.4)
     $$

     $m=n\geq2$:
     $$
     f=x^1\xi^1+\ldots+x^{m-1}\xi^{m-1}+x^m\xi^m,
     \eqno(9.5)
     $$
     $$
     f=x^1\xi^1+\ldots+x^{m-1}\xi^{m-1}+x^m+\xi^m;
     \eqno(9.6)
     $$

     $m=n-1\geq2$:
     $$
     f=x^1\xi^1+\ldots+x^m\xi^m+\xi^{m+1}.
     \eqno(9.7)
     $$

     For all the other pairs of values of the natural numbers $m$ and $n$, satisfying
     the above-said condition of $n\geq m\geq 1$, no physical structures of rank
     $(n+1,m+1)$ exist.}

     \vspace{10mm}

     The phenomenological symmetry of the geometries of two sets (physical structures)
     that are defined by the metric functions (9.2)--(9.7) above is, correspondingly, naturally
     expressed by the following equations:

     $$
     f(i\alpha)-f(i\beta)-f(j\alpha)+f(j\beta)=0;
     \eqno(9.2')
     $$
$$
$$
     $$
     \left|
     \begin{array}{ccc}
     f(i\alpha ) & f(i\beta ) & 1 \\
     f(j\alpha ) & f(j\beta ) & 1 \\
     f(k\alpha ) & f(k\beta ) & 1
     \end{array}
     \right| =0;
     \eqno(9.3')
     $$
$$
$$

\newpage

$$
\left|\begin{array}{cccc}
f(i\alpha) & f(i\beta) & f(i\alpha)f(i\beta) & 1 \\
f(j\alpha) & f(j\beta) & f(j\alpha)f(j\beta) & 1 \\
f(k\alpha) & f(k\beta) & f(k\alpha)f(k\beta) & 1 \\
f(l\alpha) & f(l\beta) & f(l\alpha)f(l\beta) & 1
\end{array}\right|=0;
\eqno(9.4')
$$
$$
$$
$$
\left|\begin{array}{cccc}
f(i\alpha) & f(i\beta) & \dotsc & f(i\tau)  \\
f(j\alpha) & f(j\beta) & \dotsc & f(j\tau)  \\
\dotsc & \dotsc & \dotsc & \dotsc \\
f(v\alpha) & f(v\beta) & \dotsc & f(v\tau)
\end{array}\right|=0,
\eqno(9.5')
$$
$$
$$
$$
\left|\begin{array}{ccccc}
0 & 1 & 1 & \dotsc & 1 \\
1 & f(i\alpha) & f(i\beta) & \dotsc & f(i\tau)  \\
1 & f(j\alpha) & f(j\beta) & \dotsc & f(j\tau)  \\
\dotsc & \dotsc & \dotsc & \dotsc & \dotsc \\
1 & f(v\alpha) & f(v\beta) & \dotsc & f(v\tau)
\end{array}\right|=0;
\eqno(9.6')
$$
$$
$$
$$
\left|\begin{array}{ccccc}
f(i\alpha) & f(i\beta) & \dotsc & f(i\tau) & 1 \\
f(j\alpha) & f(j\beta) & \dotsc & f(j\tau) & 1 \\
f(k\alpha) & f(k\beta) & \dotsc & f(k\tau) & 1 \\
\dotsc & \dotsc & \dotsc & \dotsc & \dotsc  \\
f(v\alpha) & f(v\beta) & \dotsc & f(v\tau) & 1
\end{array}\right|=0.
\eqno(9.7')
$$
$$
$$

Под движением в геометрии двух множеств мы понимаем такую пару
гладких локально обратимых преобразований многообразий
$\mathfrak{M}$ и $\mathfrak{N}$:
$$
x'=\lambda(x), \ \xi'=\sigma(\xi),
\eqno(9.8)
$$
при которых функция (9.1) сохраняется:
$$
f(\lambda(x),\sigma(\xi))=f(x,\xi).
\eqno(9.9)
$$

Если метрическая функция $f$ задана в ее явном координатном
представлении (9.1), то равенство (9.9) является функциональным
уравнением относительно двух преобразований (9.8), решая которое
можно найти группу движений и установить число ее непрерывных
параметров. Ниже в настоящем параграфе приводятся также полные
локаль-

\newpage

     $$
     \left|\begin{array}{cccc}
     f(i\alpha) & f(i\beta) & f(i\alpha)f(i\beta) & 1 \\
     f(j\alpha) & f(j\beta) & f(j\alpha)f(j\beta) & 1 \\
     f(k\alpha) & f(k\beta) & f(k\alpha)f(k\beta) & 1 \\
     f(l\alpha) & f(l\beta) & f(l\alpha)f(l\beta) & 1
     \end{array}\right|=0;
     \eqno(9.4')
     $$
     $$
     $$
     $$
     \left|\begin{array}{cccc}
     f(i\alpha) & f(i\beta) & \dotsc & f(i\tau)  \\
     f(j\alpha) & f(j\beta) & \dotsc & f(j\tau)  \\
     \dotsc & \dotsc & \dotsc & \dotsc \\
     f(v\alpha) & f(v\beta) & \dotsc & f(v\tau)
     \end{array}\right|=0,
     \eqno(9.5')
     $$
     $$
     $$
     $$
     \left|\begin{array}{ccccc}
     0 & 1 & 1 & \dotsc & 1 \\
     1 & f(i\alpha) & f(i\beta) & \dotsc & f(i\tau)  \\
     1 & f(j\alpha) & f(j\beta) & \dotsc & f(j\tau)  \\
     \dotsc & \dotsc & \dotsc & \dotsc & \dotsc \\
     1 & f(v\alpha) & f(v\beta) & \dotsc & f(v\tau)
     \end{array}\right|=0;
     \eqno(9.6')
     $$
     $$
     $$
     $$
     \left|\begin{array}{ccccc}
     f(i\alpha) & f(i\beta) & \dotsc & f(i\tau) & 1 \\
     f(j\alpha) & f(j\beta) & \dotsc & f(j\tau) & 1 \\
     f(k\alpha) & f(k\beta) & \dotsc & f(k\tau) & 1 \\
     \dotsc & \dotsc & \dotsc & \dotsc & \dotsc  \\
     f(v\alpha) & f(v\beta) & \dotsc & f(v\tau) & 1
     \end{array}\right|=0.
     \eqno(9.7')
     $$
     $$
     $$

     By a motion in a geometry of two sets we understand such a pair of smooth locally
     invertible transformations of the manifolds $\mathfrak{M}$ and $\mathfrak{N}$:
     $$
     x'=\lambda(x), \ \xi'=\sigma(\xi),
     \eqno(9.8)
     $$
     under which the function (9.1) is preserved:
     $$
     f(\lambda(x),\sigma(\xi))=f(x,\xi).
     \eqno(9.9)
     $$

     If the metric function $f$ is defined in its explicit coordinate representation
     (9.1), then the equality (9.9) is the functional equation with respect to the
     two transformations (9.8), the solution of which gives the group of motions
     and helps establish the number of its continuous parameters. Further in this
     paragraph we also give the full local groups of local motions for each of
     the

     \newpage

\noindent
ные группы локальных движений для каждой из шести метрических
функций (9.2)--(9.7), которые могут быть найдены (см. [24], \S2)
как общие решения соответствующих уравнений (9.9), причем на
функции $\lambda(x)$ и $\sigma(\xi)$ преобразований (9.8), кроме
гладкости и локальной обратимости, никакие дополнительные
ограничения (например, линейность) не налагаются.

\vspace{5mm}
{\bf Теорема 1.} {\it Группа движений $(9.8)$
феноменологически симметричной геометрии двух множеств(физической структуры)
ранга $(n+1,m+1)$, задаваемой одной из метрических
функций $(9.2)-(9.7)$, представляется следующими преобразованиями
многообразий $\mathfrak{M}$ и $\mathfrak{N}$:

для метрической функции $(9.2)$:
$$
x'=x+a, \ \xi'=\xi-a;
\eqno(9.10)
$$

для метрической функции $(9.3)$:
$$          x'=ax+b, \ \xi'=\xi/a, \ \eta'=\eta-b\xi/a,
\eqno(9.11)
$$
где $a\neq0$;

для метрической функции $(9.4)$:
$$
\left.\begin{array}{c}
x'=(ax+b)/(cx+d), \ \xi'=(d\xi-c\eta)/(d-c\vartheta), \\
\eta'=(a\eta-b\xi)/(d-c\vartheta), \ \vartheta'=(a\vartheta-b)/(d-c\vartheta),
\end{array}\right\}
\eqno(9.12)
$$
где $ad-bc=\pm1$;

для метрической функции $(9.5)$:
$$
\left.\begin{array}{c}
 x^{\prime\mu}=a^{\mu1}x^1+\ldots+a^{\mu m}x^m, \\
\xi^{\prime\mu}=\tilde{a}^{1\mu}\xi^1+\ldots+\tilde{a}^{m\mu}\xi^m,
\end{array}\right\}
\eqno(9.13)
$$
где $\mu=1,\ldots,m$ \ и \ $a$ -- квадратная невырожденная матрица порядка $m$, \
$\tilde{a}$ -- обратная к ней матрица;

для метрической функции $(9.6)$:
$$
\left.\begin{array}{c}
x^{\prime\nu}=a^{\nu1}x^1+\ldots+a^{\nu,m-1}x^{m-1}+b^{\nu}, \\
x^{\prime m}=x^m+c^1x^1+\ldots+c^{m-1}x^{m-1}+b^m, \\
\xi^{\prime\nu}=\tilde{a}^{1\nu}(\xi^1-c^1)+\ldots+
\tilde{a}^{m-1,\nu}(\xi^{m-1}-c^{m-1}), \\
\xi^{\prime m}=\xi^m-(b^1\tilde{a}^{11}+\ldots+b^{m-1}\tilde{a}^{1,m-1})
(\xi^1-c^1)-\ldots- \\
-(b^1\tilde{a}^{m-1,1}+\ldots+b^{m-1}\tilde{a}^{m-1,m-1})
(\xi^{m-1}-c^{m-1})-b^m,
\end{array}\right\}
\eqno(9.14)
$$}

\newpage

      \noindent
      six metric functions (9.2)--(9.7) that can be found (see [24],  \S2) as general
     solutions of the respective equations (9.9), no restrictions other than smooth- \ ness
     and local invertibility being imposed on the functions $\lambda(x)$ and $\sigma(\xi)$
     of the transformations (9.8) (such as necessity of being linear etc.).

     \vspace{10mm}
     {\bf Theorem 1.} {\it The group of motions $(9.8)$ of the phenomenologically
     sym- \ metric geometry of two sets (physical structure) of rank $(n+1,m+1)$
     defined by one of the metric functions $(9.2)-(9.7)$ is represented by the
     transformations of the manifolds $\mathfrak{M}$ and $\mathfrak{N}$ as
     follows:

     for the metric function $(9.2)$:
     $$
     x'=x+a, \ \xi'=\xi-a;
     \eqno(9.10)
     $$

     for the metric function $(9.3)$:
     $$
      x'=ax+b, \ \xi'=\xi/a, \ \eta'=\eta-b\xi/a,
     \eqno(9.11)
     $$
     where $a\neq0$;

     for the metric function $(9.4)$:

     $$
     \left.\begin{array}{c}
     x'=(ax+b)/(cx+d), \ \xi'=(d\xi-c\eta)/(d-c\vartheta), \\
     \eta'=(a\eta-b\xi)/(d-c\vartheta), \ \vartheta'=(a\vartheta-b)/(d-c\vartheta),
     \end{array}\right\}
     \eqno(9.12)
     $$
     where $ad-bc=\pm1$;

     for the metric function $(9.5)$:
     $$
     \left.\begin{array}{c}
      x^{\prime\mu}=a^{\mu1}x^1+\ldots+a^{\mu m}x^m, \\
     \xi^{\prime\mu}=\tilde{a}^{1\mu}\xi^1+\ldots+\tilde{a}^{m\mu}\xi^m,
     \end{array}\right\}
     \eqno(9.13)
     $$
     where $\mu=1,\ldots,m$ \ and \ $a$ is a quadratic nondegenerate matrix of
     degree $m$, \ and $\tilde{a}$ is its reciprocal matrix;

     for the metric function $(9.6)$:
     $$
     \left.\begin{array}{c}
     x^{\prime\nu}=a^{\nu1}x^1+\ldots+a^{\nu,m-1}x^{m-1}+b^{\nu}, \\
     x^{\prime m}=x^m+c^1x^1+\ldots+c^{m-1}x^{m-1}+b^m, \\
     \xi^{\prime\nu}=\tilde{a}^{1\nu}(\xi^1-c^1)+\ldots+
     \tilde{a}^{m-1,\nu}(\xi^{m-1}-c^{m-1}), \\
     \xi^{\prime m}=\xi^m-(b^1\tilde{a}^{11}+\ldots+b^{m-1}\tilde{a}^{1,m-1})

     (\xi^1-c^1)-\ldots- \\
     -(b^1\tilde{a}^{m-1,1}+\ldots+b^{m-1}\tilde{a}^{m-1,m-1})
     (\xi^{m-1}-c^{m-1})-b^m,
     \end{array}\right\}
     \eqno(9.14)
     $$ }

     \newpage

\noindent
{\it где $\nu=1,\ldots,m-1$ \ и \ $a$ -- квадратная невырожденная
матрица порядка $m-1$, \ $\tilde{a}$ -- обратная к ней матрица;

для метрической функции $(9.7)$:
$$
\left.\begin{array}{c}
x^{\prime\mu}=a^{\mu1}x^1+\ldots+a^{\mu m}x^m+b^{\mu}, \\
\xi^{\prime\mu}=\tilde{a}^{1\mu}\xi^1+\ldots+\tilde{a}^{m\mu}\xi^m, \\
\xi^{\prime m+1}=\xi^{m+1}-(b^1\tilde{a}^{11}+\ldots+b^m\tilde{a}^{1m})
\xi^1- \\
-\ldots-(b^1\tilde{a}^{m1}+\ldots+b^m\tilde{a}^{mm})\xi^m,
\end{array}\right\}
\eqno(9.15)
$$
где $\mu=1,\ldots,m$ \ и \ $a$ -- квадратная невырожденная матрица порядка $m$,
\ $\tilde{a}$ -- обратная к ней матрица.}
\vspace{5mm}

Все перечисленные в теореме 1 группы движений зависят от конечного
числа непрерывных параметров, число которых в соответствии с
теоремой 2 из предыдущего \S8 равно $mn$, то есть произведению
размерностей $m$ и $n$ многообразий $\mathfrak{M}$ и
$\mathfrak{N}$. Для сравнения заметим, что в $n$-мерной
феноменологически симметричной геометрии ранга $n+2$ на одном
множестве $\mathfrak{M}$ это число равно $n(n+1)/2$. Отметим
также, что не для всякой метрической функции (9.1) уравнение (9.9)
имеет нетривиальное решение, то есть полная группа движений может
состоять только из тождественных преобразований многообразий
$\mathfrak{M}$ и $\mathfrak{N}$. Нетрудно, например, установить,
что для метрической функции $f(x,\xi)= x\xi+\xi^3$ уравнение (9.9)
имеет только тривиальное решение: $\lambda(x)=x, \
\sigma(\xi)=\xi$, и потому полная группа движений соответствующей
геометрии двух множеств содержит только тождественные
преобразования $x'=x$ и $\xi'=\xi$ одномерных многообразий
$\mathfrak{M}$ и $\mathfrak{N}$. Согласно теореме 3 из \S8
наделенная такой тривиальной групповой симметрией геометрия двух
множеств, задаваемая на одномерных многообразиях $\mathfrak{M}$ и
$\mathfrak{N}$ этой метрической функцией, не является физической
структурой ранга (2,2).

\vspace{5mm}

Рассмотрим более подробно феноменологически симметричную геометрию
двух множеств (физическую структуру) ранга (3,3), существующую в двух
вариантах, задаваемых на двумерных многообразиях метрическими функциями
$$
f=x\xi+y\eta,
\eqno(9.16)
$$

\newpage

      \noindent
     {\it where $\nu=1,\ldots,m-1$ \ and \ $a$ is a quadratic nondegenerate matrix of degree $m-1$,
     and \ $\tilde{a}$ is its reciprocal matrix;

     for the metric function $(9.7)$:
     $$
     \left.\begin{array}{c}
     x^{\prime\mu}=a^{\mu1}x^1+\ldots+a^{\mu m}x^m+b^{\mu}, \\
     \xi^{\prime\mu}=\tilde{a}^{1\mu}\xi^1+\ldots+\tilde{a}^{m\mu}\xi^m, \\
     \xi^{\prime m+1}=\xi^{m+1}-(b^1\tilde{a}^{11}+\ldots+b^m\tilde{a}^{1m})
     \xi^1- \\
     -\ldots-(b^1\tilde{a}^{m1}+\ldots+b^m\tilde{a}^{mm})\xi^m,
     \end{array}\right\}
     \eqno(9.15)
     $$
     where $\mu=1,\ldots,m$ \ and \ $a$ is a quadratic nondegenerate matrix of degree
      $m$, and \ $\tilde{a}$ is its reciprocal matrix.}
     \vspace{7mm}

     All the groups of motions represented in Theorem 1 depend on the finite
     number of continuous parameters which number, according to Theorem
     2 of \S8 is equal to $mn$, i.e. the direct product of the dimensionalities $m$
     and $n$ of the manifolds $\mathfrak{M}$ and $\mathfrak{N}$. For the sake of
     comparison, we shall note that in the $n$-dimensional phenomenologically symmetric
     geometry of rank $n+2$ on one set $\mathfrak{M}$ that number is equal to $n(n+1)/2$.
     We shall also note that not for every metric function (9.1) the equation (9.9) has a
     nontrivial solution, i.e. the full group of motions may only consist of identical
     transformations of the manifolds $\mathfrak{M}$ and $\mathfrak{N}$. For example,
     it is easy to establish that for the metric function $f(x,\xi)= x\xi+\xi^3$ the
     equation (9.9) has only a trivial solution: $\lambda(x)=x, \sigma(\xi)=\xi$,
     and so the full group of motions of the respective geometry of two sets contains
     only identical transformations $x'=x$ and $\xi'=\xi$ of the unimetric manifolds
     $\mathfrak{M}$ and $\mathfrak{N}$. Under Theorem 3 of \S8, a geometry of two
     sets defined on the unimetric manifolds $\mathfrak{M}$ and $\mathfrak{N}$ by
     that metric function and endowed with such trivial group symmetry is not a physical
     structure of rank (2,2).

     \vspace{7mm}

     Let us scrutinize in greater detail the phenomenologically symmetric geometry
     of two sets (physical structure) of rank (3,3), which exists in two variants defined
     on two-dimensional manifolds by the metric functions
     $$
     f=x\xi+y\eta,
     \eqno(9.16)
     $$

\newpage

$$
f=x\xi+y+\eta,
\eqno(9.17)
$$
координатные представления которых получаются из выражений (9.5), (9.6)
для случая $m=n=2$ при введении безиндексных обозначений координат:
$x=x^1, \ y=x^2, \ \xi=\xi^1, \ \eta=\xi^2$.

Легко убедиться в том, что их феноменологическая симметрия выражается
уравнениями
$$
\left|
\begin{array}{ccc}
f(i\alpha ) & f(i\beta ) & f(i\gamma) \\
f(j\alpha ) & f(j\beta ) & f(j\gamma) \\
f(k\alpha ) & f(k\beta ) & f(k\gamma)
\end{array}
\right| =0,
\eqno(9.16')
$$
$$
$$
$$
\left|
\begin{array}{cccc}
0 & 1 & 1 & 1 \\
1 & f(i\alpha ) & f(i\beta ) & f(i\gamma) \\
1 & f(j\alpha ) & f(j\beta ) & f(j\gamma) \\
1 & f(k\alpha ) & f(k\beta ) & f(k\gamma)
\end{array}
\right| =0,
\eqno(9.17')
$$ \\
соответственно.

Установим, прежде всего, что эти две физические структуры неэквивалентны.

\vspace{5mm}

{\bf Теорема 2.} {\it Ни при каких заменах координат и масштабных
преобразованиях метрические функции (9.16) и (9.17) не переходят
друг в друга.}

\vspace{5mm}

Доказательство теоремы 2 проведем методом от
противного, предположив, что при некоторых
гладких обратимых заменах координат $\lambda(x,y)\to x, \ \sigma(x,y)\to y$
\ и \ $\rho(\xi,\eta)\to\xi, \ \tau(\xi,\eta)\to\eta$ в многообразиях
$\mathfrak{M}$ и $\mathfrak{N}$, а также масштабном преобразовании
$\chi(f)\to f$ одна из метрических функций (9.16), (9.17)
переходит в другую, например:
$$
\lambda(x,y)\rho(\xi,\eta)+\sigma(x,y)\tau(\xi,\eta)=\chi(x\xi+y+\eta),
\eqno(9.18)
$$
где $\partial(\lambda,\sigma)/\partial(x,y)\neq0$,
$\partial(\rho,\tau)/\partial(\xi,\eta)\neq0$ и $\chi'\neq0$. Теорема 2
будет верна, если функциональное уравнение (9.18) не имеет решения.

\newpage

     $$
     f=x\xi+y+\eta,
     \eqno(9.17)
     $$
     whose coordinate representations are obtained from the expressions (9.5) and (9.6)
     for the case of $m=n=2$ by way of introducing non-index designations of the coordinates:
     $x=x^1, \ y=x^2, \ \xi=\xi^1, \ \eta=\xi^2$.

     It is easy to make sure that their phenomenological symmetry is expressed by the
     equations
     $$
     \left|
     \begin{array}{ccc}
     f(i\alpha ) & f(i\beta ) & f(i\gamma) \\
     f(j\alpha ) & f(j\beta ) & f(j\gamma) \\
     f(k\alpha ) & f(k\beta ) & f(k\gamma)
     \end{array}
     \right| =0,
     \eqno(9.16')
     $$
$$
$$
     $$
     \left|
     \begin{array}{cccc}
     0 & 1 & 1 & 1 \\
     1 & f(i\alpha ) & f(i\beta ) & f(i\gamma) \\
     1 & f(j\alpha ) & f(j\beta ) & f(j\gamma) \\
     1 & f(k\alpha ) & f(k\beta ) & f(k\gamma)
     \end{array}
     \right| =0,
     \eqno(9.17')
     $$ \\
     respectively.

     We shall establish, in the first place, that these two physical structures are not equivalent.

     \vspace{5mm}

     {\bf Theorem 2.} {\it Under no changes of coordinates and no scaling transforma- \ tions
     may the metric functions (9.16) and (9.17) be transformed one into the
     other.}

     \vspace{5mm}

     We shall prove Theorem 2 using the method of proof by contradiction supposing
     that with some smooth invertible changes of coordinates $\lambda(x,y)\to x, \
     \sigma(x,y)\to y$ \ and \ $\rho(\xi,\eta)\to\xi, \ \tau(\xi,\eta)\to\eta$ in the
     manifolds $\mathfrak{M}$ and $\mathfrak{N}$, and a scaling transformation
     $\chi(f)\to f$ one of the metric functions (9.16), (9.17) is transformed into the other,
     for example:
     $$
     \lambda(x,y)\rho(\xi,\eta)+\sigma(x,y)\tau(\xi,\eta)=\chi(x\xi+y+\eta),
     \eqno(9.18)
     $$
     where $\partial(\lambda,\sigma)/\partial(x,y)\neq0$, $\partial(\rho,\tau)/\partial(\xi,\eta)\neq0$
     and $\chi'\neq0$. Theorem 2 will be true if the functional equation (9.18) has no
     solution.

\newpage

Продифференцируем уравнение (9.18) по переменным $\xi$, $\eta$ и
разделим один результат дифференцирования на другой:
$\lambda\rho_\xi+\sigma\tau_\xi=
(\lambda\rho_\eta+\sigma\tau_\eta)x$, откуда, фиксируя переменные
$\xi,\eta$, получаем связь:
$$
\sigma(x,y)=A(x)\lambda(x,y),
\eqno(9.19)
$$
где $A(x)=(ax+b)/(cx+d)$ -- дробно-линейная функция с отличной от нуля
производной: $A'(x)=(ad-bc)/(cx+d)^2\neq0$, так как функции $\lambda$ и
$\sigma$ независимы.

Совершенно аналогично, дифференцируя уравнение (9.18) по переменным $x,y$,
получаем вторую связь:
$$
\tau(\xi,\eta)=B(\xi)\rho(\xi,\eta),
\eqno(9.20)
$$
где $B(\xi)=(k\xi+l)/(m\xi+n)$ -- дробно-линейная функция с
отличной от нуля производной: $B'(\xi)=(kn-lm)/(m\xi+n)^2\neq0$,
так как независимы функции $\rho$ и $\tau$.

Полученные две связи (9.19), (9.20) подставим в исходное
функциональное уравнение (9.18):
$$
\lambda(x,y)\rho(\xi,\eta)(1+A(x)B(\xi))=\chi(x\xi+y+\eta)
\eqno(9.21)
$$
и продифференцируем его по переменным $y,\eta$, после чего исключим
производную $\chi'$. Разделяя далее
переменные, получаем дифференциальные уравнения $\lambda_y/\lambda=
\rho_\eta/\rho=h\neq0$, откуда после интегрирования:
$$
\lambda(x,y)=C(x)\exp{hy}, \ \ \ \rho(\xi,\eta)=D(\xi)\exp{h\eta},
\eqno(9.22)
$$
где, очевидно, $C(x)\neq0, \ D(\xi)\neq0$.

Перепишем уравнение (9.21) с функциями (9.22):
$$
(1+A(x)B(\xi))C(x)D(\xi)\exp{h(y+\eta)}=\chi(x\xi+y+\eta).
\eqno(9.23)
$$

Полагая $x=0, \ \xi=0$ и вводя переменную $z=y+\eta$, из уравнения
(9.23) получаем выражение $\chi(z)=E\exp{hz}$, где $E\neq0$, с
которым оно значительно упростится:
$$
(1+A(x)B(\xi))C(x)D(\xi)=E\exp{hx\xi}
$$

Логарифмируем это уравнение:

\newpage

     We shall differentiate the equation (9.18) with respect to the variables $\xi$, and
     $\eta$ and divide one result of the differentiation by the other:
     $\lambda\rho_\xi+\sigma\tau_\xi=(\lambda\rho_\eta+\sigma\tau_\eta)x$, wherefrom,
     fixing the variables $\xi, and \eta$, we get the relation:
     $$
     \sigma(x,y)=A(x)\lambda(x,y),
     \eqno(9.19)
     $$
     where $A(x)=(ax+b)/(cx+d)$ is a homographic function with the derivative unequal to
     zero: $A'(x)=(ad-bc)/(cx+d)^2\neq0$, as the functions $\lambda$ and $\sigma$ are
     independent.

     Quite similarly, differentiating the equation (9.18) with respect to the variables $x, and y$,
     we get the second relation:
     $$
     \tau(\xi,\eta)=B(\xi)\rho(\xi,\eta),
     \eqno(9.20)
     $$
     where $B(\xi)=(k\xi+l)/(m\xi+n)$ is a homographic function with the derivative
     unequal to zero: $B'(\xi)=(kn-lm)/(m\xi+n)^2\neq0$, as the functions $\rho$ and
     $\tau$ are independent.

     We shall substitute the two relations, (9.19) and (9.20), obtained into the initial
     functional equation (9.18):
     $$
     \lambda(x,y)\rho(\xi,\eta)(1+A(x)B(\xi))=\chi(x\xi+y+\eta)
     \eqno(9.21)
     $$
     and differentiate it with respect to the variables $y and \eta$, whereafter
     eliminate the variable $\chi'$. Dividing, further, the variables, we get the differential
     equation $\lambda_y/\lambda=\rho_\eta/\rho=h\neq0$, wherefrom, after
     integrating:
     $$
     \lambda(x,y)=C(x)\exp{hy}, \ \ \ \rho(\xi,\eta)=D(\xi)\exp{h\eta},
     \eqno(9.22)
     $$
     where, obviously, $C(x)\neq0, \ D(\xi)\neq0$.

     We shall rewrite the equation (9.21) with the functions (9.22):
     $$
     (1+A(x)B(\xi))C(x)D(\xi)\exp{h(y+\eta)}=\chi(x\xi+y+\eta).
     \eqno(9.23)
     $$

     Setting $x=0, \ \xi=0$ and introducing the variable $z=y+\eta$, we obtain from the
     equation (9.23) the expression $\chi(z)=E\exp{hz}$, where $E\neq0$, with which it
     becomes much simpler:
     $$
     (1+A(x)B(\xi))C(x)D(\xi)=E\exp{hx\xi}
     $$

     We find the logarithm of that equation:

    \newpage

$$
\ln{(1+A(x)B(\xi))}+\ln{C(x)}+\ln{D(\xi)}=\ln E+hx\xi,
$$
дифференцируя затем его по переменным $x$, $\xi$ и приводя к общему
знаменателю:
$$
A'(x)B'(\xi)=h(1+A(x)B(\xi))^2.
$$

Повторим предыдущие операции по отношению к последнему результату:
$$
A'(x)B'(\xi)=0,
$$
что, очевидно, противоречит установленному выше необращению в нуль
производных функций $A(x)$ и $B(\xi)$, входящих в соотношения (9.19) и (9.20).
Полученное противоречие означает, что исходное функциональное уравнение
(9.18) не имеет решения и потому метрические функции (9.16) и (9.17)
неэквивалентны. Теорема 2 полностью  доказана.

\vspace{5mm}

Установим теперь групповую симметрию физической структуры ранга (3,3), степень
которой по теореме 2 из \S8 должна быть равна четырем.

\vspace{5mm}

{\bf Теорема 3.} {\it Группа движений феноменологически симметричной
геометрии двух множеств (физической структуры) ранга $(3,3)$, задаваемой на
двумерных многообразиях метрической функцией $(9.16)$: $f=x\xi+y\eta$,
представляется следующими уравнениями:
$$
\left.\begin{array}{c}
x'=ax+ by, \ \ y'=cx+dy, \\
\xi'=(d\xi-c\eta)/\Delta, \ \ \eta'=(-b\xi+a\eta)/\Delta,
\end{array}\right\}
\eqno(9.24)
$$
где $\Delta=ad-bc\neq0.$}

\vspace{5mm}

Движение в этой геометрии можно записать следующими уравнениями:
$$
\left.\begin{array}{c}
x'=\lambda(x,y), \ \ y'=\sigma(x,y), \\
\xi'=\rho(\xi,\eta), \ \ \eta'=\tau(\xi,\eta),
\end{array}\right\}
\eqno(9.25)
$$
где $\partial(\lambda,\sigma)/\partial(x,y)\neq0$,
$\partial(\rho,\tau)/\partial(\xi,\eta)\neq0$, так как
соответствующие преобразования многообразий $\mathfrak{M}$ и
$\mathfrak{N}$ в движении должны быть локально обратимыми.
Поскольку движение (9.25) сохраняет метрическую функцию (9.16),
для него получаем функциональное уравнение

\newpage

     $$
     \ln{(1+A(x)B(\xi))}+\ln{C(x)}+\ln{D(\xi)}=\ln E+hx\xi,
     $$
     next differentiating it with respect to the variables $x$ and $\xi$ performing next
     reduction to a common denominator:
     $$
     A'(x)B'(\xi)=h(1+A(x)B(\xi))^2.
     $$

     With respect to the latter result, the same actions are repeated:
     $$
     A'(x)B'(\xi)=0,
     $$
     which is, obviously, in contradiction with the non-vanishing into zero of the derivatives
     of the functions $A(x)$ and $B(\xi)$ that are part of the relations (9.19) and (9.20) that
     we established above. The contradiction we arrive at means that the initial functional
     equation (9.18) has no solution, and so the metric functions (9.16) and (9.17) are
     nonequivalent. The proof of Theorem 2 is complete.

     \vspace{5mm}

     Let us now find the group symmetry of the physical structure of rank (3,3) whose
     degree under Theorem 2 of \S8 must be equal to four.

     \vspace{5mm}

      {\bf Theorem 3.} {\it The group of motions of the phenomenologically symmetric
     geometry of two sets (physical structure) of rank $(3,3)$ defined on two-dimensional
     manifolds by the metric function $(9.16)$: $f=x\xi+y\eta$, is represented by the
     equations as follows:
     $$
     \left.\begin{array}{c}
     x'=ax+ by, \ \ y'=cx+dy, \\
     \xi'=(d\xi-c\eta)/\Delta, \ \ \eta'=(-b\xi+a\eta)/\Delta,
     \end{array}\right\}
     \eqno(9.24)
     $$
     where $\Delta=ad-bc\neq0.$}

     \vspace{5mm}

     Motion in such geometry may be written using equations as follows:
     $$
     \left.\begin{array}{c}
     x'=\lambda(x,y), \ \ y'=\sigma(x,y), \\
     \xi'=\rho(\xi,\eta), \ \ \eta'=\tau(\xi,\eta),
     \end{array}\right\}
     \eqno(9.25)
     $$
     where $\partial(\lambda,\sigma)/\partial(x,y)\neq0$, $\partial(\rho,\tau)/\partial(\xi,\eta)\neq0$, as
     the respective transforma- \ tions of the two-dimensional manifolds $\mathfrak{M}$ and $\mathfrak{N}$
     in the motions must be invertible. Since the motion (9.25) preserves the metric function
     (9.16), we have for it the functional equation

    \newpage

$$
\lambda(x,y)\rho(\xi,\eta)+\sigma(x,y)\tau(\xi,\eta)=x\xi+y\eta,
\eqno(9.26)
$$
которое выполняется тождественно по четырем координатам $x,y$ и
$\xi,\eta$.

Продифференцируем уравнение (9.26) по переменным $\xi$, $\eta$:
$$
\lambda\rho_\xi+\sigma\tau_\xi=x, \ \
\lambda\rho_\eta+\sigma\tau_\eta=y
$$
и разрешим полученные равенства относительно функций $\lambda,\sigma$:
$$
\lambda(x,y)=\frac{x\tau_\eta-
y\tau_\xi}{\rho_\xi\tau_\eta-\rho_\eta\tau_\xi}, \
\ \sigma(x,y)=\frac{-x\rho_\eta+
y\rho_\xi}{\rho_\xi\tau_\eta-\rho_\eta\tau_\xi}.
$$

Дифференцируя полученные для функций $\lambda$ и $\sigma$
выражения по переменным $x,y$, убеждаемся в том, что коэффициенты
при них являются константами. Введя для них соответствующие
обозначения, получаем первую пару уравнений (9.24), которыми
определяется преобразование двумерного многообразия $\mathfrak{M}$
в движении (9.25), причем из их обратимости, очевидно, вытекает
условие $\Delta\neq0$ . Вторая пара уравнений (9.24), определяющая
преобразование другого двумерного многообразия $\mathfrak{N}$ в
движении (9.25), легко получается из функционального уравнения
(9.26) при подстановке в него первой пары.

Множество движений (9.24) зависит от четырех непрерывных
параметров $a,b,c$, $d$, на которые наложено условие
$\Delta=ad-bc\neq0$. Легко убедиться в том, что это множество по
композиции движений является группой. Для этого запишем, например,
преобразования первого многообразия $\mathfrak{M}$ в множестве
движений (9.24) в матричной форме:
$$
\left\|\begin{array}{cc} x' \\ y' \end{array}\right\|=
\left\|\begin{array}{cc} a & b \\ c & d \end{array}\right\|
\left\|\begin{array}{cc} x \\ y \end{array}\right\|.
$$
То есть каждому такому преобразованию однозначно сопоставляется
квадратная невырожденная матрица второго порядка, а композиции
двух преобразований -- их матричное умножение по правилу "строка
на столбец". Хорошо известно, что множество всех невырожденных
квадратных матриц по операции их обычного умножения является
группой и потому группой является и множество преобразований
многообразия $\mathfrak{M}$ в множестве движений (9.24).

\newpage

     $$
     \lambda(x,y)\rho(\xi,\eta)+\sigma(x,y)\tau(\xi,\eta)=x\xi+y\eta,
     \eqno(9.26)
     $$
     that is satisfied identically with respect to all the coordinates $x,y$ and $\xi,\eta$.

     We shall differentiate this equation with respect to the variables $\xi$, $\eta$:
     $$
     \lambda\rho_\xi+\sigma\tau_\xi=x, \ \
     \lambda\rho_\eta+\sigma\tau_\eta=y
     $$
     and solve the equalities obtained with respect to the functions $\lambda,\sigma$:
     $$
     \lambda(x,y)=\frac{x\tau_\eta-
     y\tau_\xi}{\rho_\xi\tau_\eta-\rho_\eta\tau_\xi}, \
     \ \sigma(x,y)=\frac{-x\rho_\eta+
     y\rho_\xi}{\rho_\xi\tau_\eta-\rho_\eta\tau_\xi}.
     $$

     Differentiating the expressions obtained for the functions $\lambda$ and $\sigma$
     with respect to the variables $x,y$, we make sure that the coefficients with them
     are constants. Introducing proper designations for them, we get the first pair of equations
     (9.24), which define the transformation of the two-dimensional manifold $\mathfrak{M}$
     in the motion (9.25), their invertibility obviously having, as the corollary of it, the
     condition of $\Delta\neq0$. The other pair of equations (9.24), which defines the
     transformation of the two-dimensional manifold $\mathfrak{N}$ in the motion (9.25), is
     easily come by from the functional equation (9.26) by way of substituting the former
     pair of equations into it.

     The set of motions (9.24) depends on the four continuous parameters $a,b,c$, $d$, with
     the condition imposed upon them of $\Delta=ad-bc\neq0$. It is easy to make sure that
     that set, by its composition of motions, is a group. To do it, let us write, for example, the
     transformations of the former manifold, $\mathfrak{M}$, in the set of motions (9.24) in
     the matrix form:
     $$
     \left\|\begin{array}{cc} x' \\ y' \end{array}\right\|=
     \left\|\begin{array}{cc} a & b \\ c & d \end{array}\right\|
     \left\|\begin{array}{cc} x \\ y \end{array}\right\|.
     $$
     That is, every such transformation is univocally assigned a quadratic nondege- \ nerate
     matrix of second order, and the composition of the two transformations their matrix
     multiplication according to the \ "row by column" \ rule. It is well-known that the set of
     all nondegenerate quadratic matrices in the operation of their ordinary multiplication
     is a group, so the set of transformations of the manifold $\mathfrak{M}$ in the set of
     motions (9.24) is a group too.

\newpage

Преобразованиям многообразия $\mathfrak{N}$ в движениях (9.24)
сопоставляются обратные транспонированные матрицы, множество
которых также составляет группу, которая изоморфна группе прямых
матриц. Следовательно, и все множество движений (9.24) как
совокупность двух изоморфных групп преобразований различных
многообразий является группой, определяющей групповую симметрию
физической структуры ранга (3,3), задаваемой метрической функцией
(9.16). Степень этой симметрии равна четырем, так как группа
движений (9.24) зависит от четырех непрерывных и независимых
параметров. Теорема 3 доказана.

\vspace{5mm}

{\bf Теорема 4.} {\it Группа движений феноменологически симметричной
геометрии двух множеств (физической структуры) ранга $(3,3)$, задаваемой на
двумерных многообразиях метрической функцией $(9.17)$: $f=x\xi+y+\eta$,
представляется следующими уравнениями:
$$
\left.\begin{array}{c}
x'=ax+ b, \ \ y'=y+cx+d, \\
\xi'=(\xi-c)/a, \ \ \eta'=\eta -b\xi/a-(ad-bc)/a,
\end{array}\right\}
\eqno(9.27)
$$
где $a\neq0.$}

\vspace{5mm}

Запишем функциональное уравнение на множество движений (9.25) для метрической
функции (9.17):
$$
\lambda(x,y)\rho(\xi,\eta)+\sigma(x,y)+\tau(\xi,\eta)=x\xi+y+\eta
\eqno(9.28)
$$
и продифференцируем его по переменным $\xi,\eta$: \
$\lambda\rho_\xi+\tau_\xi=x, \ \ \lambda\rho_\eta+\tau_\eta=1,$
откуда находим: $\lambda(x,y)=(x\tau_\eta-
\tau_\xi)/(\rho_\xi\tau_\eta-\rho_\eta\tau_\xi).$ Зафиксируем в
правой части переменные $\xi,\eta$ и введем удобные обозначения
постоянных коэффициентов: $\lambda(x,y)=ax+b$, где $a\neq0$, так
как $\lambda(x,y)\neq const$. Подставляя это выражение для функции
$\lambda$ в исходное функциональное уравнение (9.28) и снова
фиксируя переменные $\xi,\eta$ получаем выражение для другой
функции $\sigma(x,y)= y+cx+d$. Тем самым получены преобразования
многообразия $\mathfrak{M}$ в множестве движений (9.27).
Преобразования второго многообразия $\mathfrak{N}$ находятся из
функционального уравнения (9.28) при подстановке в него только что
найденных выражений для функций $\lambda$  и $\sigma$. В
результате получаем все множество движений (9.27).

\newpage

      The transformations of the manifold $\mathfrak{N}$ in
     motions  (9ю24) are assigned the reciprocal transposes of matrices, whose set is also a
     group, which is isomorphic with respect to the group of the initial matrices. Therefore,
     the whole set of motions (9.24), as a congregate of isomorphic groups of transformations
     of different manifolds is a group that gives the group symmetry of the physical structure
     of rank (3,3) defined by the metric function (9.16). The degree of that symmetry is equal
     to 4, because the group of motions (9.24) depends on four continuous and independent
     parame- \ ters. Theorem 3 is proved.

     \vspace{5mm}

     {\bf Theorem 4.} {\it The group of motions of a phenomenologically symmetric
     geometry of two sets (physical structure) of rank $(3,3)$ defined on two-dimensional
     manifolds by the metric function $(9.17)$: $f=x\xi+y+\eta$, is represented by the
     following equations:
     $$
     \left.\begin{array}{c}
     x'=ax+ b, \ \ y'=y+cx+d, \\
     \xi'=(\xi-c)/a, \ \ \eta'=\eta -b\xi/a-(ad-bc)/a,
     \end{array}\right\}
     \eqno(9.27)
     $$
     where $a\neq0.$}

     \vspace{5mm}

     We shall write the functional equation on the set of motions (9.25) for the metric function
     (9.17):
     $$
     \lambda(x,y)\rho(\xi,\eta)+\sigma(x,y)+\tau(\xi,\eta)=x\xi+y+\eta
     \eqno(9.28)
     $$
     and differentiate it with respect to the variables $\xi$ and
     $\eta$: \
     $\lambda\rho_\xi+\tau_\xi=x, \ \ \lambda\rho_\eta+\tau_\eta=1,$
     which yields: $\lambda(x,y)=(x\tau_\eta-\tau_\xi)/(\rho_\xi\tau_\eta-\rho_\eta\tau_\xi)$.
     We shall fix in the right-hand member the variables $\xi,\eta$ and introduce suitable
     designation for the constant coefficients: $\lambda(x,y)=ax+b$, where $a\neq0$, as
     $\lambda(x,y)\neq const$. By substituting that expression for the function $\lambda$ in
     the initial functional equation (9.28) and fixing again the variables $\xi,\eta$ we get an
     expression for the other function $\sigma(x,y)= y+cx+d$. Thus we have obtained
     transformations for the manifold $\mathfrak{M}$ in the set of motions (9.27). The
     transformations for the second manifold, $\mathfrak{N}$, are found from the functional
     equation (9.28) by way of substituting the expressions for the functions $\lambda$ and
     $\sigma$ that we have just found. Which gives us the full set of motions (9.27).

    \newpage

Для того, чтобы убедиться в том, что множество движений (9.27)
является группой, запишем, например, преобразование множества
$\mathfrak{M}$ из него в следующей матричной форме:
$$
\left\|\begin{array}{ccc} x' \\ y' \\ 1 \end{array}\right\|=
\left\|
\begin{array}{ccc}
a & 0 & b \\
c & 1 & d \\
0 & 0 & 1
\end{array}
\right\|
\left\|\begin{array}{ccc} x \\ y \\ 1 \end{array}\right\|,
$$
откуда видно, что каждому такому преобразованию сопоставляется
\linebreak невырожденная матрица третьего порядка, структура
которой, очевидно, сохраняется при обычном матричном умножении по
правилу "строка на столбец"$,$ причем композиции двух
преобразований сопоставляется произведение соответствующих матриц.
Множество невырожденных матриц подобной структуры по операции их
умножения является группой и потому группой является множество
преобразований многообразия $\mathfrak{M}$ в движениях (9.27).
Нетрудно сообразить, что преобразованию второго многообразия
$\mathfrak{N}$ в этих движениях сопоставляется транспонированная
обратная матрица, множество которых также является группой,
изоморфной группе прямых матриц. Таким образом, все множество
движений (9.27) является группой, которая определяет групповую
симметрию геометрии двух множеств ранга (3,3), задаваемой на
двумерных многообразиях метрической функцией (9.17). Степень
групповой симметрии равна четырем, так как группа движений (9.27)
зависит от четырех непрерывных и независимых параметров. Теорема 4
доказана.

\vspace{5mm}

{\bf Теорема 5.} {\it Двухточечный инвариант группы преобразований $(9.24)$
совпадает с метрической функцией $(9.16)$ с тoчностью до масштабного
преобразования.}

\vspace{5mm}

Тождественному преобразованию в группе (9.24) соответствуют
параметры $a=1, \ b=0, \ c=0, \ d=1$. Введем параметры бесконечно
малого (инфинитезимального) преобразования $\alpha, \beta, \gamma,
\delta$, полагая $a=1+\alpha, \ b=\beta, \ c=\gamma, \
d=1+\delta$. C точностью до величин первого порядка малости
преобразования (9.24) запишутся в следующем виде:

\newpage

    In order to make certain that the set of motions (9.27) is a group, let us write, for example,
     the transformation of a set $\mathfrak{M}$ from it in the matrix form as follows:

     $$
     \left\|\begin{array}{ccc} x' \\ y' \\ 1 \end{array}\right\|=
     \left\|
     \begin{array}{ccc}
     a & 0 & b \\
     c & 1 & d \\
     0 & 0 & 1
     \end{array}
     \right\|
     \left\|\begin{array}{ccc} x \\ y \\ 1 \end{array}\right\|,
     $$  \\
     which demonstrates that every such transformation is assigned a nondegene- \ rate
     matrix of third order, whose structure is obviously preserved under usual matrix
     "row by column" \ multiplication, the composition of the two transformations being
     assigned the product of the matrices. The set of nondegenerate matrices of such
     structure is a group under the operation of their multiplication, and so the set of
     transformations of the manifold $\mathfrak{M}$ in the motions (9.27) is also a group.
     It is not difficult to see that the transformations of the manifold $\mathfrak{N}$ in
     these motions are assigned the transposed reciprocal matriceswhose set is also
     aё
     group, isomorphic to the group of the initial matrices. Thus, the whole set of motions
     (9.27) is a group that determines the symmetry of the geometry of two sets of rank
     (3,3) that is defined on two-dimensional manifolds by the metric function (9.17). The
     degree of the group symmetry equals 4, as the group of motions (9.27) depends on the
     four continuous and independent parameters. Theorem 4 is proved.

     \vspace{7mm}

     {\bf Theorem 5.} {\it The two-point invariant of the group of transformations $(9.24)$
     coincides with the metric function $(9.16)$ with an accuracy up to a scaling transformation.}

     \vspace{7mm}

     The parameters of the identity transformation in the group (9.24) are $a=1, \ b=0, \ c=0, \ d=1$.
     We shall introduce the parameters of the indefinitely small (infinitesimal) transformation
     $\alpha, \beta, \gamma, \delta$, setting $a=1+\alpha, \ b=\beta, \ c=\gamma, \ d=1+\delta$.
     Then, with an accuracy up to values of the first order of smallness the transformations (9.24)
     will be as follows:

    \newpage

$$
\left.\begin{array}{c}
x'=x+\alpha x+\beta y, \ \ y'=y+\gamma x+\delta y, \\
\xi'=\xi-\alpha\xi-\gamma\eta, \ \ \eta'=\eta-\beta\xi-\delta\eta.
\end{array}\right\}
\eqno(9.29)
$$

Бесконечно малым преобразованиям (9.29) можно сопоставить две системы
четырех линейных дифференциальных операторов:
$$
\left.\begin{array}{c}
X_1=x\partial_x, \ X_2=y\partial_x, \ X_3=x\partial_y, \ X_4=y\partial_y, \\
\Xi_1=-\xi\partial_\xi, \ \Xi_2=-\xi\partial_\eta, \
\Xi_3=-\eta\partial_\xi, \ \Xi_4=-\eta\partial_\eta,
\end{array}\right\}
\eqno(9.30)
$$
где, например, $\partial_x=\partial/\partial x$, которые составляют
естественные координатные базисы двух изоморфных с точностью до
совпадения структурных констант четырехмерных алгебр Ли преобразований
двумерных многообразий $\mathfrak{M}$ и $\mathfrak{N}$.

При известных преобразованиях (9.24) двухточечный инвариант
$f=f(x,y,\xi,\eta)$ является решением функционального уравнения
$$
f(x',y',\xi',\eta')=f(x,y,\xi,\eta).
\eqno(9.31)
$$
Если в функциональное уравнение (9.31) подставить бесконечно малые
преобразования (9.29), затем продифференцировать его по каждому из
четырех параметров $\alpha, \beta, \gamma, \delta$ и придать им
нулевые значения, то относительно двухточечного инварианта
получается система четырех дифференциальных уравнений
$$
X_\omega f+\Xi_\omega f=0,
\eqno(9.32)
$$
где $\omega=1,2,3,4$, с операторами (9.30).

Поскольку дифференциальные уравнения (9.32) линейные однородные в
частных производных первого порядка, их можно решать методом
характеристик. Для первого и четвертого уравнений системы
соответствующие уравнения характеристик $dx/x=-d\xi/\xi, \
dy/y=-d\eta/\eta$ имеют интегралы $x\xi=$ const, $y\eta=$ const.
Общее решение $f=\theta(x\xi,y\eta)$ первого и четвертого
уравнений системы (9.32), где $\theta(u,v)$ -- произвольная
функция двух переменных, подставим в ее второе и третье уравнения:
$\theta_u-\theta_v=0$. Это уравнение также решается методом
характеристик и его общее решение задается выражением
$\theta(u,v)=\chi(u+v)$, где $\chi$ -- произвольная функция уже
только одной переменной с отличной от нуля производной \ $\chi'$.
\ Таким образом, двухточечный инвариант, как

\newpage

     $$
     \left.\begin{array}{c}
     x'=x+\alpha x+\beta y, \ \ y'=y+\gamma x+\delta y, \\
     \xi'=\xi-\alpha\xi-\gamma\eta, \ \ \eta'=\eta-\beta\xi-\delta\eta.
     \end{array}\right\}
     \eqno(9.29)
     $$

     The infinitely small transformations (9.29) may be assigned two systems of four linear
     differential operators:
     $$
     \left.\begin{array}{c}
     X_1=x\partial_x, \ X_2=y\partial_x, \ X_3=x\partial_y, \ X_4=y\partial_y, \\
     \Xi_1=-\xi\partial_\xi, \ \Xi_2=-\xi\partial_\eta, \
     \Xi_3=-\eta\partial_\xi, \ \Xi_4=-\eta\partial_\eta,
     \end{array}\right\}
     \eqno(9.30)
     $$
     where, for example, $\partial_x=\partial/\partial x$ which comprise natural coordinate
     bases of the two isomorphic with an accuracy up to the coincidence of the structural
     constants four-dimensional Lie algebras of the transformations of the two-dimensional
     manifolds $\mathfrak{M}$ and $\mathfrak{N}$.

     With the known transformations (9.24), the two-point invariant $f=f(x,y,\xi,\eta)$ is the
     solution of the functional equation
     $$
     f(x',y',\xi',\eta')=f(x,y,\xi,\eta).
     \eqno(9.31)
     $$
     If we substitute in the functional equation (9.31) the infinitely small transfor- \ mations (9.29),
     differentiate it with respect to each of the four parameters $\alpha, \beta, \gamma, \delta$
     and assign zero values to them, then with respect to the two-point invariant a system of
     four differential equations
     $$
     X_\omega f+\Xi_\omega f=0,
     \eqno(9.32)
     $$
     appears where $\omega=1,2,3,4$, with the operators (9.30).

     Since the differential equations (9.32) are ones linear homogeneous in the partial
     derivatives of the second order, they may be solved by the method of characteristics.
     For the first and the fourth equations of the system, the respective equations of
     characteristics are $dx/x=-d\xi/\xi, \ dy/y=-d\eta/\eta$ and have the integrals
     $x\xi=$ const, $y\eta=$ const. We shall substitute the general solution $f=\theta(x\xi,y\eta)$
     of the first and the fourth equations of the system (9.32), where $\theta(u,v)$ is an
     arbitrary function of two variables, into the second and third equations: $\theta_u-\theta_v=0$.
     This equation is also solved by the method of characteristics, and its general solution is
     the expression $\theta(u,v)=\chi(u+v)$, where $\chi$ is an arbitrary function of only one
     variable with the derivative $\chi'$ which is unequal to zero. Thus, the two-point invariant,
     as

    \newpage

\noindent
решение системы дифференциальных уравнений (9.32) с операторами
(9.30) задается выражением
$$
f=\chi(x\xi+y\eta),
\eqno(9.33)
$$
которое масштабным преобразованием $\chi^{-1}(f)\to f$ с обратной функцией
$\chi^{-1}$ переводится в
метрическую функцию (9.16). Теорема 5 доказана.

\vspace{5mm}

{\bf Теорема 6.} {\it Двухточечный инвариант группы преобразований $(9.27)$
совпадает с метрической функцией $(9.17)$ с тoчностью до масштабного
преобразования.}

\vspace{5mm}

Доказательства теоремы 6 в общих чертах повторяет доказательство
предыдущей теоремы, хотя в деталях, конечно же, от него
отличается. Тождественным преобразованием в группе (9.27) будет
преобразование с параметрами $a=1, \ b=0, \ c=0, \ d=0$. Полагая
$a=1+\alpha, \ b=\beta, \ c=\gamma, \ d=\delta$, с точностью до
малых величин первого порядка из уравнений (9.27) получаем
уравнения для бесконечно малых (инфинитезимальных) преобразований:
$$
\left.\begin{array}{c}
x'=x+\alpha x+\beta, \ \ y'=y+\gamma x+\delta, \\
\xi'=\xi-\alpha\xi -\gamma, \ \ \eta'=\eta -\beta\xi-\delta,
\end{array}\right\}
\eqno(9.34)
$$
которым соответствуют две системы четырех линейных
дифференциальных операторов:
$$
\left.\begin{array}{c}
X_1=x\partial_x, \ X_2=\partial_x, \ X_3=x\partial_y, \ X_4=\partial_y, \\
\Xi_1=-\xi\partial_\xi, \ \Xi_2=-\xi\partial_\eta, \
\Xi_3=-\partial_\xi, \ \Xi_4=-\partial_\eta,
\end{array}\right\}
\eqno(9.35)
$$
которые составляют естественные координатные базисы двух
изоморфных с точностью до совпадения структурных констант
четырехмерных алгебр Ли преобразований (9.27) двумерных
многообразий  $\mathfrak{M}$ и $\mathfrak{N}$.

Подставим в функциональное уравнение (9.31) для двухточечного
инварианта инфинитезимальные преобразования (9.34),
продифференцируем его по каждому из четырех параметров
$\alpha,\beta,\gamma,\delta$ и в результатах дифференцирования
придадим им нулевые значения, соответствующие тождественным
преобразованиям. В итоге на двухточечный инвариант
$f=f(x,y,\xi,\eta)$ возникает система четырех дифференциальных
уравнений (9.32) с операторами (9.35). \ Как и в предыдущем
случае, \ эти

\newpage

\noindent
 the solution of the system of differential equations (9.32)
with the operators (9.30) is

     $$
     f=\chi(x\xi+y\eta),
     \eqno(9.33)
     $$ \\
     which is transformed by the scaling transformation $\chi^{-1}(f)\to f$ with the inverse
     $\chi^{-1}$ into the metric function (9.16). Theorem 5 is proved.

     \vspace{5mm}

     {\bf Theorem 6.} {\it The two-point invariant of the group of transformations $(9.27)$
     coincides with the metric function $(9.17)$ with an accuracy up to a scaling transformation.}
     \vspace{5mm}

     The proof of Theorem 6 mainly repeats the proof of Theorem 5, differing from it, grosso
     modo, in some small detail. The identity transformation in the group (9.27) will be the
     transformation with the parameters $a=1, \ b=0, \ c=0, \ d=0$. Setting $a=1+\alpha,
     \ b=\beta, \ c=\gamma, \ d=\delta$, with an accuracy up to the terms of the first order
     of smallness from the equation (9.27) we get equations for infinitely small (infinitesimal)
     transformations:
     $$
     \left.\begin{array}{c}
     x'=x+\alpha x+\beta, \ \ y'=y+\gamma x+\delta, \\
     \xi'=\xi-\alpha\xi -\gamma, \ \ \eta'=\eta -\beta\xi-\delta,
     \end{array}\right\}
     \eqno(9.34)
     $$
     which correspond two systems of four linear differential operators:
     $$
     \left.\begin{array}{c}
     X_1=x\partial_x, \ X_2=\partial_x, \ X_3=x\partial_y, \ X_4=\partial_y, \\
     \Xi_1=-\xi\partial_\xi, \ \Xi_2=-\xi\partial_\eta, \
     \Xi_3=-\partial_\xi, \ \Xi_4=-\partial_\eta,
     \end{array}\right\}
     \eqno(9.35)
     $$
     which comprise the natural coordinate bases of two isomorphic with an accuracy
     up to the coincidence of the structural constants four-dimensional Lie algebras of
     the transformations (9.27) of the two-dimensional manifolds $\mathfrak{M}$ and
     $\mathfrak{N}$.

     We shall substitute into the functional equation (9.31) for the two-point invariant
     the infinitesimal transformations (9.34), differentiate it with respect to each of the
     four parameters$\alpha,\beta,\gamma,\delta$ and in the results of the differentiation
     assign them zero values corresponding the identity transformations. That yields for
     the two-point invariant $f=f(x,y,\xi,\eta)$ the system of the four differential
     equations (9.32) with the operators (9.35). As is in the
     previous

    \newpage

\noindent
уравнения решаются методом характеристик. Для первого и четвертого
уравнений системы (9.32) соответствующие уравнения характеристик
$dx/x=-d\xi/\xi, \ dy=-d\eta$ легко интегрируются: $x\xi=$ const,
\ $y+\eta=$ const, поэтому их общее решение запишется в следующем
виде: $f=\theta(x\xi,y+\eta)$, где $\theta(u,v)$ -- произвольная
функция двух переменных. После подстановки этого выражения во
второе и третье уравнения системы получаем дифференциальное
уравнение $\theta_u-\theta_v=0$, решение которого
$\theta(u,v)=\chi(u+v)$ записывается через произвольную функцию
$\chi$ от одной только переменной, причем $\chi'\neq0$. В
результате для двухточечного инварианта $f$ получаем выражение
$$
f=\chi(x\xi+y+\eta),
\eqno(9.36)
$$
которое переходит в метрическую функцию (9.17) при масштабном преобразовании
$\chi^{-1}(f)\to f$ с обратной функцией $\chi^{-1}$.
Теорема 6 доказана.

\vspace{5mm}

Заметим, что две четырехмерные алгебры Ли (9.30) в соответствующих
базисах имеют одинаковые структурные константы. Очевидный переход
к другому базису и тривиальная замена координат переводят один
базис в другой, что говорит о слабой эквивалентности этих алгебр.
Однако никакая только замена координат не переведет эти базисы
один в другой. Отмеченное обстоятельство означает, что
соответствующие им две группы Ли преобразований (9.24), как
различные действия в двумерном многообразии одной и той же
четырехмерной группы Ли, подобны, но не эквивалентны. То есть
некоторый автоморфизм в группе и замена координат переведут одну
группу преобразований в другую (подобие или слабая
эквивалентность), но никакая замена координат без автоморфизма
этого не сможет сделать (неэквивалентность в сильном смысле).
Аналогичное замечание можно сделать и в отношении двух
четырехмерных алгебр Ли (9.35), соответствующих группам Ли
преобразований (9.27).

\newpage

     \noindent
    case, the equations
     are solved with the method of characteristics. For the first and the fourth equations of
     the system (9.32) the respective equations of characteristics, $dx/x=-d\xi/\xi,
     \ dy=-d\eta$, are easily integrable: $x\xi=$ const, \ $y+\eta=$ const, and their general
     solution will be written as follows: $f=\theta(x\xi,y+\eta)$, where $\theta(u,v)$ is an
     arbitrary function of two variables. The substitution of that expression into the second
     and third equations of the system yields the differential equation $\theta_u-\theta_v=0$
     whose solution $\theta(u,v)=\chi(u+v)$ is written via the arbitrary function $\chi$ of
     only one variable, $\chi'$ being unequal to zero. As result, for the two-point invariant
     $f$ we get the expression
     $$
     f=\chi(x\xi+y+\eta),
     \eqno(9.36)
     $$
     that is transformed into the metric function (9.17) through the scaling transformation
     $\chi^{-1}(f)\to f$ with the inverse function $\chi^{-1}$. Theorem 6 is proved.
     \vspace{5mm}

     We shall note that the two four-dimensional Lie algebras (9.30) in the respective
     bases have the same structural constants. The obvious transition to the other basis
     coupled with a trivial change of coordinates transforms one basis into the other,
     which means the weak equivalence of the algebras. However, no change of
     coordinates {\it alone} transforms the bases one into the other. That circumstance
     implies that the two corresponding groups of Lie transformations (9.24), as
     different actions in the two-dimensional manifold of one and the same four-dimensional
     Lie group, are similar but not equivalent. That is, some automorphism in the group
     accompanied by the change of coordinates {\it will} transform one group of
     transformations into the other (similarity, or weak equivalence), but no change of
     coordinates {\it without} automorphism can do the same (nonequivalence in the strong
     sense). The same is to be said as concerns the two four-dimensional Lie algebras
     (9.35) that correspond the Lie transformations (9.27).

     \newpage

\begin{center}
{\bf \large \S10. Двуметрические и триметрические \\ физические структуры}
\end{center}

Полная классификация двуметрических физических структур (феноменологически
симметричных геометрий двух множеств) построена только для ранга $(n+1,2)$.
Краткое же их определение получается из общего
определения 1 полиметрических физических структур ранга $(n+1,m+1)$,
данного в начале \S8, если в нем положить $s=2$ и $m=1$.

Пусть имеются два множества $\mathfrak{M}$ и $\mathfrak{N}$, являющиеся
2-мерным и $2n$-мерным многообразиями соответственно, где $n$ -- натуральное
число. Обозначим локальные координаты в этих многообразиях через $x=(x^1,x^2)$
\ и \ $\xi=(\xi^1,\ldots,\xi^{2n})$. Пусть также имеется функция $f$ с открытой и
плотной в \ $\mathfrak{M\times N}$ \ областью определения $\mathfrak{S}_f$,
сопоставляющая каждой паре из нее два вещественных числа, то есть
$f:\mathfrak{S}_f\to R^2$. Двухточечную
двухкомпонентную функцию $f=(f^1,f^2)$ будем
называть {\it метрической}. Предполагается, что ее локальное координатное
представление задается достаточно гладкой невырожденной
функцией
$$
f=f(x,\xi)=f(x^1,x^2,\xi^1,\dots,\xi^{2n}),
\eqno(10.1)
$$
выражение для которой получается из выражения (8.2) при $s=2$ и $m=1$.
Невырожденность метрической функции (10.1) понимается в смысле
аксиомы III из \S8 и,
вообще говоря, в отличие от случая $s=1$, то есть однометрических физических
структур, означает нечто большее, чем просто ее существенную зависимость
от координат $x=(x^1,x^2)$ \ и \ $\xi=(\xi^1,\dots,\xi^{2n})$.
А именно, должны быть отличны
от нуля якобианы \ $\partial f(i\alpha)/\partial x_i$ \ и \
$\partial(f(i_1\alpha),\dots,f(i_n\alpha))/\partial\xi_\alpha$ \
для плотных множеств пар
$\langle  i\alpha  \rangle$ $\in$ $\mathfrak{M\times N}$ и кортежей
$\langle  i_1\ldots i_n,\alpha  \rangle \ \in \mathfrak{M}^n\times\mathfrak{N}$ длины
$n+1$.

Далее строим функцию $F$ с естественной в
$\mathfrak{M}^{n+1}\times\mathfrak{N}^2$ областью определения
$\mathfrak{S}_F$, сопоставляя каждому кортежу длины $n+3$ из
$\mathfrak{S}_F$ все $4(n+1)$ возможные по метрической функции
$f=(f^1,f^2)$ расстояния. Будем говорить, что двухкомпонентная
функция $f$ с локальным координатным представлением (10.1) задает
на 2-мерном и $2n$-мерном многообразиях $\mathfrak{M}$ и
$\mathfrak{N}$ {\it двуметрическую физическую структуру}
(феноменологически симметричную геометрию двух множеств) {\it
ранга} $(n+1,2)$, если локально множество значений
$F(\mathfrak{S}_F)$ в $R^{4(n+1)}$ принадлежит мно-

\newpage

     \begin{center}
     {\bf \large \S10. Dimetric and trimetric \\ physical structures}
     \end{center}

     The full classification of the dimetric physical structures (phenomenologi- \ cally
     symmetric geometries of two sets) has been only built for the rank $(n+1,2)$.
     A brief definition of theirs is derived from Definition 1 of the polymetric physical
     structures of rank $(n+1,m+1)$, that was given at the beginning of \S8, if we set
     $s=2$ and $m=1$ in it.

     Suppose there are two sets, $\mathfrak{M}$ and $\mathfrak{N}$, that are a two-
     dimensional  and a $2n$-dimensional manifolds respectively, where $n$ is a natural
     number. We shall designate the local coordinates in the manifolds as $x=(x^1,x^2)$
     \ and \ $\xi=(\xi^1,\ldots,\xi^{2n})$. Suppose there is also a function $f$ with the
     domain $\mathfrak{S}_f$ open and dense in \ $\mathfrak{M\times N}$ \ that assigns
     to every pair of it two real numbers, i.e. $f:\mathfrak{S}_f\to R^2$. We shall call the
     two-point two-component function $f=(f^1,f^2)$ a {\it metric} one. It is supposed that
     its local coordinate representation is defined by a sufficiently smooth nondegenerate
     function

     $$
     f=f(x,\xi)=f(x^1,x^2,\xi^1,\dots,\xi^{2n}),
     \eqno(10.1)
     $$ \\
     the expression for which is obtained from the expression (8.2) with $s=2$ and $m=1$.
     The nondegeneracy of the metric function (10.1) is understood in the sense of Axiom III
     of \S8 and, generally speaking, and in contrast to the case of  $s=1$, i.e. that of unimetric
     physical structures, means somewhat more than its mere essential dependence on the
     coordinates $x=(x^1,x^2)$ \ and \ $\xi=(\xi^1,\dots,\xi^{2n})$. And that is the necessary
     nonzero quality of the Jacobians \ $\partial f(i\alpha)/\partial x_i$ \ and \
     $\partial(f(i_1\alpha),\dots,f(i_n\alpha))/\partial\xi_\alpha$ \ for the dense sets of pairs
     $\langle  i\alpha  \rangle$ $\in$ $\mathfrak{M\times N}$ and corteges
     $\langle  i_1\ldots i_n,\alpha  \rangle \ \in \mathfrak{M}^n\times\mathfrak{N}$ of length
     $n+1$.
     Further, we build the function $F$ with the natural in $\mathfrak{M}^{n+1}\times\mathfrak{N}^2$
     domain $\mathfrak{S}_F$ by assigning to every cortege $n+3$ from $\mathfrak{S}_F$
     all the $4(n+1)$ distances possible in the metric function $f=(f^1,f^2)$. We shall say that
     the two-component function $f$ with the local coordinate representation (10.1) gives on
     a two-dimensional manifold $\mathfrak{M}$ and a $2n$-dimensional manifold $\mathfrak{N}$
     a {\it dimetric physical structure} (phenomenologically symmetric geometry of two sets) of
     {\it rank} $(n+1,2)$, \ if locally the set of values $F(\mathfrak{S}_F)$ \ in
     $R^{4(n+1)}$

\newpage

\noindent
жеству нулей некоторой достаточно гладкой двухкомпонентной функции
$\Phi= (\Phi_1,\Phi_2)$ от $4(n+1)$ переменных с независимыми
компонентами $\Phi_1$ и $\Phi_2$, то есть имеет место уравнение
$$
\Phi(f(i\alpha),f(i\beta),f(j\alpha),f(j\beta),\dots,f(v\alpha),f(v\beta))=0
\eqno(10.2)
$$
для всех кортежей $\langle  ijk\ldots v,\alpha\beta  \rangle$ из некоторого плотного и
открытого в $\mathfrak{S}_F\subseteq\mathfrak{M}^{n+1}\times\mathfrak{N}^2$
множества. Таким образом, локально множество $F(\mathfrak{S}_F)$ принадлежит
некоторой регулярной коразмерности 2 поверхности в $R^{4(n+1)}$, не
обязательно совпадая с ней.

Заметим, что не всякая двухкомпонентная функция $f=(f^1,f^2)$
может задавать двуметрическую физическую структуру и потому
основной задачей теории является полная классификация таких
функций, которая, как обычно, проводится с точностью до
масштабного преобразования, в данном случае двумерного, и
возможности выбора в многообразиях $\mathfrak{M}$ и $\mathfrak{N}$
любых допустимых систем локальных координат.

\vspace{5mm}

{\bf Теорема 1.} {\it Двуметрические физические структуры (феноменологически
симметричные геометрии двух множеств) ранга $(n+1,2)$
существуют только для $n=1,2,3,4$, то есть ранга
$(2,2), \ (3,2), \ (4,2),$ $(5,2)$,
и не существуют для $n\geq5$, то есть ранга $(6,2), \ (7,2)$ и т.д. С точностью
до масштабного преобразования двухкомпонентная метрическая
функция $f=(f^1,f^2)$, задающая
на $2$-мерном и $2n$-мерном многообразиях $\mathfrak{M}$ и $\mathfrak{N}$
двуметрическую физическую структуру ранга $(n+1,2)$, в надлежаще выбранных в
них системах локальных координат $x=(x^1,x^2)=(x,y)$ и \
$\xi=(\xi^1,\xi^2,\xi^3,\xi^4,\ldots)=(\xi,\eta,\mu,\nu,\ldots)$ определяется
следующими каноническими выражениями:

для $n=1$, то есть ранга $(2,2)$:
$$
f^1=x+\xi, \ f^2=y+\eta,
\eqno(10.3)
$$
$$
f^1=(x+\xi)y, \ f^2=(x+\xi)\eta;
\eqno(10.4)
$$

для $n=2$, то есть ранга $(3,2)$:
$$
f^1=x\xi+\varepsilon y\eta+\mu, \ f^2=x\eta+y\xi+\nu, \ \varepsilon=0,\pm1,
\eqno(10.5)
$$
$$
f^1=x\xi+\mu, \ f^2=x\eta+y\xi^c+\nu, \ c\neq1,
\eqno(10.6)
$$ }

\newpage

      \noindent
      belongs
     to the set of zeros of some sufficiently smooth two-component function $\Phi=(\Phi_1,\Phi_2)$
     of $4(n+1)$ variables with the independent components $\Phi_1$ and $\Phi_2$, i.e. an
     equation

     $$
     \Phi(f(i\alpha),f(i\beta),f(j\alpha),f(j\beta),\dots,f(v\alpha),f(v\beta))=0
     \eqno(10.2)
     $$ \\
     takes place for all the corteges $\langle  ijk\ldots v,\alpha\beta  \rangle$ from a set dense and
     open in $\mathfrak{S}_F\subseteq\mathfrak{M}^{n+1}\times\mathfrak{N}^2$.
     Thus, the set $F(\mathfrak{S}_F)$ locally belongs to some regular surface in $R^{4(n+1)}$,
     of codimension 2, not necessarily coinciding with it.

     We shall note that not every two-component function $f=(f^1,f^2)$ may give a
     dimetric physical structure, so the principle task for the theory is their complete
     classification that is to be carried out with an accuracy, as usual, up to a scaling
     transformation, two-dimensional in this case, and the possibility of choosing in the
     manifold $\mathfrak{M}$ and $\mathfrak{N}$ of any allowable systems of local
     coordinates.

     \vspace{5mm}

     {\bf Theorem 1.} {\it Dimetric physical structures (phenomenologically symmetric
     geometries of two sets) of rank $(n+1,2)$ exist only for $n=1,2,3,4$, that is rank $(2,2), \
     (3,2), \ (4,2),$ $(5,2)$, and do not exist for $n\geq5$, i.e. for rank $(6,2), \ (7,2)$ etc. With
     an accuracy up to a scaling transformation the two-component metric function $f=(f^1,f^2)$
     that defines on a $2$-dimensional and a $2n$-dimensional manifolds $\mathfrak{M}$ and
     $\mathfrak{N}$ a dimetric physical structure of rank $(n+1,2)$ in systems of local coordinates
     $x=(x^1,x^2)=(x,y)$ and $\xi=(\xi^1,\xi^2,\xi^3,\xi^4,\ldots)=(\xi,\eta,\mu,\nu,\ldots)$
     suitably chosen in them is defined with the following canonical expressions:

     for $n=1$, that is for rank $(2,2)$:
     $$
     f^1=x+\xi, \ f^2=y+\eta,
     \eqno(10.3)
     $$
     $$
     f^1=(x+\xi)y, \ f^2=(x+\xi)\eta;
     \eqno(10.4)
     $$

     for $n=2$, i.e. for rank $(3,2)$:
     $$
     f^1=x\xi+\varepsilon y\eta+\mu, \ f^2=x\eta+y\xi+\nu, \ \varepsilon=0,\pm1,
     \eqno(10.5)
     $$
     $$
     f^1=x\xi+\mu, \ f^2=x\eta+y\xi^c+\nu, \ c\neq1,
     \eqno(10.6)
     $$ }

\newpage

 {\it $$
f^1=x\xi+\mu, \ f^2=x\eta+y\xi^2+x^2\xi^2\ln\xi+\nu,
\eqno(10.7)
$$
$$
f^1=x\xi+y\mu, \ f^2=x\eta+y\nu;
\eqno(10.8)
$$

для $n=3$, то есть ранга $(4,2)$:
$$
\left.\begin{array}{c}
f^1=\displaystyle\frac{(x\xi+\varepsilon y\eta+\mu)(x+\rho)-
\varepsilon(x\eta+y\xi+\nu)(y+\tau)}{(x+\rho)^2-\varepsilon(y+\tau)^2}, \\
\phantom{aaaaa} \\
f^2=\displaystyle\frac{(x\xi+\varepsilon y\eta +\mu)(y+\tau)-
(x\eta+y\xi+\nu)(x+\rho)}{(x+\rho)^2-\varepsilon(y+\tau)^2},
\end{array}\right\}
\eqno(10.9)
$$
где $\varepsilon=0,\pm1$,
$$
f^1=\frac{x\xi+\mu}{x+\rho}, \ f^2=\frac{x\eta+y\nu+\tau}{x+\rho},
\eqno(10.10)
$$
$$
f^1=x\xi+y\mu+\rho, \ f^2=x\eta+y\nu+\tau;
\eqno(10.11)
$$

для $n=4$, то есть ранга $(5,2)$:
$$
f^1=\frac{x\xi+y\mu+\rho}{x\varphi+y+\omega}, \
f^2=\frac{x\eta+y\nu+\tau}{x\varphi+y+\omega}.
\eqno(10.12)
$$}

\vspace{5mm}

Доказательство теоремы 1 можно найти в \S7 монографии автора [24] и в его
работе [26].

\vspace{5mm}

Обратимся теперь к уравнению (10.2), которое выражает
феноменологическую симметрию двуметрической физической структуры ранга
$(n+1,2)$. Выпишем его явно для каждой из метрических функций (10.3) -- (10.12)
соответственно:

для метрической функции (10.3):
$$
\left.\begin{array}{c}
f^1(i\alpha)-f^1(i\beta)-f^1(j\alpha)+f^1(j\beta)=0, \\
f^2(i\alpha)-f^2(i\beta)-f^2(j\alpha)+f^2(j\beta)=0;
\end{array}\right\}
$$

для метрической функции (10.4):
$$
\left.\begin{array}{c}
$$
\left|
\begin{array}{cc}
f^1(i\alpha)-f^1(i\beta) & f^1(i\alpha)f^2(j\alpha) \\
f^1(j\alpha)-f^1(j\beta) & f^1(j\alpha)f^2(i\alpha)
\end{array}
\right|=0,
$$ \\
\phantom{aaaaa} \\
$$
\phantom{ab} \left|
\begin{array}{cc}
f^2(i\alpha)-f^2(j\alpha) & f^2(i\alpha)f^1(i\beta) \\
f^2(i\beta)-f^2(j\beta) &  f^2(i\beta)f^1(i\alpha)
\end{array}
\right|=0;
$$
\end{array}\right\}
$$

\newpage

     {\it    $$
     f^1=x\xi+\mu, \ f^2=x\eta+y\xi^2+x^2\xi^2\ln\xi+\nu,
     \eqno(10.7)
     $$
     $$
     f^1=x\xi+y\mu, \ f^2=x\eta+y\nu;
     \eqno(10.8)
     $$

     for $n=3$, i.e. for rank $(4,2)$:
     $$
     \left.\begin{array}{c}
     f^1=\displaystyle\frac{(x\xi+\varepsilon y\eta+\mu)(x+\rho)-
     \varepsilon(x\eta+y\xi+\nu)(y+\tau)}{(x+\rho)^2-\varepsilon(y+\tau)^2}, \\
     \phantom{aaaaa} \\
     f^2=\displaystyle\frac{(x\xi+\varepsilon y\eta +\mu)(y+\tau)-
     (x\eta+y\xi+\nu)(x+\rho)}{(x+\rho)^2-\varepsilon(y+\tau)^2},
     \end{array}\right\}
     \eqno(10.9)
     $$
     where $\varepsilon=0,\pm1$,
     $$
     f^1=\frac{x\xi+\mu}{x+\rho}, \ f^2=\frac{x\eta+y\nu+\tau}{x+\rho},
     \eqno(10.10)
     $$
     $$
     f^1=x\xi+y\mu+\rho, \ f^2=x\eta+y\nu+\tau;
     \eqno(10.11)
     $$

     for $n=4$, i.e. for rank $(5,2)$:
     $$
     f^1=\frac{x\xi+y\mu+\rho}{x\varphi+y+\omega}, \
     f^2=\frac{x\eta+y\nu+\tau}{x\varphi+y+\omega}.
     \eqno(10.12)
     $$}

     \vspace{5mm}

     The proof of Theorem 1 may be found in \S7 of the author's monograph [24] and in his
     note [26].

     \vspace{5mm}

     Let us now take the equation (10.2), that expresses the phenomenological symmetry
     of the dimetric physical structure of rank$(n+1,2)$. We shall write it explicitly for each
     of the metric functions (10.3) to (10.12) respectively:

     for the metric function (10.3):
     $$
     \left.\begin{array}{c}
     f^1(i\alpha)-f^1(i\beta)-f^1(j\alpha)+f^1(j\beta)=0, \\
     f^2(i\alpha)-f^2(i\beta)-f^2(j\alpha)+f^2(j\beta)=0;
     \end{array}\right\}
     $$

     for the metric function (10.4):
     $$
     \left.\begin{array}{c}
     $$
     \left|
     \begin{array}{cc}
     f^1(i\alpha)-f^1(i\beta) & f^1(i\alpha)f^2(j\alpha) \\
     f^1(j\alpha)-f^1(j\beta) & f^1(j\alpha)f^2(i\alpha)
     \end{array}
     \right|=0,
     $$ \\
     \phantom{aaaaa} \\
     $$
     \phantom{ab} \left|
     \begin{array}{cc}
     f^2(i\alpha)-f^2(j\alpha) & f^2(i\alpha)f^1(i\beta) \\
     f^2(i\beta)-f^2(j\beta) &  f^2(i\beta)f^1(i\alpha)
     \end{array}
     \right|=0;
     $$
     \end{array}\right\}
     $$

     \newpage

для метрической функции (10.5):
$$
\left.\begin{array}{c}
$$
\left|
\begin{array}{ccc}
f^1(i\alpha) & f^1(i\beta) & 1 \\
f^1(j\alpha) & f^1(j\beta) & 1 \\
f^1(k\alpha) & f^1(k\beta) & 1
\end{array}
\right|+\varepsilon \left|
\begin{array}{ccc}
f^2(i\alpha) & f^2(i\beta) & 1 \\
f^2(j\alpha) & f^2(j\beta) & 1 \\
f^2(k\alpha) & f^2(k\beta) & 1
\end{array}
\right|=0,
$$ \\
\phantom{aaaaa} \\
$$
\phantom{abc} \left|
\begin{array}{ccc}
f^1(i\alpha) & f^2(i\beta) & 1 \\
f^1(j\alpha) & f^2(j\beta) & 1 \\
f^1(k\alpha) & f^2(k\beta) & 1
\end{array}
\right|+\left|
\begin{array}{ccc}
f^2(i\alpha) & f^1(i\beta) & 1 \\
f^2(j\alpha) & f^1(j\beta) & 1 \\
f^2(k\alpha) & f^1(k\beta) & 1
\end{array}
\right|=0;
$$
\end{array}\right\}
$$

для метрической функции (10.6):
$$
{\bf \hat{R}}(\alpha\beta) \frac{\left| \begin{array}{ccc}
f^1(i\alpha) & f^2(i\alpha) & 1 \\
f^1(j\alpha) & f^2(j\alpha) & 1 \\
f^1(k\alpha) & f^2(k\alpha) & 1
\end{array} \right| }{(f^1(i\alpha)-f^1(j\alpha))^{c+1} }=0, \ \
\left|\begin{array}{ccc}
f^1(i\alpha) & f^1(i\beta) & 1 \\
f^1(j\alpha) & f^1(j\beta) & 1 \\
f^1(k\alpha) & f^1(k\beta) & 1
\end{array}\right|=0,
$$
где ${\bf \hat{R}}(\alpha\beta)$ -- оператор альтернирования (антисимметризации) по
элементам $\alpha,\beta$, то есть ${\bf \hat{R}}(\alpha\beta)\varphi(\alpha\beta)=
\varphi(\alpha\beta)-\varphi(\beta\alpha)$;

для метрической функции (10.7):
$$
\left.\begin{array}{c}
\left| \begin{array}{ccc}
f^1(i\alpha) & f^1(i\beta) & 1 \\
f^1(j\alpha) & f^1(j\beta) & 1 \\
f^1(k\alpha) & f^1(k\beta) & 1
\end{array} \right|=0,\\
\phantom{aaaaa} \\
{\bf\hat{R}}(\alpha\beta)\displaystyle\frac{1}{(f^1(i\alpha)-f^1(j\alpha))^3}
\{\left| \begin{array}{ccc}
f^1(i\alpha) & f^2(i\alpha) & 1 \\
f^1(j\alpha) & f^2(j\alpha) & 1 \\
f^1(k\alpha) & f^2(k\alpha) & 1
\end{array}\right|- \\
\phantom{aaaaa} \\
-\left| \begin{array}{ccc}
f^1(i\alpha) & (f^1(i\alpha))^2 & 1 \\
f^1(j\alpha) & (f^1(j\alpha))^2 & 1 \\
f^1(k\alpha) & (f^1(k\alpha))^2 & 1
\end{array}\right|\ln[f^1(i\alpha)-f^1(j\alpha)]\}=0;
\end{array}\right\}
$$

\newpage

     for the metric function (10.5):
     $$
     \left.\begin{array}{c}
     $$
     \left|
     \begin{array}{ccc}
     f^1(i\alpha) & f^1(i\beta) & 1 \\
     f^1(j\alpha) & f^1(j\beta) & 1 \\
     f^1(k\alpha) & f^1(k\beta) & 1
     \end{array}
     \right|+\varepsilon \left|
     \begin{array}{ccc}
     f^2(i\alpha) & f^2(i\beta) & 1 \\
     f^2(j\alpha) & f^2(j\beta) & 1 \\
     f^2(k\alpha) & f^2(k\beta) & 1
     \end{array}
     \right|=0,
     $$ \\
     \phantom{aaaaa} \\
     $$
     \phantom{abc} \left|
     \begin{array}{ccc}
     f^1(i\alpha) & f^2(i\beta) & 1 \\
     f^1(j\alpha) & f^2(j\beta) & 1 \\
     f^1(k\alpha) & f^2(k\beta) & 1
     \end{array}
     \right|+\left|
     \begin{array}{ccc}
     f^2(i\alpha) & f^1(i\beta) & 1 \\
     f^2(j\alpha) & f^1(j\beta) & 1 \\
     f^2(k\alpha) & f^1(k\beta) & 1
     \end{array}
     \right|=0;
     $$
     \end{array}\right\}
     $$

     for the metric function (10.6):
     $$
     {\bf \hat{R}}(\alpha\beta) \frac{\left| \begin{array}{ccc}
     f^1(i\alpha) & f^2(i\alpha) & 1 \\
     f^1(j\alpha) & f^2(j\alpha) & 1 \\
     f^1(k\alpha) & f^2(k\alpha) & 1
     \end{array} \right| }{(f^1(i\alpha)-f^1(j\alpha))^{c+1} }=0, \ \
     \left|\begin{array}{ccc}
     f^1(i\alpha) & f^1(i\beta) & 1 \\
     f^1(j\alpha) & f^1(j\beta) & 1 \\
     f^1(k\alpha) & f^1(k\beta) & 1
     \end{array}\right|=0,
     $$
     where ${\bf \hat{R}}(\alpha\beta)$ is the operator of alternation (antisymmetrization)
     with respect to the elements $\alpha,\beta$, i.e. ${\bf \hat{R}}(\alpha\beta)\varphi(\alpha\beta)=
     \varphi(\alpha\beta)-\varphi(\beta\alpha)$;

     for the metric function (10.7):
     $$
     \left.\begin{array}{c}
     \left| \begin{array}{ccc}
     f^1(i\alpha) & f^1(i\beta) & 1 \\
     f^1(j\alpha) & f^1(j\beta) & 1 \\
     f^1(k\alpha) & f^1(k\beta) & 1
     \end{array} \right|=0,\\
     \phantom{aaaaa} \\
     {\bf\hat{R}}(\alpha\beta)\displaystyle\frac{1}{(f^1(i\alpha)-f^1(j\alpha))^3}
     \{\left| \begin{array}{ccc}
     f^1(i\alpha) & f^2(i\alpha) & 1 \\
     f^1(j\alpha) & f^2(j\alpha) & 1 \\
     f^1(k\alpha) & f^2(k\alpha) & 1
     \end{array}\right|- \\
     \phantom{aaaaa} \\
     -\left| \begin{array}{ccc}
     f^1(i\alpha) & (f^1(i\alpha))^2 & 1 \\
     f^1(j\alpha) & (f^1(j\alpha))^2 & 1 \\
     f^1(k\alpha) & (f^1(k\alpha))^2 & 1
     \end{array}\right|\ln[f^1(i\alpha)-f^1(j\alpha)]\}=0;
     \end{array}\right\}
     $$

\newpage

для метрической функции (10.8):

$$
\left.\begin{array}{c} {\bf\hat{R}}(ij) \left|\begin{array}{cc}
f^1(i\alpha) & f^1(i\beta) \\
f^1(k\alpha) & f^1(k\beta)
\end{array}\right|\times
\left|\begin{array}{cc}
f^2(j\alpha) & f^2(j\beta) \\
f^2(k\alpha) & f^2 (k\beta)
\end{array}\right|=0, \\
\phantom{aaaaa} \\
\phantom{ab}{\bf \hat{R}}(ik)
\left|\begin{array}{cc}
f^1(i\alpha) & f^1(i\beta) \\
f^1(j\alpha) & f^1(j\beta)
\end{array}\right|\times
\left|\begin{array}{cc}
f^2(k\alpha) & f^2(k\beta) \\
f^2(j\alpha) & f^2(j\beta)
\end{array}\right|=0;
\end{array}\right\}
$$ \\

для метрической функции (10.9) уравнение (10.2) можно получить комплексификацией
уравнения

$$
\left|\begin{array}{cccc}
f(i\alpha) & f(i\beta) & f(i\alpha)f(i\beta) & 1 \\
f(j\alpha) & f(j\beta) & f(j\alpha)f(j\beta) & 1 \\
f(k\alpha) & f(k\beta) & f(k\alpha)f(k\beta) & 1 \\
f(l\alpha) & f(l\beta) & f(l\alpha)f(l\beta) & 1
\end{array}\right|=0,
$$ \\
полагая в нем $f=f^1+ef^2$, где
$e^2=\varepsilon=0,\pm1$, и отделяя затем реальную и мнимую части;

для метрической функции (10.10):

$$
\left.\begin{array}{c}
\hat R(\alpha\beta)\dfrac{(f^1(j\alpha)-f^1(l\alpha))(f^1(i\alpha)-f^1(k\alpha))}
{(f^1(i\alpha)-f^1(l\alpha))(f^1(j\alpha)-f^1(k\alpha))}=0,
\\
\\
\hat R(\alpha\beta)\dfrac{f^1(j\alpha)-f^1(l\alpha)}{f^1(i\alpha)-f^1(l\alpha)}\times
\\
\\
\times\dfrac{\begin{vmatrix}
f^1(i\alpha) & f^1(k\alpha)-f^1(l\alpha)\\
f^2(i\alpha) & f^2(k\alpha)-f^2(l\alpha)\\
\end{vmatrix}
+
\begin{vmatrix}
f^1(k\alpha) & f^1(l\alpha)\\
f^2(k\alpha) & f^2(l\alpha)\\
\end{vmatrix}
}
{
\begin{vmatrix}
f^1(j\alpha) & f^1(k\alpha)-f^1(l\alpha)\\
f^2(j\alpha) & f^2(k\alpha)-f^2(l\alpha)\\
\end{vmatrix}
+
\begin{vmatrix}
f^1(k\alpha) & f^1(l\alpha)\\
f^2(k\alpha) & f^2(l\alpha)\\
\end{vmatrix}}=0;
\end{array}\right\}
$$

     \newpage

     for the metric function (10.8):

     $$
     \left.\begin{array}{c}
     {\bf\hat{R}}(ij)
     \left|\begin{array}{cc}
     f^1(i\alpha) & f^1(i\beta) \\
     f^1(k\alpha) & f^1(k\beta)
     \end{array}\right|\times
     \left|\begin{array}{cc}
     f^2(j\alpha) & f^2(j\beta) \\
     f^2(k\alpha) & f^2 (k\beta)
     \end{array}\right|=0, \\
     \phantom{aaaaa} \\
     \phantom{ab}{\bf \hat{R}}(ik)
     \left|\begin{array}{cc}
     f^1(i\alpha) & f^1(i\beta) \\
     f^1(j\alpha) & f^1(j\beta)
     \end{array}\right|\times
     \left|\begin{array}{cc}
     f^2(k\alpha) & f^2(k\beta) \\
     f^2(j\alpha) & f^2(j\beta)
     \end{array}\right|=0;
     \end{array}\right\}
     $$ \\

     for the metric function (10.9) the equation (10.2) may be obtained by way of
     complexification of the equation

     $$
     \left|\begin{array}{cccc}
     f(i\alpha) & f(i\beta) & f(i\alpha)f(i\beta) & 1 \\
     f(j\alpha) & f(j\beta) & f(j\alpha)f(j\beta) & 1 \\
     f(k\alpha) & f(k\beta) & f(k\alpha)f(k\beta) & 1 \\
     f(l\alpha) & f(l\beta) & f(l\alpha)f(l\beta) & 1
     \end{array}\right|=0,
     $$ \\
     by setting $f=f^1+ef^2$, where $e^2=\varepsilon=0,\pm1$, and separating the
     real and the imaginary part;

     for the metric function (10.10):

     $$
     \left.\begin{array}{c}
     \hat R(\alpha\beta)\dfrac{(f^1(j\alpha)-f^1(l\alpha))(f^1(i\alpha)-f^1(k\alpha))}
     {(f^1(i\alpha)-f^1(l\alpha))(f^1(j\alpha)-f^1(k\alpha))}=0,
     \\
     \\
     \hat R(\alpha\beta)\dfrac{f^1(j\alpha)-f^1(l\alpha)}{f^1(i\alpha)-f^1(l\alpha)}\times
     \\
     \\
     \times\dfrac{\begin{vmatrix}
     f^1(i\alpha) & f^1(k\alpha)-f^1(l\alpha)\\
     f^2(i\alpha) & f^2(k\alpha)-f^2(l\alpha)\\
     \end{vmatrix}
     +
     \begin{vmatrix}
     f^1(k\alpha) & f^1(l\alpha)\\
     f^2(k\alpha) & f^2(l\alpha)\\
     \end{vmatrix}
     }
     {
     \begin{vmatrix}
     f^1(j\alpha) & f^1(k\alpha)-f^1(l\alpha)\\
     f^2(j\alpha) & f^2(k\alpha)-f^2(l\alpha)\\
     \end{vmatrix}
     +
     \begin{vmatrix}
     f^1(k\alpha) & f^1(l\alpha)\\
     f^2(k\alpha) & f^2(l\alpha)\\
     \end{vmatrix}}=0;
     \end{array}\right\}
     $$ \\

\newpage

для метрической функции (10.11):

$$
\left.\begin{array}{c}
{\bf\hat{R}}(ij)
\left|\begin{array}{ccc}
f^1(i\alpha) & f^1(i\beta) & 1 \\
f^1(k\alpha) & f^1(k\beta) & 1 \\
f^1(l\alpha) & f^1(l\beta) & 1
\end{array}\right| \times
\left|\begin{array}{ccc}
f^2(j\alpha) & f^2(j\beta) & 1 \\
f^2(k\alpha) & f^2(k\beta) & 1 \\
f^2(l\alpha) & f^2(l\beta) & 1
\end{array}\right|=0, \\
\phantom{aaaaa} \\
\phantom{a}{\bf \hat{R}}(kl)
\left|\begin{array}{ccc}
f^1(i\alpha) & f^1(i\beta) & 1 \\
f^1(j\alpha) & f^1(j\beta) & 1 \\
f^1(k\alpha) & f^1(k\beta) & 1
\end{array}\right| \times
\left|\begin{array}{ccc}
f^2(i\alpha) & f^2(i\beta) & 1 \\
f^2(j\alpha) & f^2(j\beta) & 1 \\
f^2(l\alpha) & f^2(l\beta) & 1
\end{array}\right|=0;
\end{array}\right\}
$$ \\

для последней метрической функции (10.12), задающей единственную
физическую структуру ранга (5,2):

$$
\left.\begin{array}{c}
\hat R(\alpha\beta)
\dfrac{
\begin{vmatrix}
f^1(i\alpha) & f^2(i\alpha) & 1\\
f^1(k\alpha) & f^2(k\alpha) & 1\\
f^1(l\alpha) & f^2(l\alpha) & 1\\
\end{vmatrix}}
{\begin{vmatrix}
f^1(j\alpha) & f^2(j\alpha) & 1\\
f^1(k\alpha) & f^2(k\alpha) & 1\\
f^1(l\alpha) & f^2(l\alpha) & 1\\
\end{vmatrix}}\times
\dfrac{
\begin{vmatrix}
f^1(j\alpha) & f^2(j\alpha) & 1\\
f^1(k\alpha) & f^2(k\alpha) & 1\\
f^1(m\alpha) & f^2(m\alpha) & 1\\
\end{vmatrix}}
{\begin{vmatrix}
f^1(i\alpha) & f^2(i\alpha) & 1\\
f^1(k\alpha) & f^2(k\alpha) & 1\\
f^1(l\alpha) & f^2(l\alpha) & 1\\
\end{vmatrix}}=0,
\\
\\
\hat R(\alpha\beta)
\dfrac{
\begin{vmatrix}
f^1(i\alpha) & f^2(i\alpha) & 1\\
f^1(k\alpha) & f^2(k\alpha) & 1\\
f^1(l\alpha) & f^2(l\alpha) & 1\\
\end{vmatrix}}
{\begin{vmatrix}
f^1(j\alpha) & f^2(j\alpha) & 1\\
f^1(k\alpha) & f^2(k\alpha) & 1\\
f^1(l\alpha) & f^2(l\alpha) & 1\\
\end{vmatrix}}\times
\dfrac{
\begin{vmatrix}
f^1(j\alpha) & f^2(j\alpha) & 1\\
f^1(l\alpha) & f^2(l\alpha) & 1\\
f^1(m\alpha) & f^2(m\alpha) & 1\\
\end{vmatrix}}
{\begin{vmatrix}
f^1(i\alpha) & f^2(i\alpha) & 1\\
f^1(l\alpha) & f^2(l\alpha) & 1\\
f^1(m\alpha) & f^2(m\alpha) & 1\\
\end{vmatrix}}=0.
\end{array}\right\}
$$

\vspace{5mm}

Оказывается, связь метрической функции (10.1), задающей
двуметрическую физическую структуру ранга $(n+1,2)$ и уравнения (10.2),
выражающего ее феноменологическую симметрию, может стать  более прозрачной,
о чем говорит следующая, доказанная Р.М. Мурадовым [32, \S18], теорема:

     \newpage

     for the metric function (10.11):

     $$
     \left.\begin{array}{c}
     {\bf\hat{R}}(ij)
     \left|\begin{array}{ccc}
     f^1(i\alpha) & f^1(i\beta) & 1 \\
     f^1(k\alpha) & f^1(k\beta) & 1 \\
     f^1(l\alpha) & f^1(l\beta) & 1
     \end{array}\right| \times
     \left|\begin{array}{ccc}
     f^2(j\alpha) & f^2(j\beta) & 1 \\
     f^2(k\alpha) & f^2(k\beta) & 1 \\
     f^2(l\alpha) & f^2(l\beta) & 1
     \end{array}\right|=0, \\
     \phantom{aaaaa} \\
     \phantom{a}{\bf \hat{R}}(kl)
     \left|\begin{array}{ccc}
     f^1(i\alpha) & f^1(i\beta) & 1 \\
     f^1(j\alpha) & f^1(j\beta) & 1 \\
     f^1(k\alpha) & f^1(k\beta) & 1
     \end{array}\right| \times
     \left|\begin{array}{ccc}
     f^2(i\alpha) & f^2(i\beta) & 1 \\
     f^2(j\alpha) & f^2(j\beta) & 1 \\
     f^2(l\alpha) & f^2(l\beta) & 1
     \end{array}\right|=0;
     \end{array}\right\}
     $$ \\

     for the metric function (10.12), the one that gives an only physical
     structure of rank (5,2):

     $$
     \left.\begin{array}{c}
     \hat R(\alpha\beta)
     \dfrac{
     \begin{vmatrix}
     f^1(i\alpha) & f^2(i\alpha) & 1\\
     f^1(k\alpha) & f^2(k\alpha) & 1\\
     f^1(l\alpha) & f^2(l\alpha) & 1\\
     \end{vmatrix}}
     {\begin{vmatrix}
     f^1(j\alpha) & f^2(j\alpha) & 1\\
     f^1(k\alpha) & f^2(k\alpha) & 1\\
     f^1(l\alpha) & f^2(l\alpha) & 1\\
     \end{vmatrix}}\times
     \dfrac{
     \begin{vmatrix}
     f^1(j\alpha) & f^2(j\alpha) & 1\\
     f^1(k\alpha) & f^2(k\alpha) & 1\\
     f^1(m\alpha) & f^2(m\alpha) & 1\\
     \end{vmatrix}}
     {\begin{vmatrix}
     f^1(i\alpha) & f^2(i\alpha) & 1\\
     f^1(k\alpha) & f^2(k\alpha) & 1\\
     f^1(l\alpha) & f^2(l\alpha) & 1\\
     \end{vmatrix}}=0,
     \\
     \\
     \hat R(\alpha\beta)
     \dfrac{
     \begin{vmatrix}
     f^1(i\alpha) & f^2(i\alpha) & 1\\
     f^1(k\alpha) & f^2(k\alpha) & 1\\
     f^1(l\alpha) & f^2(l\alpha) & 1\\
     \end{vmatrix}}
     {\begin{vmatrix}
     f^1(j\alpha) & f^2(j\alpha) & 1\\
     f^1(k\alpha) & f^2(k\alpha) & 1\\
     f^1(l\alpha) & f^2(l\alpha) & 1\\
     \end{vmatrix}}\times
     \dfrac{
     \begin{vmatrix}
     f^1(j\alpha) & f^2(j\alpha) & 1\\
     f^1(l\alpha) & f^2(l\alpha) & 1\\
     f^1(m\alpha) & f^2(m\alpha) & 1\\
     \end{vmatrix}}
     {\begin{vmatrix}
     f^1(i\alpha) & f^2(i\alpha) & 1\\
     f^1(l\alpha) & f^2(l\alpha) & 1\\
     f^1(m\alpha) & f^2(m\alpha) & 1\\
     \end{vmatrix}}=0.
     \end{array}\right\}
     $$ \\

     It turns out that the relation of the metric function (10.1), that gives a dimetric
     physical structure of rank $(n+1,2)$, and the equation (10.2), that expresses its
     phenomenological symmetry, may be made more transparent, which is demonstrated by the
     following theorem proved by R.M. Muradov [32, \S18]:

\newpage

{\bf Теорема 2.} \emph{Если двухкомпонентная метрическая функция}
$$
f=f(x,y,\xi,\eta,\mu,\nu,\ldots)
$$
\emph{задает на 2-мерном и} $2n$-\emph{мерном многообразиях} $\mathfrak{M}$
\emph{и} $\mathfrak{N}$ \emph{двуметрическую физическую структуру
(феноменологически симметричную геометрию двух множеств) ранга}
$(n+1,2)$, \emph{то с точностью до масштабного преобразования
и замены координат в многообразиях она определяет в $R^{2n}$ такую
квазигрупповую операцию с правой единицей,
что правый обратный элемент совпадает с исходным
и в уравнении, выражающем феноменологическую симметрию,
под оператором альтернирования} $\hat{R}{(\alpha\beta)}$ \emph{стоит выражение,
подобное метрической функции}:
$$
\hat R(\alpha\beta)
f(f^1(i\alpha),f^2(i\alpha),f^1(j\alpha),f^2(j\alpha),f^1(k\alpha),f^2(k\alpha),
\ldots)=0;
$$

\emph{для}  $n=1$, \emph{то есть ранга} $(2,2)$:

$$
f^1=x-\xi, \ \ \ f^2=y-\eta,
\eqno(10.3')
$$
$$
f^1=(x-\xi)\eta, \ \ \ f^2=y/\eta;
\eqno(10.4')
$$ \\

\emph{для} $n=2$, \emph{то есть ранга} $(3,2)$:

$$
f^1=\dfrac{
\begin{vmatrix}
x & \xi-\mu\\
\mu & \xi-\mu\\
\end{vmatrix}
-\varepsilon
\begin{vmatrix}
y & \eta-\nu\\
\nu & \eta-\nu\\
\end{vmatrix}}
{(\xi-\mu)^2-\varepsilon(\eta-\nu)^2}, \ \ \
f^2=\dfrac{
\begin{vmatrix}
y & x & 1\\
\eta & \xi & 1\\
\nu & \mu & 1\\
\end{vmatrix}}
{(\xi-\mu)^2-\varepsilon(\eta-\nu)^2},
\eqno(10.5')
$$ \\
\emph{где} $\varepsilon=0,\pm1$;

$$
f^1=\dfrac{x-\mu}{\xi-\mu}, \ \ \
f^2=\dfrac{
\begin{vmatrix}
y & x & 1\\
\eta & \xi & 1\\
\nu & \mu & 1\\
\end{vmatrix}}
{(\xi-\mu)^{c+1}},
\eqno(10.6')
$$ \\
\emph{где} $c\neq 1$;

\newpage

     {\bf Theorem 2.} \emph{If a two-component metric function}
     $$
     f=f(x,y,\xi,\eta,\mu,\nu,\ldots)
     $$
     \emph{gives on a two-dimensional and} a $2n$-\emph{dimensional manifolds} $\mathfrak{M}$
     \emph{and} $\mathfrak{N}$ \emph{a dimetric physical structure(a phenomenologically
     symmetric geometry of two sets) of rank}$(n+1,2)$, \emph{then with an accuracy up to a
     scaling transforma- \ tion and change of coordinates in the manifolds it defines in $R^{2n}$
     such a quasigroup operation with a right identity that the right inverse coincides with the parent
     element, and in the equation expressing the phenomenologi- \ cal symmetry under the operator
     of alternation} $\hat{R}{(\alpha\beta)}$ \emph{there stands an expression similar to the metric
     function itself}:
     $$
     \hat R(\alpha\beta)
     f(f^1(i\alpha),f^2(i\alpha),f^1(j\alpha),f^2(j\alpha),f^1(k\alpha),f^2(k\alpha),
     \ldots)=0;
     $$

     \emph{for}  $n=1$, \emph{i.e. for rank} $(2,2)$:

     $$
     f^1=x-\xi, \ \ \ f^2=y-\eta,
     \eqno(10.3')
     $$
     $$
     f^1=(x-\xi)\eta, \ \ \ f^2=y/\eta;
     \eqno(10.4')
     $$ \\

     \emph{for} $n=2$, \emph{i.e. for rank} $(3,2)$:

     $$
     f^1=\dfrac{
     \begin{vmatrix}
     x & \xi-\mu\\
     \mu & \xi-\mu\\
     \end{vmatrix}
     -\varepsilon
     \begin{vmatrix}
     y & \eta-\nu\\
     \nu & \eta-\nu\\
     \end{vmatrix}}
     {(\xi-\mu)^2-\varepsilon(\eta-\nu)^2}, \ \ \
     f^2=\dfrac{
     \begin{vmatrix}
     y & x & 1\\
     \eta & \xi & 1\\
     \nu & \mu & 1\\
     \end{vmatrix}}
     {(\xi-\mu)^2-\varepsilon(\eta-\nu)^2},
     \eqno(10.5')
     $$ \\
     \emph{where} $\varepsilon=0,\pm1$;
     $$
     f^1=\dfrac{x-\mu}{\xi-\mu}, \ \ \
     f^2=\dfrac{
     \begin{vmatrix}
     y & x & 1\\
     \eta & \xi & 1\\
     \nu & \mu & 1\\
     \end{vmatrix}}
     {(\xi-\mu)^{c+1}},
     \eqno(10.6')
     $$ \\
     \emph{where} $c\neq 1$;

\newpage

 $$
f^1=\dfrac{x-\mu}{\xi-\mu}, \ \ \
f^2=\dfrac{
\begin{vmatrix}
y & x & 1\\
\eta & \xi & 1\\
\nu & \mu & 1\\
\end{vmatrix}-
\ln(\xi-\mu)
\begin{vmatrix}
x^2 & x & 1\\
\xi^2 & \xi & 1\\
\mu^2 & \mu & 1\\
\end{vmatrix}}{(\xi-\mu)^3},
\eqno(10.7')
$$ \\
$$
f^1=\dfrac{x\nu-y\mu}{\xi\nu-\eta\mu}, \ \ \
f^2=\dfrac{x\eta-y\xi}{\xi\nu-\eta\mu};
\eqno(10.8')
$$ \\

\emph{для} $n=3$, \emph{то есть ранга} $(4,2)$:

$$
\left.\begin{array}{c}
f^1=\dfrac{
\begin{vmatrix}
\dfrac{(x-\mu)(\xi-\rho)+\varepsilon(y-\nu)(\eta-\tau)}
{(x-\rho)(\eta-\nu)+(y-\tau)(\xi-\mu)} & 1\\
\dfrac{\varepsilon((x-\mu)(\eta-\tau)+(y-\nu)(\xi-\rho))}
{(x-\rho)(\xi-\mu)+\varepsilon(y-\tau)(\eta-\nu)} & 1\\
\end{vmatrix}}
{
\begin{vmatrix}
\dfrac{(x-\rho)(\xi-\mu)+\varepsilon(y-\tau)(\eta-\nu)}
{(x-\rho)(\eta-\nu)+(y-\tau)(\xi-\mu)} & 1\\
\dfrac{\varepsilon((x-\rho)(\eta-\nu)+(y-\tau)(\xi-\mu))}
{(x-\rho)(\xi-\mu)+\varepsilon(y-\tau)(\eta-\nu)} & 1\\
\end{vmatrix}},
\\
\\
f^2=\dfrac{
\begin{vmatrix}
x^2 & y & x & 1\\
\xi^2 & \eta & \xi & 1\\
\mu^2 & \nu & \mu & 1\\
\rho^2 & \tau & \rho & 1\\
\end{vmatrix}
-\varepsilon
\begin{vmatrix}
y^2 & y & x & 1\\
\eta^2 & \eta & \xi & 1\\
\nu^2 & \nu & \mu & 1\\
\tau^2 & \tau & \rho & 1\\
\end{vmatrix}}
{((x-\rho)^2-\varepsilon(y-\tau)^2)((\xi-\mu)^2-\varepsilon(\eta-\nu)^2)},
\end{array}\right\}
\eqno(10.9')
$$ \\
\emph{где} $\varepsilon=0,\pm1$;

$$
f^1=\dfrac{(x-\mu)(\xi-\rho)}{(x-\rho)(\xi-\mu)}, \ \ \
f^2=\dfrac{
(\mu-\rho)(\xi-\rho)}{(x-\rho)(\xi-\mu)}\cdot
\dfrac{\begin{vmatrix}
y & x & 1\\
\eta & \xi & 1\\
\nu & \mu & 1\\
\end{vmatrix}}
{\begin{vmatrix}
\xi & \eta & 1\\
\mu & \nu & 1\\
\rho & \tau & 1\\
\end{vmatrix}},
\eqno(10.10')
$$

\newpage

     $$
     f^1=\dfrac{x-\mu}{\xi-\mu}, \ \ \
     f^2=\dfrac{
     \begin{vmatrix}
     y & x & 1\\
     \eta & \xi & 1\\
     \nu & \mu & 1\\
     \end{vmatrix}-
     \ln(\xi-\mu)
     \begin{vmatrix}
     x^2 & x & 1\\
     \xi^2 & \xi & 1\\
     \mu^2 & \mu & 1\\
     \end{vmatrix}}{(\xi-\mu)^3},
     \eqno(10.7')
     $$ \\
     $$
     f^1=\dfrac{x\nu-y\mu}{\xi\nu-\eta\mu}, \ \ \
     f^2=\dfrac{x\eta-y\xi}{\xi\nu-\eta\mu};
     \eqno(10.8')
     $$ \\

     \emph{for} $n=3$, \emph{i.e. for rank} $(4,2)$:

     $$
     \left.\begin{array}{c}
     f^1=\dfrac{
     \begin{vmatrix}
     \dfrac{(x-\mu)(\xi-\rho)+\varepsilon(y-\nu)(\eta-\tau)}
     {(x-\rho)(\eta-\nu)+(y-\tau)(\xi-\mu)} & 1\\
     \dfrac{\varepsilon((x-\mu)(\eta-\tau)+(y-\nu)(\xi-\rho))}
     {(x-\rho)(\xi-\mu)+\varepsilon(y-\tau)(\eta-\nu)} & 1\\
     \end{vmatrix}}
     {
     \begin{vmatrix}
     \dfrac{(x-\rho)(\xi-\mu)+\varepsilon(y-\tau)(\eta-\nu)}
     {(x-\rho)(\eta-\nu)+(y-\tau)(\xi-\mu)} & 1\\
     \dfrac{\varepsilon((x-\rho)(\eta-\nu)+(y-\tau)(\xi-\mu))}
     {(x-\rho)(\xi-\mu)+\varepsilon(y-\tau)(\eta-\nu)} & 1\\
     \end{vmatrix}},
     \\
     \\
     f^2=\dfrac{
     \begin{vmatrix}
     x^2 & y & x & 1\\
     \xi^2 & \eta & \xi & 1\\
     \mu^2 & \nu & \mu & 1\\
     \rho^2 & \tau & \rho & 1\\
     \end{vmatrix}
     -\varepsilon
     \begin{vmatrix}
     y^2 & y & x & 1\\
     \eta^2 & \eta & \xi & 1\\
     \nu^2 & \nu & \mu & 1\\
     \tau^2 & \tau & \rho & 1\\
     \end{vmatrix}}
     {((x-\rho)^2-\varepsilon(y-\tau)^2)((\xi-\mu)^2-\varepsilon(\eta-\nu)^2)},
     \end{array}\right\}
     \eqno(10.9')
     $$ \\
     \emph{where} $\varepsilon=0,\pm1$;
     $$
     f^1=\dfrac{(x-\mu)(\xi-\rho)}{(x-\rho)(\xi-\mu)}, \ \ \
     f^2=\dfrac{
     (\mu-\rho)(\xi-\rho)}{(x-\rho)(\xi-\mu)}\cdot
     \dfrac{\begin{vmatrix}
     y & x & 1\\
     \eta & \xi & 1\\
     \nu & \mu & 1\\
     \end{vmatrix}}
     {\begin{vmatrix}
     \xi & \eta & 1\\
     \mu & \nu & 1\\
     \rho & \tau & 1\\
     \end{vmatrix}},
     \eqno(10.10')
     $$

\newpage

$$
f^1=\dfrac{
\begin{vmatrix}
x & y & 1\\
\mu & \nu & 1\\
\rho & \tau & 1\\
\end{vmatrix}}
{\begin{vmatrix}
\xi & \eta & 1\\
\mu & \nu & 1\\
\rho & \tau & 1\\
\end{vmatrix}}, \ \ \
f^2=\dfrac{
\begin{vmatrix}
x & y & 1\\
\xi & \eta & 1\\
\rho & \tau & 1\\
\end{vmatrix}}
{\begin{vmatrix}
\xi & \eta & 1\\
\mu & \nu & 1\\
\rho & \tau & 1\\
\end{vmatrix}};
\eqno(10.11')
$$ \\

\emph{для} $n=4$, \emph{то есть ранга} $(5,2)$:

$$
f^1=\dfrac{
\begin{vmatrix}
x & y & 1\\
\mu & \nu & 1\\
\varphi & \omega & 1\\
\end{vmatrix}}
{\begin{vmatrix}
\xi & \eta & 1\\
\mu & \nu & 1\\
\varphi & \omega & 1\\
\end{vmatrix}}\cdot
\dfrac{
\begin{vmatrix}
\xi & \eta & 1\\
\rho & \tau & 1\\
\varphi & \omega & 1\\
\end{vmatrix}}
{\begin{vmatrix}
x & y & 1\\
\rho & \tau & 1\\
\varphi & \omega & 1\\
\end{vmatrix}},\ \ \
f^2=\dfrac{
\begin{vmatrix}
x & y & 1\\
\xi & \eta & 1\\
\rho & \tau & 1\\
\end{vmatrix}}
{\begin{vmatrix}
\xi & \eta & 1\\
\mu & \nu & 1\\
\rho & \tau & 1\\
\end{vmatrix}}\cdot
\dfrac{
\begin{vmatrix}
\mu & \nu & 1\\
\rho & \tau & 1\\
\varphi & \omega & 1\\
\end{vmatrix}}
{\begin{vmatrix}
x & y & 1\\
\rho & \tau & 1\\
\varphi & \omega & 1\\
\end{vmatrix}},
\eqno(10.12')
$$

\vspace{5mm}

Полная классификация триметрических физических структур (феноменологически
симметричных геометрий двух множеств) построена
только для ранга (2,2). Такая структура, согласно общему определению 1 из \S8,
в котором надо положить $s=3, \ m=1, \ n=1$, задается
трехкомпонентной функцией $f=(f^1,f^2,f^3)$ на 3-мерных
многообразиях $\mathfrak{M}$ и $\mathfrak{N}$. Обозначим локальные координаты
в этих многообразиях через \ $x,y,z$ \ и \ $\xi,\eta,\vartheta$.\ Тогда координатное
представление метрической функции $f$ запишется в следующем виде:
$$
f=f(x,y,z,\xi,\eta,\vartheta),
\eqno(10.13)
$$
причем для конкретной пары $\langle  i\alpha  \rangle$ из области ее определения
$\mathfrak{S}_f$ будем иметь:
$$
f(i\alpha)=f(x_i,y_i,z_i,\xi_\alpha,\eta_\alpha,\vartheta_\alpha),
$$ \\
а ее невырожденность означает отличие от нуля следующих якобианов:

\newpage

     $$
     f^1=\dfrac{
     \begin{vmatrix}
     x & y & 1\\
     \mu & \nu & 1\\
     \rho & \tau & 1\\
     \end{vmatrix}}
     {\begin{vmatrix}
     \xi & \eta & 1\\
     \mu & \nu & 1\\
     \rho & \tau & 1\\
     \end{vmatrix}}, \ \ \
     f^2=\dfrac{
     \begin{vmatrix}
     x & y & 1\\
     \xi & \eta & 1\\
     \rho & \tau & 1\\
     \end{vmatrix}}
     {\begin{vmatrix}
     \xi & \eta & 1\\
     \mu & \nu & 1\\
     \rho & \tau & 1\\
     \end{vmatrix}};
     \eqno(10.11')
     $$ \\

     \emph{for} $n=4$, \emph{i.e. for rank} $(5,2)$:

      $$
     f^1=\dfrac{
     \begin{vmatrix}
     x & y & 1\\
     \mu & \nu & 1\\
     \varphi & \omega & 1\\
     \end{vmatrix}}
     {\begin{vmatrix}
     \xi & \eta & 1\\
     \mu & \nu & 1\\
     \varphi & \omega & 1\\
     \end{vmatrix}}\cdot
     \dfrac{
     \begin{vmatrix}
     \xi & \eta & 1\\
     \rho & \tau & 1\\
     \varphi & \omega & 1\\
     \end{vmatrix}}
     {\begin{vmatrix}
     x & y & 1\\
     \rho & \tau & 1\\
     \varphi & \omega & 1\\
     \end{vmatrix}},\ \ \
     f^2=\dfrac{
     \begin{vmatrix}
     x & y & 1\\
     \xi & \eta & 1\\
     \rho & \tau & 1\\
     \end{vmatrix}}
     {\begin{vmatrix}
     \xi & \eta & 1\\
     \mu & \nu & 1\\
     \rho & \tau & 1\\
     \end{vmatrix}}\cdot
     \dfrac{
     \begin{vmatrix}
     \mu & \nu & 1\\
     \rho & \tau & 1\\
     \varphi & \omega & 1\\
     \end{vmatrix}}
     {\begin{vmatrix}
     x & y & 1\\
     \rho & \tau & 1\\
     \varphi & \omega & 1\\
     \end{vmatrix}},
     \eqno(10.12')
     $$
     \vspace{5mm}

     The complete classification of the trimetric physical structures (phenomeno- \ logically
     symmetric geometries of two sets) has only been built for rank (2,2). Such a structure,
     according to general Definition 1 of \S8 where we must set $s=3, \ m=1, \ n=1$, is
     defined by a three-component function $f=(f^1,f^2,f^3)$ on three-dimensional
     manifolds $\mathfrak{M}$ and $\mathfrak{N}$. We shall designate the local
     coordinates in these manifolds by \ $x,y,z$ \ and \ $\xi,\eta,\vartheta$.\
     Then the coordinate representation of the metric function $f$ is written as follows:
     $$
     f=f(x,y,z,\xi,\eta,\vartheta),
     \eqno(10.13)
     $$
     for any pair $\langle  i\alpha  \rangle$ from the domain $\mathfrak{S}_f$ of it there taking
     place an expression:

    $$
     f(i\alpha)=f(x_i,y_i,z_i,\xi_\alpha,\eta_\alpha,\vartheta_\alpha),
     $$ \\
     and its nondegeneracy meaning the nonzero quality of the two Jacobians:

\newpage

$$
\frac{\partial(f^1(i\alpha),f^2(i\alpha),f^3(i\alpha))}
{\partial(x_i,y_i,z_i)}\neq0, \ \ \
\frac{\partial(f^1(i\alpha),f^2(i\alpha),f^3(i\alpha))}
{\partial(\xi_\alpha,\eta_\alpha,\vartheta_\alpha)}\neq0
\eqno(10.14)
$$ \\
для плотных множеств пар $\langle  i\alpha  \rangle \ \in \mathfrak{M\times N}$.

Феноменологическая симметрия рассматриваемой  триметрической
геометрии двух множеств выражается уравнением
$$
\Phi(f(i\alpha),f(i\beta),f(j\alpha),f(j\beta))=0,
\eqno(10.15)
$$
в котором независимы все три компоненты функции $\Phi=(\Phi_1,\Phi_2,\Phi_3)$.
А это означает, что множество значений функции $F:\mathfrak{S}_F\to R^{12}$,
где $\mathfrak{S}_F\subseteq\mathfrak{M}^3\times \mathfrak{N}^3$ -- естественная
область ее определения, локально принадлежит девятимерной поверхности в $R^{12}$,
задаваемой тремя уравнениями $\Phi=0$.

По теореме 2 из \S8 функция (10.13),
задающая на 3-мерных многообразиях $\mathfrak{M}$ и $\mathfrak{N}$
триметрическую физическую структуру ранга (2,2), допускает трехмерную группу
движений, состоящую из двух действий группы $G^3$  в них.
Выпишем явно действия этой группы в $\mathfrak{M}$:
$$
\left.\begin{array}{c}
x'=\lambda(x,y,z;a^1,a^2,a^3), \\
y'=\sigma(x,y,z;a^1,a^2,a^3), \\
z'=\tau(x,y,z;a^1,a^2,a^3),
\end{array}\right\}
\eqno(10.16)
$$
где $(a^1,a^2,a^3)\in G^3$. Ее действие во втором многообразии
$\mathfrak{N}$ записывается аналогично:
$$
\left.\begin{array}{c}
\xi'=\tilde{\lambda}(\xi,\eta,\vartheta;a^1,a^2,a^3), \\
\eta'=\tilde{\sigma}(\xi,\eta,\vartheta;a^1,a^2,a^3), \\
\vartheta'=\tilde{\tau}(\xi,\eta,\vartheta;a^1,a^2,a^3),
\end{array}\right\}
$$
причем функции $\tilde{\lambda},\tilde{\sigma},\tilde{\tau}$,
задающие это действие, не
обязательно совпадают с функциями $\lambda,\sigma,\tau$ в действии (10.16). Но
если эти действия эквивалентны, то
всегда можно найти такие системы координат в многообразиях $\mathfrak{M}$ и
$\mathfrak{N}$, для которых $\lambda=\tilde{\lambda}, \
\sigma=\tilde{\sigma}, \ \tau=\tilde{\tau}$ при соответствующей перестановке
координат многообразий.

Инвариантность метрической функции (10.13) относительно группы движений
означает ее сохранение согласно уравнению

\newpage

     $$
     \frac{\partial(f^1(i\alpha),f^2(i\alpha),f^3(i\alpha))}
     {\partial(x_i,y_i,z_i)}\neq0, \ \ \
     \frac{\partial(f^1(i\alpha),f^2(i\alpha),f^3(i\alpha))}
     {\partial(\xi_\alpha,\eta_\alpha,\vartheta_\alpha)}\neq0
     \eqno(10.14)
     $$ \\
     for dense sets of pairs $\langle  i\alpha  \rangle \ \in \mathfrak{M\times N}$.

     The phenomenological symmetry of the trimetric geometry of two sets in
     question is expressed by the equation
     $$
     \Phi(f(i\alpha),f(i\beta),f(j\alpha),f(j\beta))=0,
     \eqno(10.15)
     $$
     in which all the three components of the function $\Phi=(\Phi_1,\Phi_2,\Phi_3)$ are
     independent. And that implies that the set of values of the function $F:\mathfrak{S}_F\to R^{12}$,
     where $\mathfrak{S}_F\subseteq\mathfrak{M}^3\times \mathfrak{N}^3$ is its
     natural domain, belongs locally to  the nine-dimensional surface in $R^{12}$ that is
     defined by three equations of $\Phi=0$.

     Under Theorem 2 of \S8 the function (10.13) that gives on the three-dimensional
     manifolds $\mathfrak{M}$ and $\mathfrak{N}$ a trimetric physical structure of rank
     (2,2) allows a three-dimensional group of motions that consists of two actions of the
     group $G^3$ in them. We shall write the actions of that group in $\mathfrak{M}$ in
     the explicit form:

     $$
     \left.\begin{array}{c}
     x'=\lambda(x,y,z;a^1,a^2,a^3), \\
     y'=\sigma(x,y,z;a^1,a^2,a^3), \\
     z'=\tau(x,y,z;a^1,a^2,a^3),
     \end{array}\right\}
     \eqno(10.16)
     $$
     where $(a^1,a^2,a^3)\in G^3$. Its action in the other manifold, manifold $\mathfrak{N}$,
     is written similarly:
     $$
     \left.\begin{array}{c}
     \xi'=\tilde{\lambda}(\xi,\eta,\vartheta;a^1,a^2,a^3), \\
     \eta'=\tilde{\sigma}(\xi,\eta,\vartheta;a^1,a^2,a^3), \\
     \vartheta'=\tilde{\tau}(\xi,\eta,\vartheta;a^1,a^2,a^3),
     \end{array}\right\}
     $$
     the functions $\tilde{\lambda},\tilde{\sigma},\tilde{\tau}$, which define that action, not
     necessarily coinciding with the functions $\lambda,\sigma,\tau$ in the action (10.16). But
     if those actions are equivalent, then we can always find systems of coordinates in the
     manifolds $\mathfrak{M}$ and $\mathfrak{N}$ such that $\lambda=\tilde{\lambda}, \
     \sigma=\tilde{\sigma}, \ \tau=\tilde{\tau}$ in them with the corresponding permutation
     of coordinates in the manifolds.

     The invariance of the metric function (10.13) with respect to the group of motions
     implies its being preserved according to the equation

\newpage

$$
f(x',y',z',\xi',\eta',\vartheta')=f(x,y,z,\xi,\eta,\vartheta),
\eqno(10.17)
$$ \\
которое для  ее компонент $f^1,f^2,f^3$ выполняется тождественно
по координатам \ $x,y,z$ \ и \ $\xi,\eta,\vartheta$ \ точек
многообразий $\mathfrak{M}$ и $\mathfrak{N}$, а также параметрам
$a^1,a^2,a^3$ действующей в них группы $G^3$.

\vspace{5mm}

{\bf Теорема 3.} {\it С точностью до масштабного преобразования
трехкомпонентная метрическая функция $f=(f^1,f^2,f^3)$, задающая
на $3$-мерных многообразиях $\mathfrak{M}$ и $\mathfrak{N}$
триметрическую физическую структуру (феноменологически
симметричную геометрию двух множеств) ранга $(2,2)$, в надлежаще
выбранных в них системах локальных координат \ $x,y,z$ \ и \
$\xi,\eta,\vartheta$ \ определяется следующими одиннадцатью
каноническими выражениями: }
$$
f^1=x+\xi, \ f^2=y+\eta, \ f^3=z+\vartheta;
\eqno(10.18)
$$
$$
$$
$$
f^1=y-\eta, \ f^2=(x+\xi)y+z+\vartheta, \ f^3=(x+\xi)\eta+z+\vartheta;
\eqno(10.19)
$$
$$
$$
$$
\left.\begin{array}{c}
f^1=(x+\xi)^2\exp(2\displaystyle\frac{y+\eta}{x+\xi}), \\
\phantom{aaaaa} \\
f^2=(x+\xi)z, \ f^3=(x+\xi)\vartheta;
\end{array}\right\}
\eqno(10.20)
$$
$$
$$
$$
f^1=\frac{x+\xi}{y+\eta}, \ f^2=(x+\xi)z, \ f^3=(x+\xi)\vartheta;
\eqno(10.21)
$$
$$
$$
$$
f^1=(x+\xi)(y+\eta), \ f^2=(x+\xi)z, \ f^3=(x+\xi)\vartheta;
\eqno(10.22)
$$
$$
$$
$$
f^1=y+\eta, \ f^2=(x+\xi)z, \ f^3=(x+\xi)\vartheta;
\eqno(10.23)
$$
$$
$$
$$
f^1=\frac{(x+\xi)^p}{y+\eta}, \ f^2=(x+\xi)z, \ f^3=(x+\xi)\vartheta;
\eqno(10.24)
$$

\newpage

     $$
     f(x',y',z',\xi',\eta',\vartheta')=f(x,y,z,\xi,\eta,\vartheta),
     \eqno(10.17)
     $$ \\
     which is identically satisfied for its components $f^1,f^2,f^3$ with respect
     to the coordinates \ $x,y,z$ \ and \ $\xi,\eta,\vartheta$ \ of the points of the manifolds
     $\mathfrak{M}$ and $\mathfrak{N}$, as well as to the parameters $a^1,a^2,a^3$ of the
     group $G^3$ acting in them.

     \vspace{5mm}

     {\bf Theorem 3.} {\it With an accuracy up to a scaling transformation
     the three-component metric function $f=(f^1,f^2,f^3)$ that gives on
     $3$-dimensional manifolds $\mathfrak{M}$ and $\mathfrak{N}$ a trimetric
     physical structure (a phenomenologically symmetric geometry of two sets) of
     rank $(2,2)$ is, in a suitably chosen systems of local coordinates \ $x,y,z$ \
     and \ $\xi,\eta,\vartheta$ \, defined by the following canonical
     expressions: }

     $$
     f^1=x+\xi, \ f^2=y+\eta, \ f^3=z+\vartheta;
     \eqno(10.18)
     $$
     $$
     $$
     $$
     f^1=y-\eta, \ f^2=(x+\xi)y+z+\vartheta, \ f^3=(x+\xi)\eta+z+\vartheta;
     \eqno(10.19)
     $$
     $$
     $$
     $$
     \left.\begin{array}{c}
     f^1=(x+\xi)^2\exp(2\displaystyle\frac{y+\eta}{x+\xi}), \\
     \phantom{aaaaa} \\
     f^2=(x+\xi)z, \ f^3=(x+\xi)\vartheta;
     \end{array}\right\}
     \eqno(10.20)
     $$
     $$
     $$
     $$
     f^1=\frac{x+\xi}{y+\eta}, \ f^2=(x+\xi)z, \ f^3=(x+\xi)\vartheta;
     \eqno(10.21)
     $$
     $$
     $$
     $$
     f^1=(x+\xi)(y+\eta), \ f^2=(x+\xi)z, \ f^3=(x+\xi)\vartheta;
     \eqno(10.22)
     $$
     $$
     $$
     $$
     f^1=y+\eta, \ f^2=(x+\xi)z, \ f^3=(x+\xi)\vartheta;
     \eqno(10.23)
     $$
     $$
     $$
     $$
     f^1=\frac{(x+\xi)^p}{y+\eta}, \ f^2=(x+\xi)z, \ f^3=(x+\xi)\vartheta;
     \eqno(10.24)
     $$

\newpage

 $$ \left.\begin{array}{c}
f^1=(x+\xi)^2+(y+\eta)^2, \\
\phantom{aaaaa} \\
f^2=z+\text{arctg}\displaystyle\frac{y+\eta}{x+\xi}, \ \
f^3=\vartheta+\text{arctg}\displaystyle\frac{y+\eta}{x+\xi};
\end{array}\right\}
\eqno(10.25)
$$
$$
$$
$$
\left.\begin{array}{c}
f^1=((x+\xi)^2+(y+\eta)^2)\exp(2\gamma \text{arctg}\displaystyle\frac{y+\eta}{x+\xi}), \\
\phantom{aaaaa} \\
f^2=z+\text{arctg}\displaystyle\frac{y+\eta}{x+\xi}, \ \
f^3=\vartheta+\text{arctg}\displaystyle\frac{y+\eta}{x+\xi};
\end{array}\right\}
\eqno(10.26)
$$
$$
$$
$$
\left.\begin{array}{c}
f^1=\sin y\sin\eta\cos(x+\xi)+\cos y\cos\eta, \\
\phantom{aaaaa} \\
f^2=z+\arcsin\displaystyle\frac{\sin(x+\xi)\sin \eta}{\sqrt{1-(f^1)^2}}, \\
\phantom{aaaaa} \\
f^3=\vartheta+\arcsin\displaystyle\frac{\sin(x+\xi)\sin y}{\sqrt{1-(f^1)^2}};
\end{array}\right\}
\eqno(10.27)
$$
$$
$$
$$
f^1=(x+\xi)y\eta, \ \
f^2=z+\displaystyle\frac{1}{(x+\xi)y^2}, \ \
f^3=\vartheta+\displaystyle\frac{1}{(x+\xi)\eta^2},
\eqno(10.28)
$$
{\it  где $0<|p|<1$ \ и \ $0<\gamma<\infty$.}

\vspace{5mm}

Доказательство этой теоремы можно найти в \S8 монографии [24].

\vspace{5mm}

Перейдем теперь к уравнению (10.15), выражающему феноменологическую симметрию
триметрических физических структур ранга (2,2), задаваемых на трехмерных
многообразиях метрическими функциями (10.18)--(10.28).
Как и в случае двуметрических физических структур (см. теорему 2)
сделаем связь метрической функции
(10.13) и уравнения (10.15) более прозрачной, воспользовавшись следующей
теоремой, доказанной Р.М. Мурадовым [32, \S19]:

\vspace{5mm}

{\bf Теорема 4.} \emph{Если трехкомпонентная метрическая функция}
$$
f=f(x,y,z,\xi,\eta,\vartheta)
$$
\emph{задает на 3-мерных многообразиях} $\mathfrak{M}$ \emph{и}
$\mathfrak{N}$ \emph{триметрическую физическую структуру ранга}
$(2,2)$, \emph{то с точностью до масштабного преоб- }

\newpage

     $$
     \left.\begin{array}{c}
     f^1=(x+\xi)^2+(y+\eta)^2, \\
     \phantom{aaaaa} \\
     f^2=z+\arctan\displaystyle\frac{y+\eta}{x+\xi}, \ \
     f^3=\vartheta+\arctan\displaystyle\frac{y+\eta}{x+\xi};
     \end{array}\right\}
     \eqno(10.25)
     $$
     $$
     $$
     $$
     \left.\begin{array}{c}
     f^1=((x+\xi)^2+(y+\eta)^2)\exp(2\gamma \arctan\displaystyle\frac{y+\eta}{x+\xi}), \\
     \phantom{aaaaa} \\
     f^2=z+\arctan\displaystyle\frac{y+\eta}{x+\xi}, \ \
     f^3=\vartheta+\arctan\displaystyle\frac{y+\eta}{x+\xi};
     \end{array}\right\}
     \eqno(10.26)
     $$
     $$
     $$
     $$
     \left.\begin{array}{c}
     f^1=\sin y\sin\eta\cos(x+\xi)+\cos y\cos\eta, \\
     \phantom{aaaaa} \\
     f^2=z+\arcsin\displaystyle\frac{\sin(x+\xi)\sin \eta}{\sqrt{1-(f^1)^2}}, \\
     \phantom{aaaaa} \\
     f^3=\vartheta+\arcsin\displaystyle\frac{\sin(x+\xi)\sin y}{\sqrt{1-(f^1)^2}};
     \end{array}\right\}
     \eqno(10.27)
     $$
     $$
     $$
     $$
     f^1=(x+\xi)y\eta, \ \
     f^2=z+\displaystyle\frac{1}{(x+\xi)y^2}, \ \
     f^3=\vartheta+\displaystyle\frac{1}{(x+\xi)\eta^2},
     \eqno(10.28)
     $$
    {\it     where $0<|p|<1$ \ and \ $0<\gamma<\infty$.}

     \vspace{5mm}

     The proof of the theorem is in \S8 of the monograph [24].

     \vspace{5mm}

     Let us now take the equation (10.15) that expresses the phenomenological symmetry
     of trimetric physical structures of rank (2,2) defined on three-dimensional manifolds
     by the metric functions (10.18) to (10.28). Just as in the case of the dimetric physical
     structures (see Theorem 2), we shall make clearer the relation of the metric function
     (10.13) and the equation (10.15), for which sake we shall use the following theorem
     proved by R.M. Muradov [32, \S19]:

     \vspace{5mm}

     {\bf Theorem 4.} \emph{If a three-component metric function}
     $$
     f=f(x,y,z,\xi,\eta,\vartheta)
     $$
     \emph{gives on 3-dimensional manifolds} $\mathfrak{M}$ \emph{and} $\mathfrak{N}$
     \emph{a trimetric physical structure of rank} $(2,2)$, \emph {then, with an accuracy up to
     a scaling transformation and }

\newpage

\noindent {\it разования и замены координат в многообразиях  она \
определяет \ в $R^3$ такую квазигрупповую операцию с правым
единичным элементом и правым обратным, совпадающим с исходным, что
в уравнении, выражающем феноменологическую симметрию
соответствующей геометрии двух множеств, под знаком оператора
альтернирования} $\hat{R}{(\alpha\beta)}$ \emph{стоит выражение,
подобное самой метрической функции и получаемое из нее при
подстановках $x\to f^1(i\alpha), \ y\to f^2(i\alpha), \ z\to
f^3(i\alpha), \ \xi\to f^1(j\alpha), \ \eta\to f^2(j\alpha),
\vartheta\to f^3(j\alpha):$}

$$
\hat R(\alpha\beta)
f(f^1(i\alpha),f^2(i\alpha),f^3(i\alpha),
f^1(j\alpha),f^2(j\alpha),f^3(j\alpha))=0;
$$

$$
f^1=x-\xi,\ \ f^2=y-\eta, \ \ f^3=z-\vartheta;
\eqno(10.18')
$$

$$
f^1=x-\xi,\ \ f^2=y-\eta,\ \ f^3=(x-\xi)\eta+z-\vartheta;
\eqno(10.19')
$$

$$
f^1=(x-\xi)\vartheta,\ \ f^2=(y-\eta-(x-\xi)\ln\vartheta)\vartheta,\
f^3=z/\vartheta;
\eqno(10.20')
$$

$$
f^1=(x-\xi)\vartheta,\ \ f^2=(y-\eta)\vartheta,\ \ f^3=z/\vartheta;
\eqno(10.21')
$$

$$
f^1=(x-\xi)\vartheta,\ \ f^2=(y-\eta)/\vartheta,\ \ f^3=z/\vartheta;
\eqno(10.22')
$$

$$
f^1=(x-\xi)\vartheta,\ \ f^2=y-\eta,\ \ f^3=z/\vartheta;
\eqno(10.23')
$$

$$
f^1=(x-\xi)\vartheta,\ \ f^2=(y-\eta)\vartheta^p,\ \ f^3=z/\vartheta;
\eqno(10.24')
$$
{\it где} $0<|p|<1$;

$$
\left.\begin{array}{c}
f^1=(x-\xi)\cos\vartheta-(y-\eta)\sin\vartheta,\\
\phantom{aaaaa} \\
f^2=(x-\xi)\sin\vartheta+(y-\eta)\cos\vartheta,\\
\phantom{aaaaa} \\
f^3=z-\vartheta;
\end{array}\right\}
\eqno(10.25')
$$

\newpage

     \noindent
    {\it  a change of coordinates in the manifolds, it defines in $R^3$
     a quazigroup operation with a right identity and its right inverse that coincides with
     the parent element such that in the equation expressing the phenomenological symmetry of the
     respective geometry of two sets under the sign of the operator of alternation} $\hat{R}{(\alpha\beta)}$
     \emph{there stands an expression similar to the metric function itself and derived from it
     under the substitutions of  $x\to f^1(i\alpha), \ y\to f^2(i\alpha), \ z\to f^3(i\alpha), \ \xi\to
     f^1(j\alpha), \ \eta\to f^2(j\alpha), \vartheta\to f^3(j\alpha):$}

     $$
     \hat R(\alpha\beta)
     f(f^1(i\alpha),f^2(i\alpha),f^3(i\alpha),
     f^1(j\alpha),f^2(j\alpha),f^3(j\alpha))=0;
     $$

     $$
     f^1=x-\xi,\ \ f^2=y-\eta, \ \ f^3=z-\vartheta;
     \eqno(10.18')
     $$

     $$
     f^1=x-\xi,\ \ f^2=y-\eta,\ \ f^3=(x-\xi)\eta+z-\vartheta;
     \eqno(10.19')
     $$

     $$
     f^1=(x-\xi)\vartheta,\ \ f^2=(y-\eta-(x-\xi)\ln\vartheta)\vartheta,\
     f^3=z/\vartheta;
     \eqno(10.20')
     $$

     $$
     f^1=(x-\xi)\vartheta,\ \ f^2=(y-\eta)\vartheta,\ \ f^3=z/\vartheta;
     \eqno(10.21')
     $$

     $$
     f^1=(x-\xi)\vartheta,\ \ f^2=(y-\eta)/\vartheta,\ \ f^3=z/\vartheta;
     \eqno(10.22')
     $$

     $$
     f^1=(x-\xi)\vartheta,\ \ f^2=y-\eta,\ \ f^3=z/\vartheta;
     \eqno(10.23')
     $$

     $$
     f^1=(x-\xi)\vartheta,\ \ f^2=(y-\eta)\vartheta^p,\ \ f^3=z/\vartheta;
     \eqno(10.24')
     $$
     {\it where} $0<|p|<1$;

    $$
     \left.\begin{array}{c}
     f^1=(x-\xi)\cos\vartheta-(y-\eta)\sin\vartheta,\\
     \phantom{aaaaa} \\
     f^2=(x-\xi)\sin\vartheta+(y-\eta)\cos\vartheta,\\
     \phantom{aaaaa} \\
     f^3=z-\vartheta;
     \end{array}\right\}
     \eqno(10.25')
     $$

\newpage

$$
\left.\begin{array}{c}
f^1=\dfrac{(x-\xi)\cos\vartheta-(y-\eta)\sin\vartheta}{\exp(\gamma\vartheta)},\\
\phantom{aaaaa} \\
f^2=\dfrac{(x-\xi)\sin\vartheta+(y-\eta)\cos\vartheta}{\exp(\gamma\vartheta)},\\
\phantom{aaaaa} \\
f^3=z-\vartheta;
\end{array}\right\}
\eqno(10.26')
$$ \\
{\it где} $0<\gamma<\infty$;

$$
\left.\begin{array}{c}
f^1=x\sqrt{1-\xi^2-\eta^2-\vartheta^2}-\xi\sqrt{1-x^2-y^2-z^2}+y\vartheta-z\eta,\\
\phantom{aaaaa} \\
f^2=y\sqrt{1-\xi^2-\eta^2-\vartheta^2}-\eta\sqrt{1-x^2-y^2-z^2}+z\xi-x\vartheta,\\
\phantom{aaaaa} \\
f^1=z\sqrt{1-\xi^2-\eta^2-\vartheta^2}-\vartheta\sqrt{1-x^2-y^2-z^2}+x\eta-y\xi;
\end{array}\right\}
\eqno(10.27')
$$
\\
$$
\left.\begin{array}{c}
f^1=\dfrac{(x-\xi)\eta^2}{1-(x-\xi)\vartheta\eta^2},\\
\phantom{aaaaa} \\
f^2=\dfrac{(1-(x-\xi)\vartheta\eta^2)y}{\eta},\\
\phantom{aaaaa} \\
f^3=z-\dfrac{\vartheta\eta^2}{(1-(x-\xi)\vartheta\eta^2)y^2}.
\end{array}\right\}
\eqno(10.28')
$$

`
\vspace{5mm}

В заключение отметим, что В.А.Кыров в работе [18] одновременно с
классификацией четыреметрических феноменологически симметричных геометрий
ранга 3 на одном множестве, приведенной в конце \S3, построил также классификацию четыреметрических
физических структур (феноменологически симметричных геометрий двух множеств)
ранга (2,2).

\newpage

     $$
     \left.\begin{array}{c}
     f^1=\dfrac{(x-\xi)\cos\vartheta-(y-\eta)\sin\vartheta}{\exp(\gamma\vartheta)},\\
     \phantom{aaaaa} \\
     f^2=\dfrac{(x-\xi)\sin\vartheta+(y-\eta)\cos\vartheta}{\exp(\gamma\vartheta)},\\
     \phantom{aaaaa} \\
     f^3=z-\vartheta;
     \end{array}\right\}
     \eqno(10.26')
     $$ \\
     {\it where} $0<\gamma<\infty$;

     $$
     \left.\begin{array}{c}
     f^1=x\sqrt{1-\xi^2-\eta^2-\vartheta^2}-\xi\sqrt{1-x^2-y^2-z^2}+y\vartheta-z\eta,\\
     \phantom{aaaaa} \\
     f^2=y\sqrt{1-\xi^2-\eta^2-\vartheta^2}-\eta\sqrt{1-x^2-y^2-z^2}+z\xi-x\vartheta,\\
     \phantom{aaaaa} \\
     f^1=z\sqrt{1-\xi^2-\eta^2-\vartheta^2}-\vartheta\sqrt{1-x^2-y^2-z^2}+x\eta-y\xi;
     \end{array}\right\}
     \eqno(10.27')
     $$
     \\
     $$
     \left.\begin{array}{c}
     f^1=\dfrac{(x-\xi)\eta^2}{1-(x-\xi)\vartheta\eta^2},\\
     \phantom{aaaaa} \\
     f^2=\dfrac{(1-(x-\xi)\vartheta\eta^2)y}{\eta},\\
     \phantom{aaaaa} \\
     f^3=z-\dfrac{\vartheta\eta^2}{(1-(x-\xi)\vartheta\eta^2)y^2}.
     \end{array}\right\}
     \eqno(10.28')
     $$
     \vspace{5mm}

     We shall note in conclusion that V.A. Kyrov, in his note [18], simultaneous- \ ly with the
     classification of the four-metric phenomenologically symmetric geometries of rank 3
     on one set, which is at the end of \S3, built the classification of the four-metric physical
     structures (phenomenologically symmetric geometries of two sets) of rank (2,2).

\newpage

\begin{center}
{\bf \large \S11. Групповая симметрия произвольных \\ физических структур}
\end{center}

Бинарные \ физические структуры как \ феноменологически
симметричные геометрии естественно определяются на одном и двух
множествах. Двухточечная функция, задающая такую геометрию,
допускает нетривиальную группу движений с конечным числом
непрерывных параметров, которое было названо степенью групповой
симметрии. При определенном соотношении между рангом физической
структуры, числом существенных параметров группы движений и
размерностью множеств групповая и феноменологическая симметрии
соответствующей геометрии оказываются эквивалентными. Эти
соотношения были заложены в определение физической структуры, ее
феноменологической и групповой симметрий. Естественно возникает
вопрос об их происхождении и обосновании. Кроме того, имеется
много возможностей обобщения и развития понятия физической
структуры, одна из которых была реализована в \S1 и в \S8, когда
двум точкам сопоставлялось несколько действительных чисел. Другая
возможность обобщения, реализованная в \S5, состоит в определении
тернарных физических структур, когда метрическая функция, задающая
структуру, сопоставляет число не паре точек, а трем точкам.
Тернарные физические структуры естественно определяются на одном,
двух и трех множествах и случаи минимального ранга для них были
рассмотрены автором в работах [5], [33], [34]. Однако уже
предварительное их исследование показало, что тернарные структуры,
в отличие от бинарных, не наделяются групповой симметрией, то есть
исходная трехточечная функция не допускает нетривиальной группы
движений. Этот результат в какой-то мере объясняет, почему столь
богаты и содержательны в физическом и математическом смыслах
бинарные структуры, в то время как тернарные существуют только в
случае наименьшего возможного ранга. Поэтому возникает еще вопрос
о внутренних причинах такого различия между бинарными и тернарными
физическими структурами.

Для полного ответа на вопрос о соотношении между рангом физи-
\\ческой \ структуры, степенью групповой симметрии и размерностью

\newpage

     \begin{center}
     {\bf \large \S11. The group symmetry of arbitrary \\ physical structures}
     \end{center}

     Binary physical structures as phenomenologically symmetric geometries are naturally
     defined on one and two sets. A two-point function that defines such a geometry
     allows a nontrivial group of motions with a finite number of continuous parameters
     which number we named the degree of the group symmetry. Under certain
     relationship among the rank of the physical struc- \ ture, the number of the
     essential parameters of the group of motions, and the dimensionality of the sets, the
     group and phenomenological symmetries of the respective geometry turn out to be
     equivalent. Those relationships were built into the definition of the physical structure
     and the phenomenological and group symmetries of it. The question naturally suggests
     itself of their origin and rationale. Besides, a good many potentialities spring up for
     generalization and development of the notion of a physical structure, one of which was
     realized in \S1 and in \S8, when two points were assigned more than one real number.
     Another potentiality of generalization, realized in \S5, consists in defining the ternary
     physical structures, i.e. such that the metric function defining the physical structure does
     not assigns a number to two but to three points. The ternary physical structures are naturally
     defined on one, two, and three sets, and the cases of their minimal rank are treated by
     the author in his notes [5], [33], and [34]. However, as early as on the stage of the
     preliminary investigation it was already found that the ternary structures, in contrast to the
     binary ones, are not endowed with a group symmetry, i.e. the initial three-point function
     does not allow a nontrivial group of motions. Such a result throws into relief the rich
     potentialities of the binary structures, which are in such a contrast with the ternary ones,
     which only exist in case of the smallest possible rank. So a question also suggests itself
     of the intrinsic causes of such a difference.

     To give the final answer to the question  of the relationship of the rank of a physical structure,
     the degree of the group symmetry, and the dimensionality

\newpage

\noindent
 множеств (многообразий), на которых она задана, а также
о причинах различия бинарных и полиарных (в частности, тернарных)
структур, необходимо исходить из более общего определения
физической структуры. Тогда можно будет установить, при каких
соотношениях между основными характеристиками структуры она может
быть наделена групповой симметрией, а при каких нет. Естественно
предположить, что только те структуры содержательны в физическом и
математическом смыслах, группы движений которых нетривиальны. Для
краткости последующего изложения определение произвольных
физических структур будет дано в самом общем виде, достаточном,
однако, для проведения доказательных рассуждений.

Пусть имеются $p$ множеств $\mathfrak{M}_1, \ldots,\mathfrak{M}_p$
произвольной природы, каждое из которых представляет собой гладкое
многообразие размерности $m_1,\ldots,m_p$ соответственно. Пусть
также имеется функция
$$
f:\mathfrak{S}_f\to R^s,
\eqno(11.1)
$$
где $\mathfrak{S}_f\subseteq\mathfrak{M}^{q_1}_1\times\ldots\times
\mathfrak{M}^{q_p}_p$, сопоставляющая каждому кортежу длины $q=q_1+\ldots+
q_p$ из $\mathfrak{S}_f$ некоторую точку из $R^s$, то есть $s$ действительных
чисел. Предполагается, что область определения $\mathfrak{S}_f$ функции $f$
открыта и плотна в $q$-арном прямом произведении $\mathfrak{M}^{q_1}_1\times
\ldots\times\mathfrak{M}^{q_p}_p$ исходных множеств $\mathfrak{M}_1,\ldots,
\mathfrak{M}_p$, а ее координатное представление достаточно гладкое.
Числовой кортеж $(q_1,\ldots,q_p)$ назовем {\it кратностью},
число $q=q_1+\ldots+q_p$ -- {\it арностью}, а
функцию (11.1) {\it метрической}.

Пусть $M_1,\ldots,M_p$ -- произвольные целые числа, такие что $M_1>q_1,\ldots,
M_p>q_p$. Построим отображение
$$
F:\mathfrak{S}_F\to R^{sC^{q_1}_{M_1}\times\ldots\times C^{q_p}_{M_p}},
\eqno(11.2)
$$
где $\mathfrak{S}_F\subseteq\mathfrak{M}^{M_1}_1\times\ldots\times
\mathfrak{M}^{M_p}_p$, сопоставляя каждому кортежу длины $M_1+\ldots+M_p$
из $\mathfrak{S}_F$ упорядоченную по нему
совокупность $sC^{q_1}_{M_1}\times\ldots\times
C^{q_p}_{M_p}$ чисел, соответствующих всем кортежам длины
$q=q_1+\ldots+q_p$, которые являются проекциями исходного кортежа на область
$\mathfrak{S}_f$. Область определения $\mathfrak{S}_F$ функции (11.2) будет,
очевидно, открытой и плотной в $\mathfrak{M}^{M_1}_1\times\ldots\times
\mathfrak{M}^{M_p}_p$. Аналогично построим второе отображение

\newpage

     \noindent
      of the sets (manifolds) where the
     structure is defined, as well as the question of the difference between binary and polyary
     (ternary, in particular) struct- \ ures, it is necessary to proceed from a more general definition
     of a physical structure. Then we shall be able to establish what are the relations among the
     principle characteristics of a structure that make it able to be endowed with a group symmetry,
     and what are those that render it unable to be so endowed. It is natural to suppose that only structures
     whose groups of motions are nontrivial may have real physical and mathematical meaning. For
     the sake of brevity of the further exposition, the definition of arbitrary physical structures will be
     given in the most general form, sufficient, however, for the exposition to be cogent.

     Suppose there are $p$ sets $\mathfrak{M}_1, \ldots,\mathfrak{M}_p$ of arbitrary
     nature, each being a smooth manifold of dimension $m_1,\ldots,m_p$,
     respectively. Suppose there is also a function
     $$
     f:\mathfrak{S}_f\to R^s,
     \eqno(11.1)
     $$
     where $\mathfrak{S}_f\subseteq\mathfrak{M}^{q_1}_1\times\ldots\times
     \mathfrak{M}^{q_p}_p$ that assigns to each cortege of length $q=q_1+\ldots+
     q_p$ from $\mathfrak{S}_f$ a point from $R^s$, i.e. $s$ real numbers. It is supposed
     that the domain $\mathfrak{S}_f$ of the function $f$ is open and dense in $q$-ary direct
     product $\mathfrak{M}^{q_1}_1\times\ldots\times\mathfrak{M}^{q_p}_p$ of the
     assumed sets $\mathfrak{M}_1,\ldots, \mathfrak{M}_p$, and its coordinate representation
     is sufficiently smooth. We shall call the numerical cortege $(q_1,\ldots,q_p)$ a {\it multiplicity },
     the number $q=q_1+\ldots+q_p$ -- an {\it arity}, and the function (11.1) a {\it metric} one.

     Let $M_1,\ldots,M_p$ be arbitrary integers, such that $M_1>q_1,\ldots, M_p>q_p$. We
     shall build a mapping
     $$
     F:\mathfrak{S}_F\to R^{sC^{q_1}_{M_1}\times\ldots\times C^{q_p}_{M_p}},
     \eqno(11.2)
     $$
     where $\mathfrak{S}_F\subseteq\mathfrak{M}^{M_1}_1\times\ldots\times
     \mathfrak{M}^{M_p}_p$, by assigning to every cortege of length $M_1+\ldots+M_p$
     from $\mathfrak{S}_F$ a collection ordered with respect to it of $sC^{q_1}_{M_1}\times\ldots\times
     of C^{q_p}_{M_p}$ numbers corresponding all the corteges of length $q=q_1+\ldots+q_p$
     which are the projections of the initial cortege onto the region $\mathfrak{S}_f$. The
     domain $\mathfrak{S}_F$ of the function (11.2) will obviously be open and dense in
     $\mathfrak{M}^{M_1}_1\times\ldots\times \mathfrak{M}^{M_p}_p$. Quite similarly,
     we shall build another mapping

\newpage

$$
F':\mathfrak{S}_{F'}\to R^{sC^{q_1}_{M'_1}\times\ldots\times
C^{q_p}_{M'_p}},
\eqno(11.2')
$$
где $\mathfrak{S}_{F'}\subseteq\mathfrak{M}^{M'_1}_1\times\ldots\times
\mathfrak{M}^{M'_p}_p$ и $M'_1\geq M_1,\ldots,M'_p\geq M_p$. Проекцию
отображения $F'$ получим, опуская из области его определения
$\mathfrak{S}_{F'}$ некоторую совокупность кортежей длины $q=q_1+\ldots+q_p$,
а из области его значений все числа, соответствующие им по функции (11.1).

{\bf Определение 1.} Будем говорить, что функция (11.1) задает на $m_1,\ldots,
m_p$-мерных многообразиях $\mathfrak{M}_1,\ldots,\mathfrak{M}_p \ q$-арную
полиметрическую физическую структуру ранга $(M_1,\ldots,M_p)$ и кратности
$(q_1,\ldots,q_p)$, если на плотном в $\mathfrak{S}_F$ множестве ранг
отображения $F$ равен $s(C^{q_1}_{M_1}\times\ldots\times C^{q_p}_{M_p}-1)$, а
ранг любой проекции отображения $F'$, область определения которой
не включает в себя какую-либо область
отображения $F$, максимален на плотном в $\mathfrak{S}_{F'}$ множестве.

Другими словами, локально множество значений отображения $F$ является
подмножеством множества нулей системы $s$ независимых функций $\Phi=
(\Phi_1,\ldots,\Phi_s)$ от $sC^{q_1}_{M_1}\times\ldots\times C^{q_p}_{M_p}$
переменных, причем $s$ функциональных связей
$$
\Phi=(\Phi_1,\ldots,\Phi_s)=0
\eqno(11.3)
$$
являются порождающими в том смысле, что любые другие нетривиальные связи будут
только их следствием.

{\bf Определение 2.} Будем говорить, что определенная выше
физическая структура наделена групповой симметрией конечной
степени $r\geq1$, если заданы такие эффективные гладкие локальные
действия некоторой $r$-мерной локальной группы Ли $G^r$ в
многообразиях $\mathfrak{M}_1,\ldots,\mathfrak{M} _p$, что для их
взаимного расширения на прямое произведение
$\mathfrak{M}^{q_1}_1\times\ldots\times\mathfrak{M}^{q_p}_p$
метрическая функция (11.1), задающая структуру, является
\linebreak$q$-точечным инвариантом, где $q=q_1+\ldots+q_p$.

Поскольку преобразуемые многообразия конечномерны, естественно
условие в определении 2, что конечно максимальное
число $r$ непрерывных параметров группы движений, которая  поэтому
является конечномерной локальной группой Ли специальных преобразований
многообразия $\mathfrak{M}^{q_1}_1\times\ldots\times\mathfrak{M}^{q_p}_p$
размерности $q_1m_1+\ldots+q_pm_p$, являющихся  взаимным расширением
преобразований многообразий $\mathfrak{M}_1,\ldots,\mathfrak{M}_p$.

\newpage

     $$
     F':\mathfrak{S}_{F'}\to R^{sC^{q_1}_{M'_1}\times\ldots\times
     C^{q_p}_{M'_p}},
     \eqno(11.2')
     $$
     where $\mathfrak{S}_{F'}\subseteq\mathfrak{M}^{M'_1}_1\times\ldots\times \mathfrak{M}^{M'_p}_p$
     and $M'_1\geq M_1,\ldots,M'_p\geq M_p$. The projection of the mapping $F'$ may be
     obtained by way of dropping a collection of corteges of length $q=q_1+\ldots+q_p$ from
     its domain $\mathfrak{S}_{F'}$, together with dropping all the numbers that correspond
     those corteges with respect to the function (11.1) from its range of values.

     {\bf Definition 1.} We shall say that the function (11.1) gives on $m_1,\ldots, m_p$-
     dimensional manifolds $\mathfrak{M}_1,\ldots,\mathfrak{M}_p a \ q$-ary polymetric
     physical structure of rank $(M_1,\ldots,M_p)$ and of arity $(q_1,\ldots,q_p)$, if on the
     set dense in $\mathfrak{S}_F$ the rank of the mapping $F$ is equal to $s(C^{q_1}_{M_1}\times
     \ldots\times C^{q_p}_{M_p}-1)$, and the rank of any projection of the mapping $F'$
     whose domain does not include any region of the mapping $F$ is maximal on a set dense
     in $\mathfrak{S}_{F'}$.

     In other words, the set of values of the mapping $F$ is locally a subset of the set of
     zeros of the system of $s$ independent functions of $\Phi=(\Phi_1,\ldots,\Phi_s)$ of
     $sC^{q_1}_{M_1}\times\ldots\times C^{q_p}_{M_p}$ variables, $s$ functional relations
     $$
     \Phi=(\Phi_1,\ldots,\Phi_s)=0
     \eqno(11.3)
     $$
     being generating ones, in the sense that any other nontrivial relations will be nothing more
     than their corollaries.

     {\bf Definition 2.} We shall say that a physical structure we have so defined is endowed
     with a group symmetry of finite degree $r\geq1$ if there are also effective smooth local
     actions defined of some $r$-dimensional local Lie group $G^r$ in the manifolds
     $\mathfrak{M}_1,\ldots,\mathfrak{M} _p$, such that for their mutual expansion onto the
     direct product $\mathfrak{M}^{q_1}_1\times\ldots\times\mathfrak{M}^{q_p}_p$
     the metric function (11.1) is a $q$-point invariant, $q$ being $q_1+\ldots+q_p$.

     Since the manifolds being transformed are finite dimensional, the condition in Definition
     2 of the maximal number $r$ of the continuous parameters of the group of motions
     being also finite is quite natural, and the group itself is therefore a finite dimensional
     local Lie group of special transformations of the manifold
     $\mathfrak{M}^{q_1}_1\times\ldots\times\mathfrak{M}^{q_p}_p$
     of dimension $q_1m_1+\ldots+q_pm_p$ which transformations are a mutual expansion
     of the transformations of the mani- \ folds $\mathfrak{M}_1,\ldots,\mathfrak{M}_p$.

\newpage

Запишем систему $sC^{q_1}_{M'_1}\times\ldots\times C^{q_p}_{M'_p}$
уравнений сохранения метрической функции (11.1):
$$
Df|_{F'}=0
\eqno(11.4)
$$
относительно $M'_1m_1+\ldots+M'_pm_p$ дифференциалов координат точек кортежа
из $\mathfrak{S}_{F'}$. Если физическая структура наделена групповой
симметрией, то однородная система (11.4), с одной стороны, должна иметь хотя
бы одно ненулевое решение, а с другой, число ее линейно независимых
ненулевых решений для любых чисел $M'_1,\ldots,M'_p$ не должно превышать
некоторого конечного значения, равного степени групповой симметрии. Число
таких решений равно, как известно, числу неизвестных в системе минус ранг ее
матрицы. Но матрица системы уравнений (11.4)
есть функциональная матрица для системы
функций $f$, соответствующих всем проекциям области определения
$\mathfrak{S}_{F'}$ отображения $(11.2')$ на область определения
$\mathfrak{S}_f$ исходной функции (11.1). Ранг этой матрицы,
очевидно, не изменится, если из системы функций $f|_{F'}$ исключить функции,
зависимые по связи (11.3). Исключив их, получим максимальную проекцию
отображения $(11.2')$, не содержащую в себе отображения (11.2). Обозначим
число функций $f$ в этой максимальной проекции через $N(M'_1,\ldots,M'_p)$.
Тогда по определению 1 физической структуры ранг матрицы системы уравнений
(11.4) будет равен
$$
\min(M'_1m_1+\ldots+M'_pm_p; \ N(M'_1,\ldots,M'_p)).
$$

Если \ найдутся \ такие \ значения \ чисел \ $M'_1,\ldots,M'_p$, \ для \
которых \ $M'_1m_1+\ldots+M'_pm_p\leq N(M'_1,\ldots,M'_p)$,
то ранг матрицы системы уравнений
(11.4) для них будет равен $M'_1m_1+\ldots+M'_pm_p$, то есть числу
неизвестных в ней. Но тогда эта система будет иметь только нулевое решение,
что означает отсутствие нетривиальной групповой симметрии у рассматриваемой
физической структуры. Если же для любых значений $M'_1,\ldots,M'_p$
выполняется строгое неравенство $N(M'_1,\ldots,M'_p)<M'_1m_1+\ldots+M'_pm_p$,
то ранг матрицы системы (11.4) будет равен $N(M'_1,\ldots,M'_p)$ и
число ее линейно независимых ненулевых решений окажется равным
$$
r'=M'_1m_1+\ldots+M'_pm_p-N(M'_1,\ldots,M'_p)>0.
$$

\newpage

     Let us write a system of $sC^{q_1}_{M'_1}\times\ldots\times C^{q_p}_{M'_p}$ equations
     of conservation of the metric function (11.1):
     $$
     Df|_{F'}=0
     \eqno(11.4)
     $$
     with respect to $M'_1m_1+\ldots+M'_pm_p$ differentials of the coordinates of the
     points of a cortege from $\mathfrak{S}_{F'}$. In case the physical structure is endowed
     with a group symmetry, the homogeneous system (11.4) must, on the one hand, have at
     least one nonzero solution, and the number of its linearly independent nonzero solutions
     for any numbers $M'_1,\ldots,M'_p$ must not, on the other hand, be bigger than some
     finite value equal to the degree of the group symmetry. It is known that the number of
     such solutions is equal to the number of the unknowns in the system minus the rank of
     the matrix. But the matrix of the system of equations (11.4) is a functional matrix for the
     system of functions $f$ which correspond all the projections of the domain $\mathfrak{S}_{F'}$
     of the mapping $(11.2')$ onto the domain $\mathfrak{S}_f$ of the assumed function
     (11.1). Obviously, the rank of the matrix will not change if we eliminate from the system
     of functions $f|_{F'}$ the functions dependent with respect to the relation (11.3). Their
     elimination yields the maximal projection of the mapping $(11.2')$, which projection does
     not contain the mapping (11.2). We shall designate the number of functions $f$ in that
     maximal projection by $N(M'_1,\ldots,M'_p)$. Then, under Definition 1 of a physical
     structure the rank of the matrix of the system of equations (11.4) will be equal to
     $$
     \min(M'_1m_1+\ldots+M'_pm_p; \ N(M'_1,\ldots,M'_p)).
     $$

     If values of the numbers \ $M'_1,\ldots,M'_p$, \  may be found such that
     \ $M'_1m_1+\ldots+M'_pm_p\leq N(M'_1,\ldots,M'_p)$, then the rank of the matrix
     of the system of equations  (11.4) for them will be equal to $M'_1m_1+\ldots+M'_pm_p$,
     i.e. to the number of the unknowns in the matrix. But then the system may only have a
     zero solution which means the absence of a nontrivial group symmetry of the physical
     structure in question. And in case for any values of $M'_1,\ldots,M'_p$ a strict inequality
    $N(M'_1,\ldots,M'_p)<M'_1m_1 + \ldots+M'_pm_p$ takes place the rank of the matrix of
     the system (11.4) is equal to $N(M'_1,\ldots,M'_p)$ and the number of its linearly independent
     zero solutions turns out to be
     $$
     r'=M'_1m_1+\ldots+M'_pm_p-N(M'_1,\ldots,M'_p)>0.
     $$

\newpage

Число $r'$, как было отмечено выше, в случае наделения физической
структуры групповой симметрией согласно определению 2, не должно
превышать некоторого конечного значения. Из такого условия
установим, при каких соотношениях между размерностью множеств,
кратностью и рангом физическая структура, задаваемая метрической
функцией (11.1), может быть наделена групповой симметрией и
определить степень $r$ этой симметрии.

Бинарные физические структуры $(q=2)$ могут быть заданы функцией (11.1)
на одном и на двух множествах. В случае одного множества (p=1) они были
подробно рассмотрены в \S5 под наименованием бинарных
феноменологически симметричных
геометрий. Выведенные там соотношения (5.7) и (5.8) устанавливали связь
размерности многообразия с рангом феноменологической симметрии и степенью
групповой симметрии. Собственно физические структуры определялись
первоначально как феноменологически симметричные геометрии двух множеств
(p=2) и к их рассмотрению, дополняющему результаты \S5, мы сейчас перейдем.

Бинарная физическая структура ранга $(M,N)$ и кратности
(1,1), где $M\geq2$ и $N\geq2$, задается метрической функцией (11.1)
на двух многообразиях $\mathfrak{M}$ и $\mathfrak{N}$ размерности $m$ и $n$
соответственно, где $\mathfrak{S}_f\subseteq\mathfrak
{M\times N}$. По определению 1 ранг отображения
$F:\mathfrak{S}_F\to R^{sMN}$, где $\mathfrak{S}_F\subseteq\mathfrak{M}^M\times
\mathfrak{N}^N$, равен $s(MN-1)$. Число зависимых в системе $sM'N'$
функций отображения $F':\mathfrak{S}_{F'}\to R^{sM'N'}$, где $\mathfrak{S}_{F'}
\subseteq\mathfrak{M}^{M'}\times\mathfrak{N}^{N'}$ и $M'\geq M, \ N'\geq N$,
определяется способом
наложения на матрицу пар для кортежа
длины $M'+N'$ из области $\mathfrak{S}_{F'}$ матрицы пар для кортежа длины
$M+N$ из области $\mathfrak{S}_F$, аналогичным описанному в \S5.
Число это находится достаточно просто и
равно $s(M'-M+1)(N'-N+1)$. Поэтому ранг функциональной
матрицы системы функций
$f|_{F'}$, а следовательно, и системы уравнений (11.4) по определению 1
будет равным
$$
\min(M'm+N'n; \ sM'N'-s(M'-M+1)(N'-N+1)).
$$

Если $m<s(N-1)$ или $n<s(M-1)$, то для некоторых значений $M'$ и
$N'$ ранг матрицы системы уравнений (11.4) будет равен числу
неизвестных $M'm+N'n$ в ней и для них она имеет только нулевое ре-

\newpage

     As mentioned above, the number $r'$ must not exceed some finite value if a physical
     structure is endowed with a group symmetry under Definition 2. Let us establish at which
     correlations of the dimensionality of the sets the arity and the rank the physical structure
     defined by the metric function (11.1) thus restricted may have a group symmetry and let
     us find the degree $r$ of that symmetry.

     Binary physical structures $(q=2)$ may be defined by the function (11.1) on one or two sets.
     The case of one set (that of p=1), where we named them binary phenomenologically
     symmetric geometries, was discussed in detail in \S5. The relationships (5.7) and (5.8) we
     deduced in \S5 established the relation between the dimension of the manifold on the one
     hand and the rank of the phenomenological symmetry and the degree of the group symmetry
     on the other. Physical structures were defined as phenomenologi- \ cally symmetric geometries
     of two sets (those with p=2), and we are going to discuss them now, whereby we are going to
     add more detail to what was said in \S5.

     Binary physical structure of rank $(M,N)$ and arity (1,1) where $M\geq2$ and $N\geq2$ is
     defined by the metric function (11.1) on two manifolds, $\mathfrak{M}$ and $\mathfrak{N}$
     of dimensions $m$ and $n$ respectively where $\mathfrak{S}_f\subseteq\mathfrak{M\times N}$.
     Under Definition 1, the rank of the mapping $F:\mathfrak{S}_F\to R^{sMN}$, where
     $\mathfrak{S}_F\subseteq\mathfrak{M}^M\times\mathfrak{N}^N$, is equal to $s(MN-1)$.
     The number of the dependents in the system of $sM'N'$ functions of the mapping
     $F':\mathfrak{S}_{F'}\to R^{sM'N'}$, where $\mathfrak{S}_{F'} \subseteq\mathfrak{M}^{M'}\times\mathfrak{N}^{N'}$ and
     $M'\geq M, \ N'\geq N$, is determined by way of superposing the matrix of pairs for the
     cortege of length $M+N$ from the region $\mathfrak{S}_F$ on the matrix of pairs for the
     cortege of length $M'+N'$ from the region $\mathfrak{S}_{F'}$, in a way similar to that
     described in \S5. It is very easy to find that it is equal to $s(M'-M+1)(N'-N+1)$. So, the rank
     of the system of functions $f|_{F'}$, and of the system of equations (11.4) is according to
     Definition 1 is equal to
     $$
     \min(M'm+N'n; \ sM'N'-s(M'-M+1)(N'-N+1)).
     $$

     If $m<s(N-1)$ or $n<s(M-1)$, then for some values of $M'$ and $N'$ the rank of the
     matrix of the system of equations (11.4) will be equal to the number of the unknowns
     $M'm+N'n$ in it, and for them it only has a zero

    \newpage

\noindent
шение, что и означает отсутствие групповой симметрии у
рассматриваемой физической структуры. Если же $m\geq s(N-1)$ и
$n\geq s(M-1)$, то ранг матрицы системы уравнений (11.4) для любых
значений $M'$ и $N'$ равен $sM'N'-s(M'-M+1)(N'-N+1)$ и она потому
имеет
$$
r'=M'm+N'n-sM'(N-1)-sN'(M-1)+s(M-1)(N-1)
$$
линейно независимых ненулевых решений. При $m>s(N-1)$ или $n>s(M-1)$ с ростом
$M'$ и $N'$ число $r'$ таких решений может стать сколь угодно большим, что
противоречит предположению о конечности степени групповой симметрии.
Поэтому бинарная физическая структура ранга
$(M,N)$, задаваемая на $m$-мерном и $n$-мерном многообразиях $\mathfrak{M}$
и $\mathfrak{N}$ \ метрической функцией (11.1), будет наделена групповой
симметрией только при выполнении следующих соотношений:
$$
{\bf m=s(N-1), \ n=s(M-1).}
\eqno(11.5)
$$

Степень $r$ групповой симметрии, то есть число независимых и существенных
параметров группы движений, равно числу $r'$ линейно независимых ненулевых
решений системы уравнений (11.4) при соотношениях (11.5):
$$
{\bf r=s(M-1)(N-1)=mn/s.}
\eqno(11.6)
$$

Соотношения (11.5) и (11.6) были использованы в основных определениях работы
автора [35] и \S8 настоящей монографии. При сопоставлении надо учесть только
следующую сдвижку обозначений $M\to n+1, \ N\to m+1, \ m\to sm, \ n\to sn$. То
есть на $sm$-мерном и $sn$-мерном многообразиях $\mathfrak{M}$ и
$\mathfrak{N}$ \ метрическая функция (11.1) задает физическую структуру ранга
$(n+1,m+1)$, которая наделена групповой симметрией степени $r=smn$.

Тернарные физические структуры (q=3) могут быть заданы функцией
(11.1) на одном, двух и трех множествах. В случае одного множества
(p=1) они уже были рассмотрены в \S5 под наименованием тернарных
феноменологически симметричных геометрий. Оказалось, что такие
геометрии не могут быть наделены групповой симметрией конечной
степени. Естественно ожидать, что в случае двух множеств (p=2) и
трех множеств (p=3) вывод окажется тем же самым. Покажем это.

\newpage

      \noindent
     solution, which implies the absence of
     any group symmetry of the physical structure in question. And if $m\geq s(N-1)$ and
     $n\geq s(M-1)$, then the rank of the matrix of the system of equations (11.4) for any
     values $M'$ and $N'$ is equal to $sM'N'-s(M'-M+1)(N'-N+1)$, and so it has
     $$
     r'=M'm+N'n-sM'(N-1)-sN'(M-1)+s(M-1)(N-1)
     $$
     linearly independent nonzero solutions. With either $m>s(N-1)$ or $n>s(M-1)$, increasing
     $M'$ and $N'$ may result in the number $r'$ of such solutions becoming arbitrarily large,
     which is in contradiction with the assumption of the finiteness of the degree of the group.
     That's why the binary physical structure of rank $(M,N)$ defined on an $m$-dimensional
     and an $n$-dimensional manifolds $\mathfrak{M}$ and $\mathfrak{N}$ \ by the metric
     function (11.1) is only endowed with a group symmetry if the conditions are satisfied as
     follows:
     $$
     {\bf m=s(N-1), \ n=s(M-1).}
     \eqno(11.5)
     $$

     The degree $r$ of the group symmetry, i.e. the number of independent and essential
     parameters of the group of motions, is equal to the number $r'$ of the linearly independent
     nonzero solutions of the system of equations (11.4) with the relations (11.5):
     $$
     {\bf r=s(M-1)(N-1)=mn/s.}
     \eqno(11.6)
     $$

     The relations (11.5) and (11.6) were used in the principle definitions of the author's
     note [35] and in \S8 of this monograph. It is only necessary to keep in mind some
     small change of designation: $M\to n+1, \ N\to m+1, \ m\to sm, \ n\to sn$. That is,
     on an $sm$-dimensional and an $sn$-dimensional manifolds $\mathfrak{M}$ and
     $\mathfrak{N}$ \ the metric function (11.1) gives a physical structure of rank $(n+1,m+1)$
     endowed with a group symmetry of degree $r=smn$.

     The ternary physical structures (with q=3) may be defined by the function (11.1) on
     one, two, or three sets. The case of one set (p=1) was already discussed in \S5,
     where they were called ternary phenomenologically symmetric geometries, where
     they turned out unable to have a group symmetry of finite degree. It is natural to
     expect that we shall have the same result in case of two (p=2) and three (p=3) sets.
     Let us establish that.

\newpage

Тернарные физические структуры ранга $(M,N)$ и кратности (2,1),
где $M\geq3, \ N\geq2$, задаются на $m$-мерном и $n$-мерном
многообразиях $\mathfrak{M}$ и $\mathfrak{N}$ метрической функцией
(11.1), где
$\mathfrak{S}_f\subseteq\mathfrak{M}^2\times\mathfrak{N}$. По
определению 1 ранг отображения $F:\mathfrak{S}_F\to
R^{sM(M-1)N/2}$, где $\mathfrak{S}_F\subseteq\mathfrak{M}^M
\times\mathfrak{N}^N$, равен $sM(M-1)N/2-s$. Ранг отображения
$F':\mathfrak{S}_{F'}\to R^{sM'(M'-1)N'/2}$, где
$\mathfrak{S}_{F'}\subseteq\mathfrak{M}^{M'}\times
\mathfrak{N}^{N'}$ и $M'\geq M, \ N'\geq N$,  то есть ранг матрицы
системы уравнений (11.4), найдем, налагая матрицу троек для
кортежа длины $M+N$ из области $\mathfrak{S}_F$ на матрицу троек
для кортежа длины $M'+N'$ из области $\mathfrak{S}_{F'}$:
\begin{eqnarray*}
\min(M'm+N'n; \ sM'(M'-1)N'/2- \phantom{aaaa} \\
\phantom{aaaa} \mbox{}-s(M'-M+1)(M'-M+2)(N'-N+1)/2).
\end{eqnarray*}
Поскольку $M>2$ и $N>1$, для достаточно больших значений $M'$ и $N'$ этот ранг
равен $M'm+N'n$, то есть числу неизвестных в системе уравнений (11.4), которая
будет иметь только нулевое решение. Таким образом, тернарные физические
структуры на двух множествах не могут быть наделены групповой симметрией
конечной степени.

Для тернарной физической структуры ранга $(M,N,L)$ и кратности
(1,1,1), где $M\geq2, \ N\geq2, \ L\geq2$, задаваемой на трех
многообразиях $\mathfrak{M,N,L}$ размерности $m,n,l$
соответственно метрической функцией (11.1), где
$\mathfrak{S}_f\subseteq\mathfrak{M\times N\times L}$, ранг
отображения $F:\mathfrak{S}_F\to R^{sMNL}$, где $\mathfrak{S}_F
\subseteq\mathfrak{M}^M\times\mathfrak{N}^N\times\mathfrak{L}^L$,
по определению 1 равен $sMNL-s$. Ранг же отображения
$F':\mathfrak{S}_{F'}\to R^{sM'N'L'}$, где
$\mathfrak{S}_{F'}\subseteq\mathfrak{M}^{M'}\times
\mathfrak{N}^{N'}\times\mathfrak{L}^{L'}$ и $M'\geq M, \ N'\geq N,
\ L'\geq L$, то есть ранг матрицы системы уравнений (11.4),
нетрудно найти методом наложения, использованным выше:
\begin{eqnarray*}
\min(M'm+N'n+L'l; \ sM'N'L'- \phantom{aaa} \\
\phantom{aaa} \mbox{}-s(M'-M+1)(N'-N+1)(L'-L+1)).
\end{eqnarray*}
Поскольку $M>1, \ N>1, \ L>1$, для достаточно больших значений $M',N',L'$ ранг
отображения $F'$ равен $M'm+N'n+L'l$, то есть числу неизвестных в системе
уравнений (11.4), которая для них имеет только нулевое
решение. Таким образом, и на трех множествах тернарные физические структуры не
могут быть наделены групповой симметрией конечной степени.

\newpage

     Ternary physical structures of rank $(M,N)$ and multiplicity (2,1), where
     $M\geq3, \ N\geq2$, are defined on an $m$-dimensional and an $n$-dimensional
     manifolds $\mathfrak{M}$ and $\mathfrak{N}$ by the metric function (11.1), where
     $\mathfrak{S}_f\subseteq\mathfrak{M}^2\times\mathfrak{N}$. Under Definition 1,
     the rank of the mapping $F:\mathfrak{S}_F\to R^{sM(M-1)N/2}$, where$\mathfrak{S}_F\subseteq\mathfrak{M}^M
     \times\mathfrak{N}^N$, is equal to $sM(M-1)N/2-s$. The rank of the mapping
     $F':\mathfrak{S}_{F'}\to R^{sM'(M'-1)N'/2}$,
     where $\mathfrak{S}_{F'}\subseteq\mathfrak{M}^{M'}\times\mathfrak{N}^{N'}$
     and $M'\geq M, \ N'\geq N$, that is the rank of the matrix of
     the system of equations (11.4), may be found by way of superposition of the matrix
     of triples for a cortege of length $M+N$ from the region $\mathfrak{S}_F$ on the
     matrix of triples for a cortege of length $M'+N'$ from the region $\mathfrak{S}_{F'}$:
     \begin{eqnarray*}
     \min(M'm+N'n; \ sM'(M'-1)N'/2- \phantom{aaaa} \\
     \phantom{aaaa} \mbox{}-s(M'-M+1)(M'-M+2)(N'-N+1)/2).
     \end{eqnarray*}
     Since $M>2$ and $N>1$, for sufficiently big values of $M'$ and $N'$ that rank is
     equal to $M'm+N'n$, i.e. to the number of the unknowns in the system of equations
     (11.4), which system may have only a nonzero solution. Thus, the ternary physical
     structures on two sets may not be endowed with a group symmetry of finite degree.

     For the ternary physical structure of rank $(M,N,L)$ and multiplicity (1,1,1), where
     $M\geq2, \ N\geq2, \ L\geq2$, defined on three manifolds $\mathfrak{M,N,L}$ of
     dimensions $m,n,l$ respectively by the metric function (11.1), where $\mathfrak{S}_f
     \subseteq\mathfrak{M\times N\times L}$, the rank of the mapping  $F: \mathfrak{S}_F\to
     R^{sMNL}$, where $\mathfrak{S}_F \subseteq\mathfrak{M}^M\times\mathfrak{N}^N\times\mathfrak{L}^L$,
     under Definition 1, is equal to $sMNL-s$ And the rank of the mapping
     $F':\mathfrak{S}_{F'}\to R^{sM'N'L'}$, where $\mathfrak{S}_{F'}\subseteq\mathfrak{M}^{M'}\times
     \mathfrak{N}^{N'}\times\mathfrak{L}^{L'}$ and $M'\geq M, \ N'\geq N, \ L'\geq L$,
     i.e. the rank of the matrix of equations (11.4), is as easily found by the same method
     of superposition:
     \begin{eqnarray*}
     \min(M'm+N'n+L'l; \ sM'N'L'- \phantom{aaa} \\
     \phantom{aaa} \mbox{}-s(M'-M+1)(N'-N+1)(L'-L+1)).
     \end{eqnarray*}
     Since $M>1, \ N>1, \ L>1$, the rank of the mapping $F'$ for sufficiently great values
     of $M',N',L'$  is equal to $M'm+N'n+L'l$, i.e. to the number of the unknowns in the
     system of equations (11.4), which system for them has only a nonzero solution. Thus,
     ternary physical structures cannot have a group symmetry of finite degree on three
     sets either.

\newpage

Физические структуры ранга $(M_1,\ldots,M_p)$ и кратности
$(q_1,\ldots,q_p)$, где $M_1>q_1,\ldots,M_p>q_p$, задаются на $p$
многообразиях $\mathfrak{M}_1,\ldots,\mathfrak{M}_p$ размерности
$m_1,\ldots,m_p$ соответственно метрической функцией (11.1). Ранг
отображения (11.2) по определению 1 равен
$s(C^{q_1}_{M_1}\times\ldots\times C^{q_p}_{M_p}-1)$, а ранг
отображения $(11.2')$, то есть ранг матрицы системы уравнений
(11.4), можно найти, налагая на матрицу кортежей длины
$q=q_1+\ldots+q_p$ для кортежа длины $M'_1+\ldots+ M'_p$ из
области $\mathfrak{S}_{F'}$ матрицу кортежей той же длины $q$ для
кортежа длины $M_1+\ldots+M_p$ из области $\mathfrak{S}_F$:
\begin{eqnarray*}
\phantom{aa} \min(M'_1m_1+\ldots+M'_pm_p; \
sC^{q_1}_{M'_1}\times\ldots\times C^{q_p}_{M'_p}- \phantom{aaaaabbbbbcc} \\
\phantom{aaaaabbbbbcccccdd} \mbox{}-sC^{q_1}_{M'_1-M_1+q_1}\times\ldots\times
C^{q_p}_{M'_p-M_p+q_p}).  \phantom{aaaaabbbbbcc}
\end{eqnarray*}

 Поскольку бинарные $(q=2)$ и тернарные $(q=3)$
физические структуры выше были рассмотрены, будем предполагать,
что их арность $q>3$. Число неизвестных в системе уравнений (11.4)
от $M'_1,\ldots,M'_p$ зависит линейным образом. В то же время
разность, входящая во вторую половину последнего выражения,
содержит по тем же переменным $M'_1,\ldots,M'_p$ члены порядка
$q-1>2$, которые неограниченно возрастают, так как
$M_1>q_1,\ldots,M_p>q_p$. А это означает, что для достаточно
больших значений $M'_1,\ldots,M'_p$ ранг матрицы системы уравнений
(11.4) станет равен числу неизвестных в ней, и потому она для них
будет иметь только нулевое решение. Таким образом, $q$-арные
физические структуры, задаваемые на множествах
$\mathfrak{M}_1,\ldots,\mathfrak{M}_p$ функцией (11.1), и в случае
$q>3$ не могут быть наделены групповой симметрией конечной
степени.

Окончательный вывод, к которому мы приходим по результатам проведенного выше
исследования выражает следующая

\vspace{5mm}

{\bf Теорема.}  {\it Групповой симметрией конечной степени могут
быть наделены только бинарные физические структуры на одном и двух
множествах, в то время как для $q$-арных физических структур с
$q\geq3$ метрическая функция $(11.1)$ не допускает никаких
нетривиальных локальных движений.}

\vspace{3mm}

Групповая симметрия бинарных физических структур, которым было
уделено основное внимание в монографиях автора [10] и [24],
явля-

\newpage

     The physical structures of rank $(M_1,\ldots,M_p)$ and multiplicity $(q_1,\ldots,q_p)$,
     where $M_1>q_1,\ldots,M_p>q_p$, are defined on $p$ manifolds $\mathfrak{M}_1,\ldots,\mathfrak{M}_p$
     of dimensions $m_1,\ldots,m_p$ respectively by the metric function (11.1). The rank
     of the mapping (11.2) is, under Definition 1, equal to $s(C^{q_1}_{M_1}\times\ldots\times C^{q_p}_{M_p}-1)$,
     and the rank of the mapping $(11.2')$, i.e. the rank of the system of equations (11.4),
     may be found by superposing on the matrix of corteges of length $q=q_1+\ldots+q_p$
     for a cortege of length $M'_1+\ldots+ M'_p$ from the region $\mathfrak{S}_{F'}$ that
     of corteges of the same length $q$ for a cortege of length $M_1+\ldots+M_p$
     from the region $\mathfrak{S}_F$:
     \begin{eqnarray*}
     \phantom{aa} \min(M'_1m_1+\ldots+M'_pm_p; \
     sC^{q_1}_{M'_1}\times\ldots\times C^{q_p}_{M'_p}- \phantom{aaaaabbbbbcc} \\
     \phantom{aaaaabbbbbcccccdd} \mbox{}-sC^{q_1}_{M'_1-M_1+q_1}\times\ldots\times
     C^{q_p}_{M'_p-M_p+q_p}).  \phantom{aaaaabbbbbcc}
     \end{eqnarray*}

     As binary $(q=2)$ and ternary $(q=3)$ physical structures have been investigated, we
     shall assume that their arity $q>3$. The number of the unknowns in the system of
     equations (11.4) linearly depends on $M'_1,\ldots,M'_p$. At the same time, the residual
     that is part of the second half of the latter expression contains, with respect to the same
     variables $M'_1,\ldots,M'_p$, members of order $q-1>2$ whose number becomes
     unrestrictedly large because $M_1>q_1,\ldots,M_p>q_p$. And that means that
     for sufficiently great values of $M'_1,\ldots,M'_p$ the rank of the matrix of the system
     of equations (11.4) will become equal to the number of the unknowns in it, and therefore
     it will only have a zero solution for them. Thus, $q$-ary physical structures defined on
     sets $\mathfrak{M}_1,\ldots,\mathfrak{M}_p$ by the function (11.1) cannot be endowed
    with a group symmetry of finite degree in the case of $q>3$ either.

     The final conclusion we arrive at under the results of the exposition above is
     expressed by the following theorem.

     \vspace{5mm}
     {\bf Theorem.}  {\it It is only binary physical structures on one or two sets that can
     be endowed with a group symmetry of finite degree, while for $q$-ary physical
     structures with $q\geq3$ the metric function $(11.1)$ does not allow any nontrivial
     local motions.}
     \vspace{5mm}

     The group symmetry of binary physical structures, which were the principle object of
     investigation in monographs [10] and [24] be the author, is
     the de-

    \newpage

\noindent
ется определяющей. То есть функция $f:\mathfrak{S}_f\to R^s$, где
$\mathfrak{S}_f\subseteq\mathfrak{M}^2$ или
$\mathfrak{S}_f\subseteq \mathfrak{M\times N}$, будет задавать
физическую структуру в том и только в том случае, если она
допускает нетривиальную конечномерную группу движений. Условие
наделения физических структур групповой симметрией конечной
степени определяет эту степень, устанавливая ее связь с
размерностью множеств и рангом структуры соотношениями (5.7),(5.8)
и (11.5),(11.6). С другой стороны, без предположения о групповой
симметрии конечной степени даже соотношения (5.7) и (11.5),
устанавливающие связь размерности множеств с рангом структуры и не
содержащие степень групповой симметрии, должны в исходных аксиомах
оговариваться дополнительно без достаточно убедительного
обоснования этой связи.

\vspace{3mm}

В заключение отметим, что результаты настоящего параграфа
опубликованы автором в работе [36].

\vspace{15mm}

\begin{center}
{\bf \large \S12. Функциональные уравнения в теории \\ физических структур}
\end{center}

В математическом аппарате теории физических структур (ТФС)
\linebreak функциональные уравнения играют ключевую роль, причем
феноменологическая и групповая симметрии приводят к различному их
типу. В настоящем параграфе будут рассмотрены функциональные
уравнения для физических структур на двух множествах. Напомним,
что для геометрических физических структур на одном множестве
соответствующие функциональные уравнения были рассмотрены в \S6.

Физическая структура ранга $(n+1,m+1)$ задается невырожденной
$s$-компонентной метрической функцией
$$
f=f(x,\xi)=
f(x^1,\ldots,x^{sm}, \ \xi^1,\ldots,\xi^{sn}),
\eqno(12.1)
$$
где $m,n,s\geq1$, на $sm$- и $sn$-мерных многообразиях
$\mathfrak{M}$ и $\mathfrak{N}$. Ее феноменологическая симметрия
означает, что для каждого кортежа \linebreak$\langle ijk\ldots
v,\alpha\beta\gamma\ldots\tau \rangle$ из некоторой окрестности
$U\subset\mathfrak{S}_F\subseteq\mathfrak{M}^{n+1}\times\mathfrak{N}^{m+1}$
плотного в $\mathfrak{S}_F$ множества кортежей выполняется
тождество

\newpage

      \noindent
      termining one. That is,
     the function $f:\mathfrak{S}_f\to R^s$ where either $\mathfrak{S}_f\subseteq\mathfrak{M}^2$
     or $\mathfrak{S}_f\subseteq\mathfrak{M}\times{N}$ defines a physical structure if and
     only if it allows a nontrivial finite dimensional group of motions. The condition of physical
     structures having a group symmetry of finite degree determines that degree, establishing
     whereby its relationship with the dimensionality of the sets and the rank of the structure
     by the relations (5.7), (5.8) and (11.5), (11.6). On the other hand, without the assumption
     of a group symmetry of finite degree even the relations (5.7) and (11.5), establishing the
     relationship of the dimensionality of the sets with the rank of the structure and containing
     no degree of a group symmetry, must be additionally stipulated in the initial axioms without
     any sufficient validation of the presence of any such relation.

    \vspace{3mm}

     We shall note in conclusion that the results of this paragraph were published
     by the author in his note [36].

     \vspace{15mm}

     \begin{center}
     {\bf \large \S12. Functional equations in the theory \\ of physical structures}
     \end{center}

     In the mathematical apparatus of the theory of physical structures (TPS) functional
     equations are of key importance, and it is worth noting that the phenomenological
     and group symmetry yield different types of such equations. In this paragraph we are
     going to discuss the functional equations for physical structures on two sets. We remind
     that for geometric physical structures on one set the respective functional equations
     were discussed in \S6.

     The physical structure of rank $(n+1,m+1)$ is defined by the nondegenerate
     $s$-component metric function
     $$
     f=f(x,\xi)=
     f(x^1,\ldots,x^{sm}, \ \xi^1,\ldots,\xi^{sn}),
     \eqno(12.1)
     $$
     where $m,n,s\geq1$, on an $sm$- and an $sn$-dimensional manifolds $\mathfrak{M}$
     and $\mathfrak{N}$. Its phenomenological symmetry means that for every cortege
     $\langle ijk\ldots v, \\ \alpha\beta\gamma\ldots\tau \rangle$ from some neighbourhood
     $U\subset\mathfrak{S}_F\subseteq\mathfrak{M}^{n+1}\times\mathfrak{N}^{m+1}$
     of a set of corteges dense in $\mathfrak{S}_F$ the identity is satisfied as follows:

\newpage

$$
\Phi(f(i\alpha),f(i\beta),\ldots,f(v\tau))=0,
\eqno(12.2)
$$
в котором функция $\Phi$, как и метрическая функция (12.1), имеет $s$ компонент,
которые независимы.

Тождество (12.2) есть, с одной стороны, аналитическое выражение
принципа феноменологической симметрии, а с другой, представляет
собой основное функциональное уравнение в ТФС. В общем случае,
неизвестными в уравнении (12.2) являются и функция
$f=(f^1,\ldots,f^s)$, задающая физическую структуру, и функция
$\Phi=(\Phi_1,\ldots,\Phi_s)$, выражающая этим уравнением ее
феноменологическую симметрию. Понимаемое именно в таком смысле,
функциональное уравнение (12.2) решено только в некоторых случаях.
Полностью для $s=1$ (см. \S9), частично для $s=2$ (см. \S10) и
$s=3, 4$ (см. \S10). Более простым является тот вариант уравнения
(12.2), когда в нем из двух функций $f$ и $\Phi$ какая-то известна
заранее. Чаще всего известной оказывается метрическая функция $f$,
поэтому сначала рассмотрим именно такой вариант уравнения (12.2).

Простейшие случаи уравнения (12.2) с известной функцией $f$ возникают при
анализе второго закона Ньютона и закона Ома.
Если закон Ньютона записать обычной формулой
$$
a=F/m,
\eqno(12.3)
$$
где $a$ -- функция ускорения тела массы $m$ под действием ускорителя силы $F$,
то для любых двух материальных тел $i,j$ и любых двух ускорителей $\alpha,
\beta$ легко находится связь, в которую входят только четыре измеряемые
в опыте ускорения:
$$
a_{i\alpha}a_{j\beta}-a_{i\beta}a_{j\alpha}=0.
\eqno(12.4)
$$

Феноменологически симметричная форма (12.4) второго закона Ньютона является
решением функционального уравнения
$$
\Phi(a_{i\alpha},a_{i\beta},a_{j\alpha},a_{j\beta})=0,
\eqno(12.5)
$$
в котором функция ускорения $a$ известна по обычной форме (12.3)
этого закона. Функция $a$ задает на множестве материальных тел
$\mathfrak{M}$ и множестве ускорителей $\mathfrak{N}$ физическую
структуру минимального ранга (2,2),

\newpage

     $$
     \Phi(f(i\alpha),f(i\beta),\ldots,f(v\tau))=0,
     \eqno(12.2)
     $$
     in which the function $\Phi$, as well as the metric function (12.1), has $s$ compo- \ nents
     that are independent.

     The identity (12.2) is, on the one hand, an analytical expression of the principle of
     phenomenological symmetry, and on the other, is a functional equation in the TPS.
     In the general case, the unknowns in the equation (12.2) are both the function
     $f=(f^1,\ldots,f^s)$ that gives the physical structure and the function
     $\Phi=(\Phi_1,\ldots,\Phi_s)$ that expresses via that equation the phenomenological
     symmetry of the physical structure. Understood in that sense, the functional equation
     (12.2) has been solved only for some particular cases. It is solved completely for $s=1$
     (see \S9), and partly for $s=2$ (see \S10) and $s=3, 4$ (see \S10). A simpler variant of
     the equation (12.2) is that with one of the functions $f$ and $\Phi$ known. More often it
     is the function $f$ that is known, so that is the case we take into consideration first.

     The simplest cases of the equation (12.2) with the known function $f$ take place in
     the analysis of the Second law of Newton and Ohm's law. In the traditional Newtonian
     law
     $$
     a=F/m,
     \eqno(12.3)
     $$ \\
     where $a$ is a function of the acceleration of a body of weight $m$ under the impact of
     the accelerator $F$, it is easy to find for any two bodies $i,j$ and any two accelerators
     $\alpha, \beta$ the relation that only comprises the four accelerations measured in
     experiment:
     $$
     a_{i\alpha}a_{j\beta}-a_{i\beta}a_{j\alpha}=0.
     \eqno(12.4)
     $$

\vspace{3mm}

     The phenomenologically symmetric form (12.4) of Newton's 2nd law is the solution of
     the functional equation
     $$
     \Phi(a_{i\alpha},a_{i\beta},a_{j\alpha},a_{j\beta})=0,
     \eqno(12.5)
     $$
     in which the function of acceleration $a$ is known from the ordinary formula (12.3) of
     the law. The function $a$ gives on the set of material bodies $\mathfrak{M}$ and
     the set of accelerators $\mathfrak{N}$ a physical structure of minimal rank (2,2), and
     so

    \newpage

 \noindent
поэтому и уравнение (12.5) оказалось достаточно простым. Ясно, что
это уравнение есть частный случай общего уравнения (12.2) для
случая $m=n=s=1$, если в нем положить  $f=a$.

Перейдем теперь к закону Ома:
$$
I=\mathcal{E}/(R+r),
\eqno(12.6)
$$
где $I$ -- функция тока, измеряемая амперметром в замкнутой цепи, содержащей
проводник с сопротивлением $R$ и источник тока с электродвижущей силой
$\mathcal{E}$ и внутренним сопротивлением $r$. Для любых трех проводников $i,
j,k$ и любых двух источников тока $\alpha,\beta$ шесть возможных значений
тока связаны соотношением, в котором отсутствуют их характеристики:
$$
\left|\begin{array}{ccc}
(I_{i\alpha})^{-1} & (I_{i\beta})^{-1} & 1 \\
(I_{j\alpha})^{-1} & (I_{j\beta})^{-1} & 1 \\
(I_{k\alpha})^{-1} & (I_{k\beta})^{-1} & 1
\end{array}\right|=0.
\eqno(12.7)
$$

Соотношение (12.7) задает феноменологически симметричную форму закона Ома,
являющуюся решением функционального уравнения
$$
\Phi(I_{i\alpha},I_{i\beta},I_{j\alpha},I_{j\beta},I_{k\alpha},I_{k\beta})=0,
\eqno(12.8)
$$
в котором функция тока $I$ известна по обычной форме (12.6) этого
закона. Заметим, что функция $I$ задает на множестве проводников
$\mathfrak{M}$ и множестве источников тока $\mathfrak{N}$
физическую структуру ранга (3,2), а уравнение (12.8) получается из
общего уравнения (12.2) при $n=2,m=s=1$, если в нем положить
$f=I$.

В \S9 приведены все возможные выражения (9.2)--(9.7) функции $f$,
задающей на $m$-мерном и $n$-мерном многообразиях $\mathfrak{M}$ и
$\mathfrak{N}$ однометрическую физическую структуру ранга
$(n+1,m+1)$. Для каждого из них, кроме, может быть, выражения
(9.4), решение функционального уравнения (12.2) находится
сравнительно просто и чисто алгебраически: из $(m+1)(n+1)$
значений функции $f$, соответствующих всем парам точек кортежа
$\langle ijk\ldots v,\alpha\beta\gamma\ldots\tau \rangle \
\in\mathfrak{S}_F\subseteq
\mathfrak{M}^{n+1}\times\mathfrak{N}^{m+1}$ исключаются координаты
точек этого кортежа. Получающиеся при этом соотношения
$(9.2')$--$(9.7')$ и являются соответствующими решениями
функционального уравнения (12.2).

\newpage

      \noindent
     the equation (12.5) turns out to be simple enough. It is quite obvious that that equation
     is a special case of the general equation (12.2) for the case of $m=n=s=1$ if we set $f=a$
     in it.

     Now let us take Ohm's law:
     $$
     I=\mathcal{E}/(R+r),
     \eqno(12.6)
     $$
     where $I$ is the function of current measured by ammeter in a closed circuit that
     contains a conductor with resistance $R$ and a source of current that has electromotive
     force $\mathcal{E}$ and internal resistance $r$. For any three conductors $i,j,k$ and any
     two sources of current $\alpha,\beta$, the six possible values of current are tied by the
     relationship that does not contain the characteristics of theirs:
     $$
     \left|\begin{array}{ccc}
     (I_{i\alpha})^{-1} & (I_{i\beta})^{-1} & 1 \\
     (I_{j\alpha})^{-1} & (I_{j\beta})^{-1} & 1 \\
     (I_{k\alpha})^{-1} & (I_{k\beta})^{-1} & 1
     \end{array}\right|=0.
     \eqno(12.7)
     $$

     The relation (12.7) defines the phenomenologically symmetric form of Ohm's law that
     is the solution of the functional equation
     $$
     \Phi(I_{i\alpha},I_{i\beta},I_{j\alpha},I_{j\beta},I_{k\alpha},I_{k\beta})=0,
     \eqno(12.8)
     $$
     in which the function of the current $I$ is known from the ordinary form (12.6) of the
     same law. We shall note that the function $I$ gives on the set of conductors
     $\mathfrak{M}$ and the set of current sources $\mathfrak{N}$ a physical structure of
     rank (3,2), and the equation (12.8) is derived from the general equation (12.2) if
     $n=2,m=s=1$, and if we set $f=I$ in it.

     In \S9 we gave all the possible expressions (9.2)--(9.7) of the function $f$ giving on an
     $m$-dimensional and an $n$-dimensional manifolds $\mathfrak{M}$ and $\mathfrak{N}$
     a unimetric physical structure of rank $(n+1,m+1)$. For each of them, except perhaps for
     the expression (9.4), the solution of the functional equation (12.2) is relatively easy to find
     and it is done in a purely algebraic way at that: by way of excluding from the $(m+1)(n+1)$
     values of the function $f$ that correspond to all the pairs of points of the cortege  $\langle ijk\ldots
     v, \alpha\beta\gamma \ldots \\  \ldots\tau \rangle \ \in\mathfrak{S}_F\subseteq \mathfrak{M}^{n+1}\times
    \mathfrak{N}^{m+1}$ the coordinates of the points of that cortege. The relations $(9.2')$--$(9.7')$
     that that artifice yields are the respective solutions of the functional equation (12.2).

\newpage

В \S10 приведены все возможные выражения (10.3)--(10.12) для
двухкомпонентной функции $f=(f^1,f^2)$, задающей на двумерном и
$2n$ -мерном многообразиях $\mathfrak{M}$ и $\mathfrak{N}$
физическую структуру ранга $(n+1,2)$, где $n\geq1$. Функциональное
уравнение (12.2) для двуметрической физической структуры ранга
$(n+1,2)$ записывается в следующем виде:
$$
\Phi(f(i\alpha),f(i\beta),f(j\alpha),f(j\beta),\ldots,f(v\alpha),f(v\beta))=0,
\eqno(12.9)
$$
причем надо помнить, что функции $f$ и $\Phi$ двухкомпонентные, то есть
$f=(f^1,f^2)$ и $\Phi=(\Phi_1,\Phi_2)$.
Решениями этого уравнения для каждой из функций (10.3)--(10.12) являются
следующие в \S10 за теоремой 1 соотношения. Теорема 2
определяет решения того же функционального уравнения (12.9), но с другими
эквивалентными выражениями $(10.3')$--$(10.12'$) для метрической функции.

В \S10 приведена также полная классификация
(10.18)--(10.28) трехкомпонентных функций
$f=(f^1,f^2,f^3)$, задающих на трехмерных многообразиях $\mathfrak{M}$ и
$\mathfrak{N}$ физическую структуру ранга (2,2). Функциональное уравнение
(12.2) для них записывается в следующем виде:
$$
\Phi(f(i\alpha),f(i\beta),f(j\alpha),f(j\beta))=0,
\eqno(12.10)
$$
которое есть система трех функциональных уравнений, так как в нем
функция $\Phi=(\Phi_1,\Phi_2,\Phi_3)$ тоже трехкомпонентная.
Решения уравнения (12.10) для эквивалентных выражений
$(10.18')$--$(10.28')$ определяются теоремой 4 из \S10.

Другой вариант функционального уравнения (12.2), когда при
известной функции $\Phi=(\Phi_1,\ldots,\Phi_s)$ необходимо найти
метрическую функцию  $f=(f^1,\ldots,f^s)$, задающую на
многообразиях $\mathfrak{M}$ и $\mathfrak{N}$ размерности $sm$ и
$sn$ физическую структуру ранга $(n+1,m+1)$, решается следующим
образом. В прямых произведениях $\mathfrak{M}^n$ и
$\mathfrak{N}^m$ фиксируются кортежи $\langle j_0k_0\ldots v_0 \rangle$ и
$\langle \beta_0\gamma_0\ldots\tau_0 \rangle$ длины $n$ и $m$ соответственно.
Точки этих кортежей выбираются такие, чтобы уравнение (12.2),
записанное для кортежа $\langle ij_0k_0\ldots
v_0,\alpha\beta_0\gamma_0\ldots\tau_0 \rangle \ \in \mathfrak{S}_F$,
могло быть однозначно разрешено относительно $f(i\alpha)$. Затем
удобным образом вводятся локальные координаты $x^1_i,x^2_i,\ldots,
x^{sm}_i$ и $\xi^1_\alpha,\xi^2_\alpha,\ldots,\xi^{sn}_\alpha$,
через которые и выражается метрическая функция (12.1). Если
полученное выражение невырожде-

\newpage

     In \S10 we gave all the possible expressions (10.3)--(10.12) for for the two-component
     function $f=(f^1,f^2)$ that defines on a two-dimensional and a $2n$-dimensional manifolds
     $\mathfrak{M}$ and $\mathfrak{N}$ a physical structure of rank $(n+1,2)$, where $n\geq1$.
     The functional equation (12.2) for the dimetric physical structure of rank $(n+1,2)$ is as
     follows:
     $$
     \Phi(f(i\alpha),f(i\beta),f(j\alpha),f(j\beta),\ldots,f(v\alpha),f(v\beta))=0,
     \eqno(12.9)
     $$
     \vspace{3mm}
     and we are to keep in mind that the functions $f$ and $\Phi$ are two-component ones,
     i.e. $f=(f^1,f^2)$ and $\Phi=(\Phi_1,\Phi_2)$. The solutions of that equation for each of
     the functions (10.3)--(10.12) are the expressions that follow Theorem 1 of \S10. Theorem 2
     determines the solutions of the same functional equation (12.9), but with different
     equivalent expressions $(10.3')$--$(10.12'$) for the metric function.

     \S10 also gives the complete classification (10.18)--(10.28) of the three-component
     functions $f=(f^1,f^2,f^3)$ that give on three-dimensional mani- \ folds $\mathfrak{M}$ and
     $\mathfrak{N}$ a physical structure of rank (2,2). The functional equation (12.2) for them
     is as follows:
     $$
     \Phi(f(i\alpha),f(i\beta),f(j\alpha),f(j\beta))=0,
     \eqno(12.10)
     $$
     \vspace{3mm}
     which is a system of three functional equations, as the function $\Phi=(\Phi_1,\Phi_2,\Phi_3)$
     it contains is also a three-component one. The solutions of the equation (12.10) for the
    equivalent expressions $(10.18')$--$(10.28')$ are determined by Theorem 4 of \S10.

     Another case, namely that of the functional equation (12.2) where the function
     $\Phi=(\Phi_1,\ldots,\Phi_s)$ is known and the metric function $f=(f^1,\ldots,f^s)$
     defining on the manifolds $\mathfrak{M}$ and $\mathfrak{N}$ of dimensions $sm$
     and $sn$ a physical structure of rank $(n+1,m+1)$ is to be found is solved as follows.
     In the direct products $\mathfrak{M}^n$ and $\mathfrak{N}^m$ corteges $\langle j_0k_0\ldots v_0 \rangle$
     and $\langle \beta_0\gamma_0\ldots\tau_0 \rangle$ are fixed of lengths $n$ and $m$ respectively.
     Points of the corteges are selected such that the equation (12.2) written for the
     cortege $\langle ij_0k_0\ldots v_0,\alpha\beta_0\gamma_0\ldots\tau_0 \rangle \ \in \mathfrak{S}_F$
     might be solved uniquely with respect to $f(i\alpha)$. Next, local coordinates $x^1_i,
     x^2_i,\ldots,x^{sm}_i$ and $\xi^1_\alpha,\xi^2_\alpha,\ldots,\xi^{sn}_\alpha$, are introduced
     in a suitable way, and it is via them that the metric function (12.1) is expressed. \
     If the expression

    \newpage

\noindent
 но и при его подстановке в исходное уравнение (12.2)
последнее превращается в тождество по всем координатам точек
кортежа $\langle ijk\ldots v,\alpha\beta\gamma \\ \ldots \tau
\rangle$, то найденная таким образом метрическая функция
действительно задает физическую структуру ранга $(n+1,m+1)$.

Проиллюстрируем описанный выше метод решения функционального уравнения (12.2)
на примере однометрической и двуметрической физических структур минимального
ранга (2,2).

Запишем феноменологически симметричное соотношение
\vspace{2mm}
$$
f(i\alpha)-f(i\beta)-f(j\alpha)+f(j\beta)=0
$$
\vspace{3mm}
 для четверки $\langle ij_0,\alpha\beta_0 \rangle$:
\vspace{2mm}
$$
f(i\alpha)-f(i\beta_0)-f(j_0\alpha)+f(j_0\beta_0)=0,
$$
\vspace{3mm}
 после чего разрешим его относительно $f(i\alpha)$:
\vspace{2mm}
$$
f(i\alpha)=f(i\beta_0)+f(j_0\alpha)-f(j_0\beta_0).
$$
\vspace{3mm}
 Вводя координаты $x_i=f(i\beta_0)-f(j_0\beta_0)/2$ \
и \ $\xi_\alpha=f(j_0\alpha)-f(j_0\beta_0)/2$, где
$f(j_0\beta_0)$, очевидно, константа, получаем координатное
представление метрической функции (9.2), которое при подстановке в
исходное соотношение обращает его в тождество, подтверждая тем
самым, что эта функция задает на одномерных многообразиях
$\mathfrak{M}$ и $\mathfrak{N}$ однометрическую физическую
структуру ранга (2,2).

Перейдем к более сложному
феноменологически симметричному соотношению
$$
\left.\begin{array}{c}
$$
\left|
\begin{array}{cc}
f^1(i\alpha)-f^1(i\beta) & f^1(i\alpha)f^2(j\alpha) \\
f^1(j\alpha)-f^1(j\beta) & f^1(j\alpha)f^2(i\alpha)
\end{array}
\right|=0,
$$ \\
\phantom{aaaaa} \\
$$
\phantom{ab} \left|
\begin{array}{cc}
f^2(i\alpha)-f^2(j\alpha) & f^2(i\alpha)f^1(i\beta) \\
f^2(i\beta)-f^2(j\beta) &  f^2(i\beta)f^1(i\alpha)
\end{array}
\right|=0,
$$
\end{array}\right\}
$$ \\
которое раcсмотрим как систему двух функциональных уравнений и
сначала запишем их для четверки $\langle j_0i,\beta_0\alpha \rangle$:

\newpage

      \noindent
     obtained is nondegenerate and its substitution into the initial equation
     (12.2) yields an identity with respect to all the coordinates of the points of the cortege
     $\langle ijk\ldots v,\alpha\beta\gamma\ldots\tau \rangle$, then the metric function found does give
     a physical structure of rank $(n+1,m+1)$.

     We shall illustrate the method we have described of solving the functional equation (12.2)
     with an example of a unimetric and dimetric physical struc- \ tures of minimal rank (2,2).

    We shall write the phenomenologically symmetric relation
     \vspace{2mm}
     $$
     f(i\alpha)-f(i\beta)-f(j\alpha)+f(j\beta)=0
     $$
     \vspace{3mm}
     for a quadruple $\langle ij_0,\alpha\beta_0 \rangle$:
     \vspace{2mm}
     $$
     f(i\alpha)-f(i\beta_0)-f(j_0\alpha)+f(j_0\beta_0)=0,
     $$
     \vspace{3mm}
     and then solve it with respect to $f(i\alpha)$:
     \vspace{2mm}
     $$
     f(i\alpha)=f(i\beta_0)+f(j_0\alpha)-f(j_0\beta_0).
     $$
     \vspace{3mm}
     By introducing coordinates $x_i=f(i\beta_0)-f(j_0\beta_0)/2$ \ and \
     $\xi_\alpha=f(j_0\alpha)-f(j_0\beta_0)/2$, where $f(j_0\beta_0)$ is, obviously, a
     constant, we get the coordinate representation of the metric function (9.2) that,
     substituted into the initial relation, turns it into an identity confirming whereby that that
     function does give on one-dimensional manifolds $\mathfrak{M}$ and $\mathfrak{N}$
     a unimetric physical structure of rank (2,2).

     Let us take a more complicated phenomenologically symmetric relation
     $$
     \left.\begin{array}{c}
     $$
     \left|
     \begin{array}{cc}
     f^1(i\alpha)-f^1(i\beta) & f^1(i\alpha)f^2(j\alpha) \\
     f^1(j\alpha)-f^1(j\beta) & f^1(j\alpha)f^2(i\alpha)
     \end{array}
     \right|=0,
     $$ \\
     \phantom{aaaaa} \\
     $$
     \phantom{ab} \left|
     \begin{array}{cc}
     f^2(i\alpha)-f^2(j\alpha) & f^2(i\alpha)f^1(i\beta) \\
     f^2(i\beta)-f^2(j\beta) &  f^2(i\beta)f^1(i\alpha)
     \end{array}
     \right|=0,
     $$
     \end{array}\right\}
     $$
     which we shall consider as a system of two functional equations, and shall write them
     first for the quadruple $\langle j_0i,\beta_0\alpha \rangle$:

\newpage

$$
\left.\begin{array}{c}
\left|\begin{array}{cc}
f^1(j_0\beta_0)-f^1(j_0\alpha) & f^1(j_0\beta_0)f^2(i\beta_0) \\
f^1(i\beta_0)-f^1(i\alpha) & f^1(i\beta_0)f^2(j_0\beta_0)
\end{array}\right|=0, \\
\phantom{aaaaaabbbbbccccc} \\
\left|\begin{array}{cc}
f^2(j_0\beta_0)-f^2(i\beta_0) & f^2(j_0\beta_0)f^1(j_0\alpha) \\
f^2(j_0\alpha)-f^2(i\alpha) & f^2(j_0\alpha)f^1(j_0\beta_0)
\end{array}\right|=0,
\end{array}\right\}
$$ \\
после чего разрешим относительно $f(i\alpha)=(f^1(i\alpha),f^2(i\alpha))$:

$$
\left.\begin{array}{c}
f^1(i\alpha)=[f^2(i\beta_0)f^1(j_0\beta_0)+f^1(j_0\alpha)f^2(j_0\beta_0)- \\
\phantom{aaaaabbbbb}
-f^1(j_0\beta_0)f^2(j_0\beta_0)]f^1(i\beta_0)/f^2(i\beta_0)f^1(j_0\beta_0), \\
\phantom{aaaaabbbbbccccc} \\
f^2(i\alpha)=[f^2(i\beta_0)f^1(j_0\beta_0)+f^1(j_0\alpha)f^2(j_0\beta_0)- \\
\phantom{aaaaabbbbbc}
-f^1(j_0\beta_0)f^2(j_0\beta_0)]f^2(j_0\alpha)/f^1(j_0\alpha)f^2(j_0\beta_0).
\end{array}\right\}
$$ \\
Вводя удобные координаты
$x_i=f^2(i\beta_0)f^1(j_0\beta_0), \
y_i=f^1(i\beta_0)/f^2(i\beta_0)\times$ \linebreak $\times f^1(j_0\beta_0)$ \ и \
$\xi_\alpha=(f^1(j_0\alpha)-f^1(j_0\beta_0))f^2(j_0\beta_0), \
\eta_\alpha=f^2(j_0\alpha)/f^1(j_0\alpha)\times$ \linebreak $\times f^2(j_0\beta_0)$ в двумерных
многообразиях $\mathfrak{M}$ и $\mathfrak{N}$, получаем координатное
представление (10.4) метрической функции,
задающей на этих многообразиях двуметрическую физическую
структуру ранга (2,2), так как ее подстановка в исходное соотношение
обращает его в тождество.

Все другие функциональные уравнения (12.2) с известной функцией $\Phi$
решаются аналогично, хотя для некоторых из них возникают значительные
чисто технические трудности его разрешения относительно переменной
$f(i\alpha)$ и рационального введения систем локальных
координат в многообразиях $\mathfrak{M}$ и $\mathfrak{N}$.

Описанный выше метод решения функционального уравнения (12.2) может быть
применен к любой наперед заданной функции $\Phi$. Но если равенство $\Phi=0$
не определяет феноменологически симметричное соотношение для физической
структуры, то полученное координатное представление
функции $f(i\alpha)$ при подстановке в уравнение (12.2) не обращает его в
тождество по всем координатам точек кортежа \linebreak$\langle ijk\ldots v,\alpha\beta\gamma
\ldots\tau \rangle$. Приведем интересный в этом смысле пример.

Обобщая феноменологически симметричное соотношение $(9.4')$ для однометрической
физической структуры ранга (4,2), естественно было

\newpage

     $$
     \left.\begin{array}{c}
     \left|\begin{array}{cc}
     f^1(j_0\beta_0)-f^1(j_0\alpha) & f^1(j_0\beta_0)f^2(i\beta_0) \\
     f^1(i\beta_0)-f^1(i\alpha) & f^1(i\beta_0)f^2(j_0\beta_0)
     \end{array}\right|=0, \\
     \phantom{aaaaaabbbbbccccc} \\
     \left|\begin{array}{cc}

     f^2(j_0\beta_0)-f^2(i\beta_0) & f^2(j_0\beta_0)f^1(j_0\alpha) \\
     f^2(j_0\alpha)-f^2(i\alpha) & f^2(j_0\alpha)f^1(j_0\beta_0)
     \end{array}\right|=0,
     \end{array}\right\}
     $$ \\
     solving them whereafter with respect to $f(i\alpha)=(f^1(i\alpha),f^2(i\alpha))$:

     $$
     \left.\begin{array}{c}

     f^1(i\alpha)=[f^2(i\beta_0)f^1(j_0\beta_0)+f^1(j_0\alpha)f^2(j_0\beta_0)- \\
     \phantom{aaaaabbbbb}
     -f^1(j_0\beta_0)f^2(j_0\beta_0)]f^1(i\beta_0)/f^2(i\beta_0)f^1(j_0\beta_0), \\
     \phantom{aaaaabbbbbccccc} \\
     f^2(i\alpha)=[f^2(i\beta_0)f^1(j_0\beta_0)+f^1(j_0\alpha)f^2(j_0\beta_0)- \\
     \phantom{aaaaabbbbbc}
     -f^1(j_0\beta_0)f^2(j_0\beta_0)]f^2(j_0\alpha)/f^1(j_0\alpha)f^2(j_0\beta_0).
     \end{array}\right\}
     $$ \\
     By way of introducing suitable coordinates
     $x_i=f^2(i\beta_0)f^1(j_0\beta_0), \
     y_i=f^1(i\beta_0)/f^2(i\beta_0)\times f^1(j_0\beta_0)$ \ and \
     $\xi_\alpha=(f^1(j_0\alpha)-f^1(j_0\beta_0))f^2(j_0\beta_0), \
     \eta_\alpha=f^2(j_0\alpha)/f^1(j_0\alpha)\times f^2(j_0\beta_0)$ in
     two-dimensional manifolds $\mathfrak{M}$ and $\mathfrak{N}$, we get the coordinate
     representation (10.4) for the metric function that gives on these manifolds a dimetric
     physical structure of rank (2,2), as the substitution of it into the initial relation yields an
     identity.

     All the other functional equations (12.2) with the known function $\Phi$ are solved
     similarly, though for some of them some difficulties, of purely technical nature, arise
     of solving them with respect to the variable $f(i\alpha)$ and of finding a rational way
     of introducing systems of local coordinates in the manifolds $\mathfrak{M}$ and
     $\mathfrak{N}$.

     The described method of solution of the functional equation (12.2) may be used with
     any preassigned function $\Phi$. However, unless the equality $\Phi=0$ defines a
     phenomenologically symmetric relation for the physical structure, the coordinate
     representation obtained of the function $f(i\alpha)$, being substituted into
     the equation (12.2), does not yield an identity with respect to all the points of the
     cortege $\langle ijk\ldots v,\alpha\beta\gamma \ldots\tau \rangle$. We shall give an interesting
     example.

     While generalizing the phenomenologically symmetric relationship $(9.4')$ for the
     unimetric physical structure of rank (4,2), it was natural to
     suppose

    \newpage

\noindent
предположить, что феноменологически симметричное соотношение для
физической структуры ранга (5,3) должно записываться в виде
равенства нулю следующего определителя пятого порядка:
$$
\left|\begin{array}{ccccc}
1 & f(i\alpha) & f(i\beta) & f(i\gamma) & f(i\alpha)f(i\beta)f(i\gamma) \\
1 & f(j\alpha) & f(j\beta) & f(j\gamma) & f(j\alpha)f(j\beta)f(j\gamma) \\
1 & f(k\alpha) & f(k\beta) & f(k\gamma) & f(k\alpha)f(k\beta)f(k\gamma) \\
1 & f(l\alpha) & f(l\beta) & f(l\gamma) & f(l\alpha)f(l\beta)f(l\gamma) \\
1 & f(q\alpha) & f(q\beta) & f(q\gamma) & f(q\alpha)f(q\beta)f(q\gamma)
\end{array}\right|=0.
$$

Запишем это соотношение для кортежа $\langle ij_0k_0l_0q_0,
\alpha\beta_0\gamma_0 \rangle$, разрешим его относительно
переменной $f(i\alpha)$ и введем удобным образом координаты
$x_i,y_i$ в двумерном многообразии $\mathfrak{M}$ и
$\xi_\alpha,\eta_\alpha,\mu_\alpha$, $\nu_\alpha$ в четырехмерном
многообразии $\mathfrak{N}$. В результате для функции $f(i\alpha)$
получаем следующее локальное координатное представление:
$$
f(i\alpha)=(x_i\xi_\alpha+y_i\eta_\alpha+\mu_\alpha)/(x_iy_i+\nu_\alpha).
$$
Однако подстановка найденной функции в исходное соотношение не обращает его в
тождество, что было установлено с помощью компьютерной программы "Maple".
Этот результат можно было предвидеть
заранее, так как согласно приведенной в начале \S9 классификации
однометрическая физическая структура ранга (5,3) не существует.

Описанная в \S8 эквивалентность феноменологической и групповой симметрий,
согласно которой функция $f$, задающая на двух множествах $\mathfrak{M}$ и
$\mathfrak{N}$ физическую структуру, является
двухточечным инвариантом некоторой группы их преобразований,
приводит к функциональному уравнению (8.6),
принципиально отличному от рассмотренного выше уравнения (12.2), хотя решения
этих уравнений для функции $f$ должны совпадать.

Ниже удобно будет в уравнении (8.6) опустить явное указание точек $i$ и
$\alpha$ многообразий $\mathfrak{M}$ и $\mathfrak{N}$, записывая его в
следующем виде:
$$
f(\lambda(x),\sigma(\xi))=f(x,\xi),
\eqno(12.11)
$$
где $\lambda:\mathfrak{M\to M}, \ \sigma:\mathfrak{N\to N}$ -- локальные
обратимые преобразования  многообразий, а
$x=(x^1,\ldots$, $x^{sm})$, \ $\xi=(\xi^1,\ldots,\xi^{sn})$ --
локальные координаты в них.

\newpage

     \noindent
     that the
     phenomenologically symmetric relationship for the physical structure of rank (5,3)
     must be written as the equality to zero of the following determinant of 5th order:
     $$
     \left|\begin{array}{ccccc}
     1 & f(i\alpha) & f(i\beta) & f(i\gamma) & f(i\alpha)f(i\beta)f(i\gamma) \\
     1 & f(j\alpha) & f(j\beta) & f(j\gamma) & f(j\alpha)f(j\beta)f(j\gamma) \\
     1 & f(k\alpha) & f(k\beta) & f(k\gamma) & f(k\alpha)f(k\beta)f(k\gamma) \\
     1 & f(l\alpha) & f(l\beta) & f(l\gamma) & f(l\alpha)f(l\beta)f(l\gamma) \\
     1 & f(q\alpha) & f(q\beta) & f(q\gamma) & f(q\alpha)f(q\beta)f(q\gamma)
     \end{array}\right|=0.
     $$

     We shall write that relationship for the cortege $\langle ij_0k_0l_0q_0, \alpha\beta_0\gamma_0 \rangle$,
     solve it with respect to the variable $f(i\alpha)$, and introduce coordinates $x_i,y_i$
     in a suitable way in a two-dimensional manifold $\mathfrak{M}$ and coordinates
     $\xi_\alpha,\eta_\alpha,\mu_\alpha$, $\nu_\alpha$ in a four-dimensional manifold
     $\mathfrak{N}$. As result, we have for the function $f(i\alpha)$ the following
     local coordinate representation:
     $$
     f(i\alpha)=(x_i\xi_\alpha+y_i\eta_\alpha+\mu_\alpha)/(x_iy_i+\nu_\alpha).
     $$
     But the substitution of the function obtained into the initial relationship does not
     turn it into an identity, which fact we established by using the "Maple" computing
     package. However, that result could have been anticipated in advance, as, according
     to the classification described at the beginning of \S9, no unimetric physical structure
     of rank (5,3) exists.

     The equivalence of the phenomenological and group symmetries described in \S8,
     under which the function $f$ defining on two sets, $\mathfrak{M}$ and $\mathfrak{N}$,
     a physical structure is a two-point invariant of some group of their transformations,
     yields the functional equation (8.6), which is basically different from the equation (12.2)
     we have discussed, though the solutions of both equations for the function $f$ must
     coincide.

     In further discussion, it will be suitable to drop in the equation (8.6) the explicit
     inclusion of the points $i$ and $\alpha$ of the manifolds $\mathfrak{M}$ and
     $\mathfrak{N}$, writing it as follows:
     $$
     f(\lambda(x),\sigma(\xi))=f(x,\xi),
     \eqno(12.11)
     $$
     where $\lambda:\mathfrak{M\to M}, \ \sigma:\mathfrak{N\to N}$ are locally
     invertible transformations of the manifolds, and $x=(x^1,\ldots$, $x^{sm})$,
     and $\xi=(\xi^1,\ldots,\xi^{sn})$ are local coordinates in them.

    \newpage

В общем случае функциональное уравнение (12.11), как и уравнение
(12.2), допускает два толкования. В первом случае неизвестны как
метрическая функция $f$, так и функции $\lambda,\sigma$,
определяющие преобразования многообразий $\mathfrak{M,N}$. Тогда,
зная по итоговой теореме 3 из \S8 размерность группы движений
геометрии двух множеств, задаваемой функцией $f$, \ сначала
проводим полную с точностью до эквивалентности (замены локальных
координат) классификацию $smn$-мерных групп преобразований
многообразий $\mathfrak{M}$ и $\mathfrak{N}$ размерности $sm$ и
$sn$, после чего находим по уравнению (12.11) невырожденные
двухточечные инварианты. Но решение такого варианта уравнения для
больших размерностей многообразий $\mathfrak{M,N}$ и группы их
преобразований наталкивается на значительные технические
трудности, связанные с классификацией этих групп, и может быть
проведено до конца только для малых их размерностей. Во втором
случае, когда известна метрическая функция $f$ или известны
действия $\lambda$ и $\sigma$ группы в многообразиях
$\mathfrak{M}$ и $\mathfrak{N}$, функциональное уравнение (12.11)
может быть решено сведением его к системе дифференциальных
уравнений в частных производных. Ниже будут рассмотрены некоторые
примеры решения этого уравнения для второго случая.

Для функции (9.2): $f=x+\xi$, задающей физическую структуру ранга (2,2) на
одномерных многообразиях $\mathfrak{M}$ и $\mathfrak{N}$, функциональное
уравнение (12.11):
$$
\lambda(x)+\sigma(\xi)=x+\xi
$$
имеет решение $\lambda(x)=x+a, \ \sigma(\xi)=\xi-a$, определяющее
однопараметрическую группу движений (9.10) для этой функции, а двухточечный
инвариант полученной группы преобразований $x'=x+a, \ \xi'=\xi-a$ находится
как решение того же функционального уравнения (12.11):
$$
f(x+a,\xi-a)=f(x,\xi),
$$
совпадая с исходной метрической функцией с точностью до
масштабного преобразования: $f=\chi(x+\xi)$.

Для функции (9.3): $f=x\xi+\eta$, задающей на одномерном и
двумерном многообразиях $\mathfrak{M}$ и $\mathfrak{N}$ физическую
структуру ранга (3,2), функциональное уравнение (12.11):

\newpage

     In the general case, the functional equation (12.11), just as the equation
     (12.2), allows two interpretations. Either both the metric function $f$ and the
     functions $\lambda,\sigma$ defining the transformations of the manifolds
     $\mathfrak{M,N}$ are unknown. Then, knowing, from Theorem 3 of \S8 the
     dimension of the group of motions of the geometry of two sets defined by the
     function $f$, \ we perform full, with an accuracy up to equivalence (change of
     local coordinates), classification of the $smn$-dimensional groups of transformations
     of the mani- \ folds $\mathfrak{M}$ and $\mathfrak{N}$ of the dimensions $sm$ and
     $sn$, and then via the equation (12.11) come by the nondegenerate two-point
     invariants. However, for large dimensionalities of the manifolds $\mathfrak{M,N}$
     and the group of their transformations, solution of that type of equations is
     encountered with technical difficulties and can only be done for small dimensionalities.
     In the latter case, i.e. that when the metric function $f$ is known, or known are the
     actions $\lambda$ and $\sigma$ of the group of manifolds $\mathfrak{M}$
     and $\mathfrak{N}$, the functional equation (12.11) may be solved by way of
     its reduction to a system of differential equations in partial derivatives. Further, we
     shall discuss some examples of that latter sort.

     For the function (9.2): $f=x+\xi$, which gives a physical structure of rank (2,2) on
     one-dimensional manifolds $\mathfrak{M}$ and $\mathfrak{N}$, the functional
     equation (12.11):
     $$
     \lambda(x)+\sigma(\xi)=x+\xi
     $$ \\
     has the solution $\lambda(x)=x+a, \ \sigma(\xi)=\xi-a$ that defines the one-parameter
     group of motions (9.10) for that function, while the two-point invariant of the group
     of transformations $x'=x+a, \ \xi'=\xi-a$ obtained is to be found as the solution of the
     same functional equation:

     $$
     f(x+a,\xi-a)=f(x,\xi),
     $$ \\
     and that two-point invariant coincides with the initial metric function with an accuracy
     up to a scaling transformation: $f=\chi(x+\xi)$.

     For the function (9.3): $f=x\xi+\eta$, which gives a physical structure of rank (3,2) on a one-
     and a two-dimensional manifolds $\mathfrak{M}$ and $\mathfrak{N}$, the functional
    equation (12.11):

    \newpage

$$
\lambda(x)\sigma(\xi,\eta)+\rho(\xi,\eta)=x\xi+\eta
$$
 имеет решение $\lambda(x)=ax+b, \
\sigma(\xi,\eta)=\xi/a, \ \rho(\xi,\eta)=\eta-b\xi/a$, где
$a\neq0$, определяющее двухпараметрическую группу движений (9.11)
для этой функции. Сама же функция (9.3) может быть найдена как
двухточечный инвариант этой группы по тому же функциональному
уравнению (12.11):
$$
f(ax+b,\xi/a,\eta-b\xi/a)=f(x,\xi,\eta)
$$
 с точностью до масштабного преобразования:
$f=\chi(x\xi+\eta)$.

Для функции (9.4): $f=(x\xi+\eta)/(x+\vartheta)$, задающей физическую
структуру ранга (4,2) на одномерном и трехмерном многообразиях $\mathfrak{M}$
и $\mathfrak{N}$, функциональное уравнение (12.11):
$$
\frac{\lambda(x)\sigma(\xi,\eta,\vartheta)+\rho(\xi,\eta,\vartheta)}
{\lambda(x)+\tau(\xi,\eta,\vartheta)}=\frac{x\xi+\eta}{x+\vartheta}
$$
имеет решение, определяющее трехпараметрическую группу движений
(9.12). Сама же метрическая функция (9.4) может быть найдена как двухточечный
инвариант этой группы по функциональному уравнению (12.11):
$$
f\left(\frac{ax+b}{cx+d},\frac{d\xi-c\eta}{d-c\vartheta},
\frac{a\eta-b\xi}{d-c\vartheta},\frac{a\vartheta-b}{d-c\vartheta}\right)=
f(x,\xi,\eta,\vartheta).
$$
 с точностью до масштабного преобразования $\chi(f)\to f$.

Для двухкомпонентной функции (10.3): $f^1=x+\xi, \ f^2=y+\eta$,
задающей двуметрическую физическую структуру ранга (2,2) на
двумерных многообразиях $\mathfrak{M}$ и $\mathfrak{N}$, функциональное
уравнение (12.11):
$$
\left.\begin{array}{c}
\lambda^1(x,y)+\sigma^1(\xi,\eta)=x+\xi, \\
\lambda^2(x,y)+\sigma^2(\xi,\eta)=y+\eta
\end{array}\right\}
$$
имеет решение $\lambda^1(x,y)=x+a, \ \lambda^2(x,y)=y+b, \
\sigma^1(\xi,\eta)=\xi-a, \ \sigma^2(\xi,\eta)=\eta-b$, определяющее
двухпараметрическую группу движений для этой функции:
$x'=x+a, \ y'=y+b, \ \xi'=\xi-a, \ \eta'=\eta-b$. Сама же функция (10.3)
находится как двухточечный инвариант этой группы
по функциональному уравнению (12.11):

\newpage

     $$
     \lambda(x)\sigma(\xi,\eta)+\rho(\xi,\eta)=x\xi+\eta
     $$
     has the solution $\lambda(x)=ax+b, \ \sigma(\xi,\eta)=\xi/a, \\rho(\xi,\eta)=\eta-b\xi/a$,
     with $a\neq0$, that defines the one-parameter group of motions (9.11) for that
     function. The function (9.3) itself may be found as the two-point invariant of that
     group via the same functional equation (12.11):
     $$
     f(ax+b,\xi/a,\eta-b\xi/a)=f(x,\xi,\eta)
     $$ \\
     with an accuracy up to a scaling transformation: $f=\chi(x\xi+\eta)$.

     For the function (9.4): $f=(x\xi+\eta)/(x+\vartheta)$, which gives a physical structure
     of rank (4,2) on a one-dimensional and a three-dimensional mani- \ folds $\mathfrak{M}$
     and $\mathfrak{N}$, the functional equation (12.11):
     $$
     \frac{\lambda(x)\sigma(\xi,\eta,\vartheta)+\rho(\xi,\eta,\vartheta)}
     {\lambda(x)+\tau(\xi,\eta,\vartheta)}=\frac{x\xi+\eta}{x+\vartheta}
     $$
     has the solution that defines the three-parameter group of motions (9.12). As to the
     metric function (9.4) itself, it may be found as the two-point invariant of that group via the
     functional equation (12.11):
     $$
     f\left(\frac{ax+b}{cx+d},\frac{d\xi-c\eta}{d-c\vartheta},
     \frac{a\eta-b\xi}{d-c\vartheta},\frac{a\vartheta-b}{d-c\vartheta}\right)=
     f(x,\xi,\eta,\vartheta).
     $$
     with an accuracy up to a scaling transformation $\chi(f)\to f$.

     For the two-component function (10.3): $f^1=x+\xi, \ f^2=y+\eta$, which gives a
     dimetric physical structure of rank (2,2) on two-dimensional manifolds $\mathfrak{M}$
     and $\mathfrak{N}$, the functional equation (12.11):
     $$
     \left.\begin{array}{c}
     \lambda^1(x,y)+\sigma^1(\xi,\eta)=x+\xi, \\
     \lambda^2(x,y)+\sigma^2(\xi,\eta)=y+\eta
     \end{array}\right\}
     $$
     has the solution $\lambda^1(x,y)=x+a, \ \lambda^2(x,y)=y+b, \
     \sigma^1(\xi,\eta)=\xi-a, \ \sigma^2(\xi,\eta)=\eta-b$ that defines a two-parameter
     group of motions for that function: $x'=x+a, \ y'=y+b, \ \xi'=\xi-a, \ \eta'=\eta-b$.
     The function (10.3) is found as the two-point invariant of that group via the
     functional equation (12.11):

\newpage

$$
f(x+a,y+b,\xi-a,\eta-b)=f(x,y,\xi,\eta).
$$
с точностью до двумерного масштабного преобразования:
$f^1=\chi^1(x+\xi,y+\eta), \  f^2=\chi^2(x+\xi,y+\eta)$.

Для второй функции (10.4): $f^1=(x+\xi)y, \ f^2=(x+\xi)\eta$,
задающей на двумерных многообразиях другую двуметрическую
физическую структуру ранга (2,2), функциональное
уравнение (12.11):
$$
\left.\begin{array}{c}
(\lambda^1(x,y)+\sigma^1(\xi,\eta))\lambda^2(x,y)=(x+\xi)y, \\
(\lambda^1(x,y)+\sigma^1(\xi,\eta))\sigma^2 (\xi,\eta)=(x+\xi)\eta
\end{array}\right\}
$$
имеет решение $\lambda^1(x,y)=ax+b, \ \lambda^2(x,y)=y/a, \
\sigma^1(\xi,\eta)=a\xi-b, \ \sigma^2(\xi,\eta)=\eta/a$, где $a\neq0$,
определяющее двухпараметрическую группу движений для этой функции: $x'=ax+b, \
y'=y/a, \ \xi'=a\xi-b, \ \eta'=\eta/a$. Сама же метрическая функция находится
как двухточечный инвариант группы движений решением
функционального уравнения (12.11):
$$
f(ax+b,y/a,a\xi-b,\eta/a)=f(x,y,\xi,\eta)
$$
с \ точностью \ до \ масштабного \ преобразования \
$\chi^1(f^1,f^2) \ \to \ f^1$, \\ $\chi^2(f^1,f^2) \ \to \ f^2$.

Для всех остальных функций (10.5)--(10.12), задающих двуметрические физические
структуры ранга (3,2), (4,2) и (5,2),
а также функций (10.18)--(10.28), задающих
триметрические физические структуры ранга (2,2), функциональное уравнение
(12.11) рассматривается аналогично.

Функциональные уравнения естественно появляются и в теории групп
преобразований, с которой теория физических структур тесно связана.
Для групп преобразований $G^r(\lambda)$  и  $H^r(\sigma)$ многообразий
$\mathfrak{M}$  и  $\mathfrak{N}$ с действиями $x'=\lambda(x,a)$  и
$\xi'=\sigma(\xi,\alpha)$, где $a\in G^r$  и  $\alpha\in H^r$, с законами
умножения $ab=\varphi(a,b)$  и  $\alpha\beta=\psi(\alpha,\beta)$ в
соответствующих параметрических группах $G^r$  и  $H^r$ их изоморфизм
устанавливается по решению функционального уравнения
$$
u(\varphi(a,b))=\psi(u(a),u(b))
\eqno(12.12)
$$
относительно обратимого отображения $u:G^r\to H^r$, а их подобие
по решению системы двух функциональных уравнений: (12.12) и

\newpage

     $$
     f(x+a,y+b,\xi-a,\eta-b)=f(x,y,\xi,\eta).
     $$
     with an accuracy up to a two-dimensional scaling transformation:
     $f^1=\chi^1(x+\xi,y+\eta), \  f^2=\chi^2(x+\xi,y+\eta)$.

     For the second function (10.4): $f^1=(x+\xi)y, \ f^2=(x+\xi)\eta$, which gives
     another physical structure of rank (2,2), the functional equation (12.11):
     $$
     \left.\begin{array}{c}
     (\lambda^1(x,y)+\sigma^1(\xi,\eta))\lambda^2(x,y)=(x+\xi)y, \\
     (\lambda^1(x,y)+\sigma^1(\xi,\eta))\sigma^2 (\xi,\eta)=(x+\xi)\eta
     \end{array}\right\}
     $$
     has the solution $\lambda^1(x,y)=ax+b, \ \lambda^2(x,y)=y/a, \
     \sigma^1(\xi,\eta)=a\xi-b, \ \sigma^2(\xi,\eta)=\eta/a$, with $a\neq0$,
     that defines a two-parameter group of motions for that function: $x'=ax+b, \
     y'=y/a, \ \xi'=a\xi-b, \ \eta'=\eta/a$. The metric function itself is come by as the
     two-point invariant of the group of motions via solving the functional equation (12.11):
     $$
     f(ax+b,y/a,a\xi-b,\eta/a)=f(x,y,\xi,\eta)
     $$
     with \ an \ accuracy \ up \ to \ a \ scaling \ transformation \ $\chi^1(f^1,f^2) \ \to \ f^1$,  \\
     $\chi^2(f^1,f^2) \ \to \ f^2$.

     For all the other functions, (10.5) to (10.12), that give dimetric physical structures of
     ranks (3,2), (4,2) and (5,2), as well as for the functions (10.18)--(10.28), that give trimetric
     physical structures of rank (2,2), the functional equation (12.11) is considered similarly.

     Functional equations appear quite naturally in the theory of groups of transformations too,
     for that theory is inherently related with that of physical structures. For the groups of
     transformations $G^r(\lambda)$ and $H^r(\sigma)$ of the manifolds $\mathfrak{M}$ and
     $\mathfrak{N}$ with the actions $x'=\lambda(x,a)$ and $\xi'=\sigma(\xi,\alpha)$, where
     $a\in G^r$ and $\alpha\in H^r$, and the rules of multiplication in the parameter groups
     $G^r$ and $H^r$ being $ab=\varphi(a,b)$ and $\alpha\beta=\psi(\alpha,\beta)$ respectively,
     their isomorphism is established by the solution of the functional equation
     $$
     u(\varphi(a,b))=\psi(u(a),u(b))
     \eqno(12.12)
     $$
     with respect to the invertible mapping $u:G^r\to H^r$, while their similarity is established
     by the solution of the system of two functional equations: (12.12) and

\newpage

     $$
     v(\lambda(x,a))=\sigma(v(x),u(a))
     \eqno(12.13)
     $$
относительно обратимых отображений $u:G^r\to H^r$ и $v:\mathfrak{M}\to
\mathfrak{N}$. Ясно, что подобие возможно только при совпадении размерностей
многообразий $\mathfrak{M}$ и $\mathfrak{N}$.

Слабая эквивалентность групп преобразований $G^r(\lambda)$ и
$G^r(\sigma)$ с действиями $x'=\lambda(x,a)$ и
$\xi'=\sigma(\xi,a)$, имеющих одну и ту же параметрическую группу
$G^r$, устанавливается по решению системы функциональных уравнений
(12.12), (12.13), где $\psi=\varphi$, относительно автоморфизма
$u:G^r\to G^r$ и обратимого отображения
$v:\mathfrak{M}\to\mathfrak{N}$, а их сильная эквивалентность --
по решению функционального уравнения
$$
w(\lambda(x,a))=\sigma(w(x),a)
\eqno(12.14)
$$
относительно обратимого отображеня $w:\mathfrak{M}\to\mathfrak{N}$.

Заметим, что вполне возможен случай, когда система функциональных
уравнений (12.12), (12.13) с
$\psi=\varphi$ имеет решение, в то время как функциональное уравнение
(12.14) решения не имеет, то есть группы преобразований $G^r(\lambda)$ и
$G^r(\sigma)$, будучи слабо эквивалентными,
не эквивалентны в сильном смысле.

\vspace{6mm}

\begin{center}
{\bf \large \S13. Интерпретации физических структур}
\end{center}

Физические структуры как математические формы могут быть наполнены
разным содержанием, то есть имеют разнообразные физические и
геометрические интерпретации. Приведем некоторые примеры.

Второй закон Ньютона: $F=ma$, рассмотренный во Введении (см. уравнения (В.17)
и (В.18)), запишем в мультипликативной канонической форме:
$$
f=x\xi, \ \ f(i\alpha)f(j\beta)-f(i\beta)f(j\alpha)=0,
\eqno(13.1)
$$
где, например, $f(i\alpha)=x_i\xi_\alpha$, введя следующие единые
обозначения функций и координат:
$$
f=a, \ \ x=1/m, \ \ \xi=F.
\eqno(13.2)
$$

\newpage

     $$
     v(\lambda(x,a))=\sigma(v(x),u(a))
     \eqno(12.13)
     $$
     with respect to the invertible mappings $u:G^r\to H^r$ and $v:\mathfrak{M}\to
     \mathfrak{N}$. It is evident that similarity is only possible if there is coincidence of the
     dimensions of the manifolds $\mathfrak{M}$ and $\mathfrak{N}$.

     The weak equivalence of the groups of transformations $G^r(\lambda)$ and $G^r(\sigma)$,
     which have one and the same parameter group $G^r$, with the actions $x'=\lambda(x,a)$ and
     $\xi'=\sigma(\xi,a)$ is established by the solution of the system of the functional
     equations (12.12) and (12.13), where $\psi=\varphi$, with respect to the automorphism
     $u:G^r\to G^r$ and the invertible mapping $v:\mathfrak{M}\to\mathfrak{N}$, and their
     strong equivalence is established by the solution of the functional equation
     $$
     w(\lambda(x,a))=\sigma(w(x),a)
     \eqno(12.14)
     $$
     with respect to the invertible mapping $w:\mathfrak{M}\to\mathfrak{N}$.

     We shall note that a case is quite possible when the system of functional equations
     (12.12) and (12.13) with $\psi=\varphi$ {\it does} have a solution, while the functional
     equation (12.14) has no solution, i.e. the groups of transformations $G^r(\lambda)$
     and $G^r(\sigma)$ while equivalent weakly have at the same time no strong equivalence.

     \vspace{15mm}

     \begin{center}
     {\bf \large \S13. Interpretations of physical structures}
     \end{center}

     Physical structures, as mathematical forms, may have various meanings, i.e. they may
     have various physical and geometric interpretations. Now we shall give examples.

     We shall write Newton's 2nd law: $F=ma$, considered in the Introduction (see the
     equations (T.17) and (T.18)), in the multiplicative canonical form:
     $$
     f=x\xi, \ \ f(i\alpha)f(j\beta)-f(i\beta)f(j\alpha)=0,
     \eqno(13.1)
     $$
     where, for example, $f(i\alpha)=x_i\xi_\alpha$, and introduce a single notation
     for the functions and the coordinates:
     $$
     f=a, \ \ x=1/m, \ \ \xi=F.
     \eqno(13.2)
     $$

\newpage

 Уравнения канонической формы (13.1) представляют собой чисто
математические соотношения, которые можно наполнить разным
физическим содержанием. Для второго закона Ньютона согласно
обозначениям (13.2) функция $f$ есть измеряемое в опыте ускорение
$a$ тела под действием ускорителя, координата $x$ задает величину
обратную массе $m$ тела, а координата $\xi$ совпадает с силой $F$
ускорителя.

Такой подход к канонической форме (13.1) оправдан тем, что она может быть
наполнена и другим физическим содержанием, то есть к ней приводится не только
второй закон Ньютона в механике, но и многие другие физические законы.

Рассмотрим, например, еще закон преломления в оптике для того случая,
когда луч света падает
из вакуума в среду, известная формула которого: $\sin\varphi/\sin\psi=n$
прочитывается следующим образом: {\it отношение синуса угла падения к синусу
угла преломления равно показателю преломления среды.}

Легко понять, что в законе преломления, также как и во втором
законе Ньютона, между собой связаны разные по своей природе
физические величины. В самом деле, угол падения $\varphi$
характеризует только падающий луч света, а показатель преломления
$n$ только среду. Но угол преломления $\psi$, непосредственно
измеряемый в опыте, характеризует одновременно и падающий луч и
оптическую среду, определяя их взаимодействие.

Подчеркнем отмеченное обстоятельство, введя множество падающих
лучей света $\mathfrak{M}=\{i,j,k,...\}$ и множество оптических
сред $\mathfrak{N}=\{\alpha,\beta, \\ \gamma,...\}$. Тогда для
произвольного луча $i \in \mathfrak{M}$ с углом падения
$\varphi_i$ и произвольной среды $\alpha \in \mathfrak{N}$  с
показателем преломления $n_\alpha$ формула закона преломления
примет следующий вид:
$$
\sin\varphi_i/\sin\psi_{i\alpha}=n_\alpha,
\eqno(13.3)
$$
откуда видим, что величины $\varphi, n$ и $\psi$ имеют еще и
различную {\it математическую} природу, так как первые две
величины одноиндексные и характеризуют падающий луч и оптическую
среду, а третья -- двухиндексная и характеризует уже
взаимодействие падающего луча и среды.

Ключевую роль в законе преломления (13.3) играет, очевидно, угол

\newpage

     The canonical equations (13.1) are purely mathematical relationships that can be filled
     with different physical contents. For Newton's 2nd law, according to the designations
     (13.2), the function $f$ is the acceleration $a$ of a body under the impact of the accelerator,
     the coordinate $x$ gives the value that is the inverse of the mass $m$ of the body, and
     the coordinate $\xi$ coincides with the force $F$ of the accelerator.

     The justification of such an approach to the canonical form (13.1) is in the rich
     opportunities it presents of different physical interpretations, i.e. it is not only the
     Newton's Second Law of Mechanics that can be reduced to it, but many other laws of
     physics.

     For example, let us consider the optical Law of Refraction, for the case of a beam of
     light falling from the vacuum into some refracting medium, whose formula is known
     to be $\sin\varphi/\sin\psi=n$, that reads:
    {\it the ratio of the sine of the angle of incidence to the sine of the
     angle of refraction is equal to the index of refraction of the medium.}

     It is easy to see that the Law of Refraction, just as Newton's Second Law, relates
     physical quantities of different natures. Indeed, the angle of incidence $\varphi$ only
     characterizes the beam, while the refractive index $n$ characterizes the medium. But
     the angle of the refraction $\psi$, directly measured by experiment, characterizes
     simultaneously the beam and the optical medium, establishing whereby their interaction.

     To stress that circumstance, we shall introduce a set of incident beams $\mathfrak{M}=\{i,j,k,...\}$
     and a set of optical mediums $\mathfrak{N}=\{\alpha,\beta,\gamma,...\}$.
     Then, for an arbitrary beam $i \in \mathfrak{M}$ with the incidence angle $\varphi_i$
     and an arbitrary medium $\alpha \in \mathfrak{N}$ with the refractive index $n_\alpha$,
     the formula of the law of refraction is as follows:
     $$
     \sin\varphi_i/\sin\psi_{i\alpha}=n_\alpha,
     \eqno(13.3)
     $$
     where we can see that the {\it mathematical} natures of the quantities $\varphi, n$
     and $\psi$ are different too, as the two former quantities are one-index ones and
     characterize the incident beam and the optical medium, while the third is a two-index
     one and characterizes the interaction of the incident beam and the medium.

The critical part in the law of refraction (13.3) is obviously
that played

\newpage

\noindent
преломления, и потому естественно переписать этот закон
в феноменологически симметричной форме, содержащей только
измеряемые в опыте углы преломления. Для этого, как и в случае
второго закона Ньютона, необходимо взять по два элемента из
каждого множества, то есть два луча $i,j$ из множества падающих
лучей $\mathfrak{M}$ и две среды $\alpha,\beta$ из множества
оптических сред $\mathfrak{N}$. Между четырьмя возможными углами
преломления $\psi_{i\alpha}, \psi_{i\beta}, \psi_{j\alpha},
\psi_{j\beta}$, используя формулу (13.3), легко находим связь
$$
\sin \psi_{i\alpha}\sin \psi_{j\beta}-\sin \psi_{i\beta}\sin \psi_{j\alpha}=0,
\eqno(13.3')
$$ \\
уравнение которой
задает закон преломления в феноменологически симметричной форме. Заметим, что
два уравнения (13.3) и (13.3') закона преломления приводятся к мультипликативной
канонической форме (13.1), если положить
$$
f=\sin\psi, \ \ x=\sin\varphi, \ \ \xi=1/n.
\eqno(13.4)
$$

\vspace{2mm}
 Таким образом, каноническая форма (13.1) может быть
наполнена различным физическим содержанием, если точно указать, из
каких физических объектов состоят множества $\mathfrak{M}$ и
$\mathfrak{N}$, а так же какой измерительной процедурой двум
объектам из этих множеств сопоставляется число, характеризующее их
взаимодействие. Математический объект, для которого выполняются
уравнения (13.1), назван физической структурой, так как он имеет,
как было показано выше, различные физические интерпретации.
Говорят также (см. \S8), что функция $f$ задает на двух множествах
$\mathfrak{M}$ и $\mathfrak{N}$ {\it физическую структуру ранга
(2,2)}, поскольку вторым уравнением (13.1) задается функциональная
связь значений этой функции для любых двух элементов $ i,j$ из
первого множества и любых двух элементов $\alpha,\beta$ из
второго.

Закон Ома для замкнутой цепи: $I=\mathcal{E}/(R+r)$ подробно был
рассмотрен во Введении (см. формулы (В.19) и (В.20)). Придадим ему
каноническую форму, полученную автором в работе [37], введя
следующие удобные обозначения: $R=x, \ 1/\mathcal{E}=\xi, \
r/\mathcal{E}=\eta, \ 1/I=f$:

\newpage

\noindent
     by the angle of
     refraction, and so it is natural to rewrite that law in the phenomenologically symmetric
     form that should only contain the angles of refraction measured by experiment. To do it
     it is necessary, just as in the case of Newton's Second Law, to take two elements from each
     set, that is two beams $i,j$ of the set of incident beams $\mathfrak{M}$ and two mediums
     $\alpha,\beta$ of the set of optical mediums $\mathfrak{N}$. The relation among the four
     possible angles of refraction $\psi_{i\alpha}, \psi_{i\beta}, \psi_{j\alpha}, \psi_{j\beta}$ is
     easily found by using the formula (13.3):

  $$
     \sin \psi_{i\alpha}\sin \psi_{j\beta}-\sin \psi_{i\beta}\sin \psi_{j\alpha}=0,
     \eqno(13.3')
     $$ \\
     which equation gives the Law of Refraction in the phenomenologically symmetric form.
     We shall note that the equations (13.3) and (13.3') of the law of refraction are reduced
     to the multiplicative canonical form (13.1) if we set

     $$
     f=\sin\psi, \ \ x=\sin\varphi, \ \ \xi=1/n.
     \eqno(13.4)
     $$

     \vspace{3mm}
     Thus, the canonical form (13.1) may be filled with different physical contents if we
     point out precisely what kind of physical objects the sets $\mathfrak{M}$ and
     $\mathfrak{N}$ comprise, and what measurement procedure assigns two objects
     of these sets the number that characterizes their interaction. We call a mathematical
     object for which the equations (13.1) are satisfied a physical structure, for it has, as was
     demonstrated above, various physical interpreta- \ tions. It is also said (see \S8) that the
     function $f$ defines on the sets $\mathfrak{M}$ and $\mathfrak{N}$ a{\it physical
     structure of rank (2,2)} because the second equation (13.1) defines the functional
     relation of the values of that function for any two elements $i,j$ of the former set
     and any two elements $\alpha,\beta$ of the latter.

     Ohm's law for a closed circuit: $I=\mathcal{E}/(R+r)$ was discussed in detail in the
     Introduction (see the expressions (В.19) and (В.20)). Let us endow it with the canonical
     form, derived by the author in the article [37], introducing suitable designation as follows: $R=x, \ 1/\mathcal{E}=\xi, \ r/\mathcal{E}=\eta,
     \ 1/I=f$:

\newpage

$$
\left.\begin{array}{c}
f=x\xi+\eta,
\\
\phantom{aaaaaaaaaaa}
\\
\left|
\begin{array}{ccc}
f(i\alpha ) & f(i\beta ) & 1 \\
f(j\alpha ) & f(j\beta ) & 1 \\
f(k\alpha ) & f(k\beta ) & 1
\end{array}
\right| =0,
\end{array}\right\}
\eqno(13.5)
$$ \\
где, например, $f(i\alpha)=x_i\xi_\alpha+\eta_\alpha$.

Оказывается, что каноническая форма (13.5) может быть наполнена и
другим физическим содержанием. Рассмотрим закон линейного
теплового расширения твердых тел: $L=L_0(1+Et),$ где $L$ -- длина
стержня при данной температуре $t$ в градусах по Цельсию, $L_0$ --
его длина при нулевой температуре и $E$ -- коэффициент теплового
расширения. В этом законе, также как и в законе Ома, связаны между
собой физически разнородные величины. Действительно, температура
$t$ характеризует тот термостат, в котором проводится измерение
длины стержня, а начальная длина $L_0$ и коэффициент теплового
расширения $E$ характеризуют стержень. Длина $L$ зависит и от
стержня и от того термостата, в котором стержень находится.

Подчеркнем это различие, введя множество
термостатов \linebreak$\mathfrak{M}=\{i,j,k,...\}$ и множество стержней $\mathfrak{N}=
\{\alpha,\beta,\gamma, ...\}$. Термостат $i$ характеризуется своей
температурой $t_i$, измеряемой термометром, а стержень
$\alpha$ характеризуется своей начальной длиной $L_{0\alpha}$
при нулевой температуре и
коэффициентом объемного расширения $E_\alpha$, который в линейном
приближении считается
постоянным. Измеряемая же в опыте длина $L_{i\alpha}$ стержня
$\alpha$, находящегося в термостате $i$, должна быть двухиндексной величиной.
Закон теплового расширения теперь запишется в таком
виде:
$$
L_{i\alpha}=L_{0\alpha}(1+E_\alpha t_i),
\eqno(13.6)
$$ \\
в котором явно указана физическая и математическая разнородность
величин, в него входящих.

Закон теплового расширения (13.6) можно записать в единой с
законом Ома канонической форме (13.5), \ если ввести следующие
обозна-

\newpage

$$
     \left.\begin{array}{c}
     f=x\xi+\eta,
     \\

     \phantom{aaaaaaaaaaa}
     \\
     \left|
     \begin{array}{ccc}
     f(i\alpha ) & f(i\beta ) & 1 \\
     f(j\alpha ) & f(j\beta ) & 1 \\
     f(k\alpha ) & f(k\beta ) & 1
     \end{array}
     \right| =0,
     \end{array}\right\}
     \eqno(13.5)
     $$ \\
     where, for example, $f(i\alpha)=x_i\xi_\alpha+\eta_\alpha$.

     It appears that the canonical form (13.5) may have still another physical meaning. Now
     we shall consider the law of the linear thermal expansion of solid bodies: $L=L_0(1+Et),$
     where $L$ is the length of a bar at the given temperature $t$, in degrees of centigrade,
     $L_0$ is its length at the zero temperature, and $E$ is the thermal expansion coefficient.
     In that law, as in Ohm's Law, physically different quantities are related one with another.
     Indeed, the temperature $t$ characterizes the thermostat where the measuring of the
     length of a bar is performed, while the original length $L_0$ and the thermal expansion
     coefficient $E$ characterize the bar. The length $L$ depends on both the bar and the
     thermostat where the bar is placed.

     We shall stress that difference, introducing a set of thermostats $\mathfrak{M}=\{i,j,k,...\}$
     and a set of bars $\mathfrak{N}= \{\alpha,\beta,\gamma, ...\}$. The thermostat $i$ is
     characte- \ rized by its temperature $t_i$ of it measured by a thermometer, and the bar $\alpha$
     is characterized with its original length $L_{0\alpha}$ at the zero temperature and the
     coefficient of volume expansion $E_\alpha$, which is, in the linear approximation,
     considered to be constant, the length $L_{i\alpha}$ of the bar $\alpha$ placed into the
     thermostat $i$ measured by experiment being a two-index quantity. Then, the law of
     thermal expansion acquires the form as follows:

     $$
     L_{i\alpha}=L_{0\alpha}(1+E_\alpha t_i),
     \eqno(13.6)
     $$ \\
     which demonstrates clearly the physical and mathematical heterogeneity of the
     quantities it includes.

     The law of thermal expansion (13.6) and Ohm's law may be written in the single
     canonical form (13.5), if we introduce the designation as follows:

\newpage

\noindent
 чения: $t=x, \ EL_0=\xi, \ L_0=\eta, \ L=f$. Тогда
феноменологически симметричной формой этого закона, как и закона
Ома, будет, очевидно, функциональная связь, задаваемая вторым
уравнением из (13.5).

Единая для двух различных физических законов
каноническая форма (13.5)
может быть освобождена от всякого физического содержания и рассмотрена как
чисто математический объект, который, в силу его происхождения, назван
{\it физической структурой ранга} (3,2),
являющейся феноменологически симметричной
геометрией двух множеств того же ранга, так как метрическая функция
$f=x\xi+\eta$ двухточечная и ее значение $f(i\alpha)$
в некотором обобщенном смысле можно назвать
расстоянием между точкой $i$ и точкой $\alpha$ из разных множеств
$\mathfrak{M}$ и $\mathfrak{N}$.

Остановимся еще на интерпретациях {\it физической структуры ранга} (4,2),
каноническая форма которой приведена в \S9 (см. (9.4) и $(9.4'))$:

$$
\left.\begin{array}{c}
f=(x\xi+\eta)/(x+\vartheta),
\\
\phantom{aaaaaaaaaaa}
\\
\left|\begin{array}{cccc}
f(i\alpha) & f(i\beta) & f(i\alpha)f(i\beta) & 1 \\
f(j\alpha) & f(j\beta) & f(j\alpha)f(j\beta) & 1 \\
f(k\alpha) & f(k\beta) & f(k\alpha)f(k\beta) & 1 \\
f(l\alpha) & f(l\beta) & f(l\alpha)f(l\beta) & 1
\end{array}\right|=0,
\end{array}\right\}
\eqno(13.7)
$$ \\
где, например, $f(i\alpha)=(x_i\xi_\alpha+\eta_\alpha)/(x_i+
\vartheta_\alpha)$. В справедливости уравнения, выражающего
феноменологическую симметрию этой структуры, можно убедиться
непосредственной подстановкой в него метрической функции, применяя
метод разложения определителя по сумме в
столбце. Или, проще, с помощью программных
пакетов "Maple" \ и "Mathematica"$,$ позволяющих вычислять
определители и ранги матриц.

Рассмотрим сначала оптику толстой линзы (см. [1], стр. 506-508).
Ее формула внешне совпадает с формулой тонкой линзы:

$$
\frac{1}{a}+\frac{1}{b}=\frac{1}{F} \ ,
$$ \\
в которой $a$ -- расстояние вдоль главной оси от предмета до
центра линзы, $b$ -- соответствующее расстояние для изображения и
$F$ -- фокусное

\newpage

      \noindent
      $t=x, \ EL_0=\xi, \ L_0=\eta,
     \ L=f$. Then, the phenomenologically symmetric form of either of the laws will,
     obviously, be the functional relation defined by the second equation in (13.5).

     The canonical form (13.5), single for two different laws of physics, may be disengaged
     from any physical meaning and considered as a sheerly mathemati- \ cal object that we name,
     due to its origin, a {\it physical structure of rank} (3,2) which is also a phenomenologically
     symmetric geometry of two sets of the same rank, for the metric function $f=x\xi+\eta$ is a
     two-point one and its value $f(i\alpha)$ may, in some generalized sense, be termed as the
     distance between a point $i$ and a point $\alpha$ of different sets, $\mathfrak{M}$ and $\mathfrak{N}$.

     Let us dwell more on interpretations of the {\it physical structure of rank} (4,2), whose
     canonical form was given in \S9 (see (9.4) and $(9.4'))$:

     $$
     \left.\begin{array}{c}
     f=(x\xi+\eta)/(x+\vartheta),
     \\
     \phantom{aaaaaaaaaaa}
     \\
     \left|\begin{array}{cccc}
     f(i\alpha) & f(i\beta) & f(i\alpha)f(i\beta) & 1 \\
     f(j\alpha) & f(j\beta) & f(j\alpha)f(j\beta) & 1 \\
     f(k\alpha) & f(k\beta) & f(k\alpha)f(k\beta) & 1 \\
     f(l\alpha) & f(l\beta) & f(l\alpha)f(l\beta) & 1
     \end{array}\right|=0,
     \end{array}\right\}
     \eqno(13.7)
     $$ \\
     where, for example, $f(i\alpha)=(x_i\xi_\alpha+\eta_\alpha)/(x_i+\vartheta_\alpha)$.
     The validity of the equation that expresses the phenomenological symmetry of
     that structure may be assured by way of direct substitution into it the metric
     function, using the method of expansion of the determinant with respect to the
     sum in the column. Or, to make the whole business simpler, using computing
     packages "Maple" \  and   "Mathematica" \ that can compute determinants and ranks
     of matrices.

     First, we shall take the optics of the thick lens (see [1], pp. 506-508). Its formula
     looks quite similar to that of the thin lens:

     $$
     \frac{1}{a}+\frac{1}{b}=\frac{1}{F} \ ,
     $$ \\
     in which $a$ is the distance from the object to the centre of the lens along the
     principal axis, \ $b$ is the corresponding distance for the image and $F$ is

\newpage

\noindent
 расстояние линзы. У толстой линзы величины $a$ и $b$
имеют несколько иной смысл. Дело в том, что они отсчитываются
вдоль главной оси, но не до центра линзы, а до двух ее главных
плоскостей. Пусть $x$ -- расстояние вдоль главной оси от предмета
до ближайшей точки поверхности линзы, а $\lambda$ -- от нее до
ближайшей главной плоскости. Тогда $a=x+\lambda$. Аналогично,
пусть $u$ -- расстояние от изображения до точки другой поверхности
линзы, а $\sigma$ -- от нее до ближайшей главной плоскости. Для
наглядности на рисунке удобно изображать двояковыпуклую толстую
линзу с $F>0$, тогда все введенные величины будут положительными
по знаку. Подставим приведенные выражения для величин $a$ и $b$ в
формулу толстой линзы и разрешим ее относительно расстояния от
линзы до изображения:

$$
u=\frac{x(F-\sigma)+(\lambda+\sigma)F-\lambda\sigma}{x+\lambda-F} \ .
\eqno(13.8)
$$

\vspace{2mm}
 Рассмотрим теперь множество предметов $\mathfrak{M}$
и множество толстых линз $\mathfrak{N}$, с помощью которых
строятся их изображения. Первое множество является одномерным
многообразием, точки которого задаются координатой $x$, а второе
-- трехмерным, и его точки задаются координатами
$F,\lambda,\sigma$. В законе (13.8), выведенном из формулы толстой
линзы, связаны величины различной природы. Координата $x$
характеризует предмет, координаты $F,\lambda,\sigma$ характеризуют
линзу, в то время как величина $u$ характеризует "взаимодействие"
предмета и линзы. Чтобы подчеркнуть это обстоятельство, подставим
в закон (13.8) конкретные предмет $i \in\mathfrak{M}$ и линзу
$\alpha \in \mathfrak{N}$:

$$
u_{i\alpha}=\frac{x_i(F_\alpha-\sigma_\alpha)+
(\lambda_\alpha+\sigma_\alpha)F_\alpha-\lambda_\alpha\sigma_\alpha}
{x_i+\lambda_\alpha-F_\alpha} \ .
$$

\vspace{2mm}
 Феноменологическая симметрия закона (13.8)
обнаружится сразу, если его привести к канонической форме (13.7)
следующими очевидными заменами координат: $x\to x, \
F-\sigma\to\xi, \ (\lambda+\sigma)F-\lambda\sigma\to\eta, \
\lambda-F\to\vartheta$ и переменой обозначения измеряемой
величины: $u\to f$. Тогда феноменологически симметричной формой
закона (13.8) для толстой линзы будет уравнение из (13.7).

\newpage

\noindent
    the focal
     distance of the lens. For the case of the thick lens the values $a$ and $b$ are of
     some different meaning. The thing is, they are measured along the principal axis
     too, but not to the centre of the lens, but to the two principal planes of it. Suppose
     $x$ is the distance along the principal axis from the object to the nearest point on
     the lens surface, and $\lambda$ is that from the lens to the nearest principal plane.
     Then $a=x+\lambda$. Similarly, suppose $u$ is the distance from the image to a
     point on the other surface of the lens, and $\sigma$ is that from the lens to the
     nearest principal plane. For the sake of simplicity, it is suitable to draw a biconvex
     thick lens with $F>0$, because all the values in the formula will be positive. We
     shall substitute the expressions for the values $a$ and $b$ that we have mentioned
     into the formula of the thick lens and solve it with respect to the
     distance from the lens to the image:

     $$
     u=\frac{x(F-\sigma)+(\lambda+\sigma)F-\lambda\sigma}{x+\lambda-F} \ .
     \eqno(13.8)
     $$

\vspace{2mm}
     Now we shall consider a set of objects $\mathfrak{M}$ and a set of thick lenses
     $\mathfrak{N}$ that are used to build their images. The former of the sets is a
     one-dimensional manifold whose points are defined by the coordinate $x$, while
     the latter is a three-dimensional manifold, and the points of it are defined by the
      coordinates $F,\lambda,\sigma$. In the law (13.8), derived from the formula of
     the thick lens, it is quantities of different nature that are related with one another.
     The coordinate $x$ characterizes the object, the coordinates $F,\lambda,\sigma$
     characterize the lens, and the value $u$ characterizes the "interaction"  \ of the
     object and the lens. In order to throw that into relief, we shall substitute into the law
      (13.8) some an object $i \in\mathfrak{M}$ and a lens $\alpha \in \mathfrak{N}$:

     $$
     u_{i\alpha}=\frac{x_i(F_\alpha-\sigma_\alpha)+
     (\lambda_\alpha+\sigma_\alpha)F_\alpha-\lambda_\alpha\sigma_\alpha}
     {x_i+\lambda_\alpha-F_\alpha} \ .
     $$

\vspace{2mm}
     The phenomenological symmetry of the law (13.8) is revealed forthwith if we reduce
     it to the canonical form (13.7) by way of obvious changes of coordinates as follows:
     $x\to x, \ F-\sigma\to\xi, \ (\lambda+\sigma)F-\lambda\sigma\to\eta, \lambda-F\to\vartheta$
     together with change of designation of the quantity measured: $u\to f$. Then, the
     phenomenologically symmetric form of the law (13.8) for the thick lens will be the
     equation from (13.7).

\newpage

Геометрическая интерпретация физической структуры ранга (4,2)
строится следующим образом (см. [1], стр. 501-502). Пусть
$\mathfrak{M}$ -- однопараметрическое множество прямых на
плоскости Евклида, проходящих через начало координат. Каждая такая
прямая однозначно определяется углом $\varphi$ между ней и осью
абцисс, причем $-\pi/2<\varphi\leq+\pi/2$. Вторым пусть будет
трехпараметрическое множество $\mathfrak{N}$ прямых, проходящих
через точки $(a,b)$ под различными углами $\theta$ к оси абсцисс,
причем также $-\pi/2<\theta\leq+\pi/2$. Двум прямым из этих
множеств сопоставим величину, задаваемую выражением

$$
f=\frac{-\displaystyle\frac{a}{\cos\theta}\text{tg}\varphi+
\displaystyle\frac{b}{\cos\theta}}
{\text{tg}\varphi-\text{tg}\theta},
$$ \\
 модуль которой равен расстоянию от точки их
пересечения до точки $(a,b)$. Вводя зaмены координат:
$\text{tg}\varphi\to x, \ -a/\cos\theta\to\xi,$ \
$b/\cos\theta\to\eta, \ -\text{tg}\theta\to\vartheta$, получаем
каноническую форму метрической функции (13.7), задающей на
одномерном и трехмерном многообразиях феноменологически
симметричную геометрию двух множеств (физическую структуру) ранга
(4,2).

\newpage

     A geometric interpretation of the physical structure of rank (4,2) is built as follows
     (see [1], pp. 501-502). Suppose $\mathfrak{M}$ is a one-parameter set of straight
     lines on the Euclidean plane passing through the coordinate origin. Every such straight
     line is uniquely determined by the angle $\varphi$ between the line itself and the
     abscissa axis, and $-\pi/2<\varphi\leq+\pi/2$.
     Suppose, further, the second set is a three-parameter set $\mathfrak{N}$ of straight
     lines passing through the points $(a,b)$ at different angles $\theta$ to the abscissa
     axis, and again $-\pi/2<\theta\leq+\pi/2$. We shall assign two straight lines from these
     sets the quantity defined by the expression

     $$
     f=\frac{-\displaystyle\frac{a}{\cos\theta}\tan\varphi+
     \displaystyle\frac{b}{\cos\theta}}
     {\tan\varphi-\tan\theta},
     $$ \\
     whose modulus is equal to the distance from the point of their intersection to the
     point $(a,b)$. Introducing change of coordinates: $\tan\varphi\to x, \ -a/\cos\theta\to\xi,$
     \ $b/\cos\theta\to\eta, \ -\tan\theta\to\vartheta$, we arrive at the canonical
     form of the metric function (13.7) that gives on a one-dimensional and a
     three-dimensional manifolds a phenomenologically symmetric geometry of two sets
     (a physical structure) of rank (4,2).

\newpage

\begin{center}
{\bf \large \S14. Нерешенные задачи в теории \\ физических структур}
\end{center}

Целью написания данного параграфа является краткий обзор
математических задач теории физических
структур в надежде, что некоторые из них могут заинтересовать читателя.

Феноменологическая симметрия физической структуры согласно теореме
3 из \S8 эквивалентна ее групповой симметрии в следующем смысле:
невырожденная $s$-компонентная метрическая функция $f$ допускает
$smn$-мерную группу движений в том и только в том случае, если она
задает на $sm$- и $sn$-мерных многообразиях физическую структуру
ранга $(n+1,m+1)$. А это означает, что задача решения
функционального уравнения (12.2), в котором неизвестными являются
метрическая функция $f$ и функция $\Phi$, эквивалентна задаче
решения функционального уравнения (12.11), в котором неизвестными
являются та же метрическая функция $f$ и действия $\lambda,\sigma$
группы Ли в многообразиях.

Заметим, что методы решения уравнения (12.2) совершенно отличны от
методов решения уравнения (12.11). Но поскольку, в конечном счете,
решается одна и та же задача классификации физических структур,
эти методы дополняют и взаимозаменяют друг друга, делая полученный
классификационный результат более достоверным. Одни метрические
функции найдены только как решения функционального уравнения
(12.2), другие -- только как решения функционального уравнения
(12.11). Есть и такие, которые были найдены решением обоих
функциональных уравнений. Однако полная классификация
полиметрических физических структур, исключая однометрические (см.
\S9), еще не построена. Поэтому имеет смысл представить кратко в
виде таблицы (см. ниже) обзор всех классификационных задач для
физических структур на двух множествах. Их решение имеет смысл не
только с математической точки зрения, но и с физической, так как
результатом являются возможные формы для фундаментальных
физических законов. Автор надеется, что кому-то из читателей
удастся не только плодотворно объединить известные методы, но и
найти такие новые, которые позволят продолжить и завершить
классификацию полиметрических физичес-

\newpage

     \begin{center}
     {\bf \large \S14. The unresolved problems in the theory \\ of physical structures }
     \end{center}

      This paragraph is written with an object of giving a brief overview of the
     mathematical problems o the theory of physical structures, in the hope that
     some of them may be interesting to the reader.

     The phenomenological symmetry of a physical structure is, under Theorem 3
     of \S8, equivalent to the group symmetry of it in the sense as follows: a
     nondegenerate $s$-component metric function $f$ allows an $smn$-dimensional
     group of motions if, and only if, it defines on an $sm$- and an $sn$-dimensional
     manifolds a physical structure of rank $(n+1,m+1)$. And that means that the task
     of solving the functional equation (12.2), in which the unknowns are the metric
     function $f$ and the function $\Phi$, is equivalent to the task of solving the
     functional equation (12.11), in which the unknowns are the same metric function
     $f$ and the actions $\lambda, and \sigma$ of the Lie group in the manifolds.

     We shall note that the methods of solving the equation (12.2) and those for the
     equation (12.11) are quite different. But with respect to the principal task that is
     being solved i.e. that of classifying physical structures, those methods are mutually
     complementary, making the result of the effort at classification more reliable. Some
     metric functions are found only as solutions of the functional equation (12.2), others
     only as solutions of the functional equation(12.11). There are still others, whose solutions
     were found by way of solving both of the equations. The complete classification of the
     polymetric physical structures, however, save for the monometric ones (see \S9), has not
     been built yet. So it seems it makes sense to give a brief overview of all problems of
     classification of the physical structures on two manifolds (see the table below). Trying
     to solve them makes sense not only mathematically, but in the physical sense too, as
     the result may be possible forms of fundamen- \ tal laws of physics. The author hopes
     that some readers will succeed in not only combining the already known methods, but
     in finding such new ones that would make it possible to carry on and complete the
     work of classifying

\newpage

\noindent
 ких структур произвольного ранга.

\begin{center}
\begin{tabular}{|c|c|c|c|c|c|c|c|c|c|}\hline
\multicolumn{10}{|c|}{\bf {Полиметрические физические структуры}}
\\ \hline № & $s$ & $m$ & $n$ & $sm$ & $sn$ & $(n+1,m+1)$ & $smn$
& реш. & ист.\\ \hline 1 & 1 & $\geqslant 1$ & $\geqslant m$ & $m$
& $n$ & $(n+1,m+1)$ & $mn$ & {\bf + } &\S9 \\ \hline
2 & 2 & 1 & $\geqslant 1$ & 1 & $2n$ & $(n+1,2)$ & $2n$ & {\bf + } & \S10 \\
3 & 2 & $\geqslant 2$ & $\geqslant m$ & $2m$ & $2n$ & $(n+1,m+1)$
& $2mn$ & {\bf -- } & {\bf -- } \\ \hline
4 & 3 & 1 & 1 & 3 & 3 & (2,2) & 3 & {\bf + }  & \S10 \\
5 & 3 & 1 & $\geqslant 2$ & 3 & $3n$ & $(n+1,2)$ & $3n$ & {\bf -- } & {\bf -- } \\
6 & 3 & $\geqslant 2$ & $\geqslant m$ & $3m$ & $3n$ & $(n+1,m+1)$
& $3mn$ & {\bf -- } & {\bf -- } \\ \hline
7 & 4 & 1 & 1 & 4 & 4 & (2,2) & 4 & +  & \S10 \\
8 & 4 & 1 & $\geqslant 2$ & 4 & $4n$ & $(n+1,2)$ & $4n$ & {\bf -- } & {\bf -- } \\
9 & 4 & $\geqslant 2$ & $\geqslant m$ & $4m$ & $4n$ & $(n+1,m+1)$
& $4mn$ & {\bf -- } & {\bf -- } \\ \hline 10 & $\geqslant 5$ &
$\geqslant 1$ & $\geqslant m$ & $\geqslant 5m$ & $\geqslant 5n$ &
$(n+1,m+1)$ & $\geqslant 5mn$ & {\bf -- } & {\bf -- } \\ \hline
\end{tabular}
\end{center}

\vspace{5mm}

Напомним, что $s\geq1$ -- число компонент невырожденной
метрической функции $f=(f^1,\ldots,f^s)$, задающей на $sm$-мерном
и $sn$-мерном многообразиях $\mathfrak{M}$ и $\mathfrak{N}$
физическую структуру (феноменологически симметричную геометрию
двух множеств) ранга $(n+1,m+1)$, наделенную групповой симметрией
степени $smn$. Условие $n\geq m$ введено с целью уменьшить число
строк в таблице, так как классификационный результат симметричен
относительно перестановки натуральных чисел $m$ и $n$. В
предпоследнем столбце таблицы знаками плюс и минус отмечено, что
данная задача решена $(+)$ или не решена $(-)$. \ В последнем
столбце таблицы указан номер параграфа настоящей монографии, в
котором приведена классификация и описаны методы ее построения или
указаны источники, в которых с ними можно познакомиться более
детально.

\newpage

\noindent
    the polymetric physical structures of arbitrary rank.

    \vspace{5mm}

     \begin{center}
     \begin{tabular}{|c|c|c|c|c|c|c|c|c|c|}\hline
     \multicolumn{10}{|c|}{\bf {The polymetric physical structures}} \\ \hline
     № & $s$ & $m$ & $n$ & $sm$ & $sn$ & $(n+1,m+1)$ & $smn$ & slv. & src. \\ \hline
     1 & 1 & $\geqslant 1$ & $\geqslant m$ & $m$ & $n$ & $(n+1,m+1)$ & $mn$ &
     $+$

     &\S9 \\ \hline
     2 & 2 & 1 & $\geqslant 1$ & 1 & $2n$ & $(n+1,2)$ & $2n$ & $+$ & \S10 \\
     3 & 2 & $\geqslant 2$ & $\geqslant m$ & $2m$ & $2n$ & $(n+1,m+1)$ & $2mn$
     & $-$ & $-$ \\ \hline
     4 & 3 & 1 & 1 & 3 & 3 & (2,2) & 3 & $+$  & \S10 \\
     5 & 3 & 1 & $\geqslant 2$ & 3 & $3n$ & $(n+1,2)$ & $3n$ & $-$ & $-$ \\
     6 & 3 & $\geqslant 2$ & $\geqslant m$ & $3m$ & $3n$ & $(n+1,m+1)$ & $3mn$ &

     $-$ & $-$ \\ \hline
     7 & 4 & 1 & 1 & 4 & 4 & (2,2) & 4 & $+$  & \S10 \\
     8 & 4 & 1 & $\geqslant 2$ & 4 & $4n$ & $(n+1,2)$ & $4n$ & $-$ & $-$ \\
     9 & 4 & $\geqslant 2$ & $\geqslant m$ & $4m$ & $4n$ & $(n+1,m+1)$ & $4mn$ &
     $-$ & $-$ \\ \hline
     10 & $\geqslant 5$ & $\geqslant 1$ & $\geqslant m$ & $\geqslant 5m$ &

     $\geqslant 5n$ & $(n+1,m+1)$ & $\geqslant 5mn$ & $-$ & $-$ \\ \hline
     \end{tabular}
     \end{center}

    \vspace{5mm}

     We remind that $s\geq1$ is the number of the components of the nondegene- \ rate
     metric function $f=(f^1,\ldots,f^s)$ that gives on an $sm$-dimensional and an
     $sn$-dimensional manifolds $\mathfrak{M}$ and $\mathfrak{N}$ a physical structure
     (a phenomenolo- \ gically symmetric geometry of two sets) of rank $(n+1,m+1)$
     endowed with a group symmetry of degree $smn$. The condition $n\geq m$ is
     introduced with the purpose of decreasing the number of lines in the table, because
     the classification result is symmetrical with respect to the permutation of the natural
     numbers $m$ and $n$. In the last column but one of the table the 'plus' and 'minus'
     signs are to inform whether the problem has been solved or has not been solved
     respectively. The last column of the table gives the number of the paragraph of this
     monograph where the classification has been given and the methods used in building
     are described it, or the names of works are given where they are treated in greater detail.

\newpage

\begin{center}
{\Large \bf ЗАКЛЮЧЕНИЕ}
\end{center}

Итак, установлено, что бинарные феноменологически симметричные
геометрии одного и двух множеств наделены групповой симметрией и,
кроме того, содержательны в физическом и математическом смыслах.
Поэтому основной задачей теории физических структур (ТФС) является
их полная классификация. Ее решение еще далеко от завершения, что
дает возможность каждому исследователю, имеющему творческие
способности, применить их в новой для себя сфере научной
деятельности.

\newpage

     \begin{center}
     {\Large \bf CONCLUSION}
     \end{center}

     Thus, it has been established that binary phenomenologically symmetric
     geometries of one and two sets are endowed with a group symmetry, and,
     besides, are pregnant with essential physical and mathematical meaning. So, the
     principal task for the theory of physical structures (TPS) is that of complete
     classification of them. That work is far from being complete, which gives an
     inquisitive mathematician chance to try oneself in a field of scientific investigation
     new to them.

\newpage

\begin{center}
{\bf \Large Л и т е р а т у р а}
\end{center}

\vspace{5mm}

{\bf 1.} \ Кулаков Ю.И. Теория физических структур. М.: Доминико, 2004. \\

{\bf 2.} \ Кулаков Ю.И. Геометрия пространств постоянной кривизны
как частный случай теории физических структур // Докл. АН СССР,
1970,
Т.193, №5, С.985-987. \\

{\bf 3.} \ Клейн Ф. Сравнительное обозрение новейших
геометрических исследований ("Эрлангенская программа") // Об
основаниях
геометрии. М., 1956, С.402-434. \\

{\bf 4.} \ Гельмгольц Г. О фактах, лежащих в основании геометрии
// Об
основаниях геометрии. М., 1956. С.366-388. \\

{\bf 5.} \ Михайличенко Г.Г. Вопросы единственности решения
основного уравнения теории физических структур // Кулаков Ю.И.
Элементы
теории физических структур. Новосибирск: НГУ, 1968, С.175-226. \\

{\bf 6.} \ Михайличенко Г.Г. Двумерные геометрии // Докл. АН СССР,
1981, Т.260, №4, С.803-805 (Mikhaylitchenko G.G. Geometries a deux
\linebreak dimensions dans la theorie de structures physiques //
Comptes Rendus de L'Academie des Sciences. Paris, 16 novembre
1981, T.293. Serie 1. P.529-531). \\

{\bf 7.} \ Лев В.Х. Трехмерные геометрии в теории физических
структур // Вычислительные системы. Новосибирск: ИМ СО АН СССР,
1988, Вып.
125, С.90-103. \\

{\bf 8.} \ Пуанкаре А. Об основных гипотезах геометрии // Об
основаниях геометрии. М., 1956, С.388-398.

\newpage

    \begin{center}
     {\bf \Large B i b l i o g r a p h y}
     \end{center}

     \vspace{5mm}

    {\bf 1.} \ Yu. I. Kulakov.  {\it The theory of physical structures} \ ( Кулаков Ю.И. Теория физических
     структур.  М.: Доминико, 2004. Russian).

    {\bf 2.}  \ Yu. I. Kulakov. {\it The geometry of spaces of constant curvature as a special case
     of the theory of physical structures}. Soviet Math., Dokl, 1970, Vol. 11, No. 4, pp. 1055-1057.

     {\bf 3.} \ F. Klein. {\it A comparative review of recent researches in geometry} \
     (F. Klein. Vergleic- \ hende Betrachtungen uber neuere geometrische Forschunge.
     Erlangen, 1872; printed version: Math. Ann., 1893, Vol. 43, pp. 63-100. German).

     {\bf 4.} \ H. Helmholtz. {\it On the Facts lying at the basis of Geometry} \ (H. Helmholtz. Uber die Thatsachen, die der Geometrie zu
     Grunde liegen . Nachr.
     Konigl. Ges. Wiss. und  Georg-Augusts-Univ. Gottingen, 1868,
     19 3; reprinted in his Wissenschaftliche Abhandlungen, Band
     II, Barth, Leipzig, 1883, p. 618. German).

     {\bf 5.} \ G.G. Mikhailichenko. {\it The Problems of Uniqueness of the Solution of
     the Basic Equation of physical structures}
     (Михайличенко Г.Г. Вопросы единственности решения основного уравнения
     теории физических структур // Кулаков Ю.И.  Элементы теории физических структур.
     Новосибирск: НГУ, 1968, С. 175-226. Russian).

     {\bf 6.} \ G.G. Mikhailichenko. {\it Two-dimensional geometries}.
     Soviet Math. Dokl., 1981, Vol. 24, No. 2, pp. 346- 349 \
     (The French version: G.G. Mikhaylitchenko. {\it Geometries a deux dimensions dans la theorie de
     structures physiques}. Comptes Rendus de L'Academie des Sciences. Paris, 16
     novembre 1981, t. 293. Serie 1. pp. 529-531).

     {\bf 7.} \ V.H. Lev. {\it Three-dimensional geometries in the theory of physical
     struc- \ tures} \ (Лев. В.Х. Трехмерные геометрии в теории физических структуп // Вычислительные
     системы. Новосибирск: ИМ СО АН СССР, 1988, Вып. 125. С. 90-103.  Russian).

     {\bf 8.} \ H. Poincare. {\it On the fundamental hypotheses of geometry} \ (H. Poincare. Sur les
     hypotheses fondamentales de la geometrie. Bulletin de la Societe Mathematique de
     France, 1887, Volume: 15, pp. 203-216. French).

\newpage

{\bf 9.} \ Кулаков Ю.И. О новом виде симметрии, лежащем в
основании физических теорий феноменологического типа // Докл. АН
СССР, 1971,
Т.201, №3. С.570-572.  \\

{\bf 10.} \ Михайличенко Г.Г. Полиметрические геометрии.
Новосибирск: НГУ,
2001. \\

{\bf 11.} \  Михайличенко Г.Г. Простейшие полиметрические
геометрии. I //
Сиб. мат. журн., 1998, Т.39, №2, С.377-395. \\

{\bf 12.} \ Понтрягин Л.С. Непрерывные группы. М.: Наука, 1973.  \\

{\bf 13.} \ Овсянников Л.В. Групповой анализ дифференциальных
уравнений.
М.: Наука, 1978. \\

{\bf 14.} \ Кыров В.А. Шестимерные алгебры Ли групп движений
трехмерных феноменологически симметричных геометрий //
Михайличенко Г.Г.
Полиметрические геометрии. Новосибирск: НГУ, 2001, С.116-143. \\

{\bf 15.} \ Лев В.Х. Трехмерные и четырехмерные пространства в
теории физических структур. Автореферат диссертации на соискание
ученой
степени кандидата физико-математических наук. Минск, 1990. \\

{\bf 16.} \ Lie S., Engel F. Theorie der Transformations gruppen,
Bd. 3,
Leipzig, 1893. \\

{\bf 17.} \ Михайличенко Г.Г. Трехмерные алгебры Ли локально
транзитивных преобразований пространства // Изв. вузов.
Математика, 1997,
№9(424), С.41-48. \\

{\bf 18.} \ Кыров В.А. Классификация четырехмерных транзитивных
локальных групп Ли преобразований пространства $R^4$ и их
двухточечных инвариантов // Изв. вузов. Математика, 2008, №6,
С.29-42.

\newpage

     {\bf 9.} \ Yu. I. Kulakov. {\it On a New Type of Symmetry Lying at the Basis of Physical Theories of
      Phenomenological Type} \ (Кулаков Ю.И. О новом виде симметрии, лежащем в основании
     физических теорий феноменологического типа // Докл. АН СССР, 1971, Т. 201, № 3. С.
     570-572. Russian).

     {\bf 10.} \  G.G. Mikhailichenko. {\it Polymetric geometries} \
     (Михайличенко Г.Г. Полиметрические геометрии. Новосибирск:
     НГУ  2001. Russian).

    {\bf 11.} \  G.G. Mikhailichenko. {\it The simplest polymetric geometries} I.
    Sibarian Mathematical Journal, 1998, Vol. 39, No. 2, pp. 377-395.

    {\bf 12.} \  L.S. Pontryagin. {\it Topological Groups (Classics of
     Soviet Mathematics)}. Publisher: CRC Press; 3 edition (March
     6 , 1987), ISBN-10: 2881241336.

    {\bf 13.} \  L.V. Ovsyannikov. {\it  A Group Analysis of Differential Equations} \
     (Л.В. Овсянников. Групповой анализ дифференциальных уравнений. М.:
     Наука, 1978. Russian).

    {\bf 14.} \ V.A. Kyrov. {\it Six-dimensional Lie Algebras of Groups of Motions of
     Three-Dimensional Phenomenologically Symmetric Geometries} \
     (Кыров В.А. Шестимерные алгебры Ли групп движений трехмерных
     феноменологически симметричных геометрий // Михайличенко Г.Г. Полиметрические
     геометрии. Новосибирск: НГУ, 2001, С. 116-143. Russian).

    {\bf 15.} \ V.H. Lev. {\it Three-dimensional and Four-Dimensional Spaces in the
     Theory of Physical Structures. Theses for
     the Degree of a Candidate of Physical and Mathematical Sciences} \
     (Лев В.Х. Трехмерные и четырехмерные пространства в теории
     физических структур. Автореферат диссертации на соискание ученой степени
     кандидата физико-математичес- \ ких наук. Минск, 1990. Russian).

    {\bf 16.} \ S. Lie, F. Engel. {\it The Theory of Group Transformations} \ (S. Lie, F. Engel.
     Theorie der Transformations gruppen, Bd. 3  Leipzig, 1893.
     German).

    {\bf 17.} \ G.G. Mikhailichenko. {\it The three-dimensional Lie algebras of locally transitive transformations of space}.
     Russian Mathematics (Izvestiya VUZ. Matematika), 1997, Vol. 41, No. 9, pp. 38-45.

    {\bf 18.} \ V.A. Kyrov. {\it A classification of four-dimensional transitive local groups of transformation of space \ $R^4$
     and their two-point invariants}. Russian Mathematics
     (Izvestiya VUZ. Matematika), 2008, Vol. 52, No. 6, pp. 25-36.

\newpage

{\bf 19.} \  Михайличенко Г.Г. К вопросу о симметрии расстояния в
геометрии
// Изв. вузов. Математика, 1994, №4(383), С.21-23. \\

{\bf 20.} \  Михайличенко Г.Г. Некоторые следствия гипотезы о
бинарной структуре пространства // Изв. вузов. Математика, 1991,
№6,
С.28-35. \\

{\bf 21.} \ Михайличенко Г.Г. О групповой и феноменологической
симметриях
в геометрии // Сиб. мат. журн., 1984, Т.25, №5, С.99-113. \\

{\bf 22.} \  Михайличенко Г.Г. Простейшие полиметрические
геометрии II //
Наука, культура, образование, 2001, №8/9, C.7-16. \\

{\bf 23.} \  Михайличенко Г.Г. Двумерные геометрии. Барнаул: БГПУ, 2004. \\

{\bf 24.} \  Михайличенко Г.Г. Групповая симметрия физических
структур.
Барнаул: БГПУ, 2003. \\

{\bf 25.} \  Михайличенко Г.Г. Двуметрические физические структуры
и
комплексные числа // Докл. АН СССР, 1991, Т.321, №4, С.677-680. \\

{\bf 26.} \  Михайличенко Г.Г. Двуметрические физические структуры
ранга
$(n+1,2)$ // Сиб. мат. журн., 1993. Т.34, №3, С.132--143. \\

{\bf 27.} \  Владимиров Ю.С. Реляционная теория
пространства-времени и взаимодействий, М.: Издательство
Московского университета, 1998.
\\

{\bf 28.} \  Михайличенко Г.Г. Решение функциональных уравнений в
теории физических структур // Докл. АН СССР, 1972, Т.206, №5,
С.1056--1058. \\

{\bf 29.} \  Михайличенко Г.Г. Об одном функциональном уравнении с
двухиндексными переменными // Укр. мат. журн., 1973, Т.25, №5,
С.589--598. \\

\newpage

    {\bf 19.} \  G.G. Mikhailichenko. {\it On a distance symmetry in geometry}.
     Russian Mathematics (Izvestiya VUZ. Matematika), 1994, Vol.
     38, No. 4, pp.19-20.

    {\bf 20.} \  G.G. Mikhailichenko. {\it Some consequences of the hypothesis of binary structure
     of space (within the theory of physical structures)}. Soviet Mathema- \ tics
     (Izvestiya VUZ. Matematika), 1991. Vol. 35, No. 6, pp. 28-34.

    {\bf 21.} \  G,G. Mikhailichenko. {\it On group and phenomenological symmetry in geometry}.
     Siberian Mathematical Journal, 1984, Vol. 25, No. 5, pp. 99-113.

    {\bf 22.} \  G.G. Mikhailichenko. {\it The simplest polymetric geometries} II
     (Михайличенко Г.Г. Простейшие полиметрические геометрии II // Наука, культура, образование,
      2001, № 8/9, С. 7-16. Russian).

    {\bf 23.} \  G.G. Mikhailichenko. {\it Two-dimensional geometries} \
     (Михайличенко Г.Г. Двумерные геометрии. Барнаул: БГПУ,
     2004. Russian).

    {\bf 24.} \  G.G. Mikhailichenko. {\it The group symmetry of physical structures}
     \ (Михайличенко Г.Г. Групповая симметрия физических структур.
     Барнаул, БГПУ. 2003. Russian).

    {\bf 25.} \  G.G. Mikhailichenko. {\it Dimetric physical structures
     and complex num- \ bers}. Soviet Math. Dokl., 1992, Vol. 44, No.
     3, pp. 775-778.

    {\bf 26.} \  G.G. Mikhailichenko. {\it Dimetric Physical structures of rank
     $(n+1,2)$}. Siberian Math. Journal, 1993, Vol. 34, No. 3, pp.
     513-522.

    {\bf 27.} \  Yu. S. Vladimirov. {\it A Relational Theory of Time-Space and
     Interastion} \ (Владимиров Ю.С. Реляционная теория пространства-времени и
     взаимодействий. М: МГУ, 1998. Russian).

    {\bf 28.} \  G.G. Mikhailichenko. {\it The solution of functional
     equations in the theory of physical structures}.
     Soviet Math. Dokl., 1972, Vol. 13, No. 5, pp. 1377-1380.

    {\bf 29.} \  G.G. Mikhailichenko. {\it On a functional equation with
     two-index variables}. Ukr. Mat. Zh., 1973, Vol. 25, No. 5, pp. 589-598.

\newpage

{\bf 30.} \  Михайличенко Г.Г. Об одной задаче в теории физических
структур
// Сиб. мат. журн., 1977, Т.18, №6, С.1342-1355. \\

{\bf 31.} \  Михайличенко Г.Г. Математический аппарат теории
физических
структур. Горно-Алтайск: ГАГУ, 1997. \\

{\bf 32.} \  Михайличенко Г.Г., Мурадов Р.М. Геометрия двух
множеств. {\it Основы и результаты}. Saarbrucken: LAMBERT Academic
Publishing,
2011. \\

{\bf 33.} \  Михайличенко Г.Г. Тернарная физическая структура
ранга (3,2)
// Укр. мат. журн., 1970, Т.22, №6, С.837-841. \\

{\bf 34.} \  Михайличенко Г.Г. Тернарная физическая структура
ранга (2,2,2)
// Изв. вузов. Математика, 1976, №8(171), С.60-67. \\

{\bf 35.} \  Михайличенко Г.Г. Феноменологическая и групповая
симметрии в геометрии двух множеств (теории физических структур)
// Докл. АН
СССР, 1985, Т.24, №1, С.39-41. \\

{\bf 36.} \  Михайличенко Г.Г. Групповые свойства произвольных
физических структур // Вычислительные системы. Новосибирск: ИМ,
1990, Вып.
135. С.27-39. \\

{\bf 37.} \  Михайличенко Г.Г. Феноменологически симметричная
геометрия двух множеств ранга (3,2) // Изв. вузов. Математика,
2016, №2, С.48-53.

\newpage

    {\bf 30.} \  G.G. Mikhailichenko. {\it On a problem in the theory of physical
     structures} \ (Михайличенко Г.Г. Об одной задаче в теории физических структур //
     Сиб. мат. журн. 1977, Т. 18, № 6, С. 1342-1355. Russian).

    {\bf 31.} \  G.G. Mikhailichenko. {\it The mathematical apparatus of the theory of
     physical structures} \
     (Михайличенко Г.Г. Математический аппарат теории физических структур. Горно-Алтайск:
     ГАГУ, 1997. Russian).

    {\bf 32.} \  G.G. Mikhailichenko,  \  R.M. Muradov. {\it The Geometry of two sets}.
     LAP LAMBERT Academic Publishing,
     Saarbrucken, 2013.

    {\bf 33.} \  G.G. Mikhailichenko. {\it Ternary physical structure of rank (3,2)}.
     Ukr. Mat. Zh., 1970, Vol. 22, No. 6, pp. 837-841.

    {\bf 34.} \  G.G. Mikhailichenko. {\it Ternary physical structure of
     rank (2,2,2)} \ (Михайличенко Г.Г. Тернарная физическая структура ранга (2,2,2)
      // Изв. вузов. Математика, 1976, № 8(171), С. 60-67. Russian).

    {\bf 35.} \  G.G. Mikhailichenko. {\it Phenomenological and group
     symmetry in the geometry of two sets (theory of physical
     structures)}. Soviet Math. Dokl., 1985, Vol. 32, No. 2, pp.
     371-374.

    {\bf 36.} \  G.G. Mikhailichenko. {\it Group Properties of Arbitrary Physical Structures} \
     (Михайличенко Г.Г. Групповые свойства произвольных физических
     структур. // Вычислительные системы. Новосибирск:  \ ИМ, 1990, Вып. 135, С. 27-39.
     Russian).

   {\bf 37.} \  G.G. Mikhailichenko. {\it The phenomenologically synnrtric
geometry of two sets of rank (3,2)}. \  {\bf
https://arxiv.org/pdf/1406.2070} \ .

\newpage

\begin{flushright}
{\it Приложение}
\end{flushright}

\begin{center}
\Large{{\bf Груда и физическая структура ранга (2,2)}}
\end{center}

\begin{center}
\bf{А.Н. Бородин}
\end{center}

Хорошо известно [1], что {\it грудой} называется алгебра $G$ с тернарной
операцией $\varphi:G^3\to G$, удовлетворяющей следующим тождествам:
$$
\varphi(\varphi(x,y,z),u,v)=\varphi(x,\varphi(u,z,y),v)=
\varphi(x,y,\varphi(z,u,v)),
\eqno(1)
$$
$$
\varphi(x,y,y)=\varphi(y,y,x)=x.
\eqno(2)
$$

В среднем звене тождеств (1) имеется
не вполне понятная перестановка элементов $y$ и $u$ кортежа $\langle xyzuv \rangle$.
Оказывается, что оно в определении
груды может быть опущено.

\vspace{5mm}

{\bf Лемма 1.} {\it Тождества $(1),(2)$, которым удовлетворяет тернарная
операция $\varphi$, эквивалентны тождествам}
$$
\varphi(\varphi(x,y,z),u,v)=\varphi(x,y,\varphi(z,u,v)),
\eqno(3)
$$
$$
\varphi(x,y,y) =\varphi(y,y,x)=x.
\eqno(4)
$$

\vspace{5mm}

Из леммы 1 следует, что определение груды тождествами (1),(2) по лекциям
А.Г.Куроша [1] эквивалентно ее
определению тождествами (3),(4).

\vspace{5mm}

{\bf Определение 1.} {\it Алгебра $G$ с тернарной операцией $\varphi$ называется
{\it грудой}, если эта операция удовлетворяет тождествам $(3),(4)$.}

\vspace{5mm}

{\bf Лемма 2.} {\it Тождества $(3),(4)$, которым удовлетворяет тернарная
операция $\varphi$, эквивалентны тождествам}
$$
\varphi(x,y,z)=\varphi(\varphi(x,y,s),s,z)=\varphi(x,s,\varphi(s,y,z)),
\eqno(5)
$$
$$
\varphi(x,y,y)=\varphi(y,y,x)=x.
\eqno(6)
$$

\vspace{5mm}

Детальные доказательства леммы 2 и предыдущей леммы 1 можно найти в
работе автора [2].

\newpage

     \begin{flushright}
     {\it Appendix }
     \end{flushright}

     \begin{center}
     \Large{{\bf A heap and the physical structure of rank (2,2)}}
     \end{center}

     \begin{center}
     \bf{by A.N. Borodin}
     \end{center}

     As is well known [1], a {\it heap } is an algebra $G$ with a ternary
     operation $\varphi:G^3\to G$ satisfying the following identities:
     $$
     \varphi(\varphi(x,y,z),u,v)=\varphi(x,\varphi(u,z,y),v)=
     \varphi(x,y,\varphi(z,u,v)),
     \eqno(1)
     $$
     $$
     \varphi(x,y,y)=\varphi(y,y,x)=x.
     \eqno(2)
     $$

     In medium term of the identities (1), there is a quite ambiguous permuta- tion
     present of the elements $y$ and $u$ of the cortege $\langle xyzuv \rangle$ It turns out that
     that medium term in the definition of a heap may be altogether omitted.

     \vspace{5mm}

     {\bf Lemma 1.} {\it The identities $(1) and (2)$, if satisfied with a ternary
     operation $\varphi$, are equivalent to the identities}
     $$
     \varphi(\varphi(x,y,z),u,v)=\varphi(x,y,\varphi(z,u,v)),
     \eqno(3)
     $$
     $$
     \varphi(x,y,y) =\varphi(y,y,x)=x.
     \eqno(4)
     $$

     \vspace{5mm}

     It follows from Lemma 1 that the definition of a heap by the identities (1), (2), as in
     the lectures of A.G. Kurosh [1], is equivalent to its definition by the identities (3),(4).

     \vspace{5mm}

     {\bf Definition 1.} {\it An algebra $G$ with a ternary operation $\varphi$ is a
     {\it heap}, if that operation satisfies the identities $(3), (4)$.}

     \vspace{5mm}

     {\bf Lemma 2.} {\it The identities $(3),(4)$, satisfied with the ternary operation
     $\varphi$, are equivalent to the identities}
     $$
     \varphi(x,y,z)=\varphi(\varphi(x,y,s),s,z)=\varphi(x,s,\varphi(s,y,z)),
     \eqno(5)
     $$
     $$
     \varphi(x,y,y)=\varphi(y,y,x)=x.
     \eqno(6)
     $$

     \vspace{5mm}

     Detailed proof of Lemma 2 as well as that of Lemma 1 may be found in the author's
     note [2].

\newpage

{\bf Определение 2.} {\it Алгебра $G$ с тернарной операцией
$\varphi$ называется {\it грудой}, если эта операция удовлетворяет
тождествам $(5),(6)$}.

\vspace{5mm}

{\bf Лемма 3.} {\it Определение $1$ и определение $2$ груды, как алгебры $G$
с тернарной операцией $\varphi$, удовлетворяющей тождествам $(3),(4)$ и
тождествам $(5),(6)$ соответственно, эквивалентны.}

\vspace{5mm}

Определение груды тождествами (5),(6) представляется более
естественным, так как они являются следствием принципа
феноменологической симметрии в теории физических структур [3].

\vspace{5mm}

Пусть имеются три множества $\mathfrak{M,N}$ и $G$ произвольной природы,
а также функция $f:\mathfrak{M\times N}\to G$, сопоставляющая каждой
паре $\langle i\alpha \rangle$ из прямого произведения $\mathfrak{M\times N}$ некоторый
элемент $f(i\alpha)$ из множества $G$. В отношении функции $f$ будем
предполагать выполнение следующего условия:

\vspace{5mm}

{\bf A.} {\it Для любых элементов $\beta\in\mathfrak{N}$ и $j\in\mathfrak{M}$
отображения $\mathfrak{M}\times\{\beta\}\to G$ и
$\{j\}\times\mathfrak{N}\to G$ сюръективны.}

\vspace{5mm}

Введем еще функцию $F:\mathfrak{M}^2\times\mathfrak{N}^2 \to G^4$,
сопоставляя кортежу $\langle ij,\alpha\beta \rangle \
\in \mathfrak{M}^2\times\mathfrak{N}^2$
точку $\langle f(i\alpha),f(i\beta),f(j\alpha),f(j\beta) \rangle \ \in G^4$,
координаты которой в $G^4$ есть образы соответствующих пар,
упорядоченные по исходному кортежу.

\vspace{5mm}

{\bf Определение 3.} {\it Будем говорить, что функция $f:\mathfrak{M\times N}\to
G$, удовлетворяющая условию {\bf A}, задает на множествах $\mathfrak{M}$ и
$\mathfrak{N}$ {\it физическую структуру ранга} $(2,2)$, если существует такая
тернарная алгебраическая операция $\varphi:G^3\to G$, для которой выполняется
следующее соотношение:}
$$
f(i\alpha)=\varphi(f(i\beta),f(j\beta),f(j\alpha)).
\eqno(7)
$$

\vspace{5mm}

Соотношение (7), справедливое для любого кортежа $\langle ij,\alpha\beta \rangle
\ \in \mathfrak{M}^2\times\mathfrak{N}^2$, выражает содержание
{\it принципа феноменологической симметрии} в теории физических
структур. Оно \ как \ функциональное \ уравнение

\newpage

     {\bf Definition 2.} {\it An algebra $G$ with a ternary operation $\varphi$ is a
     {\it heap}, if that operation satisfies the identities $(5),(6)$}.

     \vspace{5mm}

     {\bf Lemma 3.} {\it The definition $1$ and the definition $2$ of a heap as an
     algebra $G$ with a ternary operation $\varphi$ satisfying the identities $(3),(4)$
     and the identities $(5),(6)$ respectively are equivalent.}

     \vspace{5mm}

     To define a heap with the identities (5) and (6) deems more natural, as they
     are the corollary of the principle of phenomenological symmetry of the theory
     of physical structures [3].

     \vspace{5mm}

     Let there be three sets -- $\mathfrak{M,N}$ and $G$ -- of arbitrary nature,
     as well as a function $f:\mathfrak{M\times N}\to G$ that assigns to each pair
     $\langle i\alpha \rangle$ from the direct product $\mathfrak{M\times N}$ some element
     $f(i\alpha)$ of the set $G$. With respect to the function $f$, we shall assume
     that the condition is satisfied as follows:

     \vspace{5mm}

     {\bf A.} {\it For any elements $\beta\in\mathfrak{N}$ and $j\in\mathfrak{M}$
     the mappings $\mathfrak{M}\times\{\beta\}\to G$ and $\{j\}\times\mathfrak{N}\to
     G$ are surjective.}

     \vspace{5mm}

     We shall introduce still one more function -- $F:\mathfrak{M}^2 \times \mathfrak{N}^2
     \to G^4$, -- by assigning to the cortege $\langle ij,\alpha\beta \rangle \ \in \ \mathfrak{M}^2
     \times\mathfrak{N}^2$ a point $\langle f(i\alpha),f(i\beta), \\ f(j\alpha), f(j\beta) \rangle \ \in G^4$,
     whose coordinates in $G^4$ are the images of the correspon- \ ding pairs ordered with
     respect to the original cortege.

     \vspace{5mm}

     {\bf Definition 3.} {\it We shall say that the function $f:\mathfrak{M\times N}\to
     G$ that satisfies the condition {\bf A} gives on the sets $\mathfrak{M}$ and
     $\mathfrak{N}$ a {\it physical structure of rank} $(2,2)$, if there exists such a
     ternary algebraic operation $\varphi:G^3\to G$ for which the relation is satisfied
     as follows:}

     $$
     f(i\alpha)=\varphi(f(i\beta),f(j\beta),f(j\alpha)).
     \eqno(7)
     $$

     \vspace{5mm}

     The relation (7), true for any cortege $\langle ij,\alpha\beta \rangle \ \in
     \mathfrak{M}^2\times\mathfrak{N}^2$, expresses the essence of the {\it principle
     of phenomenological symmetry} of the theory of physical structures. \ Since it is a
     functional equation \ it \ imposes \ on \ the

\newpage

\noindent
налагает на функцию $f$ достаточно сильное
ограничение.

\vspace{5mm}
{\bf Теорема 1.} {\it Тернарная алгебраическая операция $\varphi$ из
определения $3$ физической структуры ранга $(2,2)$, устанавливающая
феноменологически симметричное соотношение $(7)$, задает на множестве $G$
груду.}
\vspace{5mm}

Положим в соотношении (7) $i=j: \ f(i\alpha)=
\varphi(f(i\beta),f(i\beta),f(i\alpha))$ и $\alpha=\beta: \ f(i\alpha)=
\varphi(f(i\alpha),f(j\alpha),f(j\alpha))$. В соответствии с условием {\bf A}
пары переменных $f(i\alpha),f(i\beta)$ и $f(i\alpha),f(j\alpha)$ независимы.
Вводя для них обозначение $x=f(i\alpha), \ y=f(i\beta)$ в первом случае и
$x=f(i\alpha), \ y=f(j\alpha)$ -- во втором, получаем тождества (6).
Возьмем в множестве
$\mathfrak{M}$ элемент $k$ и запишем
соотношение (7) для кортежей $\langle ik,\alpha\beta \rangle$ и \linebreak$\langle jk,\alpha\beta \rangle$:
$$
\left.\begin{array}{c}
f(i\alpha)=\varphi(f(i\beta),f(k\beta),f(k\alpha)), \\
f(j\alpha)=\varphi(f(j\beta),f(k\beta),f(k\alpha)).
\end{array}\right\}
\eqno(7')
$$

Из трех соотношений $(7),(7')$ легко устанавливаем равенство
$$
\varphi(f(i\beta),f(k\beta),f(k\alpha))=
\varphi(f(i\beta),f(j\beta),\varphi(f(j\beta),f(k\beta),f(k\alpha))),
$$
с независимыми по условию {\bf A} переменными $f(i\beta),
f(k\beta),f(k\alpha),f(j\beta)$. Вводя для них обозначение $x=f(i\beta), \
y=f(k\beta), \ z=f(k\alpha), \ s=f(j\beta)$, получаем одно из тождеств (5).
Возьмем, далее, элемент $\gamma$ из множества $\mathfrak{N}$ и
запишем соотношение (7) для кортежей $\langle ij,\alpha\gamma \rangle$ и
$\langle ij,\beta\gamma \rangle$:
$$
\left.\begin{array}{c}
f(i\alpha)=\varphi(f(i\gamma),f(j\gamma),f(j\alpha)), \\
f(i\beta) =\varphi(f(i\gamma),f(j\gamma),f(j\beta)).
\end{array}\right\}
\eqno(7'')
$$

Из трех соотношений $(7),(7'')$ следует равенство
$$
\varphi(f(i\gamma),f(j\gamma),f(j\alpha))=
\varphi(\varphi(f(i\gamma),f(j\gamma),f(j\beta)),f(j\beta),f(j\alpha)),
$$
с независимыми по условию {\bf A} переменными $f(i\gamma),f(j\gamma),
f(j\alpha),f(j\beta)$. Вводя для них обозначение $x=f(i\gamma), \
y=f(j\gamma), \ z=f(j\alpha), \ s=f(j\beta)$, приходим к другому из тождеств
(5). Таким образом, тождества (5) и (6), входящие в определение
2 груды, установлены, что и завершает доказательство теоремы 1.

\newpage

     \noindent
      function $f$ a strong enough restriction.

     \vspace{5mm}
     {\bf Theorem 1.} {\it The ternary algebraic operation $\varphi$ from the Definition
     $3$ of the physical structure of rank $(2,2)$ that establishes the phenomenologically
     symmetric relation $(7)$ defines a heap on the set $G$.}
     \vspace{5mm}

     We shall set in the relation (7) $i=j: \ f(i\alpha)=
     \varphi(f(i\beta),f(i\beta),f(i\alpha))$ and $\alpha=\beta: \ f(i\alpha)=
     \varphi(f(i\alpha),f(j\alpha),f(j\alpha))$. Under the condition {\bf A}, the pairs
     of the variables $f(i\alpha),f(i\beta)$ and $f(i\alpha),f(j\alpha)$ are independent.
     Introducing for them the designation $x=f(i\alpha), \ y=f(i\beta)$, for the former
     case, and $x=f(i\alpha), \ y=f(j\alpha)$, for the latter, yields the identities (6). We
     shall take an element $k$ of the set $\mathfrak{M}$ and write the relation (7)
     for the corteges $\langle ik,\alpha\beta \rangle$ and $\langle jk,\alpha\beta \rangle$:
     $$
     \left.\begin{array}{c}
     f(i\alpha)=\varphi(f(i\beta),f(k\beta),f(k\alpha)), \\
     f(j\alpha)=\varphi(f(j\beta),f(k\beta),f(k\alpha)).
     \end{array}\right\}
     \eqno(7')
     $$

     Out of the three relations $(7),(7')$, we easily establish the equality
     $$
     \varphi(f(i\beta),f(k\beta),f(k\alpha))=
     \varphi(f(i\beta),f(j\beta),\varphi(f(j\beta),f(k\beta),f(k\alpha))),
     $$
     with independent, under the condition {\bf A}, variables $f(i\beta),
     f(k\beta),f(k\alpha),f(j\beta)$. Introducing the designation $x=f(i\beta), \ y=
     f(k\beta), \ z=f(k\alpha), \ s=f(j\beta)$ for them yields one of the identities (5).
     Let us take, further, an element $\gamma$ of the set $\mathfrak{N}$ and
     write the relation (7) for the corteges $\langle ij,\alpha\gamma \rangle$ and $\langle ij,\beta\gamma \rangle$:
     $$
     \left.\begin{array}{c}
     f(i\alpha)=\varphi(f(i\gamma),f(j\gamma),f(j\alpha)), \\
     f(i\beta) =\varphi(f(i\gamma),f(j\gamma),f(j\beta)).
     \end{array}\right\}
     \eqno(7'')
     $$

     From the three relations $(7),(7'')$ there follows the equality
     $$
     \varphi(f(i\gamma),f(j\gamma),f(j\alpha))=
     \varphi(\varphi(f(i\gamma),f(j\gamma),f(j\beta)),f(j\beta),f(j\alpha)),
     $$
     with the independent, under the condition {\bf A}, variables $f(i\gamma),f(j\gamma),
     f(j\alpha), \\ f(j\beta)$. Introducing the designation $x=f(i\gamma), \ y=f(j\gamma),
     \ z=f(j\alpha), \ s=f(j\beta)$ for them yields the other of the identities (5). Thus,
     the identities(5) and (6), that are part of the Definition 2 of a heap, are established,
     which makes the proof of Theorem 1 complete.

\newpage

Обратим теперь внимание на различную роль тождеств (5) и (6).
Первые из них явно основополагающие, имеющие характер
функциональных уравнений, определяющих груду, вторые же отражают
ее частные свойства. Поэтому имеет смысл в новом определении груды
сохранить тождества (5), а тождества (6) заменить некоторым более
естественным условием, налагаемым на тернарную операцию $\varphi$.
Это условие можно получить из того же феноменологически
симметричного соотношения (7) для физической структуры ранга
(2,2).

\vspace{6mm}

{\bf Лемма 4.} {\it Тернарная алгебраическая операция $\varphi$ из
определения $3$ физической структуры ранга $(2,2)$ удовлетворяет следующему
необходимому условию:

\vspace{6mm}

${\bf B.}$ Для любых двух элементов $q,h \in G$ отображения
$x\mapsto\varphi(x,q,h)$, \ $x\mapsto\varphi(q,x,h), \ x\mapsto\varphi(q,h,x)$
сюръективны.}

\vspace{6mm}

Рассмотрим сначала первое отображение $x\mapsto\varphi(x,q,h)$. По условию
{\bf A} найдется такая пара $\langle j\alpha \rangle$, для которой $f(j\alpha)=h$. Далее, по
тому же условию {\bf A} для точек $j\in\mathfrak{M}$ и $\alpha\in\mathfrak{N}$
предыдущей пары найдутся такие точки $i\in\mathfrak{M}$ и
$\beta\in\mathfrak{N}$, для которых $f(j\beta)=q$ и $f(i\alpha)=p$, где $p$ --
произвольный элемент из $G$. Но тогда, полагая $x=f(i\beta)$, по соотношению
(7) получаем $p=\varphi(x,q,h)$. То есть у произвольного элемента $p\in G$ при
отображении $x\mapsto\varphi(x,q,h)$ имеется хотя бы один прообраз,
что и доказывает сюръективность этого отображения. Сюръективность
отображений $x\mapsto\varphi(q,x,h)$ и $x\mapsto\varphi(q,h,x)$
устанавливается совершенно аналогично. Лемма 4 доказана.

\vspace{6mm}

Условие {\bf B} имеет более привычную для алгебраистов эквивалентную форму:

\vspace{6mm}

${\bf B'.}$ {\it Для любых трех элементов $p,q,h$ из множества $G$ каждое из
уравнений
$p=\varphi(x,q,h), \ p=\varphi(q,x,h)$ и $p=\varphi(q,h,x)$ имеет решение
относительно $x$.}

\newpage

     The difference of the roles played by the identities (5) and (6) is worth noting too.
     The former are obviously basic ones, such that look like some functional relations
     that define a heap, while the latter only reflect its minor characteristics. So it makes
     sense in a new definition of a heap to retain the identities (5), and substitute for the
     identities (6) some more natural condition imposed on the ternary operation $\varphi$.
     Such condition may be obtained from the same phenomenologically
     symmetric relation (7) for the physical structure of rank (2,2).

     \vspace{6mm}

     {\bf Lemma 4.} {\it The ternary algebraic operation $\varphi$ from the definition
     $3$ of the physical structure of rank $(2,2)$ satisfies the following necessary condition:

     \vspace{6mm}

     ${\bf B.}$ For any two elements $q,h \in G$, the mappings $x\mapsto\varphi(x,q,h)$,
     \ $x\mapsto\varphi(q,x,h), \ x\mapsto\varphi(q,h,x)$ are surjective.}

     \vspace{6mm}

     Let us consider the first of the mappings, $x\mapsto\varphi(x,q,h)$. Under the condition
     {\bf A} there exists such a pair $\langle j\alpha \rangle$ for which $f(j\alpha)=h$. Next, according
     to the same condition {\bf A}, for the points $j\in\mathfrak{M}$ and $\alpha\in\mathfrak{N}$
     of the previous pair there exist points $i\in\mathfrak{M}$ and $\beta\in\mathfrak{N}$,
     for which $f(j\beta)=q$ and $f(i\alpha)=p$, where $p$ is an arbitrary element from
     $G$. But then, setting $x=f(i\beta)$ yields, with respect to the relation (7), $p=\varphi(x,q,h)$.
     I. e. an arbitrary element $p\in G$ with the mapping $x\mapsto\varphi(x,q,h)$ has at least
     one preimage, which proves the mapping being surjective. The surjectivity of the mappings
     $x\mapsto\varphi(q,x,h)$ and $x\mapsto\varphi(q,h,x)$ is established in absolutely
     the same way. Lemma 4 has been proved.

     \vspace{6mm}

     The condition {\bf B} has an equivalent form, which algebraists have been more accustomed
     to:

     \vspace{6mm}

     ${\bf B'.}$ {\it For any three elements $p,q,h$ of the set $G$ each of the equations
     $p=\varphi(x,q,h), \ p=\varphi(q,x,h)$ and $p=\varphi(q,h,x)$ has a solution with respect
     to $x$.}

\newpage

Заменим не совсем естественные тождества (6) в определении (2)
груды более естественным условием {\bf B}, которое также является
следствием феноменологической симметрии.

\vspace{5mm}

{\bf Определение 4.} Алгебра $G$ с тернарной операцией $\varphi$,
удовлетворяющей условию {\bf B} (или эквивалентному ему условию ${\bf B'}$),
называется грудой, если для нее выполняются следующие два тождества:

$$
\left.\begin{array}{c}
\varphi(x,y,z)=\varphi(\varphi(x,y,s),s,z), \\
\varphi(x,y,z)=\varphi(x,s,\varphi(s,y,z)).
\end{array}\right\}
\eqno(8)
$$

\vspace{5mm}

{\bf Лемма 5.} {\it Определение $2$ и определение $4$ груды как алгебры $G$
с тернарной операцией $\varphi$, удовлетворяющей четырем тождествам (5),(6)
или при условии {\bf B} двум тождествам $(8)$ соответственно, эквивалентны.}
\vspace{5mm}

Тождества (5) и (8) совпадают, поэтому сначала из условия {\bf B} и тождеств
(8) получим тождества (6). Запишем первое и второе из тождеств (8)
для кортежей
$\langle xyyy \rangle$ и $\langle yyxy \rangle$ соответственно: \ $\varphi(x,y,y)=
\varphi(\varphi(x,y,y),y,y)$, \ $\varphi(y,y,x)=\varphi(y,y,\varphi(y,y,x))$.
По условию {\bf B} отображения $x\mapsto\varphi(x,y,y)$ и
$x\mapsto\varphi(y,y,x)$ сюръективны. Введя соответствующие переобозначения
элементов $\varphi(x,y,y)$ и $\varphi(y,y,x)$ из $G$, получаем тождества (6).
Покажем теперь, что условие {\bf B} есть следствие тождеств (5),(6).
Предположим противное, то есть что найдутся такие три элемента $p,q,h$ из
множества $G$, что одно из трех уравнений условия ${\bf B'}$ не имеет решения.
Без ограничения общности можно предположить, что не имеет решения
уравнение $p=\varphi(x,q,h)$. Запишем первое из тождеств (5) для
кортежа $\langle phhq \rangle$: \ $\varphi(p,h,h)=\varphi(\varphi(p,h,q),q,h)$, откуда,
используя одно из тождеств (6), получаем: $p=\varphi(\varphi(p,h,q),q,h)$.
Таким образом, уравнение $p=\varphi(x,q,h)$ имеет решение $x=\varphi(p,h,q)$,
что противоречит сделанному предположению. Два других уравнения из условия
${\bf B'}$ исследуются аналогично. Устанавливаемые при этом противоречия и
показывают, что условие ${\bf B'}$ (или эквивалентное ему условие {\bf B})
является следствием тождеств (5),(6). Лемма 5 доказана.

\newpage

     We shall replace the identities (6), that do not look quite natural, in the definition (2)
     of a heap with a more natural condition {\bf B}, which is also a corollary of the
     phenomenological symmetry.
     \vspace{5mm}

     {\bf Definition 4.} The algebra $G$ with a ternary operation $\varphi$ satisfying the condition
     {\bf B} (or the condition ${\bf B'}$ that is equivalent to it) is a heap if it satisfies the two
     identities as follows:

     $$
     \left.\begin{array}{c}
     \varphi(x,y,z)=\varphi(\varphi(x,y,s),s,z), \\
     \varphi(x,y,z)=\varphi(x,s,\varphi(s,y,z)).
     \end{array}\right\}
     \eqno(8)
     $$

     \vspace{5mm}

     {\bf Lemma 5.} {\it Definition $2$ and definition $4$ of a heap as of an algebra $G$
     with a ternary operation $\varphi$ satisfying the four identities (5) and (6) or, under
     condition {\bf B}, the two identities $(8)$ respectively, are equivalent.}
     \vspace{5mm}

     The identities (5) and (8) coincide, so let us first get the identities (6) from the condition
     {\bf B} and the identities (8). We shall write the first and the second of the identities (8)
     for the corteges $\langle xyyy \rangle$ and $\langle yyxy \rangle$ respectively: \ $\varphi(x,y,y)= \varphi(\varphi(x,y,y),y,y)$,
     \ $\varphi(y,y,x)=\varphi(y,y,\varphi(y,y,x))$. Under the condition {\bf B}, the mappings
     $x\mapsto\varphi(x,y,y)$ and $x\mapsto\varphi(y,y,x)$ are surjective ones. Introducing
     corresponding redesignation of the elements $\varphi(x,y,y)$ and $\varphi(y,y,x)$ from
      $G$ yields the identities (6). Now we shall demonstrate that the condition {\bf B} is a
     corollary of the identities (5),(6). We shall assume the contrary, i.e. that there exist three
     elements $p,q,h$ from the set $G$, such that one of the three equations of the condition
     ${\bf B'}$ has no solution. Without loss of generality, it is possible to assume that it is the
     equation $p=\varphi(x,q,h)$ that has no solution. We shall write the former of the
     identities (5) for the cortege  $\langle phhq \rangle$: \ $\varphi(p,h,h)=\varphi(\varphi(p,h,q),q,h)$,
     whereof, by using one of the identities (6), we get: $p=\varphi(\varphi(p,h,q),q,h)$. Thus,
     the equation $p=\varphi(x,q,h)$ does have a solution $x=\varphi(p,h,q)$, which is in
     contradiction with the above assumption. The two other equations from the condition ${\bf B'}$
     are investigated similarly. The contradictions established in the process demonstrate that
     the condition ${\bf B'}$ (or the condition {\bf B} equivalent to it) is a corollary of the identities
     (5),(6). Lemma 5 has been proved.

\newpage

Условие {\bf B}, на первый взгляд может показаться слишком сильным, тем более,
что в доказательстве леммы 5 при получении тождеств (6) условие {\bf B}
использовалось не в полном объеме, поскольку имела значение только
сюръективность отображений $x\mapsto\varphi(x,y,y)$ и
$x\mapsto\varphi(y,y,x)$ для
произвольного элемента $y$. Сформулируем это более слабое условие:

\vspace{5mm}

{\bf C.} {\it Для любого элемента $q\in G$ сюръективны отображения,
задаваемые функциями $x\mapsto\varphi(x,q,q)$ и
$x\mapsto\varphi(q,q,x)$.}

\vspace{5mm}

Эквивалентный вариант этого условия будет следующий:

\vspace{5mm}

${\bf C'.}$ {\it Для любых двух элементов $p,q\in G$
каждое из уравнений $p=\varphi(x,q,q)$ и
$p=\varphi(q,q,x)$ имеет решение относительно $x\in G$.}

\vspace{5mm}

Заметим, однако, что слабое условие {\bf C} кажется менее естественным, чем
более сильное условие {\bf B}.

\vspace{5mm}

{\bf Определение 5.} {\it Алгебра $G$ с тернарной операцией $\varphi$,
удовлетворяющей условию {\bf C} (или эквивалентному ему условию ${\bf C'}$) и
тождествам (8), называется грудой.}

\vspace{5mm}

{\bf Лемма 6.} {\it Определение $2$ и определение $5$ груды как
алгебры $G$ с тернарной операцией $\varphi$, удовлетворяющей
четырем тождествам $(5),(6)$, или при условии {\bf C} двум
тождествам $(8)$, эквивалентны.}

\vspace{5mm}

Доказательство леммы 6 в первой его части получения тождеств (6) повторяет
соответствующую часть доказательства леммы 5, а сюръективность отображений
$x\mapsto\varphi(x,q,q)$ и $x\mapsto\varphi(q,q,x)$, требуемая условием
{\bf C}, есть непосредственное следствие тождеств (6).

\vspace{5mm}

{\bf Теорема 2.} {\it Все четыре определения груды, а именно, определения
$1, 2, 4, 5$ эквивалентны между собой.}

\vspace{5mm}

Теорема 2 является  следствием лемм 3, 5, 6,
устанавливающих транзитивную
эквивалентность пар определений 1 и 2, 2 и 4, 2 и 5.

\newpage

    The condition {\bf B} may seem too strong, and the more so because in the proof of
     Lemma 5, in obtaining the identities (6), the condition {\bf B} was not used in corpore, as
     it was only the surjectivity of the mappings $x\mapsto\varphi(x,y,y)$ and $x\mapsto\varphi(y,y,x)$
     for an arbitrary element $y$ that mattered. We shall define that weaker condition:

     \vspace{5mm}

     {\bf C.} {\it For any element $q\in G$ the mappings defined by the functions $x\mapsto\varphi(x,q,q)$
     and $x\mapsto\varphi(q,q,x)$ are surjective.}

     \vspace{5mm}

     An equivalent to it is the variant as follows:
     \vspace{5mm}

     ${\bf C'.}$ {\it For any two elements $p,q\in G$, each of the equations $p=\varphi(x,q,q)$
      and $p=\varphi(q,q,x)$ has a solution with respect to $x\in G$.}

     \vspace{5mm}
     We shall note, however, that the weak condition {\bf C} seems less natural, than
     the stronger condition {\bf B}.

     \vspace{5mm}

     {\bf Definition 5.} {\it The algebra $G$ with a ternary operation $\varphi$, which
     satisfies the condition {\bf C} (or the equivalent condition ${\bf C'}$) and the
     identities (8), is a heap.}

     \vspace{5mm}

     {\bf Lemma 6.} {\it Definition $2$ and Definition $5$ of a heap as an algebra $G$ with
     a ternary operation $\varphi$ satisfying the four identities $(5),(6)$ or, under condition
     {\bf C}, the two identities $(8)$ are equivalent.}

     \vspace{5mm}

     The proof of Lemma 6, in the first part of it, that where the identities (6) are obtained,
     repeats the respective part of the proof of Lemma 5, and the surjectivity of the mappings
     $x\mapsto\varphi(x,q,q)$ and $x\mapsto\varphi(q,q,x)$, stipulated under the condition
     {\bf C}, is a direct corollary of the identities (6).
     \vspace{7mm}

     {\bf Theorem 2.} {\it All the four definitions of a heap, i.e. Definitions $1, 2, 4, 5$ are
     equivalent one with another.}

     \vspace{7mm}

     Theorem 2 is a  corollary of Lemmas 3, 5, 6, which establish the transitive
     equivalence of the pairs of definitions 1 and 2, 2 and 4, and 2 and 5.

\newpage

Автор выражает благодарность проф. Г.Г.Михайличенко и участникам научного
семинара ФМФ ГАГУ по теории физических структур за поддержку данного
исследования и обсуждение полученных результатов.

\vspace{5mm}

\begin{center}
{\bf Литература}
\end{center}

1. Курош А.Г. Общая алгебра. Лекции 1969-1970 учебного года, М.,
1974. \\

2. Бородин А.Н. Груда и группа как физическая структура //
Михайличенко Г.Г. Групповая симметрия физических структур.
Барнаул: БГПУ, Горно-Алтайск: ГАГУ, 2003, С.195-203. \\

3. Кулаков Ю.И. Теория физических структур. М.: Доминико, 2004.

\newpage

     The author expresses gratitude to Professor G.G. Mikhailichenko and the
     participants of the academic Theory of physical structures seminar of FMF GAGU
     for the support of that research and the discussion of the results obtained.

     \vspace{5mm}

     \begin{center}
     {\bf Bibliography}
     \end{center}

     1. A.G. Kurosh. {\it The general algebra. Lectures 1969-1970 school
     year}. \ (Курош А.Г. Общая алгебра. Лекции 1969-1970 учебного
     гола. М., 1974. Russian).

     2. A.N. Borodin. {\it A heap and a group as the physical structure}
     \ (Бородин А.Н. Груда и группа как физическая структура //
     Михайличенко Г.Г. Групповая симметрия физических структур.
     Барнаул: БГПУ, Горно-Алтайск: ГАГУ, 2003, С. 195-203. Russian).

     3. Yu. I. Kulakov. {\it The theory of physical structures} \
     (Кулаков Ю.И. Теормя физических структур. М: Доминико, 2004.
     Russian).

\newpage

\begin{flushright}
{\it Сведения об авторах}
\end{flushright}

\begin{center}
{\bf Михайличенко
Геннадий Григорьевич}
\end{center}
\noindent
 1942 года рождения. В 1967 окончил физический факультет
Новосибирского госуниверситета, в 1970 аспирантуру при нем. До
1994 доцент кафедры теоретической физики Новосибирского
педуниверситета, с 1994  профессор кафедры физики Горно-Алтайского
университета. Имеет ученую степень доктора физико-математических
наук (1994) и ученое звание профессора (2000). Автор более
шестидесяти работ и пяти монографий: \ {\bf 1.} {\it
Математический аппарат теории физическмх структур} (1997) $-$
[31]; \ {\bf 2.} {\it Полиметрическме геометрии} (2001) $-$ [10];
\ {\bf 3.} {\it Групповая симметрия физических структур} (2003)
$-$ [24]; \ {\bf 4.} {\it Двумерные геометрии  } (2004) $-$ [23];
\ {\bf 5.} {\it Геометрия двух множеств   } (2011) $-$ [32]. \
Настоящая его шестая монография {\it Математические основы и
результаты теории физических структур} подводит итог научной
деятельности ее автора.

\vspace{5mm}

\noindent
{\bf Г.Г. Михайличенко} \\
Профессор, кафедра физики,  \\
Горно-Алтайский государственный университет, \\
 ул. Ленкина, д. 1, г. Горно-Алтайск, 649000, Россия. \\
 Раб. e-mail: \  {\bf mikhailichenko@gasu.ru} \ , \\
 дом. e-mail: \  {\bf michnat1956@yandex.ru}

\vspace{7mm}

\begin{center}
{\bf Бородин Александр Николаевич}
\end{center}

\noindent 1971 года рождения. В 1993 окончил физико-математический
факультет Горно-Алтайского госпединститута. Установил связь
физической структуры ранга (2,2) с такими алгебраическими
структурами как груда и группа (2000). Научные интересы лежат в
области универсальных алгебр, теоретической физики и аналитической
психологии К.Г. Юнга.

\vspace{5mm}

\noindent {\bf А.Н. Бородин} \\
 Дом. e-mail: \ {\bf serajsova@yandex.ru}

\newpage

     \begin{flushright}
     {\it About the authors}
     \end{flushright}

     \begin{center}
     {\bf Gennady Grigorievich
     Mikhailichenko}
     \end{center}
     \noindent
     was born in 1942. In 1967, he graduated from the Department of Physics of the Novosibirsk
    State  University, and in 1970 finished the post-graduate school at the same University. Up to 1994,
     he was an associate professor at the Chair of Theoretical Physics of the Novosibirsk Pedagogical
     University, since 1994 has been a professor at the Chair of Physics of Gorno-Altaisk University.
     A Doctor (1994, a post-Ph.D. degree under the Russian system) of Physical and Mathematical
     Sciences and a Professor (2000).
      He is the author of more than sixty notes and
     of five monographs: \
{\bf 1.} {\it The mathematical apparatus of the theory of physical
structures} (1997) $-$ [31]; \ {\bf 2.} {\it Polymetric geometries
} (2001) $-$ [10]; \ {\bf 3.} {\it The group symmetry of physical
structures  } (2003) $-$ [24]; \ {\bf 4.} {\it Two-dimensional
geometries  } (2004) $-$ [23]; \ {\bf 5.} {\it  The geometry of
two sets } (2013) $-$ [32]; \
      This monograph: {\it The Mathematical Basics and the Results of the Theory
     of physical structures}, which is his sixth monograph, is a sort of a conclusion of the author's
     academic activity.

    \vspace{3mm}

     \noindent
    {\bf G.G. Mikhailichenko} \\
    Professor of Chair of Physics, \\
    Gorno-Altaisk State University, \\
    1 Lenkin str. Gorno-Altaisk, 649000, Russia. \\
    Work e-mail: \  {\bf mikhailichenko@gasu.ru} \ , \\
    home e-mail: \  {\bf michnat1956@yandex.ru}

\vspace{6mm}

\begin{center}
{\bf Borodin Alexander Nicolaevich}
\end{center}
\noindent was born in 1971. In 1993, he graduated from the
Department of Physics and Mathematics of the Gorno-Altaisk State
Pedagogical Institute. He established relation  between the
physical structure of rank (2,2) and  algebraic structures such as
a heap and a group (2000). His academic interests lie in the
fields of universal algebras, theoretical physics and analytical
psychology by K.G. Yung.

 \vspace{3mm}

\noindent  {\bf A.N. Borodin}  \\
Home e-mail: \ {\bf serajsova@yandex.ru}

\newpage
\thispagestyle{empty}

\begin{center}
{\bf НАУЧНОЕ \ ИЗДАНИЕ}
\end{center}

\vspace{20mm}

\begin{center}
\Large{\bf МИХАЙЛИЧЕНКО \ ГЕННАДИЙ \ ГРИГОРЬЕВИЧ}
\end{center}

\vspace{10mm}

\begin{center}
\Huge{ \bf

МАТЕМАТИЧЕСКИЕ \ ОСНОВЫ \\
\vspace {2mm}
 И \ РЕЗУЛЬТАТЫ  \ ТЕОРИИ \\
 \vspace{2mm}
   ФИЗИЧЕСКИХ \ СТРУКТУР}
\end{center}

\vspace{15mm}

\begin{center}
\Large{\bf МОНОГРАФИЯ}
\end{center}

\vspace{40mm}

\begin{center}
Издательство Горно-Алтайского госуниверситета.\\
649000, г. Горно-Алтайск, ул. Ленкина, 1. \\ Российская Федерация.\\
\vspace{5mm}

Подписано в печать  24.06.2016. \ Формат $60\times 84 \ 1/8.$
\\ Уч.-изд. л. 37,1. \  Усл. печ. л. 35,4. \\ Тираж 500 экз. \ Заказ №
171. \\
Редакционно-издательский центр НГУ. \\
630090, г. Новосибирск, ул. Пирогова, 2. \\
Российская Федерация.
\end{center}

\newpage
\thispagestyle{empty}

     \begin{center}
     {\bf ACADEMIC EDITION}
     \end{center}

     \vspace{20mm}

     \begin{center}
     \Large{\bf
     GENNADY \ GRIGORIEVICH \ MIKHAILICHENKO}
     \end{center}

     \vspace{10mm}

     \begin{center}
     \Huge{ \bf

     THE \ MATHEMATICAL \ BASICS  \\
     \vspace{2mm}
     AND \ RESULTS \ OF \ THE \ THEORY \\
     \vspace{2mm}
     \ OF \ PHYSICAL \ STRUCTURES}
     \end{center}

     \vspace{15mm}

  \begin{center}
     {\bf MONOGRAPH}
     \end{center}

\vspace{40mm}

     \begin{center}

      Publishing house of Gorno-Altaisk State University.\\
     649000, Gorno-Altaisk, Lenkin Street, 1.  \\ Russian Federation \\

     \vspace{5mm}

     Sent to the press: 24.06.2016. \ Format $60\times 84 \ 1/8$.
     \\ Reg.-publ. lists  37.1. \ Cond. print. sheets 35.4. \\
         Copies printed 500. \  Order № 171. \\
    Editorial and publishing centre NSU. \\
     630090, 2 Pirogov Street, Novosibirsk. \\ Russian Federation.

     \end{center}

    \end{document}